\documentclass[12pt]{article}
\pdfoutput=1

\usepackage[usenames,dvipsnames,svgnames,table]{xcolor} 
\usepackage[obeyspaces,hyphens,spaces]{url}
\usepackage{jcapmod}
\usepackage{verbatim}
\usepackage{graphics}
\usepackage{epsfig}
\usepackage{booktabs}
\usepackage{booktabs}
\usepackage[english]{babel}
\usepackage{amsmath, amssymb, amsbsy, amstext, amsthm, simplewick}
\usepackage{hyperref}
\usepackage{graphicx}
\usepackage{amsfonts}
\usepackage{upgreek}
\usepackage{framed}
\usepackage{tensor}
\usepackage{pifont}
\usepackage{latexsym, mathrsfs}
\usepackage{array}
\usepackage{hyperref}
\usepackage{xspace}
\usepackage{longtable}
\usepackage{multirow}
\usepackage{cases}
\usepackage{empheq}

\usepackage{bm}
\usepackage{bbm}
\usepackage{float}
\usepackage{afterpage}
\usepackage{setspace}
\usepackage{caption}
\usepackage[load-configurations=astronomy, range-units=brackets, range-phrase=-, per-mode=reciprocal, mode=math]{siunitx}
\usepackage{threeparttable}
\usepackage{subfigure}
\usepackage{tcolorbox}
\allowdisplaybreaks 

\usepackage{braket}

\usepackage{ragged2e}

\usepackage{hhline}
\usepackage{array}
\newcolumntype{P}[1]{>{\centering\arraybackslash}p{#1}}
\usepackage{booktabs}
\usepackage{tikz}
\usetikzlibrary{calc,shadings,patterns,tikzmark,fadings}
\usepackage{cleveref} 
\Crefname{equation}{Eq.}{Eqs.}
\Crefname{section}{Sec.}{Secs.}
\Crefname{figure}{Fig.}{Figs.}
\Crefname{table}{Table}{Tables}

\definecolor{Blue}{rgb}{0.25, 0.41, 0.88}
\definecolor{Red}{rgb}{0.92,0.,0.}
\definecolor{darkorange}{rgb}{1.0,0.549,0.}
\definecolor{cobalt}{RGB}{44, 98, 120}
\definecolor{Mathematica1}{rgb}{0.368417, 0.506779, 0.709798}
\definecolor{Mathematica2}{rgb}{0.880722, 0.611041, 0.142051}
\definecolor{Mathematica3}{rgb}{0.560181, 0.691569, 0.194885}
\definecolor{Mathematica4}{rgb}{0.922526, 0.385626, 0.209179}
\definecolor{Mathematica5}{rgb}{0.528488, 0.470624, 0.701351}
\definecolor{Mathematica6}{rgb}{0.772079, 0.431554, 0.102387}
\definecolor{Mathematica7}{rgb}{0.363898, 0.618501, 0.782349}
\definecolor{Mathematica8}{rgb}{1, 0.75, 0}
\definecolor{Mathematica9}{rgb}{0.647624, 0.37816, 0.614037}
\definecolor{plotBlue}{RGB}{94, 130, 181}
\definecolor{plotRed}{RGB}{233, 85, 54}
\definecolor{plotGreen}{RGB}{142, 176, 50}
\definecolor{plotPurple}{RGB}{135, 120, 178}

\newcolumntype{C}[1]{>{\centering\let\newline\\\arraybackslash\hspace{0pt}}m{#1}}

\makeatletter
\newlength{\apb@width}
\newcommand{\autoparbox}[2][c]{\settowidth{\apb@width}{#2}\parbox[#1]{\apb@width}{#2}}

\makeatother

\makeatletter
\newsavebox\myboxA
\newsavebox\myboxB
\newlength\mylenA

\newcommand*\xoverline[2][0.75]{
	\sbox{\myboxA}{$\m@th#2$}%
	\setbox\myboxB\null
	\ht\myboxB=\ht\myboxA%
	\dp\myboxB=\dp\myboxA%
	\wd\myboxB=#1\wd\myboxA
	\sbox\myboxB{$\m@th\overline{\copy\myboxB}$}
	\setlength\mylenA{\the\wd\myboxA}
	\addtolength\mylenA{-\the\wd\myboxB}%
	ifdim\wd\myboxB<\wd\myboxA%
	\rlap{\hskip 0.5\mylenA\usebox\myboxB}{\usebox\myboxA}%
	\else
	\hskip -0.5\mylenA\rlap{\usebox\myboxA}{\hskip 0.5\mylenA\usebox\myboxB}%
	\fi}
\makeatother

\numberwithin{equation}{section}
\numberwithin{figure}{section}
\numberwithin{table}{section}

\def\beq{\begin{equation}}
\def\eeq{\end{equation}}

\def\bea{\begin{eqnarray}}
\def\eea{\end{eqnarray}}

\def\beq{\begin{equation}}
\def\eeq{\end{equation}}
\def\bea{\begin{eqnarray}}
\def\eea{\end{eqnarray}}

\numberwithin{equation}{section}
\def\beq{\begin{equation}}
\def\eeq{\end{equation}}

\def\bea{\begin{eqnarray}}
\def\eea{\end{eqnarray}}

\DeclareRobustCommand{\SkipTocEntry}[4]{}

\usepackage{colortbl}
\definecolor{blue2}{cmyk}{1, 0.1, 0.1, 0.1}

\definecolor{pyBlue}{RGB}{31, 119, 180}
\definecolor{pyRed}{RGB}{214, 39, 40}
\definecolor{pyGreen}{RGB}{44, 160, 44}
\definecolor{pyBlue2}{RGB}{0, 111, 237}
\definecolor{pyRed2}{RGB}{224, 52, 36}

\newcolumntype{P}[1]{>{\centering\arraybackslash}p{#1}}
\newcolumntype{M}[1]{>{\centering\arraybackslash}m{#1}}

\begin{document}

\tcbset{colframe=black,arc=0mm,box align=center,halign=left,valign=center,top=-10pt}

\renewcommand{\thefootnote}{\fnsymbol{footnote}}

\pagenumbering{roman}
\begin{titlepage}
	\baselineskip=2.5pt \thispagestyle{empty}
	
	
\begin{center}
 ~{\Huge 
	 {\Large \textcolor{Sepia}{\bf\sffamily Primordial Gravitational Wave Circuit Complexity}}}\\  \vspace{0.25cm}
\end{center}

		\begin{center}
	{\fontsize{12}{18}\selectfont Kiran Adhikari${}^{\textcolor{Sepia}{1}}$},
	{\fontsize{12}{18}\selectfont Sayantan Choudhury${}^{\textcolor{Sepia}{2,3,4}}$\footnote{{\sffamily \textit{ Corresponding author, E-mail}} : {\ttfamily sayantan\_ccsp@sgtuniversity.org,  sayanphysicsisi@gmail.com}}}${{}^{,}}$
	\footnote{{\sffamily \textit{ NOTE: This project is the part of the non-profit virtual international research consortium ``Quantum Aspects of Space-Time \& Matter" (QASTM)} }. }${{}^{,}}$,~
	{\fontsize{12}{18}\selectfont Hardey N. Pandya ${}^{\textcolor{Sepia}{5}}$},~
	{\fontsize{12}{18}\selectfont Rohan Srivastava${}^{\textcolor{Sepia}{6}}$},
	
\end{center}


\begin{center}
	
{ 
	\textit{\footnotesize${}^{1}$RWTH Aachen University, D-52056, Aachen, Germany}\\
	\textit{\footnotesize${}^{2}$Centre For Cosmology and Science Popularization (CCSP),
SGT University, Gurugram, Delhi- NCR, Haryana- 122505, India,}\\
		\textit{\footnotesize${}^{3}$National Institute of Science Education and Research, Bhubaneswar, Odisha - 752050, India}\\
		\textit{\footnotesize${}^{4}$Homi Bhabha National Institute, Training School Complex, Anushakti Nagar, Mumbai - 400085, India}\\	
		\textit{\footnotesize${}^{5}$  School of Technology,
Pandit Deendayal Energy University, Gandhinagar - 382355, India}\\
		\textit{\footnotesize${}^{6}$Indian Institute of Technology Jodhpur, Jodhpur 342011, India.}\\
	}
\end{center}

\vspace{1.2cm}
\hrule \vspace{0.3cm}
\noindent {\bf Abstract}\\[0.1cm]

In this article, we investigate various physical implications of quantum circuit complexity using squeezed state formalism of Primordial Gravitational Waves (PGW). Recently quantum information theoretic concepts,  such as entanglement entropy, and complexity are playing a pivotal role to understand the dynamics of quantum system even in the diverse fields such as,  high energy physics and cosmology.  This paper is devoted in studying quantum circuit complexity of PGW for various cosmological models,  such as de Sitter,  inflation,  radiation, reheating,  matter,  bouncing,  cyclic and black hole gas model etc.  We compute complexity measure using both Covariance and Nielsen's wave function method for three different choices of quantum initial vacua: Motta-Allen,  $\alpha$ and Bunch-Davies. Besides computing circuit complexity,  we have also computed Von-Neumann entanglement entropy. 
By making the comparison of complexity with entanglement entropy, we are able to probe various features regarding the dynamics of evolution for different cosmological models.  Because entanglement entropy is independent of the squeezing angle,  we are able to understand more details of the system using Nielsen's measure of complexity which is dependent on both squeezing parameter and angle.  This implies that quantum complexity could indeed be a useful probe to study quantum features in cosmological scale.  Quantum complexity is also becoming a powerful technique to understand the chaotic behaviour and random fluctuations of quantum fields.  Using the growth of complexity,  we are able to compute quantum Lyapunov exponent for various cosmological models and comment on it's chaotic nature.

\vskip10pt
\hrule
\vskip10pt

\text{Keywords:~~Quantum circuit complexity, ~Entanglement entropy, Theoretical Cosmology}

\end{titlepage}

\newpage
\setcounter{tocdepth}{2}

\tableofcontents

\newpage
	
	\clearpage
	\pagenumbering{arabic}
	\setcounter{page}{1}
	
	\renewcommand{\thefootnote}{\arabic{footnote}}

\clearpage

\section{Introduction}

Formally,  Gravitational Waves (GW) can be defined as tensor perturbations or disturbances of space-time metric.  Although many indirect evidences for existence of Gravitational waves were available,  the first direct detection of Gravitational Waves was made by Laser Interferometry Gravitational Wave Observatory (LIGO) coming from a black hole merger in 2015 \cite{LIGOScientific:2016aoc}.  It has enabled us for a completely new possibility in which we can observe our universe along with electromagnetic radiation,  neutrino astronomy and others, marking new era of Gravitational Wave Astronomy. 

Once Gravitational waves are generated,  they keep spreading in the space-time without interacting with other fields in universe.  If we can detect Gravitational Waves that were formed just after Big Bang or very early times, potentially, it can reveal information about very early universe. These gravitational waves produced at very early times after Big Bang are called Primordial Gravitational Waves (PGW).  However, because of their relatively very small amplitude,  current GW detectors are not having sufficient sensitivity for identifying PGW.  As a consequence,  some indirect ways of identifying PGW have been suggested using Cosmic Microwave Background (CMB) power spectrum.  Reader can find more detailed aspect in this review and references therein \cite{Guzzetti:2016mkm}.

Models for theory of inflation describing early universe cosmology is well developed.  Cosmic Microwave Background measured from COBE in 1989 revealed some irregularities in temperature map of CMB \cite{1990ApJ...354L..37M}.  Models of inflation can give explanation of such observation and in fact the data agrees with models of inflation.  In the context of quantum field theory of curved space-time inflation is caused by inflation field.  It is believed that the quantum fluctuations exhibited by inflation field is responsible for the observed anisotropy in CMB temperature map.  From the recent Planck 2018 observation we get the latest constraints on inflation and the related cosmological parameter estimations,  which further constrain the CMB temperature as well as polarization maps.  See refs. \cite{Planck:2018jri,Planck:2018vyg} for more details.   

Primordial quantum fluctuations are thought to be the responsible for inflation \cite{baumann2012tasi, baumann2018tasi, Mukhanov:2007zz,birrell_davies_1982}. These quantum fluctuations in early universe are believed to be responsible for structure of present day universe and also generation for PGW.  We analyse these PGW in the form of tensor perturbation in the metric for various cosmological models and then incorporating it with quantum information theoretic methods, particularly in the computation of quantum circuit complexity.  So the detection of PGW can verify theory of inflation and also the correctness of the corresponding cosmological model. 

Among all inflation models, “Slow roll” is the one widely studied. This model was studied even in the framework of quantum inverted Harmonic Oscillator system undergoing different phase transition \cite{Grishchuk:1983gpw,Guth:1985ya, Grishchuk:1989ss, Grishchuk:1990bj, Albrecht:1992kf}.
However it was studied that at late times,  PGW undergoes the phenomenon of quantum squeezing,  a phenomena widely studied in the context of quantum optics \cite{1999OptPN..10...53L}. So,  various attempts have been made for identifying the signature for detection of these squeezed PGW.  It is argued that the direct detection for such squeezed PGW is not possible and as a result,  later on,  possibility of indirect detection method using Cosmic Microwave Background fluctuations is explored.
\cite{Allen:1999xw, Bose:2001fa, Baskaran:2010zz}. The evolution of these PGW can be explained in the framework of quantum squeezing operator formalism \cite{Albrecht:1992kf}.  Since there are well-developed quantum information theoretic tools to study squeezing operation, we would like to see if we can gain new insights at the level of cosmological scale using this emerging field of quantum information. 

Quantum information theory has already became a major factor recently in various areas of physics.  Especially,  calculation of circuit complexity has been paid much attention in recent studies,  because of its relation with chaos and it's possibility to probe physics behind the horizon of black hole in the context of AdS/CFT \cite{Chapman:2016hwi,Chapman:2018dem,Chapman:2018lsv, Caceres:2019pgf,Bernamonti:2020bcf, Bai:2021ldj,Doroudiani:2019llj}. Circuit complexity has even been computed in quantum field theories \cite{Jefferson:2017sdb, Bhattacharyya:2019kvj, Guo:2018kzl, Jiang:2018nzg, Khan:2018rzm, Hackl:2018ptj,Camargo:2019isp}. Despite it being a computational framework, its application has been impactful by giving conjectures like
“Complexity = Action” and “Complexity = Volume”
\cite{Stanford:2014jda,Brown:2015bva, Brown:2015lvg, Brown:2017jil}.

In the domain of Cosmology,  this concept is recently studied to characterize quantum chaos and complexity in early universe \cite{Bhattacharyya:2020kgu, Bhattacharyya:2020rpy, Lehners:2020pem}. In bouncing cosmological framework and cosmological islands, this concept has also been explored \cite{Bhargava:2020fhl,Choudhury:2020hil}.  On the other hand, entanglement entropy is also being paid attention in quantum field theory,  gravity and cosmology \cite{Rangamani:2016dms,Calabrese:2004eu, Maldacena:2012xp, Arias:2019pzy,Iizuka:2014rua,Matsumura:2017swh}. It has also been studied to see if the entanglement Entropy has relation with quantum circuit complexity \cite{Adhikari:2021pvv,Brahma:2020zpk, Eisert:2021mjg}.

It was realized previously that zero-particle states (vacua) for de Sitter space can exist as transformation according to set of two parameters for free field\cite{Allen:1985ux,Tagirov:1972vv,Chernikov:1968zm,Mottola:1984ar}. With relation to that, in this work, we aim to analyse squeezing of PGW with respect to three quantum initial vacuum states: Bunch Davis, $\alpha$-vacuum and $\alpha-\gamma$ vacuum (also called Motta-Allen vacuum) for various cosmological backgrounds.  For that,  we first develop squeezed state formalism for tensor perturbations using which circuit complexity using Nielsen and Covariance method is then calculated.  We also give comparison of entanglement entropy and circuit complexity. Then we calculate quantum Lyapunov exponent using these complexity measures.

It is important to note that,  latest observational constraint from Planck 2018 could not able to break the degeneracy among various models of inflation,  as well as between the bounce and inflationary paradigm at the level of two-point correlation function (from the amplitude and shape of the spectrum in Fourier space) for the comoving scalar modes.  For the tensor modes only the upper bound of the tensor-to-scalar ratio is available,  which cannot able to strictly fix the two-point function and its associated power spectrum for PGW in Fourier space.  The prime difficulty is as follows,  using the present precision level of the cosmological observation the detection of tensor modes and relics of PGW is very difficult.   Also the latest observational probes cannot able to detect to signatures non-Gaussianity from the spectrum of the higher point correlations using the cosmological perturbed scalar and tensor modes with high statistical precision.  Such repeated failure from the various observations in primordial cosmology is one of the strongest motivations to revisit and rethink the overall problem from a completely different perspective.  In this work we have used the two-mode squeezed state formalism using which we have quantized the Hamiltonian of the perturbed tensor modes,  which in turn produce the relics of PGW modes in early universe.  In this formalism we parametrise everything in terms of time dependent squeezed parameters,  which are squeezing amplitude,  squeezing angle and an additional phase.  These parameters further fix the theoretical structure of two or higher point quantum correlations,  entanglement entropy and quantum circuit complexity.  Till date the direct signatures of these parameters or the mentioned derived quantities are not yet been observed in any of the past observational probes of early universe cosmology.  From the observational perspective it is really a frustrating fact.  However,  one should not loose hope by seeing the various findings of quantum information theory within the framework of primordial cosmology.  If in near future the upcoming cosmological missions instead of probing the amplitude and of the power spectrum from two point as well as from the higher point functions can able to measure the previously mentioned squeezing parameters with high statistical accuracy then one can able to explicitly measure the signatures of various quantum information theoretic quantities, such as,  quantum circuit complexity,  quantum entanglement in the cosmological correlation functions.  The most probable candidate for this purpose is out-of-time ordered correlation (OTOC),  which basically signifies a special type of correlations which can be used to probe the mentioned quantum information theoretic measures in the context of cosmology.  the most significant fact is that,  within the framework of primordial cosmological perturbation theory using canonically quantized scalar and tensor modes one can compute cosmological version of OTOC in terms of the squeezed state parameters.  Because of this specific reason one can carry forward the computation and make a clear connection among OTOC and quantum circuit complexity and quantum entanglement.  Some initial level efforts has been already made in the similar direction in refs. \cite{Choudhury:2020yaa,Bhargava:2020fhl,Choudhury:2020hil,Choudhury:2021tuu},  where the authors try to establish such connections in the various contexts.  Even if the signatures of OTOC will not be able to probe by the future cosmological missions,  then by simply measuring the squeezing parameters suffice the purpose to a great extent.  Detection of these parameters through the observational probes not only able to measure the contributions of quantum circuit complexity,  but also able to quantify the amount of quantum entanglement in the cosmological correlations.  Such detection will going to be surely help us to distinguish among various models of inflation,  as well as to break the long standing degeneracy between the bouncing and inflationary paradigm.  In this article we are going to show in the next sections that just from the PGW (tensor modes) perspective theoretically it is possible to distinguish among all of these models and paradigms using the quantum complexity measure,  which is obviously a good indication at least from the quantum information theoretic point of view.  This is because,  in the context of quantum information theory quantum circuit complexity is obviously treated as better as well as strongest measure compared to the quantum entanglement entropy.  Within the framework of primordial cosmology if we found the direct observational evidence of the chaotic signature in terms of quantum Lyapunov exponent then one can able to break the degeneracy among various inflationary models,  among inflation and bounce which one is most probable,  also can able to distinguish among reheating,  black hole solutions,  and many more things can be done. Theoretically the corresponding measure of the quantum Lyapunov exponent can be computed from the slope of the quantum complexity curves studied with respect to the scale factors computed from various cosmological solutions.  We have found from our analysis that every models can't able to show such features clearly.  Now if in near future cosmological observations can able to detect the signature of this effect then it is highly expected that using the results obtained in this work one can discard various models and distinguish among various paradigms.  In this work,  at least from the theoretical point-of-view we found the same indication,  which we will going to show in the later half of the work.  Particularly the present study with PGW (tensor modes) is extremely important compared to the scalar modes of the perturbation because finding signatures in CMB maps from PGW can able to tell us about the exact scale of the fundamental physical interactions from which inflationary/ bouncing paradigms are created.  Till date it is completely a unknown fact.  We strongly believe that the present analysis on quantum circuit complexity from the perspective of PGW generated from various cosmological models can able to address at least some of these important unknown facts.

The organization of the paper is as follows:

\begin{itemize}
    \item In Sec. \ref{sec:ChaosAndComplexity}, we provide a brief review of quantum circuit complexity and how it can be related to quantum chaos. 
    \item In Sec. \ref{sec:squeezedState}, we provide the squeezed state formalism of cosmological perturbation of PGW. We give a list of various scale factors that we are interested. As our initial state, we choose different vacua such as Motta-Allen vacua, $\alpha$ vacua and Bunch-Davies vacua. Finally, we show how the squeezed state formalism in PGW dominated primordial cosmology. 
    \item In Sec. \ref{sec:complexityMeasure}, we compute the quantum circuit complexity of PGW from squeezed states using both covariance and Nielsen's wave function approach for all three vacua.
    \item In Sec. \ref{sec:entanglementEntropy}, we compute the entanglement entropy of PGW from squeezed states for each vacua: Motta-Allen vacua, $\alpha$ vacua and Bunch-Davies vacua.
    \item In Sec. \ref{sec:Numeric}, we perform the numerical analysis for various cosmological model. In particular, we compute quantum complexity, entanglement entropy, and quantum chaos.
    \item Finally, in Sec. \ref{sec:conclusion}, we give the concluding remarks and future prospects.

\end{itemize}

\section{Chaos and Complexity: Old wine in a new glass}
\label{sec:ChaosAndComplexity}
Computationally, circuit complexity is the minimum number of elementary operations that is required to solve a certain problem. In quantum computation, we can also introduce a similar term which indicates how hard or easy it to solve a particular problem in a quantum computer. In quantum computing, such problems are solved by operations which can be represented by an unitary transformation. Quantum circuit complexity would then indicate the smallest size of the circuit that implements this unitary. 

Quantum information theoretic probes like entanglement entropy has numerous applications in physics other than quantum computing \cite{Orus:2018dya,Choudhury:2020lja}. Very recently, quantum circuit complexity is also coming as one of such tools. One can use quantum circuit complexity to detect topological phases of matter where a high-complexity quantum state indicates a topological phase. In high energy physics, quantum complexity is being used to show holographic connections of Anti-de Sitter/ Conformal Field Theory (AdS/CFT) correspondence \cite{Brown:2015lvg, Brown:2017jil, Stanford:2014jda, Hashemi:2019aop}. The "Complexity equals volume" conjecture \cite{Stanford:2014jda} in the (AdS/CFT) correspondence says that the complexity in the boundary term is proportional to the volume in bulk state. Similarly, "Complexity equals action" conjecture says that the complexity in the boundary is dual to particular space-time region's action \cite{Brown:2015bva}. Because of these wide motivations, quantum circuit complexity is now being viewed as a powerful tool to study the behavior of quantum many-body system's. However, computing complexity is a notoriously difficult open challenge. In this section, we will describe the geometric approach introduced by Nielsen \cite{Nielsen1, Nielsen2, Nielsen3, Nielsen4} to lower bound the circuit complexity. 

\subsection{Framework of quantum circuit complexity }
Let us consider a transformation $U$ which transforms a reference quantum state $\ket{\psi_{R}}$ to the target quantum state $\ket{\psi_{T}}$,  is described by the following equation:
\begin{equation}
\ket{\psi_{T}} = U \ket{\psi_{R}}
\end{equation}
In quantum computational problem, usually this unitary transformation $U$ can be written as a sequence of basic unitary gates $Q_i$ which satisfy the constraint:
 \bea U = Q_1Q_2...Q_d,\eea where $d$ physically represents depth of the circuit under consideration. The minimum number of such operations $d$ can be referred as the circuit complexity. Because, it is very difficult to achieve the perfect transformation, we can have the tolerance $\epsilon$ such that:
\begin{equation}
\label{}
||\ket{\psi_{T}} -  U \ket{\psi_{R}}||^2 \leq \epsilon
\end{equation} 

Nielsen's approach of computing the complexity is geometric where the unitary $U$ is constructed using a time-dependent Hamiltonian $H(t)$:
\begin{equation}
\label{eq:controlHamiltonian}
U = \overleftarrow{\mathcal{P}} \text{ exp}\left[ -i\int_{0}^{1}d\tau H(\tau) \right] \text{ where } H(\tau) = \sum_{I} Y^I(\tau)\mathcal{O}_I
\end{equation}
The operators $\mathcal{O}_I$ forms a basis for $H(\tau)$ and $\mathcal{P}$ is a path ordering operator which indicates the circuit is moving from left to right. We also need to define a cost function $F(U, \vec{Y}(\tau))$ which is a local functional through $U(\tau)$ and tangent vectors $\vec{Y}(\tau))$. The cost for each path is given by:
\begin{equation}
\label{eq:geodesiclength}
\mathcal{D}(U(t)) = \int_{0}^{1}dt F(U(t),\dot{U}(t))
\end{equation}
Nielsen showed that, using the similar principles of Hamiltonian control theory gives the optimal quantum circuit where we minimize the functional using variational approach. There are certain properties we would like the cost function $F$ to satisfy: 
\begin{itemize}
\item {\textbf{Continuity}:}
 Here we have $F \in C^0$,  which implies that we need to consider continuous cost function. This is expected because of the physical reason.
\item{\textbf{Positivity}}:
The continuous cost function should satisfy the constraint, 
$F(U,v) \geq 0$.  Here $F=0$ is obtained from $v = 0$. 
This also mean that when this equality condition is satisfied, the reference and target should be same..
\item{\textbf{Positive homogeneity}}:
For any positive real number $\alpha$ and any vector $v$, we get $F(\alpha v) = \alpha F(v)$. 
\item{\textbf{Triangle Inequality}}:
The continuous cost function should satisfy strictly satisfy the triangle inequality,  which is given by,  $
F(U,v+v') \leq F(U,v) + F(U,v'),$
 for all tangent vectors $v$ and $v'$.
 Now If both of these vectors belong to the same ray,  then the equality holds good perfectly.
\end{itemize}
Here it is important to note that,  the definition of the cost function is not unique and depending on this fact the quantum complexity measure also changes.  These choices are appended below:
\begin{align}
\begin{split}
  \label{eq:cost-functional}
F_1(U,Y) &= \sum_I |Y^I| \\ F_p(U,Y) &= \sum_I p_I|Y^I| \\
F_2(U,Y) &= \sqrt{\sum_I |Y^I|^2 }   \\ F_q(U,Y) &= \sqrt{\sum_I q_I|Y^I|^2 } 
\end{split}
\end{align}
Here, $F_1$,  represents the linear cost functional measure which is appearing from the concept of closest to the individual counting of gates.  Further,  $F_2$,  represents quadratic cost functional measure which is appearing from the proper distance in the manifold.  Next,  $F_{1p}$,  represents a cost function measure with penalty factors $p_I$ used to favour certain choices over other.  Depending on the problem and system we want to study, we will have to make a different choice of these cost functions.

In this connection one can also consider the following possibilities:
\begin{equation}
\begin{split}
    \label{eq:k_cost_functional}
    F_k(U,Y) &= \sum_I |Y^I|^k \\
    F_{\frac{1}{k}}(U,Y) &= \sum_I |Y^I|^{\frac{1}{k}}
\end{split}
\end{equation}
where the degree of quantity homogeneity is characterized by $k \geq 1$.
These complexities were introduced to match the results from holography such as “Complexity = Action” and “Complexity = Volume” conjectures.

\subsection{Quantum Chaos and Complexity }
Quantum chaos is a popular tool in many body quantum physics to study statistical mechanics and thermodynamics \cite{DAlessio:2015qtq}. Very recently, it was conjectured that there should be a bound on quantum chaos \cite{Maldacena:2015waa, Maldacena:2016hyu} as:
\begin{equation}
    \lambda \leq 2\pi T
\end{equation}
where, $\lambda$ is identified to be the quantum Lyapunov exponent and $T$ is the equilibrium temperature of the system. This conjecture was made in the context of holography AdS/CFT \cite{Sekino:2008he,Shenker:2013pqa}. Since, complexity of a chaotic system also grows exponentially, one could associate the growth of the complexity measure to the quantum chaos \cite{Ali:2019zcj}:
\begin{equation}
    \lambda = \frac{d\ln(C[\tau])}{d\tau}
\end{equation}
where $C[\tau]$ is the measure of complexity.

\section{Squeezed State formalism of Cosmological Perturbation Theory of PGW}
\label{sec:squeezedState}
Consider a spatially flat ($k=0$) Friedmann–Lemaître–Robertson–Walker (FLRW) metric in $3+1$ dimensions:
\bea ds^2=a^2(\tau)\left(-d\tau^2+d{\bf x}^2\right)=a^2(\tau)\left(-d\tau^2+\delta_{ij}dx^{i}dx^{j}\right),\eea
where the metric is expressed in a conformally-flat form. Here, $a(\tau)$ is the conformal time dependent scale factor which is playing the role of the conformal factor in the present context. In this context,  conformal and physical time is related via,$d\tau=\frac{dt}{a(t)},$
which is used in the above mentioned expression for the FLRW infinitesimal line element to write it in terms of the conformal time coordinate instead of the physical time coordinate. For this reason, one can treat, $t\rightarrow \tau, ~{\bf x}\rightarrow {\bf x}$ , as the coordinate transformation in the present context.

Here we start with a very simplest theory which is describe by:
\bea S=\frac{1}{2}\int d^4x\sqrt{-g}\left[R-\left(\partial\phi\right)^2-2V(\phi)\right],\eea
where we fix Planck mass $M_p=1$.  Here $\phi$ is a scalar field which is minimally coupled to gravity in FLRW space-time.   Using this set-up various types of cosmological solutions of the scale factor can be obtained by solving simultaneously the Friedmann equations and Klein-Gordon equations in spatially flat FLRW background. Depending on the constraints and structure of the effective potential all these cosmological solutions can be obtained. In this article we will not mention any particular class of effective potentials, though from the cosmological solution of the scale factors one can predict the features ans structures of these effective potentials.

In this section, instead of effective class of potentials we are actually interested in the different solutions of the conformal time dependent scale factors of Friedmann equations, which will finally produce the PGW (PGW) from the gravitational tensor mode fluctuation. The desired solutions of the scale factors are given by:
\bea
\footnotesize\displaystyle\displaystyle a(\tau)&=& \footnotesize\left\{ 
     \begin{array}{lr}
     \displaystyle  -\frac{1}{H_{*}\tau}&~ \text{\textcolor{red}{\bf De Sitter}}\\  
\displaystyle  -\frac{1}{H_{*}\tau}\left[1+\epsilon_{*}-\epsilon_{*}\ln\left(\frac{\tau}{\tau_{*}}\right)+\cdots\right]&~ \text{\textcolor{red}{\bf Inflation/Quasi De Sitter~($w\approx -1$)}}\\  
    \displaystyle~\left[\frac{(1+3w)}{3(1+w)}\tau\right]^{\frac{2}{(1+3w)}}~~~~~~~~~&~~~~~ \displaystyle\text{\textcolor{red}{\bf Reheating~($0<w<1/3$)}}\\ 
  \displaystyle  ~\frac{\tau}{2}~~~~~~~~~&\text{\textcolor{red}{\bf Radiation~($w=1/3$)}} \\ 
  \displaystyle  ~\frac{\tau^2}{9}~~~~~~~~~&\text{\textcolor{red}{\bf Matter~($w=0$)}}\\ 
  \displaystyle  ~a_{*}~\left(-\gamma\frac{\tau}{\tau_*}\right)^{\alpha_*}~~~~~~~~~&\text{\textcolor{red}{\bf Contraction~(Pre-Bounce)}} \\ 
  \displaystyle  ~a_{*}\left[1+\left(\frac{\tau}{\tau_*}\right)^2\right]~~~~~~~~~&\text{\textcolor{red}{\bf Matter~Bounce}}\\ 
  \displaystyle  ~a_{*}~{\rm sech}\left(\alpha_*(\tau-\tau_*)\right)~~~~~~~~~&\text{\textcolor{red}{\bf Sechyperbolic~Bounce}}\\ 
  \displaystyle  ~a_{*}~{\cosh}\left(\alpha_*(\tau-\tau_*)\right)~~~~~~~~~&\text{\textcolor{red}{\bf Coshyperbolic~Bounce}}\\ 
  \displaystyle  ~a_{*}~{\sinh}\left(\alpha_*(\tau-\tau_*)\right)~~~~~~~~~&\text{\textcolor{red}{\bf Sinhyperbolic~Bounce}}\\ 
  \displaystyle  ~a_{*}~{\rm cosech}\left(\alpha_*(\tau-\tau_*)\right)~~~~~~~~~&\text{\textcolor{red}{\bf Cosechyperbolic~Bounce}}\\ 
  \displaystyle  ~a_{*}~\exp\left({\rm InverseErf}\left(\alpha_*(\tau-\tau_*)\right)\right)~~~~~~~~~&\text{\textcolor{red}{\bf Exponential~Bounce}}\\ 
  \displaystyle  ~a_{*}~\left(\frac{\tau}{\tau_*}\right)^{\frac{\alpha}{1-\alpha}}~~~~~~~~~&\text{\textcolor{red}{\bf Power-Law~Bounce~$(0<\alpha<1)$}}\\ 
  \displaystyle  ~a_{*}~\sqrt{\gamma\left(\frac{\tau}{\tau_*}\right)+\delta\left(\frac{\tau}{\tau_*}\right)^{2}}~~~~~~~~~&\text{\textcolor{red}{\bf Polynomial~Bounce}}\\ 
  \displaystyle  ~a_{*}~\left(\gamma\frac{\tau}{\tau_*}\right)^{\alpha_*}~~~~~~~~~&\text{\textcolor{red}{\bf Expansion~(Post-Bounce)}}\\ 
  \displaystyle  ~a_{*}~\left[1-{\cos}\left(\alpha_*(\tau-\tau_*)\right)\right]~~~~~~~~~&\text{\textcolor{red}{\bf Matter~Cyclic}}\\ 
  \displaystyle  ~a_{*}~\sin\left(\alpha_*(\tau-\tau_*)\right)~~~~~~~~~&\text{\textcolor{red}{\bf Radiation~Cyclic}}
  \\ 
  \displaystyle  ~a^{3/2}_{*}~\exp\bigg(\frac{i\pi}{2}\bigg)~\sqrt{\frac{2}{3}}~\tau^{1/2}~~~~~~~~~&\text{\textcolor{red}{\bf Black~Hole~Gas}}\end{array}
   \right.~~~~~~~~\eea  
   Most importantly each of the scale factors carrying some information of significant physics of the early universe. We have almost tried to quote almost all of them available in the cosmology literature. The detailed study on all of these possibilities will give us a complete picture of the underlying physics within the framework of chaos and complexity in our primordial universe. In the rest of the paper our further objective is to explicitly study all of these possibilities in detail. 
\subsection{Classical perturbation due to PGW}
In this section, we will consider only the gravitational contribution which is basically described by the Einstein Hilbert term i.e.
\bea  S_{\rm grav}=\frac{1}{2}\int d^4x \sqrt{-g}~R.\eea
The production of the PGW is described by the following linearised first order perturbation in the spatially flat ($k=0$) FLRW metric:
\bea \textcolor{red}{\bf Perturbed~Linearised~Metric:}~~~ds^2=a^2(\tau)\left(-d\tau^2+\left(\delta_{ij}+\underbrace{2h_{ij}(\tau,{\bf x})}_{\textcolor{red}{\bf Linearised~PGW}}\right)dx^{i}dx^{j}\right),~~~~~~~\eea
where the linearised perturbed PGW or the spin-$2$ tensor perturbation satisfy the following constraint conditions:
\bea &&\textcolor{red}{\bf Symmetric:\Rightarrow}~~h_{ij}=h_{ji},~~~\textcolor{red}{\bf Transverse:\Rightarrow}~~\partial^{i}h_{ij}=0,~~~
\textcolor{red}{\bf Traceless:\Rightarrow}~~h^{i}_{i}=0.~~~~~~\eea
Further, substituting the above mentioned perturbed linearised metric in the Einstein Hilbert classical gravitational action the second order perturbed action is given by:
\bea \delta^{(2)}S_{\rm grav}=\frac{1}{8}\int d\tau~d^3x~a^2(\tau)~\left[(h^{'}_{ij}(\tau,{\bf x}))^2-\left(\nabla h_{ij}(\tau,{\bf x})\right)^2\right].\eea
Further, considering the above mentioned properties of PGW one can write the following ansatz:
\bea h_{ij}(\tau,{\bf x}):=\sum_{\lambda=+,\times}h^{(\lambda)}(\tau)~e^{(\lambda)}_{ij}({\bf x}),\eea
where, $e^{(\lambda)}_{ij}({\bf x})$ represents polarization tensor for PGW for two helicities, $\lambda=+,\times$, respectively.

Next, for the computational purpose it is convenient to define the following rescaled perturbation field variable for the linearised perturbation of PGW, which is given by:
\bea \frac{1}{2}a(\tau)h_{ij}(\tau,{\bf x})\equiv \frac{1}{\sqrt{2}}\begin{pmatrix}
~f_{(+)}(\tau,{\bf x})~ & ~f_{(\times)}(\tau,{\bf x})~ &~ 0~\\
~f_{(\times)}(\tau,{\bf x})~ & ~-f_{(+)}(\tau,{\bf x})~ &~ 0~\\
0 & 0 & 0\
\end{pmatrix}\eea
which is consistent with the all the required constraints on the spin-2 PGW or the tensor perturbation. Using the above mentioned convenient field redefinition the second order perturbed action for the linearised tensor modes or the PGW can be written as:
\bea \delta^{(2)}S_{\rm grav}=\frac{1}{2}\sum_{\lambda=+,\times}\int d\tau~d^3x~\left[|f^{'}_{\lambda}(\tau,{\bf x})|^2-\left|\nabla f_{\lambda}(\tau,{\bf x})\right|^2+\frac{a^{''}(\tau)}{a(\tau)}\left|f_{\lambda}(\tau,{\bf x})\right|^2\right].\eea
 
Now,  we transform the above action in the Fourier space using the following equation,
\bea f_{\lambda}(\tau,{\bf x}):=\int \frac{d^3{\bf k}}{(2\pi)^3}~f_{\lambda,{\bf k}}(\tau)~\exp(i{\bf k}.{\bf x})~~~~~\forall ~~\lambda=+,\times.\eea
Using this further we get:
\bea \delta^{(2)}S_{\rm grav}=\frac{1}{2}\sum_{\lambda=+,\times}\int d\tau~d^3{\bf k}~\left[|f^{'}_{\lambda,{\bf k}}(\tau)|^2-\omega^2(k,\tau)\left|f_{\lambda,{\bf k}}(\tau)\right|^2\right],\eea
where we define the effective frequency of the PGW as:
\bea \omega^2(k,\tau)=k^2+m^2(\tau)~~~~~~~{\rm with~effective~mass}~~~~~~m^2(\tau)=-\frac{a^{''}(\tau)}{a(\tau)}.~~~~~\eea
Further the Mukhanov Sasaki equation for PGW can be written as:
\bea f^{''}_{\lambda,{\bf k}}(\tau)+\omega^2(k,\tau)f_{\lambda,{\bf k}}(\tau)=0.\eea

Now the effective mass of the PGW can be further simplified as:
\bea m^2(\tau)&=&-\frac{a^{''}(\tau)}{a(\tau)}=-\left({\cal H}^2+{\cal H}'\right)=\left(\epsilon(\tau)-2\right)~{\cal H}^2,\eea
where we define slowly varying quantity,  $\epsilon(\tau)$, as:
\bea \epsilon(\tau)=1-\frac{{\cal H}'}{{\cal H}^2},~~~~{\rm where}~~~~{\cal H}=\frac{a'(\tau)}{a(\tau)}.\eea
On the other hand,  the mass parameter $\nu(\tau)$ for PGW can be parametrized as:
\bea  m^2(\tau)&=&-\frac{1}{\tau^2}\left(\nu^2_{\rm PGW}(\tau)-\frac{1}{4}\right),\\
{\rm where}\quad \nu_{\rm PGW}(\tau)&=&\frac{1}{2}\sqrt{1+4(\tau{\cal H})^2(2-\epsilon(\tau))}=\frac{1}{2}\sqrt{1+4(\tau{\cal H})^2\left(1+\frac{{\cal H}'}{{\cal H}^2}\right)}.\eea
Apart from this, another very crucial quantity play very significant role to determine the cosmological dynamics,  described by:
\bea {\cal D}_{{\cal H}}:=\frac{k}{a(t)H(t)}=\frac{k}{{\cal H}(\tau)}\rightarrow\left\{ 
     \begin{array}{lr}
\displaystyle   \ll 1 ~ ~~ ~~ ~~ ~&~ \text{\textcolor{red}{\bf Super-Horizon}}\\ \\
    \displaystyle =1,~ ~& \displaystyle\text{\textcolor{red}{\bf Horizon-crossing}}\\ \\
  \displaystyle  \gg 1,~ &\text{\textcolor{red}{\bf Sub-Horizon}} \end{array}
   \right.~~~~~~\eea
 In the following we will now explicitly derive the expression for the mass parameters and the above mentioned horizon determining ratio for different primordial universe models:
      \begin{enumerate}
   	\item \underline{\textcolor{red}{\bf De Sitter:}}\\

   	\bea &&\nu_{\rm PGW}(\tau)=\frac{3}{2},
   	\\
   	&&{\cal D}_{{\cal H}}=\frac{k}{{\cal H}(\tau)}=-k\tau.\eea
   	
   	\item \underline{\textcolor{red}{\bf Inflation/ Quasi De Sitter:}}\\
   	\bea &&\nu_{\rm PGW}(\tau)=\frac{3}{2}+3\epsilon(\tau)-\eta(\tau)~~~~~{\rm with}~~~~\eta(\tau)=2\epsilon(\tau)-\frac{1}{2{\cal H}}\frac{\epsilon'(\tau)}{\epsilon(\tau)},
   	\\
   	&&{\cal D}_{{\cal H}}=\frac{k}{{\cal H}(\tau)}=-k\tau.\eea
   	
   	\item \underline{\textcolor{red}{\bf Reheating~($0<w<1/3$):}}\\
   	\bea &&\nu_{\rm PGW}(\tau)=\frac{3}{2} \frac{(w-1)}{(3 w+1)},
   	\\
   	&&{\cal D}_{{\cal H}} = \frac{k}{{\cal H}(\tau)} = \frac{k a(\tau)}{a'(\tau)} =  \frac{1}{2} k \tau  (3 w+1)
   	\eea 
   	During reheating the equation of state parameter is lying within the window $0<w<1/3$,  which was discussed in detail in refs.  \cite{Munoz:2014eqa,Choudhury:2020yaa}.  However,  there is always an ambiguity of having the equation of state parameter within this prescribed window as the micro physics of the epoch of reheating is not known yet and till date it was studied from the phenomenological perspective only.  Regarding the ambiguity let us mention a ref. \cite{Liu:2015psa}, where the authors has pointed that during such epoch the equation of state parameter is $w>1/3$.  Though we have restricted the equation of state parameter within $0<w<1/3$ because at least it believed that classically reheating epoch should lie in between matter domination ($w=0$) and radiation ($w=1/3$).  The exact form of the equation of state during this epoch unknown till date.
   	\item \underline{\textcolor{red}{\bf Radiation~($w=1/3$):}}\\
   	\bea && \nu_{\rm PGW}(\tau)=-\frac{1}{2}.
   	\\&&{\cal D}_{{\cal H}}=\frac{k}{{\cal H}(\tau)}=-k\tau.\eea 
   	
   	\item \underline{\textcolor{red}{\bf Matter~($w=0$):}}\\
   	\bea &&\nu_{\rm PGW}(\tau)=-\frac{3}{2}. \\
   	&&{\cal D}_{{\cal H}}=\frac{k}{{\cal H}(\tau)}=-k\tau.\eea 
   	Bouncing cosmology is emerging as an alternative to inflationary models \cite{Brandenberger:2010dk,Brandenberger:2012zb,Bramberger:2019zez,Gao:2009wn,Koehn:2013upa,Koehn:2015vvy}. Here, the mass parameters for various bouncing models are given.
   	\item \underline{\textcolor{red}{\bf Contraction~(Pre-Bounce):}}\\
   	\bea && \nu_{\rm PGW}(\tau)=\alpha-\frac{1}{2}. \\
   	&&{\cal D}_{{\cal H}}= \frac{k\tau}{\alpha_*}.\eea 
   	
   	\item \underline{\textcolor{red}{\bf Matter~Bounce:}}\\
   	\bea && \nu_{\rm PGW}(\tau)=\frac{3}{2}~\sqrt{1-\frac{8}{9}\frac{1}{\left[1+\left(\frac{\tau}{\tau_*}\right)^2\right]}}. \\
   	&&{\cal D}_{{\cal H}}=\frac{k(\tau^2 + \tau_*^2)}{2\tau}.\eea 
   	
   	\item \underline{\textcolor{red}{\bf Sechyperbolic~Bounce:}}\\
   	\bea && \nu_{\rm PGW}(\tau)=\frac{1}{2}~\sqrt{1+4\tau^2\alpha_*^2\left[1-2~{\rm sech}^2(\alpha_*(\tau-\tau_*))\right]}. \\
   	&& {\cal D}_{{\cal H}}= \frac{-k\coth[\alpha_*(\tau-\tau_*)]}{\alpha_*}.\eea 
   	
   	\item \underline{\textcolor{red}{\bf Coshyperbolic~Bounce:}}\\
   	\bea && \nu_{\rm PGW}(\tau)=\frac{1}{2}~\sqrt{1+4\tau^2\alpha_*^2}. \\
   	&& {\cal D}_{{\cal H}}=\frac{k\coth[\alpha_*(\tau-\tau_*)]}{\alpha_*}.	\eea 
   	
   	\item \underline{\textcolor{red}{\bf Sinhyperbolic~Bounce:}}\\
   	In this case, the mass parameter can be evaluated as:
   	\bea && \nu_{\rm PGW}(\tau)=\frac{1}{2}~\sqrt{1+4\tau^2\alpha_*^2}. \\
   	 	&& {\cal D}_{{\cal H}}=\frac{k\tanh[\alpha_*(\tau-\tau_*)]}{\alpha_*}. \eea 
   	
   	\item \underline{\textcolor{red}{\bf Cosechyperbolic~Bounce:}}\\
   	\bea && \nu_{\rm PGW}(\tau)=\frac{1}{2}~\sqrt{1+4\tau^2\alpha_*^2\left[1+2~{\rm cosech}^2(\alpha_*(\tau-\tau_*))\right]}.\\
   	&& {\cal D}_{{\cal H}}=\frac{k*\text{cosech}[\alpha_*(\tau-\tau_*)]}{\alpha_*\text{cosech'}[\alpha_*(\tau-\tau_*)]}.	\eea 
   	
   	\item \underline{\textcolor{red}{\bf Exponential~Bounce:}}\\
   	\bea && \nu_{\rm PGW}(\tau)=\frac{1}{2} \sqrt{1+\pi  \alpha_* ^2 \tau^2(1+2~ \text{InverseErf}~(\alpha_*  (\tau-\tau_*)))~  \exp(2~ \text{InverseErf}~(\alpha_*  (\tau-\tau_*))^2) }.~~~~~~~~~~~\newline \\
   	&&{\cal D}_{{\cal H}} = \frac{k}{{\cal H}(\tau)} =\frac{2\exp\left( -\text{InverseErf}[\alpha_*(\tau-\tau_*)]^2 \right)k}{\sqrt{\pi}\alpha_*}. \eea

   	\item \underline{\textcolor{red}{\bf Power-Law~Bounce~($0<\alpha_*<1$):}}\\
   	\bea &&\nu_{\rm PGW}(\tau)=\frac{1}{2}~\left(\frac{3\alpha_*-1}{\alpha_*-1}\right). \\ &&{\cal D}_{{\cal H}} = \frac{(1-\alpha_* ) k \tau}{\alpha_* }
   	\eea 
   	
   	\item \underline{\textcolor{red}{\bf Polynomial~Bounce:}}\\
   	\bea \nu_{\rm PGW}(\tau)=\frac{1}{2} \frac{1}{(\tau_* \gamma +\delta  \tau)}\sqrt{\delta  \tau (2 \tau_* \gamma +\delta  \tau)}. \\ 
   	{\cal D}_{{\cal H}} = \frac{2 k \tau (\delta  \tau +\gamma  \tau_*)}{2 \delta  \tau+\gamma  \tau_*}  \eea 
   	
   	\item \underline{\textcolor{red}{\bf Matter~Cyclic:}}\\
   	\bea &&\nu_{\rm PGW}(\tau)=\frac{1}{2} \sqrt{1-2\tau^2\alpha_*^2\left[1-{\rm cot}^2\left(\frac{\alpha_*(\tau-\tau_*)}{2}\right)\right]}. \\ 
   	&&{\cal D}_{{\cal H}}=\frac{k \tan \left(\frac{ \alpha_* }{2} (\tau-\tau_*)\right)}{\alpha_* }
   	\eea 
   	
   	\item \underline{\textcolor{red}{\bf Radiation~Cyclic:}}\\
   	\bea &&\nu_{\rm PGW}(\tau)=\frac{1}{2} \sqrt{1-4\tau^2\alpha_*^2}. \\
   	&&{\cal D}_{{\cal H}}=\frac{k \tan (\alpha_*  (\tau-\tau_*))}{\alpha_* }.\eea

   	\item \underline{\textcolor{red}{\bf Black Hole Gas:}}\\
   	Black hole gas is a relatively new model which predict maximum entropy state at a very early universe \cite{Mathur:2020ivc,Mathur:2009hf,Horowitz:1996nw,Veneziano:2003sz,Fischler:1998st}.
   	In this case, the mass parameter and the horizon determining ratio can be evaluated as:
   		\bea && \nu_{\rm PGW}(\tau)=0 \\
   	&&{\cal D}_{{\cal H}} = \frac{k}{{\cal H}(\tau)} = 2 k \tau. \eea

   \end{enumerate}
   We are going to use all of these derived model dependent expressions for the mass parameters for the PGW during the numerical analysis performed in the later half of this paper. From these derived expressions it is evident that the depending on the expressions of different models the expressions for the mass parameter changes and determined in terms of the model parameters. Here we have found that for some models this parameter is exactly constant and for some class of models there are non-trivial conformal time dependence appearing. In the later half of the paper once we will perform the analysis of the squeezed state formalism, there we can able to see how these different expressions of the mass parameters will hugely change the dynamical behaviour of the quantum complexity and the related estimators of chaos.
\subsection{Classical mode function for PGW}  
For the PGW,  {\it Mukhanov-Sasaki equation} is:
\bea f^{''}_{\lambda,{\bf k}}(\tau)+\left(k^2-\frac{1}{\tau^2}\left(\nu^2_{\rm PGW}(\tau)-\frac{1}{4}\right)\right)f_{\lambda,{\bf k}}(\tau)=0.\eea
The most general analytical solution can be expressed as:
\bea f_{\lambda,{\bf k}}(\tau)=\sqrt{-\tau}\left[{\cal C}_1~{\cal H}^{(1)}_{\nu_{\rm PGW}}(-k\tau)+{\cal C}_2~{\cal H}^{(2)}_{\nu_{\rm PGW}}(-k\tau)\right].\eea
Here, ${\cal C}_1$ and ${\cal C}_2$, are arbitrary integration constants, fixed by the appropriate choice of quantum initial vacuum state which also satisfy the normalization criteria:
\bea |{\cal C}_1|^2-|{\cal C}_2|^2=1.\eea
In this solution, ${\cal H}^{(1)}_{\nu_{\rm PGW}}(-k\tau)$ and ${\cal H}^{(2)}_{\nu_{\rm PGW}}(-k\tau)$, represent the Hankel function of the first and second kind. In general one can further express these functions in terms of the Bessel function of the first kind and Newman function (which is the Bessel function of the second kind) as:
\bea &&{\cal H}^{(1)}_{\nu_{\rm PGW}}(-k\tau)={\cal J}^{(1)}_{\nu_{\rm PGW}}(-k\tau)+i{\cal Y}^{(1)}_{\nu_{\rm PGW}}(-k\tau),\\
&&{\cal H}^{(1)}_{\nu_{\rm PGW}}(-k\tau)={\cal J}^{(1)}_{\nu_{\rm PGW}}(-k\tau)-i{\cal Y}^{(1)}_{\nu_{\rm PGW}}(-k\tau),\eea
which further implies the fact that in the present context the two Hankel functions appearing are complex conjugate of each other.
If we further substitute back the above mentioned expressions in the previously obtained solution of the {\it Mukhanov-Sasaki equation} for the PGW then we get the following result:
\bea f_{\lambda,{\bf k}}(\tau)=\sqrt{-\tau}\left[{\cal D}_1~{\cal J}^{(1)}_{\nu_{\rm PGW}}(-k\tau)+{\cal D}_2~{\cal Y}^{(2)}_{\nu_{\rm PGW}}(-k\tau)\right],\eea
where we have defined two new arbitrary integration constants, ${\cal D}_1$ and ${\cal D}_2$, which are defined in terms of the above mentioned constants, ${\cal C}_1$ and ${\cal C}_2$, as:
\bea {\cal D}_1 &=& {\cal C}_1+i{\cal C}_2,\\
{\cal D}_2&=& {\cal C}_1-i{\cal C}_2.\eea
It needs to be emphasized that in general the two arbitrary integration constants, ${\cal C}_1$ and ${\cal C}_2$ might be parametrised by complex parameters. But the most important part is both the solutions are equivalent to each other and one can use any one out of them according to their preference. In this paper we will take the first solution which is expressed in terms of the Hankel functions. 

It is difficult to analyse and extract some physically meaningful information from looking at obtained solutions of rescaled field and associated canonically conjugate momentum. So, we will consider asymptotic limits in obtained solutions which will be very helpful for our analysis in various cosmological scales. We take asymptotic limits as $k\tau\rightarrow 0$ and $k\tau\rightarrow- \infty$ to decide the behaviour of Hankel functions of the first and second kind of order $\nu_{\rm PGW}$. We get expressions as follows after taking such asymptotic limits: 
\bea &&{\lim_{k\tau\rightarrow -\infty} {\cal H}^{(1)}_{\nu_{\rm PGW}}(-k\tau)=\sqrt{\frac{2}{\pi}}\frac{1}{\sqrt{-k\tau}}\exp(-ik\tau)\exp\left(-\frac{i\pi}{2}\left(\nu_{\rm PGW}+\frac{1}{2}\right)\right)},\\
&&{ \lim_{k\tau\rightarrow -\infty} {\cal H}^{(2)}_{\nu_{\rm PGW}}(-k\tau)=-\sqrt{\frac{2}{\pi}}\frac{1}{\sqrt{-k\tau}}\exp(ik\tau)\exp\left(\frac{i\pi}{2}\left(\nu_{\rm PGW}+\frac{1}{2}\right)\right)},\\
 &&{ \lim_{k\tau\rightarrow 0} {\cal H}^{(1)}_{\nu_{\rm PGW}}(-k\tau)=\frac{i}{\pi}\Gamma(\nu_{\rm PGW})\left(-\frac{k\tau}{2}\right)^{-\nu_{\rm PGW}}},\\
   &&{\lim_{k\tau\rightarrow 0} {\cal H}^{(2)}_{\nu_{\rm PGW}}(-k\tau)=-\frac{i}{\pi}\Gamma(\nu_{\rm PGW})\left(-\frac{k\tau}{2}\right)^{-\nu_{\rm PGW}}},\eea 
  which implies that:
\bea && {\lim_{k\tau\rightarrow 0} {\cal H}^{(1)}_{\nu_{\rm PGW}}(-k\tau)=-\lim_{k\tau\rightarrow 0} {\cal H}^{(2)}_{\nu_{\rm PGW}}(-k\tau)},\\
&&  {\lim_{k\tau\rightarrow -\infty} {\cal H}^{(1)}_{\nu_{\rm PGW}}(-k\tau)=- \lim_{k\tau\rightarrow -\infty} {\cal H}^{(2)}_{\nu_{\rm PGW}}(-k\tau)},
\eea
 Here,  super-horizon($k\tau\ll-1$) is  represented by   limit $k\tau\rightarrow 0$ and   sub-horizon($k\tau\gg-1$) is  represented by  limit $k\tau\rightarrow -\infty$. Transition of modes form sub-horizon to super-horizon can be represented by $k\tau=-1$, which is known as horizon exit.  

The expressions for rescaled field variable and associated canonically conjugate momentum can be obtained by placing above mentioned super-horizon ($k\tau\ll -1$) and sub-horizon  ($k\tau\gg -1$) asymptotic limits as follows:
 \bea \lim_{k\tau\rightarrow 0}f_{\lambda,{\bf k}}(\tau)&=&\sqrt{\frac{2}{k}}\frac{i}{\pi}\Gamma(\nu_{\rm PGW})\left(-\frac{k\tau}{2}\right)^{\frac{1}{2}-\nu_{\rm PGW}}\left({\cal C}_1-{\cal C}_2\right),\\
 \lim_{k\tau\rightarrow -\infty}f_{\lambda,{\bf k}}(\tau)&=&\sqrt{\frac{2}{\pi k}}\left[{\cal C}_1~\exp\left(-i\left\{k\tau+\frac{\pi}{2}\left(\nu_{\rm PGW}+\frac{1}{2}\right)\right\}\right)\right.\nonumber\\
&&\left.~~~~~~~~~~~~~~~~~~~~~~~~~~~~~~~~~~~~~~-{\cal C}_2~\exp\left(i\left\{k\tau+\frac{\pi}{2}\left(\nu_{\rm PGW}+\frac{1}{2}\right)\right\}\right)\right]. ~~~~~~~~\eea
Hence the most general solution for rescaled PGW field variable for any quantum initial vacuum state can be written as:
\bea&& f_{\lambda,{\bf k}}(\tau)=\frac{1}{\sqrt{2k}}2^{\nu_{\rm PGW}-\frac{3}{2}}(-k\tau)^{\frac{3}{2}-\nu_{\rm PGW}}\left|\frac{\Gamma(\nu_{\rm PGW})}{\Gamma\left(\frac{3}{2}\right)}\right|\nonumber\\
&&~~~~~~~~~~~~~~~~~~~\times\left[{\cal C}_1~\left(1-\frac{i}{k\tau}\right)~\exp\left(-i\left\{k\tau+\frac{\pi}{2}\left(\nu_{\rm PGW}-\frac{3}{2}\right)\right\}\right)\right.\nonumber\\
&& \left.~~~~~~~~~~~~~~~~~~~~~~~~~~~~~~+{\cal C}_2~\left(1+\frac{i}{k\tau}\right)~\exp\left(i\left\{k\tau+\frac{\pi}{2}\left(\nu_{\rm PGW}-\frac{3}{2}\right)\right\}\right)\right].~~~~~~~~~
\eea

Finally, we want to mention about a special situation, where we get a imaginary contribution, i.e. instead of having $\nu_{\rm PGW}$ we have $-i|\nu_{\rm PGW}|$ after analytic continuation. This can be very clearly understood from the following equation:
\bea \nu_{\rm PGW}(\tau)=-i|\nu_{\rm PGW}(\tau)|~~~~{\rm where}~~~~|\nu_{\rm PGW}(\tau)|=\sqrt{\frac{m^2_{\cal H}(\tau)}{{\cal H}^2}-\frac{9}{4}},\eea
where we define the heavy Hubble effective mass by the following expressions:
\bea m_{\cal H}(\tau):=\sqrt{\left(2-\tau^2\left({\cal H}^2+{\cal H}'\right)\right)}~{\cal H}\gg {\cal H},\eea
where we have the following additional constraint condition holds good in the associated conformal time scale:
\bea \tau\ll \frac{1}{\sqrt{\left({\cal H}^2+{\cal H}'\right)}}=\frac{1}{\sqrt{2-\epsilon(\tau)}~{\cal H}}.~~~~\eea
In this particular situation, the solution for the rescaled PGW field with arbitrary quantum initial vacuum can be expressed as:
\bea&& f_{\lambda,{\bf k}}(\tau)=\frac{1}{\sqrt{2k}}2^{-\left(i|\nu_{\rm PGW}|+\frac{3}{2}\right)}(-k\tau)^{\frac{3}{2}+i|\nu_{\rm PGW}|}\left|\frac{\Gamma(-i|\nu_{\rm PGW}|)}{\Gamma\left(\frac{3}{2}\right)}\right|\nonumber\\
&&~~~~~~~~~~~~~~~~~~~\times\left[{\cal C}_1~\left(1-\frac{i}{k\tau}\right)~\underbrace{\exp\left(-\frac{\pi}{2}|\nu_{\rm PGW}|\right)}_{\textcolor{red}{\bf Boltzmann ~suppression}}\exp\left(-i\left\{k\tau-\frac{3\pi}{4}\right\}\right)\right.\nonumber\\
&& \left.~~~~~~~~~~~~~~~~~~~~~~~~~~~~~~+{\cal C}_2~\left(1+\frac{i}{k\tau}\right)~\underbrace{\exp\left(\frac{\pi}{2}|\nu_{\rm PGW}|\right)}_{\textcolor{red}{\bf Boltzmann ~enhancement}}\exp\left(i\left\{k\tau-\frac{3\pi}{4}\right\}\right)\right].~~~~~~~~~
\eea
For all type of possible quantum initial vacuum state the highlighted {\it Boltzmann suppression} term play a very significant role as it is related to the fact of huge suppression in the probability distribution profile and the related spectrum of the heavy Hubble particle production. On the other hand, for excited quantum vacua states like {\it Bruce Allen vacua} and for $\alpha$ {\it vacua} an additional contribution will also appear which is also highlighted as {\it Boltzmann enhancement} contribution and this is the only term which only survives when one consider the possibility for having excited quantum vacua states as an initial choice. For the {\it Bunch-Davies vacuum state}, which is the ground quantum vacuum {\it Euclidean state}, this particular {\it Boltzmann enhancement} contribution not appears. This implies that for the excited quantum vacua there is a possibility for having enhancement in the probability distribution profile and its related spectrum of the particle production. Frankly speaking, for the excited quantum vacuum states a competition between the {\it Boltzmann suppression} term and {\it Boltzmann enhancement} term always take place, and depending on the parameter values for a specific vacuum parametrization one can decide that who wins the competition at the end of this cosmological game. To feel the cosmological essence of this competition one need to explicitly study the physical outcomes in the context of chaos and complexity.  On the other hand, for the commonly used {\it Bunch-Davies state} there is no such competition will be observed as the {\it Boltzmann suppression} term only survives. Though we will study this possibility as well in this paper to compare between the results obtained from the excited and ground quantum vacuum states.

In this paper we will further discuss about three specific choices of quantum vacuum states which are mostly appearing in the different context in the theoretical physics literature. These possibilities are appended below:
\begin{enumerate}
\item \underline{\textcolor{red}{\bf Motta-Allen ($\alpha,\gamma$)~vacua:}}\\
This is a specific choice where the arbitrary coefficients that are appearing in the classical solution of the PGW are parametrized by two real parameters $\alpha$ and $\gamma$, by:
\bea {\cal C}_1=\cosh\alpha,~~{\cal C}_2=\exp(i\gamma)~\sinh\alpha~~~\forall \alpha, \gamma \in \mathbb {R}~~~\textcolor{blue}{\bf subject~to}~~|{\cal C}_1|^2-|{\cal C}_2|^2=1.~~~~~~~\eea
This is considered as a excited $SO(1,4)$ isommetric vacuum state having oscillating feature. One can explicitly show that his type of quantum excited vacua is CPT symmetry breaking in nature and it is characterized by the real parameter $\gamma$.  Apart from having this issue to describe various physical situations the explicit role of this quantum vacuum state is extremely important. In this paper we are going to investigate the role of this two real parameters $\alpha$ and $\gamma$ in the parametrization of {\it Motta-Allen vacua} to determine the quantum complexity and to study the underlying phenomena of quantum chaos from the various cosmological primordial models of our universe using PGW perturbation. We are pretty much excited about this finding because such possibilities have not been explored. In this context, the solution for the PGW field variable for the {\it Motta-Allen} vacua can be written as:
\bea && f_{\lambda,{\bf k}}(\tau)=\frac{1}{\sqrt{2k}}2^{\nu_{\rm PGW}-\frac{3}{2}}(-k\tau)^{\frac{3}{2}-\nu_{\rm PGW}}\left|\frac{\Gamma(\nu_{\rm PGW})}{\Gamma\left(\frac{3}{2}\right)}\right|\nonumber\\
&&~~~~~~~~~~~~~~~~~~~\times\left[\cosh\alpha~\left(1-\frac{i}{k\tau}\right)~\exp\left(-i\left\{k\tau+\frac{\pi}{2}\left(\nu_{\rm PGW}-\frac{3}{2}\right)\right\}\right)\right.\nonumber\\
&& \left.~~~~~~~~~~~~~~~~~~~~~~~~~~~~~~+\sinh\alpha~\left(1+\frac{i}{k\tau}\right)~\exp\left(i\left\{\gamma+k\tau+\frac{\pi}{2}\left(\nu_{\rm PGW}-\frac{3}{2}\right)\right\}\right)\right].~~~~~~~~~
\eea
For the case of heavy field we have the following simplified expression:
\bea&& f_{\lambda,{\bf k}}(\tau)=\frac{1}{\sqrt{2k}}2^{-\left(i|\nu_{\rm PGW}|+\frac{3}{2}\right)}(-k\tau)^{\frac{3}{2}+i|\nu_{\rm PGW}|}\left|\frac{\Gamma(-i|\nu_{\rm PGW}|)}{\Gamma\left(\frac{3}{2}\right)}\right|\nonumber\\
&&~~~~~~~~~~~~~~~~~~~\times\left[\cosh\alpha~\left(1-\frac{i}{k\tau}\right)~\underbrace{\exp\left(-\frac{\pi}{2}|\nu_{\rm PGW}|\right)}_{\textcolor{red}{\bf Boltzmann ~suppression}}\exp\left(-i\left\{k\tau-\frac{3\pi}{4}\right\}\right)\right.\nonumber\\
&& \left.~~~~~~~~~~~~~~~~~+\sinh\alpha~\left(1+\frac{i}{k\tau}\right)~\underbrace{\exp\left(\frac{\pi}{2}|\nu_{\rm PGW}|\right)}_{\textcolor{red}{\bf Boltzmann ~enhancement}}\exp\left(i\left\{\gamma+k\tau-\frac{3\pi}{4}\right\}\right)\right].~~~~~~~~~ 
\eea
The limiting results for super-Hubble,  the sub-Hubble and horizon exit scale are given by:
\begin{eqnarray} 
&&\underline{\textcolor{red}{\bf I.~Sub-Hubble~limiting~solution:}}\nonumber\\
&&\underline{\textcolor{blue}{\bf A.~Massless~ \& ~Partially ~Massless~Hubble~Effective~Mass:\Longrightarrow}}\nonumber\\
&& f_{\lambda,{\bf k}}(-k\tau\gg 1)=\frac{1}{\sqrt{2k}}2^{\nu_{\rm PGW}-\frac{3}{2}}(-k\tau)^{\frac{3}{2}-\nu_{\rm PGW}}\left|\frac{\Gamma(\nu_{\rm PGW})}{\Gamma\left(\frac{3}{2}\right)}\right|\nonumber\\
&&~~~~~~~~~~~~~~~~~~~\times\left[\cosh\alpha~\exp\left(-i\left\{k\tau+\frac{\pi}{2}\left(\nu_{\rm PGW}-\frac{3}{2}\right)\right\}\right)\right.\nonumber\\
&& \left.~~~~~~~~~~~~~~~~~~~~~~~~~~~~~~+\sinh\alpha~\exp\left(i\left\{\gamma+k\tau+\frac{\pi}{2}\left(\nu_{\rm PGW}-\frac{3}{2}\right)\right\}\right)\right].~~~~~~~~~~~ 
\\
&&\underline{\textcolor{blue}{\bf B.~Heavy~Hubble~Effective~Mass:\Longrightarrow}}\nonumber\\
&&~~~~f_{\lambda,{\bf k}}(-k\tau\gg 1)=\frac{2^{-\left(i|\nu_{\rm PGW}|+\frac{3}{2}\right)}(-k\tau)^{\frac{3}{2}+i|\nu_{\rm PGW}|}}{\sqrt{2k}}\left|\frac{\Gamma(-i|\nu_{\rm PGW}|)}{\Gamma\left(\frac{3}{2}\right)}\right|\nonumber\\
&&~~~~~~~~~~~~~~~~~~~\times\left[\cosh\alpha~\underbrace{\exp\left(-\frac{\pi}{2}|\nu_{\rm PGW}|\right)}_{\textcolor{red}{\bf Boltzmann ~suppression}}\exp\left(-i\left\{k\tau-\frac{3\pi}{4}\right\}\right)\right.\nonumber\\
&& \left.~~~~~~~~~~~~~~~~~+\sinh\alpha~\underbrace{\exp\left(\frac{\pi}{2}|\nu_{\rm PGW}|\right)}_{\textcolor{red}{\bf Boltzmann ~enhancement}}\exp\left(i\left\{\gamma+k\tau-\frac{3\pi}{4}\right\}\right)\right].~~~~~~~~~
\\ \\
&&\underline{\textcolor{red}{\bf II.~Super-Hubble~limiting~solution:}}\nonumber\\
&&\underline{\textcolor{blue}{\bf A.~Massless~ \& ~Partially ~Massless~Hubble~Effective~Mass:\Longrightarrow}}\nonumber\\
&&~~~~f_{\lambda,{\bf k}}(-k\tau \ll 1)=\frac{2^{\nu_{\rm PGW}-\frac{3}{2}}(-k\tau)^{\frac{1}{2}-\nu_{\rm PGW}}}{\sqrt{2k}}\left|\frac{\Gamma(\nu_{\rm PGW})}{\Gamma\left(\frac{3}{2}\right)}\right|~\nonumber\\
&&~~~~~~~~~~~~~~~~~~~~~~~~~~~~~~~~~~\times\left[\cosh\alpha~\exp\left(-i\left\{\frac{\pi}{2}\left(\nu_{\rm PGW}-\frac{5}{2}\right)\right\}\right)\right.\nonumber\\
&&\left.~~~~~~~~~~~~~~~~~~~~~~~~~~~~~~~~~~~~~~~~~~~~~~~+\sinh\alpha~\exp\left(i\left\{\gamma+\frac{\pi}{2}\left(\nu_{\rm PGW}-\frac{5}{2}\right)\right\}\right)\right].\nonumber\\
&&\\
&&\underline{\textcolor{blue}{\bf B.~Heavy~Hubble~Effective~Mass:\Longrightarrow}}\nonumber\\ 
&&~~~~f_{\lambda,{\bf k}}(-k\tau \ll 1)=\frac{2^{-\left(i|\nu_{\rm PGW}|+\frac{3}{2}\right)}(-k\tau)^{\frac{1}{2}+i|\nu_{\rm PGW}|}}{\sqrt{2k}}\left|\frac{\Gamma(-i|\nu_{\rm PGW}|)}{\Gamma\left(\frac{3}{2}\right)}\right|~\nonumber\\
&&~~~~~~~~~~~~~~~~~~~~~~~~~~~~~~~~~~\times\left[\cosh\alpha~~\underbrace{\exp\left(-\frac{\pi}{2}|\nu_{\rm PGW}|\right)}_{\textcolor{red}{\bf Boltzmann~suppression}}~\exp\left(\frac{5\pi i}{4}\right)\right.\nonumber\\
&&\left.~~~~~~~~~~~~~~~~~~~~~~~~~~~~~~~~~~~~~~~~~~~~+\sinh\alpha~\underbrace{\exp\left(\frac{\pi}{2}|\nu_{\rm PGW}|\right)}_{\textcolor{red}{\bf Boltzmann~enhancement}}~\exp\left(i\left\{\gamma-\frac{5\pi}{4}\right\}\right)\right].\\
&&\underline{\textcolor{red}{\bf III.~Solution~at~the~point~where~boundary~condition~is~ fixed:}}\nonumber\\
&&\underline{\textcolor{blue}{\bf A.~Massless~ \& ~Partially ~Massless~Hubble~Effective~Mass:\Longrightarrow}}\nonumber\\
&&~~~~f_{\lambda,{\bf k}}(-k\tau_0=1)=\frac{2^{\nu_{\rm PGW}-1}}{\sqrt{2k}}\left|\frac{\Gamma(\nu_{\rm PGW})}{\Gamma\left(\frac{3}{2}\right)}\right|~\nonumber\\
&&~~~~~~~~~~~~~~~~~~~~~~~~~~~~\left[\cosh\alpha~\exp\left(-i\left\{\frac{\pi}{2}\left(\nu_{\rm PGW}-2\right)-1\right\}\right)\right.\nonumber\\
&&\left.~~~~~~~~~~~~~~~~~~~~~~~~~~~~~~~~~~~~~~+\sinh\alpha~\exp\left(i\left\{\gamma+\frac{\pi}{2}\left(\nu_{\rm PGW}-2\right)-1\right\}\right)\right].~~~~~~~~~~~ 
\\
&&\underline{\textcolor{blue}{\bf B.~Heavy~Hubble~Effective~Mass:\Longrightarrow}}\nonumber\\
&&~~~~f_{\lambda,{\bf k}}(-k\tau_0=1)=\frac{2^{-(i|\nu_{\rm PGW}|+1)}}{\sqrt{2k}}\left|\frac{\Gamma(-i|\nu_{\rm PGW}|)}{\Gamma\left(\frac{3}{2}\right)}\right|\left[\cosh\alpha~\underbrace{\exp\left(-\frac{\pi}{2}|\nu_{\rm PGW}|\right)}_{\textcolor{red}{\bf Boltzmann~suppression}}~\exp\left(i(\pi+1)\right)\right.\nonumber\\
&&\left.~~~~~~~~~~~~~~~~~~~~~~~~~~~~+\sinh\alpha~\underbrace{\exp\left(\frac{\pi}{2}|\nu_{\rm PGW}|\right)}_{\textcolor{red}{\bf Boltzmann~enhancement}}~\exp\left(i\left(\gamma-(\pi+1)\right)\right)\right].~~~~~~~~~~~ 
\eea
\\
\item \underline{\textcolor{red}{\bf $\alpha$~vacua:}}\\
This is a specific choice where the arbitrary coefficients that are appearing in the classical solution of the PGW are parametrized by one real parameter family $\alpha$, by:
\bea {\cal C}_1=\cosh\alpha,~~{\cal C}_2=\sinh\alpha~~~\forall \alpha\in \mathbb {R}~~~\textcolor{blue}{\bf subject~to}~~|{\cal C}_1|^2-|{\cal C}_2|^2=1.~~~~~~~\eea
This is identified as a excited $SO(1,4)$ isommetric vacuum state which is CPT symmetry preserving in nature.  The real parameter $\alpha$ of $\alpha$ {\it vacua} determine the quantum complexity and to study the underlying phenomena of quantum chaos from the various cosmological primordial models of our universe using PGW perturbation. Like the previous case this issue was not have been studied before and for this reason we are very hopeful to explore some interesting underlying physical facts from our analysis. It is important to note that by fixing $\gamma=0$ in the $(\alpha,\gamma)$ {\it vacua} or {\it Motta-Allen vacua} one can obtain the results for the $\alpha$ {\it vacua}, which in tern implies that by fixing this choice of the parameter $\gamma$ one can able to transform a CPT violating vacua to a CPT preserving vacua. In this context, we get:
\bea && f_{\lambda,{\bf k}}(\tau)=\frac{1}{\sqrt{2k}}2^{\nu_{\rm PGW}-\frac{3}{2}}(-k\tau)^{\frac{3}{2}-\nu_{\rm PGW}}\left|\frac{\Gamma(\nu_{\rm PGW})}{\Gamma\left(\frac{3}{2}\right)}\right|\nonumber\\
&&~~~~~~~~~~~~~~~~~~~\times\left[\cosh\alpha~\left(1-\frac{i}{k\tau}\right)~\exp\left(-i\left\{k\tau+\frac{\pi}{2}\left(\nu_{\rm PGW}-\frac{3}{2}\right)\right\}\right)\right.\nonumber\\
&& \left.~~~~~~~~~~~~~~~~~~~~~~~~~~~~~~+\sinh\alpha~\left(1+\frac{i}{k\tau}\right)~\exp\left(i\left\{k\tau+\frac{\pi}{2}\left(\nu_{\rm PGW}-\frac{3}{2}\right)\right\}\right)\right].~~~~~~~~~
\eea
For the case of heavy field case the PGW field can be expressed as:
\bea&& f_{\lambda,{\bf k}}(\tau)=\frac{1}{\sqrt{2k}}2^{-\left(i|\nu_{\rm PGW}|+\frac{3}{2}\right)}(-k\tau)^{\frac{3}{2}+i|\nu_{\rm PGW}|}\left|\frac{\Gamma(-i|\nu_{\rm PGW}|)}{\Gamma\left(\frac{3}{2}\right)}\right|\nonumber\\
&&~~~~~~~~~~~~~~~~~~~\times\left[\cosh\alpha~\left(1-\frac{i}{k\tau}\right)~\underbrace{\exp\left(-\frac{\pi}{2}|\nu_{\rm PGW}|\right)}_{\textcolor{red}{\bf Boltzmann ~suppression}}\exp\left(-i\left\{k\tau-\frac{3\pi}{4}\right\}\right)\right.\nonumber\\
&& \left.~~~~~~~~~~~~~~~~~~~~~~~~~~~~~~+\sinh\alpha~\left(1+\frac{i}{k\tau}\right)~\underbrace{\exp\left(\frac{\pi}{2}|\nu_{\rm PGW}|\right)}_{\textcolor{red}{\bf Boltzmann ~enhancement}}\exp\left(i\left\{k\tau-\frac{3\pi}{4}\right\}\right)\right].~~~~~~~~~
\eea
The limiting results for super-Hubble,  the sub-Hubble and horizon exit scale are given by:
\begin{eqnarray} 
&&\underline{\textcolor{red}{\bf I.~Sub-Hubble~limiting~solution:}}\nonumber\\
&&\underline{\textcolor{blue}{\bf A.~Massless~ \& ~Partially ~Massless~Hubble~Effective~Mass:\Longrightarrow}}\nonumber\\
&& f_{\lambda,{\bf k}}(-k\tau\gg 1)=\frac{1}{\sqrt{2k}}2^{\nu_{\rm PGW}-\frac{3}{2}}(-k\tau)^{\frac{3}{2}-\nu_{\rm PGW}}\left|\frac{\Gamma(\nu_{\rm PGW})}{\Gamma\left(\frac{3}{2}\right)}\right|\nonumber\\
&&~~~~~~~~~~~~~~~~~~~\times\left[\cosh\alpha~\exp\left(-i\left\{k\tau+\frac{\pi}{2}\left(\nu_{\rm PGW}-\frac{3}{2}\right)\right\}\right)\right.\nonumber\\
&& \left.~~~~~~~~~~~~~~~~~~~~~~~~~~~~~~+\sinh\alpha~\exp\left(i\left\{k\tau+\frac{\pi}{2}\left(\nu_{\rm PGW}-\frac{3}{2}\right)\right\}\right)\right].~~~~~~~~~~~ 
\\
&&\underline{\textcolor{blue}{\bf B.~Heavy~Hubble~Effective~Mass:\Longrightarrow}}\nonumber\\
&&~~~~f_{\lambda,{\bf k}}(-k\tau\gg 1)=\frac{2^{-\left(i|\nu_{\rm PGW}|+\frac{3}{2}\right)}(-k\tau)^{\frac{3}{2}+i|\nu_{\rm PGW}|}}{\sqrt{2k}}\left|\frac{\Gamma(-i|\nu_{\rm PGW}|)}{\Gamma\left(\frac{3}{2}\right)}\right|\nonumber\\
&&~~~~~~~~~~~~~~~~~~~\times\left[\cosh\alpha~\underbrace{\exp\left(-\frac{\pi}{2}|\nu_{\rm PGW}|\right)}_{\textcolor{red}{\bf Boltzmann ~suppression}}\exp\left(-i\left\{k\tau-\frac{3\pi}{4}\right\}\right)\right.\nonumber\\
&& \left.~~~~~~~~~~~~~~~~~+\sinh\alpha~\underbrace{\exp\left(\frac{\pi}{2}|\nu_{\rm PGW}|\right)}_{\textcolor{red}{\bf Boltzmann ~enhancement}}\exp\left(i\left\{k\tau-\frac{3\pi}{4}\right\}\right)\right].~~~~~~~~~
\\
&&\underline{\textcolor{red}{\bf II.~Super-Hubble~limiting~solution:}}\nonumber\\
&&\underline{\textcolor{blue}{\bf A.~Massless~ \& ~Partially ~Massless~Hubble~Effective~Mass:\Longrightarrow}}\nonumber\\
&&~~~~f_{\lambda,{\bf k}}(-k\tau \ll 1)=\frac{2^{\nu_{\rm PGW}-\frac{3}{2}}(-k\tau)^{\frac{1}{2}-\nu_{\rm PGW}}}{\sqrt{2k}}\left|\frac{\Gamma(\nu_{\rm PGW})}{\Gamma\left(\frac{3}{2}\right)}\right|~\nonumber\\
&&~~~~~~~~~~~~~~~~~~~~~~~~~~~~~~~~~~\times\left[\cosh\alpha~\exp\left(-i\left\{\frac{\pi}{2}\left(\nu_{\rm PGW}-\frac{5}{2}\right)\right\}\right)\right.\nonumber\\
&&\left.~~~~~~~~~~~~~~~~~~~~~~~~~~~~~~~~~~~~~~~~~~~~~~~+\sinh\alpha~\exp\left(i\left\{\frac{\pi}{2}\left(\nu_{\rm PGW}-\frac{5}{2}\right)\right\}\right)\right].\nonumber\\
&&\\
&&\underline{\textcolor{blue}{\bf B.~Heavy~Hubble~Effective~Mass:\Longrightarrow}}\nonumber\\ 
&&~~~~f_{\lambda,{\bf k}}(-k\tau \ll 1)=\frac{2^{-\left(i|\nu_{\rm PGW}|+\frac{3}{2}\right)}(-k\tau)^{\frac{1}{2}+i|\nu_{\rm PGW}|}}{\sqrt{2k}}\left|\frac{\Gamma(-i|\nu_{\rm PGW}|)}{\Gamma\left(\frac{3}{2}\right)}\right|~\nonumber\\
&&~~~~~~~~~~~~~~~~~~~~~~~~~~~~~~~~~~\times\left[\cosh\alpha~~\underbrace{\exp\left(-\frac{\pi}{2}|\nu_{\rm PGW}|\right)}_{\textcolor{red}{\bf Boltzmann~suppression}}~\exp\left(\frac{5\pi i}{4}\right)\right.\nonumber\\
&&\left.~~~~~~~~~~~~~~~~~~~~~~~~~~~~~~~~~~~~~~~~~~~~+\sinh\alpha~\underbrace{\exp\left(\frac{\pi}{2}|\nu_{\rm PGW}|\right)}_{\textcolor{red}{\bf Boltzmann~enhancement}}~\exp\left(-\frac{5\pi i}{4}\right)\right].\\
&&\underline{\textcolor{red}{\bf III.~Solution~at~the~point~where~boundary~condition~is~ fixed:}}\nonumber\\
&&\underline{\textcolor{blue}{\bf A.~Massless~ \& ~Partially ~Massless~Hubble~Effective~Mass:\Longrightarrow}}\nonumber\\
&&~~~~f_{\lambda,{\bf k}}(-k\tau_0=1)=\frac{2^{\nu_{\rm PGW}-1}}{\sqrt{2k}}\left|\frac{\Gamma(\nu_{\rm PGW})}{\Gamma\left(\frac{3}{2}\right)}\right|~\nonumber\\
&&~~~~~~~~~~~~~~~~~~~~~~~~~~~~\left[\cosh\alpha~\exp\left(-i\left\{\frac{\pi}{2}\left(\nu_{\rm PGW}-2\right)-1\right\}\right)\right.\nonumber\\
&&\left.~~~~~~~~~~~~~~~~~~~~~~~~~~~~~~~~~~~~~~+\sinh\alpha~\exp\left(i\left\{\frac{\pi}{2}\left(\nu_{\rm PGW}-2\right)-1\right\}\right)\right].~~~~~~~~~~~ 
\\
&&\underline{\textcolor{blue}{\bf B.~Heavy~Hubble~Effective~Mass:\Longrightarrow}}\nonumber\\
&&~~~~f_{\lambda,{\bf k}}(-k\tau_0=1)=\frac{2^{-(i|\nu_{\rm PGW}|+1)}}{\sqrt{2k}}\left|\frac{\Gamma(-i|\nu_{\rm PGW}|)}{\Gamma\left(\frac{3}{2}\right)}\right|\left[\cosh\alpha~\underbrace{\exp\left(-\frac{\pi}{2}|\nu_{\rm PGW}|\right)}_{\textcolor{red}{\bf Boltzmann~suppression}}~\exp\left(i(\pi+1)\right)\right.\nonumber\\
&&\left.~~~~~~~~~~~~~~~~~~~~~~~~~~~~+\sinh\alpha~\underbrace{\exp\left(\frac{\pi}{2}|\nu_{\rm PGW}|\right)}_{\textcolor{red}{\bf Boltzmann~enhancement}}~\exp\left(-i\left(\pi+1\right)\right)\right].~~~~~~~~~~~ 
\eea
\item \underline{\textcolor{red}{\bf Bunch-Davies~vacuum:}}\\
This is a specific choice where the arbitrary coefficients that are appearing in the classical solution of the PGW are parametrized by the following expression:
\bea {\cal C}_1=1,~~{\cal C}_2=0~~~\textcolor{blue}{\bf subject~to}~~|{\cal C}_1|^2-|{\cal C}_2|^2=1,~~~~~~~\eea
which is a $SO(1,4)$ isommetric ground state of the initial vacuum used in primordial cosmology.  In literature {\it Bunch-Davies} vacuum state is identified as {\it Cherenkov vacuum} or {\it Hartle-Hawking vacuum} state, which is actually an {\it Euclidean} state.  The general solution for PGW field variable for the Bunch-Davies vacuum is given by:
\bea && f_{\lambda,{\bf k}}(\tau)=\frac{1}{\sqrt{2k}}2^{\nu_{\rm PGW}-\frac{3}{2}}(-k\tau)^{\frac{3}{2}-\nu_{\rm PGW}}\left|\frac{\Gamma(\nu_{\rm PGW})}{\Gamma\left(\frac{3}{2}\right)}\right|\nonumber\\
&&~~~~~~~~~~~~~~~~~~~\times\left(1-\frac{i}{k\tau}\right)~\exp\left(-i\left\{k\tau+\frac{\pi}{2}\left(\nu_{\rm PGW}-\frac{3}{2}\right)\right\}\right).~~~~~~~~~
\eea
For the case of heavy field case the solution for the PGW field for the Bunch-Davies vacuum can be written as:
\bea&& f_{\lambda,{\bf k}}(\tau)=\frac{1}{\sqrt{2k}}2^{-\left(i|\nu_{\rm PGW}|+\frac{3}{2}\right)}(-k\tau)^{\frac{3}{2}+i|\nu_{\rm PGW}|}\left|\frac{\Gamma(-i|\nu_{\rm PGW}|)}{\Gamma\left(\frac{3}{2}\right)}\right|\nonumber\\
&&~~~~~~~~~~~~~~~~~~~\times\left(1-\frac{i}{k\tau}\right)~\underbrace{\exp\left(-\frac{\pi}{2}|\nu_{\rm PGW}|\right)}_{\textcolor{red}{\bf Boltzmann ~suppression}}\exp\left(-i\left\{k\tau-\frac{3\pi}{4}\right\}\right).~~~~~~~~~
\eea
The limiting results for super-Hubble,  the sub-Hubble and horizon exit scale are given by:
\begin{eqnarray} 
&&\underline{\textcolor{red}{\bf I.~Sub-Hubble~limiting~solution:}}\nonumber\\
&&\underline{\textcolor{blue}{\bf A.~Massless~ \& ~Partially ~Massless~Hubble~Effective~Mass:\Longrightarrow}}\nonumber\\
&&~~~~f_{\lambda,{\bf k}}(-k\tau\gg 1)=\frac{2^{\nu_{\rm PGW}-\frac{3}{2}}(-k\tau)^{\frac{3}{2}-\nu_{\rm PGW}}}{\sqrt{2k}}\left|\frac{\Gamma(\nu_{\rm PGW})}{\Gamma\left(\frac{3}{2}\right)}\right|~\nonumber\\
&&~~~~~~~~~~~~~~~~~~~~~~~~~~~~~~~~~~~~~~~~~~~~~~~~~~\times\exp\left(-i\left\{k\tau+\frac{\pi}{2}\left(\nu_{\rm PGW}-\frac{3}{2}\right)\right\}\right).~~~~~\nonumber\\
&&
\\
&&\underline{\textcolor{blue}{\bf B.~Heavy~Hubble~Effective~Mass:\Longrightarrow}}\nonumber\\
&&~~~~f_{\lambda,{\bf k}}(-k\tau\gg 1)=\frac{2^{-\left(i|\nu_{\rm PGW}|+\frac{3}{2}\right)}(-k\tau)^{\frac{3}{2}+i|\nu_{\rm PGW}|}}{\sqrt{2k}}\left|\frac{\Gamma(-i|\nu_{\rm PGW}|)}{\Gamma\left(\frac{3}{2}\right)}\right|~\nonumber\\
&&~~~~~~~~~~~~~~~~~~~~~~~~~~~~~~~~~~~~~~~~~~~~~~~~~~\times\underbrace{\exp\left(-\frac{\pi}{2}|\nu_{\rm PGW}|\right)}_{\textcolor{red}{\bf Boltzmann~suppression}}~\exp\left(-i\left\{k\tau-\frac{3\pi}{4}\right\}\right).~~~~~~~~~
\\
&&\underline{\textcolor{red}{\bf II.~Super-Hubble~limiting~solution:}}\nonumber\\
&&\underline{\textcolor{blue}{\bf A.~Massless~ \& ~Partially ~Massless~Hubble~Effective~Mass:\Longrightarrow}}\nonumber\\
&&~~~~f_{\lambda,{\bf k}}(-k\tau \ll 1)=\frac{2^{\nu_{\rm PGW}-\frac{3}{2}}(-k\tau)^{\frac{1}{2}-\nu_{\rm PGW}}}{\sqrt{2k}}\left|\frac{\Gamma(\nu_{\rm PGW})}{\Gamma\left(\frac{3}{2}\right)}\right|~\nonumber\\
&&~~~~~~~~~~~~~~~~~~~~~~~~~~~~~~~~~~~~~~~~~~~~~~~~~~\times\exp\left(-i\left\{\frac{\pi}{2}\left(\nu_{\rm PGW}-\frac{5}{2}\right)\right\}\right).~~~~~~
\\
&&\underline{\textcolor{blue}{\bf B.~Heavy~Hubble~Effective~Mass:\Longrightarrow}}\nonumber\\ 
&&~~~~f_{\lambda,{\bf k}}(-k\tau \ll 1)=\frac{2^{-\left(i|\nu_{\rm PGW}|+\frac{3}{2}\right)}(-k\tau)^{\frac{1}{2}+i|\nu_{\rm PGW}|}}{\sqrt{2k}}\left|\frac{\Gamma(-i|\nu_{\rm PGW}|)}{\Gamma\left(\frac{3}{2}\right)}\right|~\nonumber\\
&&~~~~~~~~~~~~~~~~~~~~~~~~~~~~~~~~~~~~~~~~~~~~~~~~~~\times~\underbrace{\exp\left(-\frac{\pi}{2}|\nu_{\rm PGW}|\right)}_{\textcolor{red}{\bf Boltzmann~suppression}}~\exp\left(\frac{5\pi i}{4}\right).~~~~~~~ 
\\
&&\underline{\textcolor{red}{\bf III.~Solution~at~the~point~where~boundary~condition~is~ fixed:}}\nonumber\\
&&\underline{\textcolor{blue}{\bf A.~Massless~ \& ~Partially ~Massless~Hubble~Effective~Mass:\Longrightarrow}}\nonumber\\
&&~~~~f_{\lambda,{\bf k}}(-k\tau_0=1)=\frac{2^{\nu_{\rm PGW}-1}}{\sqrt{2k}}\left|\frac{\Gamma(\nu_{\rm PGW})}{\Gamma\left(\frac{3}{2}\right)}\right|~\exp\left(-i\left\{\frac{\pi}{2}\left(\nu_{\rm PGW}-2\right)-1\right\}\right).~~~~~~ 
\\
&&\underline{\textcolor{blue}{\bf B.~Heavy~Hubble~Effective~Mass:\Longrightarrow}}\nonumber\\
&&~~~~f_{\lambda,{\bf k}}(-k\tau_0=1)=\frac{2^{-(i|\nu_{\rm PGW}|+1)}}{\sqrt{2k}}\left|\frac{\Gamma(-i|\nu_{\rm PGW}|)}{\Gamma\left(\frac{3}{2}\right)}\right|\nonumber\\
&&~~~~~~~~~~~~~~~~~~~~~~~~~~~~~~~~~~\times~\underbrace{\exp\left(-\frac{\pi}{2}|\nu_{\rm PGW}|\right)}_{\textcolor{red}{\bf Boltzmann~suppression}}~\exp\left(i(\pi+1)\right).~~~~~
\end{eqnarray}
\end{enumerate}

\subsection{Quantization of Hamiltonian for PGW}
In this section we quantize the Hamiltonian for PGW.  Using the sets of solutions that we have obtained for different choice of the initial vacuum state,  one can further compute the expression for the time derivative of PGW tensor mode:
\bea
 &&\nonumber f'_{\lambda,{\bf k}}(\tau)=i\sqrt{\frac{k}{2}}~2^{\nu_{\rm PGW}-\frac{3}{2}}(-k\tau)^{\frac{3}{2}-\nu_{\rm PGW}}\left|\frac{\Gamma(\nu_{\rm PGW})}{\Gamma\left(\frac{3}{2}\right)}\right|~\nonumber\\
&&~~~~~~~~~~~\times \left[{\cal C}_1~\left\{1-\left(\nu_{\rm PGW}-\frac{1}{2}\right)\frac{i}{k\tau}\left(1-\frac{i}{k\tau}\right)\right\}\exp\left(-i\left\{k\tau+\frac{\pi}{2}\left(\nu_{\rm PGW}-\frac{1}{2}\right)\right\}\right)\right.\nonumber\\
 &&\left.~~~~~~~~~~~~~-{\cal C}_2~\left\{1+\left(\nu_{\rm PGW}-\frac{1}{2}\right)\frac{i}{k\tau}\left(1+\frac{i}{k\tau}\right)\right\}\exp\left(i\left\{k\tau+\frac{\pi}{2}\left(\nu_{\rm PGW}-\frac{1}{2}\right)\right\}\right)\right].~~~~~~~~~~~\eea 
 
In the specific case of heavy field case the conformal time derivative of the PGW tensor mode can be written as:
\bea
 &&\nonumber f'_{\lambda,{\bf k}}(\tau)=i\sqrt{\frac{k}{2}}~2^{-\left(i|\nu_{\rm PGW}|+\frac{3}{2}\right)}(-k\tau)^{\frac{3}{2}+i|\nu_{\rm PGW}|}\left|\frac{\Gamma(-i|\nu_{\rm PGW}|)}{\Gamma\left(\frac{3}{2}\right)}\right|~\nonumber\\
&&~~~~~~~~~~~\times \left[{\cal C}_1~\left\{1+\left(i|\nu_{\rm PGW}|+\frac{1}{2}\right)\frac{i}{k\tau}\left(1-\frac{i}{k\tau}\right)\right\}\right.\nonumber\\
&&\left.~~~~~~~~~~~~~~~~~~~~~~~~~~~~~~~~\times\underbrace{\exp\left(-\frac{\pi}{2}|\nu_{\rm PGW}|\right)}_{\textcolor{red}{\bf Boltzmann~suppression}}\exp\left(-i\left\{k\tau+\frac{\pi}{4}\right\}\right)\right.\nonumber\\
 &&\left.~~~~~~~~~~~~~~~~~~~-{\cal C}_2~\left\{1-\left(i|\nu_{\rm PGW}|+\frac{1}{2}\right)\frac{i}{k\tau}\left(1+\frac{i}{k\tau}\right)\right\}\right.\nonumber\\
&&\left.~~~~~~~~~~~~~~~~~~~~~~~~~~~~~~~~~~~~~~\times\underbrace{\exp\left(\frac{\pi}{2}|\nu_{\rm PGW}|\right)}_{\textcolor{red}{\bf Boltzmann~enhancement}}\exp\left(i\left\{k\tau+\frac{\pi}{4}\right\}\right)\right].~~~~~~~~~~~\eea 

We further discuss about three specific choices of quantum vacuum states for which the expressions for the PGW tensor perturbed field velocities are simplified as appended below:
\begin{enumerate}
\item \underline{\textcolor{red}{\bf Motta-Allen ($\alpha,\gamma$)~vacua:}}\\
In presence of {\it Motta-Allen vacua } the PGW tensor mode velocity can be written as:
\bea
 &&\nonumber f'_{\lambda,{\bf k}}(\tau)=i\sqrt{\frac{k}{2}}~2^{\nu_{\rm PGW}-\frac{3}{2}}(-k\tau)^{\frac{3}{2}-\nu_{\rm PGW}}\left|\frac{\Gamma(\nu_{\rm PGW})}{\Gamma\left(\frac{3}{2}\right)}\right|~\nonumber\\
&&~~~~~~~~~~~\times \left[\cosh\alpha~\left\{1-\left(\nu_{\rm PGW}-\frac{1}{2}\right)\frac{i}{k\tau}\left(1-\frac{i}{k\tau}\right)\right\}\exp\left(-i\left\{k\tau+\frac{\pi}{2}\left(\nu_{\rm PGW}-\frac{1}{2}\right)\right\}\right)\right.\nonumber\\
 &&\left.~~~~~~~~~~~~~-\sinh\alpha~\left\{1+\left(\nu_{\rm PGW}-\frac{1}{2}\right)\frac{i}{k\tau}\left(1+\frac{i}{k\tau}\right)\right\}\exp\left(i\left\{\gamma+k\tau+\frac{\pi}{2}\left(\nu_{\rm PGW}-\frac{1}{2}\right)\right\}\right)\right].\nonumber\\ 
 &&\eea
 Now for the case of heavy Hubble effective mass the PGW tensor perturbed field velocity can be expressed as:
 \bea
 &&\nonumber f'_{\lambda,{\bf k}}(\tau)=i\sqrt{\frac{k}{2}}~2^{-\left(i|\nu_{\rm PGW}|+\frac{3}{2}\right)}(-k\tau)^{\frac{3}{2}+i|\nu_{\rm PGW}|}\left|\frac{\Gamma(-i|\nu_{\rm PGW}|)}{\Gamma\left(\frac{3}{2}\right)}\right|~\nonumber\\
&&~~~~~~~~~~~\times \left[\cosh\alpha~\left\{1+\left(i|\nu_{\rm PGW}|+\frac{1}{2}\right)\frac{i}{k\tau}\left(1-\frac{i}{k\tau}\right)\right\}\right.\nonumber\\
&&\left.~~~~~~~~~~~~~~~~~~~~~~~~~~~~~~~~\times\underbrace{\exp\left(-\frac{\pi}{2}|\nu_{\rm PGW}|\right)}_{\textcolor{red}{\bf Boltzmann~suppression}}\exp\left(-i\left\{k\tau-\frac{\pi}{4}\right\}\right)\right.\nonumber\\
 &&\left.~~~~~~~~~~~~~~~~~~~-\sinh\alpha~\left\{1-\left(i|\nu_{\rm PGW}|+\frac{1}{2}\right)\frac{i}{k\tau}\left(1+\frac{i}{k\tau}\right)\right\}\right.\nonumber\\
&&\left.~~~~~~~~~~~~~~~~~~~~~~~~~~~~~~~~~~~~~~\times\underbrace{\exp\left(\frac{\pi}{2}|\nu_{\rm PGW}|\right)}_{\textcolor{red}{\bf Boltzmann~enhancement}}\exp\left(i\left\{\gamma+k\tau-\frac{\pi}{4}\right\}\right)\right].~~~~~~~~~~~\eea

In the super-Hubble,  horizon crossing and sub-Hubble limit we have the following results:
\begin{eqnarray} 
&&\underline{\textcolor{red}{\bf I.~Sub-Hubble~limiting~solution:}}\nonumber\\
&&\underline{\textcolor{blue}{\bf A.~Massless~ \& ~Partially ~Massless~Hubble~Effective~Mass:\Longrightarrow}}\nonumber\\
 &&\nonumber f'_{\lambda,{\bf k}}(-k\tau\gg 1)=i\sqrt{\frac{k}{2}}~2^{\nu_{\rm PGW}-\frac{3}{2}}(-k\tau)^{\frac{3}{2}-\nu_{\rm PGW}}\left|\frac{\Gamma(\nu_{\rm PGW})}{\Gamma\left(\frac{3}{2}\right)}\right|~\nonumber\\
&&~~~~~~~~~~~\times \left[\cosh\alpha~\left\{1-\left(\nu_{\rm PGW}-\frac{1}{2}\right)\frac{i}{k\tau}\right\}\exp\left(-i\left\{k\tau+\frac{\pi}{2}\left(\nu_{\rm PGW}-\frac{1}{2}\right)\right\}\right)\right.\nonumber\\
 &&\left.~~~~~~~~~~~~~-\sinh\alpha~\left\{1+\left(\nu_{\rm PGW}-\frac{1}{2}\right)\frac{i}{k\tau}\right\}\exp\left(i\left\{\gamma+k\tau+\frac{\pi}{2}\left(\nu_{\rm PGW}-\frac{1}{2}\right)\right\}\right)\right].\nonumber\\
\\
&&\underline{\textcolor{blue}{\bf B.~Heavy~Hubble~Effective~Mass:\Longrightarrow}}\nonumber\\
&&\nonumber f'_{\lambda,{\bf k}}(\tau)=i\sqrt{\frac{k}{2}}~2^{-\left(i|\nu_{\rm PGW}|+\frac{3}{2}\right)}(-k\tau)^{\frac{3}{2}+i|\nu_{\rm PGW}|}\left|\frac{\Gamma(-i|\nu_{\rm PGW}|)}{\Gamma\left(\frac{3}{2}\right)}\right|~\nonumber\\
&&~~~~~~~~~~~\times \left[\cosh\alpha~\left\{1+\left(i|\nu_{\rm PGW}|+\frac{1}{2}\right)\frac{i}{k\tau}\right\}\right.\nonumber\\
&&\left.~~~~~~~~~~~~~~~~~~~~~~~~~~~~~~~~\times\underbrace{\exp\left(-\frac{\pi}{2}|\nu_{\rm PGW}|\right)}_{\textcolor{red}{\bf Boltzmann~suppression}}\exp\left(-i\left\{k\tau-\frac{\pi}{4}\right\}\right)\right.\nonumber\\
 &&\left.~~~~~~~~~~~~~~~~~~~-\sinh\alpha~\left\{1-\left(i|\nu_{\rm PGW}|+\frac{1}{2}\right)\frac{i}{k\tau}\right\}\right.\nonumber\\
&&\left.~~~~~~~~~~~~~~~~~~~~~~~~~~~~~~~~~~~~~~\times\underbrace{\exp\left(\frac{\pi}{2}|\nu_{\rm PGW}|\right)}_{\textcolor{red}{\bf Boltzmann~enhancement}}\exp\left(i\left\{\gamma+k\tau-\frac{\pi}{4}\right\}\right)\right].~~~~~~~~~~~
\\
&&\underline{\textcolor{red}{\bf II.~Super-Hubble~limiting~solution:}}\nonumber\\
&&\underline{\textcolor{blue}{\bf A.~Massless~ \& ~Partially ~Massless~Hubble~Effective~Mass:\Longrightarrow}}\nonumber\\
&&\nonumber f'_{\lambda,{\bf k}}(-k\tau\ll 1)=i\sqrt{\frac{k}{2}}~2^{\nu_{\rm PGW}-\frac{3}{2}}(-k\tau)^{-\left(\nu_{\rm PGW}+\frac{1}{2}\right)}\left|\frac{\Gamma(\nu_{\rm PGW})}{\Gamma\left(\frac{3}{2}\right)}\right|~\nonumber\\
&&~~~~~~~~~~~\times \left[\cosh\alpha~\left(\nu_{\rm PGW}-\frac{1}{2}\right)\exp\left(-i\left\{\frac{\pi}{2}\left(\nu_{\rm PGW}-\frac{5}{2}\right)\right\}\right)\right.\nonumber\\
 &&\left.~~~~~~~~~~~~~-\sinh\alpha~\left(\nu_{\rm PGW}-\frac{1}{2}\right)\exp\left(i\left\{\gamma+\frac{\pi}{2}\left(\nu_{\rm PGW}-\frac{5}{2}\right)\right\}\right)\right].
\\
&&\underline{\textcolor{blue}{\bf B.~Heavy~Hubble~Effective~Mass:\Longrightarrow}}\nonumber\\ 
&&\nonumber f'_{\lambda,{\bf k}}(-k\tau\ll 1)=i\sqrt{\frac{k}{2}}~2^{-\left(i|\nu_{\rm PGW}|+\frac{3}{2}\right)}(-k\tau)^{\frac{3}{2}+i|\nu_{\rm PGW}|}\left|\frac{\Gamma(-i|\nu_{\rm PGW}|)}{\Gamma\left(\frac{3}{2}\right)}\right|~\nonumber\\
&&~~~~~~~~~~~\times \left[\cosh\alpha~\left(i|\nu_{\rm PGW}|+\frac{1}{2}\right)\underbrace{\exp\left(-\frac{\pi}{2}|\nu_{\rm PGW}|\right)}_{\textcolor{red}{\bf Boltzmann~suppression}}\exp\left(\frac{i\pi}{4}\right)\right.\nonumber\\
 &&\left.~~~~~~~~~~~~~~~~~~~-\sinh\alpha\left(i|\nu_{\rm PGW}|+\frac{1}{2}\right)\underbrace{\exp\left(\frac{\pi}{2}|\nu_{\rm PGW}|\right)}_{\textcolor{red}{\bf Boltzmann~enhancement}}\exp\left(i\left\{\gamma-\frac{\pi}{4}\right\}\right)\right].~~\\
&&\underline{\textcolor{red}{\bf III.~Solution~at~the~point~where~boundary~condition~is~ fixed:}}\nonumber\\
&&\underline{\textcolor{blue}{\bf A.~Massless~ \& ~Partially ~Massless~Hubble~Effective~Mass:\Longrightarrow}}\nonumber\\
&&\nonumber f'_{\lambda,{\bf k}}(-k\tau_0=1)=i\sqrt{\frac{k}{2}}~2^{\nu_{\rm PGW}-\frac{3}{2}}\left|\frac{\Gamma(\nu_{\rm PGW})}{\Gamma\left(\frac{3}{2}\right)}\right|~\nonumber\\
&&~~~~~~\times \left[\cosh\alpha~\left\{1-\sqrt{2}\left(\nu_{\rm PGW}-\frac{1}{2}\right)\exp\left(-\frac{i\pi}{4}\right)\right\}\exp\left(-i\left\{\frac{\pi}{2}\left(\nu_{\rm PGW}-\frac{1}{2}\right)-1\right\}\right)\right.\nonumber\\
 &&\left.~~~~-\sinh\alpha~\left\{1-\sqrt{2}\left(\nu_{\rm PGW}-\frac{1}{2}\right)\exp\left(\frac{i\pi}{4}\right)\right\}\exp\left(i\left\{\gamma-1+\frac{\pi}{2}\left(\nu_{\rm PGW}-\frac{1}{2}\right)\right\}\right)\right].~~~~~~~~~
\\
&&\underline{\textcolor{blue}{\bf B.~Heavy~Hubble~Effective~Mass:\Longrightarrow}}\nonumber\\
 &&\nonumber f'_{\lambda,{\bf k}}(-k\tau_0=1)=i\sqrt{\frac{k}{2}}~2^{-\left(i|\nu_{\rm PGW}|+\frac{3}{2}\right)}\left|\frac{\Gamma(-i|\nu_{\rm PGW}|)}{\Gamma\left(\frac{3}{2}\right)}\right|~\nonumber\\
&&~~~~~~~~~~~\times \left[\cosh\alpha~\left\{1+\sqrt{2}\left(i|\nu_{\rm PGW}|+\frac{1}{2}\right)\exp\left(-\frac{i\pi}{4}\right)\right\}\right.\nonumber\\
&&\left.~~~~~~~~~~~~~~~~~~~~~~~~~~~~~~~~\times\underbrace{\exp\left(-\frac{\pi}{2}|\nu_{\rm PGW}|\right)}_{\textcolor{red}{\bf Boltzmann~suppression}}\exp\left(i\left\{\frac{\pi}{4}+1\right\}\right)\right.\nonumber\\
 &&\left.~~~~~~~~~~~~~~~~~~~-\sinh\alpha~\left\{1+\sqrt{2}\left(i|\nu_{\rm PGW}|+\frac{1}{2}\right)\exp\left(\frac{i\pi}{4}\right)\right\}\right.\nonumber\\
&&\left.~~~~~~~~~~~~~~~~~~~~~~~~~~~~~~~~~~~~~~\times\underbrace{\exp\left(\frac{\pi}{2}|\nu_{\rm PGW}|\right)}_{\textcolor{red}{\bf Boltzmann~enhancement}}\exp\left(i\left\{\gamma-1-\frac{\pi}{4}\right\}\right)\right].~~~~
\end{eqnarray}
\item \underline{\textcolor{red}{\bf $\alpha$~vacua:}}\\
In presence of $\alpha$~vacua the expression for the PGW tensor mode velocity can be simplified as: 
\bea
 &&\nonumber f'_{\lambda,{\bf k}}(\tau)=i\sqrt{\frac{k}{2}}~2^{\nu_{\rm PGW}-\frac{3}{2}}(-k\tau)^{\frac{3}{2}-\nu_{\rm PGW}}\left|\frac{\Gamma(\nu_{\rm PGW})}{\Gamma\left(\frac{3}{2}\right)}\right|~\nonumber\\
&&~~~~~~~~~~~\times \left[\cosh\alpha~\left\{1-\left(\nu_{\rm PGW}-\frac{1}{2}\right)\frac{i}{k\tau}\left(1-\frac{i}{k\tau}\right)\right\}\exp\left(-i\left\{k\tau+\frac{\pi}{2}\left(\nu_{\rm PGW}-\frac{1}{2}\right)\right\}\right)\right.\nonumber\\
 &&\left.~~~~~~~~~~~~~-\sinh\alpha~\left\{1+\left(\nu_{\rm PGW}-\frac{1}{2}\right)\frac{i}{k\tau}\left(1+\frac{i}{k\tau}\right)\right\}\exp\left(i\left\{k\tau+\frac{\pi}{2}\left(\nu_{\rm PGW}-\frac{1}{2}\right)\right\}\right)\right].\nonumber\\
 &&\eea
 Now for the case of heavy field case the PGW tensor mode velocity can be simplified as:
 \bea
 &&\nonumber f'_{\lambda,{\bf k}}(\tau)=i\sqrt{\frac{k}{2}}~2^{-\left(i|\nu_{\rm PGW}|+\frac{3}{2}\right)}(-k\tau)^{\frac{3}{2}+i|\nu_{\rm PGW}|}\left|\frac{\Gamma(-i|\nu_{\rm PGW}|)}{\Gamma\left(\frac{3}{2}\right)}\right|~\nonumber\\
&&~~~~~~~~~~~\times \left[\cosh\alpha~\left\{1+\left(i|\nu_{\rm PGW}|+\frac{1}{2}\right)\frac{i}{k\tau}\left(1-\frac{i}{k\tau}\right)\right\}\right.\nonumber\\
&&\left.~~~~~~~~~~~~~~~~~~~~~~~~~~~~~~~~\times\underbrace{\exp\left(-\frac{\pi}{2}|\nu_{\rm PGW}|\right)}_{\textcolor{red}{\bf Boltzmann~suppression}}\exp\left(-i\left\{k\tau-\frac{\pi}{4}\right\}\right)\right.\nonumber\\
 &&\left.~~~~~~~~~~~~~~~~~~~-\sinh\alpha~\left\{1-\left(i|\nu_{\rm PGW}|+\frac{1}{2}\right)\frac{i}{k\tau}\left(1+\frac{i}{k\tau}\right)\right\}\right.\nonumber\\
&&\left.~~~~~~~~~~~~~~~~~~~~~~~~~~~~~~~~~~~~~~\times\underbrace{\exp\left(\frac{\pi}{2}|\nu_{\rm PGW}|\right)}_{\textcolor{red}{\bf Boltzmann~enhancement}}\exp\left(i\left\{k\tau-\frac{\pi}{4}\right\}\right)\right].~~~~~~~~~~~\eea

In the super-Hubble,  horizon crossing and sub-Hubble limit we have the following results:
\begin{eqnarray} 
&&\underline{\textcolor{red}{\bf I.~Sub-Hubble~limiting~solution:}}\nonumber\\
&&\underline{\textcolor{blue}{\bf A.~Massless~ \& ~Partially ~Massless~Hubble~Effective~Mass:\Longrightarrow}}\nonumber\\
 &&\nonumber f'_{\lambda,{\bf k}}(-k\tau\gg 1)=i\sqrt{\frac{k}{2}}~2^{\nu_{\rm PGW}-\frac{3}{2}}(-k\tau)^{\frac{3}{2}-\nu_{\rm PGW}}\left|\frac{\Gamma(\nu_{\rm PGW})}{\Gamma\left(\frac{3}{2}\right)}\right|~\nonumber\\
&&~~~~~\times \left[\cosh\alpha~\left\{1-\left(\nu_{\rm PGW}-\frac{1}{2}\right)\frac{i}{k\tau}\right\}\exp\left(-i\left\{k\tau+\frac{\pi}{2}\left(\nu_{\rm PGW}-\frac{1}{2}\right)\right\}\right)\right.\nonumber\\
 &&\left.~~~~~~-\sinh\alpha~\left\{1+\left(\nu_{\rm PGW}-\frac{1}{2}\right)\frac{i}{k\tau}\right\}\exp\left(i\left\{k\tau+\frac{\pi}{2}\left(\nu_{\rm PGW}-\frac{1}{2}\right)\right\}\right)\right].\nonumber\\
\\
&&\underline{\textcolor{blue}{\bf B.~Heavy~Hubble~Effective~Mass:\Longrightarrow}}\nonumber\\
&&\nonumber f'_{\lambda,{\bf k}}(-k\tau\gg 1)=i\sqrt{\frac{k}{2}}~2^{-\left(i|\nu_{\rm PGW}|+\frac{3}{2}\right)}(-k\tau)^{\frac{3}{2}+i|\nu_{\rm PGW}|}\left|\frac{\Gamma(-i|\nu_{\rm PGW}|)}{\Gamma\left(\frac{3}{2}\right)}\right|~\nonumber\\
&&~~~~~~~~~~~\times \left[\cosh\alpha~\left\{1+\left(i|\nu_{\rm PGW}|+\frac{1}{2}\right)\frac{i}{k\tau}\right\}\right.\nonumber\\
&&\left.~~~~~~~~~~~~~~~~~~~~~~~~~~~~~~~~\times\underbrace{\exp\left(-\frac{\pi}{2}|\nu_{\rm PGW}|\right)}_{\textcolor{red}{\bf Boltzmann~suppression}}\exp\left(-i\left\{k\tau-\frac{\pi}{4}\right\}\right)\right.\nonumber\\
 &&\left.~~~~~~~~~~~~~~~~~~~-\sinh\alpha~\left\{1-\left(i|\nu_{\rm PGW}|+\frac{1}{2}\right)\frac{i}{k\tau}\right\}\right.\nonumber\\
&&\left.~~~~~~~~~~~~~~~~~~~~~~~~~~~~~~~~~~~~~~\times\underbrace{\exp\left(\frac{\pi}{2}|\nu_{\rm PGW}|\right)}_{\textcolor{red}{\bf Boltzmann~enhancement}}\exp\left(i\left\{k\tau-\frac{\pi}{4}\right\}\right)\right].~~~~~
\\
&&\underline{\textcolor{red}{\bf II.~Super-Hubble~limiting~solution:}}\nonumber\\
&&\underline{\textcolor{blue}{\bf A.~Massless~ \& ~Partially ~Massless~Hubble~Effective~Mass:\Longrightarrow}}\nonumber\\
&&\nonumber f'_{\lambda,{\bf k}}(-k\tau\ll 1)=i\sqrt{\frac{k}{2}}~2^{\nu_{\rm PGW}-\frac{3}{2}}(-k\tau)^{-\left(\nu_{\rm PGW}+\frac{1}{2}\right)}\left|\frac{\Gamma(\nu_{\rm PGW})}{\Gamma\left(\frac{3}{2}\right)}\right|~\nonumber\\
&&~~~~~~~~~~~\times \left[\cosh\alpha~\left(\nu_{\rm PGW}-\frac{1}{2}\right)\exp\left(-i\left\{\frac{\pi}{2}\left(\nu_{\rm PGW}-\frac{5}{2}\right)\right\}\right)\right.\nonumber\\
 &&\left.~~~~~~~~~~~~~-\sinh\alpha~\left(\nu_{\rm PGW}-\frac{1}{2}\right)\exp\left(i\left\{\frac{\pi}{2}\left(\nu_{\rm PGW}-\frac{5}{2}\right)\right\}\right)\right]. 
\\
&&\underline{\textcolor{blue}{\bf B.~Heavy~Hubble~Effective~Mass:\Longrightarrow}}\nonumber\\ 
&&\nonumber f'_{\lambda,{\bf k}}(-k\tau\ll 1)=i\sqrt{\frac{k}{2}}~2^{-\left(i|\nu_{\rm PGW}|+\frac{3}{2}\right)}(-k\tau)^{\frac{3}{2}+i|\nu_{\rm PGW}|}\left|\frac{\Gamma(-i|\nu_{\rm PGW}|)}{\Gamma\left(\frac{3}{2}\right)}\right|~\nonumber\\
&&~~~~~~~~~~~\times \left[\cosh\alpha\left(i|\nu_{\rm PGW}|+\frac{1}{2}\right)\underbrace{\exp\left(-\frac{\pi}{2}|\nu_{\rm PGW}|\right)}_{\textcolor{red}{\bf Boltzmann~suppression}}\exp\left(\frac{i \pi}{4}\right)\right.\nonumber\\
 &&\left.~~~~~~~~~~~~~~~~~~~-\sinh\alpha~\left(i|\nu_{\rm PGW}|+\frac{1}{2}\right)\underbrace{\exp\left(\frac{\pi}{2}|\nu_{\rm PGW}|\right)}_{\textcolor{red}{\bf Boltzmann~enhancement}}\exp\left(-\frac{ i\pi}{4}\right)\right].~~~~\\
&&\underline{\textcolor{red}{\bf III.~Solution~at~the~point~where~boundary~condition~is~ fixed:}}\nonumber\\
&&\underline{\textcolor{blue}{\bf A.~Massless~ \& ~Partially ~Massless~Hubble~Effective~Mass:\Longrightarrow}}\nonumber\\
&&\nonumber f'_{\lambda,{\bf k}}(-k\tau_0=1)=i\sqrt{\frac{k}{2}}~2^{\nu_{\rm PGW}-\frac{3}{2}}\left|\frac{\Gamma(\nu_{\rm PGW})}{\Gamma\left(\frac{3}{2}\right)}\right|~\nonumber\\
&&~~~~~~~~~~~\times \left[\cosh\alpha~\left\{1-\sqrt{2}\left(\nu_{\rm PGW}-\frac{1}{2}\right)\exp\left(-\frac{i\pi}{4}\right)\right\}\exp\left(-i\left\{\frac{\pi}{2}\left(\nu_{\rm PGW}-\frac{1}{2}\right)-1\right\}\right)\right.\nonumber\\
 &&\left.~~~~~~~~~~~~~-\sinh\alpha~\left\{1-\sqrt{2}\left(\nu_{\rm PGW}-\frac{1}{2}\right)\exp\left(\frac{i\pi}{4}\right)\right\}\exp\left(i\left\{\frac{\pi}{2}\left(\nu_{\rm PGW}-\frac{1}{2}\right)-1\right\}\right)\right].~~~~~~~~~
\\
&&\underline{\textcolor{blue}{\bf B.~Heavy~Hubble~Effective~Mass:\Longrightarrow}}\nonumber\\
 &&\nonumber f'_{\lambda,{\bf k}}(-k\tau_0=1)=i\sqrt{\frac{k}{2}}~2^{-\left(i|\nu_{\rm PGW}|+\frac{3}{2}\right)}\left|\frac{\Gamma(-i|\nu_{\rm PGW}|)}{\Gamma\left(\frac{3}{2}\right)}\right|~\nonumber\\
&&~~~~~~~~~~~\times \left[\cosh\alpha~\left\{1+\sqrt{2}\left(i|\nu_{\rm PGW}|+\frac{1}{2}\right)\exp\left(-\frac{i\pi}{4}\right)\right\}\right.\nonumber\\
&&\left.~~~~~~~~~~~~~~~~~~~~~~~~~~~~~~~~\times\underbrace{\exp\left(-\frac{\pi}{2}|\nu_{\rm PGW}|\right)}_{\textcolor{red}{\bf Boltzmann~suppression}}\exp\left(i\left\{\frac{\pi}{4}+1\right\}\right)\right.\nonumber\\
 &&\left.~~~~~~~~~~~~~~~~~~~-\sinh\alpha~\left\{1+\sqrt{2}\left(i|\nu_{\rm PGW}|+\frac{1}{2}\right)\exp\left(\frac{i\pi}{4}\right)\right\}\right.\nonumber\\
&&\left.~~~~~~~~~~~~~~~~~~~~~~~~~~~~~~~~~~~~~~\times\underbrace{\exp\left(\frac{\pi}{2}|\nu_{\rm PGW}|\right)}_{\textcolor{red}{\bf Boltzmann~enhancement}}\exp\left(-i\left\{\frac{\pi}{4}+1\right\}\right)\right].~~~~
\end{eqnarray}
 \\
\item \underline{\textcolor{red}{\bf Bunch-Davies~vacuum:}}\\
For $\alpha$~vacua the expression for the PGW tensor mode velocity can be written as: 
\bea
 &&\nonumber f'_{\lambda,{\bf k}}(\tau)=i\sqrt{\frac{k}{2}}~2^{\nu_{\rm PGW}-\frac{3}{2}}(-k\tau)^{\frac{3}{2}-\nu_{\rm PGW}}\left|\frac{\Gamma(\nu_{\rm PGW})}{\Gamma\left(\frac{3}{2}\right)}\right|~\nonumber\\
&&~~~~~~~~~~~\times\left\{1-\left(\nu_{\rm PGW}-\frac{1}{2}\right)\frac{i}{k\tau}\left(1-\frac{i}{k\tau}\right)\right\}\exp\left(-i\left\{k\tau+\frac{\pi}{2}\left(\nu_{\rm PGW}-\frac{1}{2}\right)\right\}\right).\nonumber\\
 &&\eea
 Now for the case of heavy field case the PGW tensor mode velocity can be simplified as:
 \bea
 &&\nonumber f'_{\lambda,{\bf k}}(\tau)=i\sqrt{\frac{k}{2}}~2^{-\left(i|\nu_{\rm PGW}|+\frac{3}{2}\right)}(-k\tau)^{\frac{3}{2}+i|\nu_{\rm PGW}|}\left|\frac{\Gamma(-i|\nu_{\rm PGW}|)}{\Gamma\left(\frac{3}{2}\right)}\right|~\nonumber\\
&&~~~~~~~~~~~\times\left\{1+\left(i|\nu_{\rm PGW}|+\frac{1}{2}\right)\frac{i}{k\tau}\left(1-\frac{i}{k\tau}\right)\right\}\nonumber\\
&&~~~~~~~~~~~~~~~~~~~~~~~~~~~~~~~~\times\underbrace{\exp\left(-\frac{\pi}{2}|\nu_{\rm PGW}|\right)}_{\textcolor{red}{\bf Boltzmann~suppression}}\exp\left(-i\left\{k\tau-\frac{\pi}{4}\right\}\right).~~~~~~~~~~~\eea

In the super-Hubble,  horizon crossing and sub-Hubble limit we have the following results:
\begin{eqnarray} 
&&\underline{\textcolor{red}{\bf I.~Sub-Hubble~limiting~solution:}}\nonumber\\
&&\underline{\textcolor{blue}{\bf A.~Massless~ \& ~Partially ~Massless~Hubble~Effective~Mass:\Longrightarrow}}\nonumber\\
 &&\nonumber f'_{\lambda,{\bf k}}(-k\tau\gg 1)=i\sqrt{\frac{k}{2}}~2^{\nu_{\rm PGW}-\frac{3}{2}}(-k\tau)^{\frac{3}{2}-\nu_{\rm PGW}}\left|\frac{\Gamma(\nu_{\rm PGW})}{\Gamma\left(\frac{3}{2}\right)}\right|\exp\left(-i\left\{k\tau+\frac{\pi}{2}\left(\nu_{\rm PGW}-\frac{1}{2}\right)\right\}\right).\nonumber\\
\\
&&\underline{\textcolor{blue}{\bf B.~Heavy~Hubble~Effective~Mass:\Longrightarrow}}\nonumber\\
&&\nonumber f'_{\lambda,{\bf k}}(-k\tau\gg 1)=i\sqrt{\frac{k}{2}}~2^{-\left(i|\nu_{\rm PGW}|+\frac{3}{2}\right)}(-k\tau)^{\frac{3}{2}+i|\nu_{\rm PGW}|}\left|\frac{\Gamma(-i|\nu_{\rm PGW}|)}{\Gamma\left(\frac{3}{2}\right)}\right|~\nonumber\\
&&~~~~~~~~~~~~~~~~~~~~~~~\times\underbrace{\exp\left(-\frac{\pi}{2}|\nu_{\rm PGW}|\right)}_{\textcolor{red}{\bf Boltzmann~suppression}}\exp\left(-i\left\{k\tau-\frac{\pi}{4}\right\}\right).
\\
&&\underline{\textcolor{red}{\bf II.~Super-Hubble~limiting~solution:}}\nonumber\\
&&\underline{\textcolor{blue}{\bf A.~Massless~ \& ~Partially ~Massless~Hubble~Effective~Mass:\Longrightarrow}}\nonumber\\
&&\nonumber f'_{\lambda,{\bf k}}(-k\tau\ll 1)=i\sqrt{\frac{k}{2}}~2^{\nu_{\rm PGW}-\frac{3}{2}}(-k\tau)^{-\left(\nu_{\rm PGW}+\frac{1}{2}\right)}\left|\frac{\Gamma(\nu_{\rm PGW})}{\Gamma\left(\frac{3}{2}\right)}\right|\nonumber\\
&&~~~~~~~~~~~~~~~~~~~~~\times~\left(\nu_{\rm PGW}-\frac{1}{2}\right)\exp\left(-i\left\{\frac{\pi}{2}\left(\nu_{\rm PGW}-\frac{5}{2}\right)\right\}\right).
\\
&&\underline{\textcolor{blue}{\bf B.~Heavy~Hubble~Effective~Mass:\Longrightarrow}}\nonumber\\ 
&&\nonumber f'_{\lambda,{\bf k}}(-k\tau\ll 1)=i\sqrt{\frac{k}{2}}~2^{-\left(i|\nu_{\rm PGW}|+\frac{3}{2}\right)}(-k\tau)^{i|\nu_{\rm PGW}|-\frac{1}{2}}\left|\frac{\Gamma(-i|\nu_{\rm PGW}|)}{\Gamma\left(\frac{3}{2}\right)}\right|~\nonumber\\
&&~~~~~~~~~~~~~~~~~~~~~~~~~\times~\left(i|\nu_{\rm PGW}|+\frac{1}{2}\right)\underbrace{\exp\left(-\frac{\pi}{2}|\nu_{\rm PGW}|\right)}_{\textcolor{red}{\bf Boltzmann~suppression}}\exp\left(\frac{i \pi }{4}\right).\\
&&\underline{\textcolor{red}{\bf III.~Solution~at~the~point~where~boundary~condition~is~ fixed:}}\nonumber\\
&&\underline{\textcolor{blue}{\bf A.~Massless~ \& ~Partially ~Massless~Hubble~Effective~Mass:\Longrightarrow}}\nonumber\\
&&\nonumber f'_{\lambda,{\bf k}}(-k\tau_0=1)=i\sqrt{\frac{k}{2}}~2^{\nu_{\rm PGW}-\frac{3}{2}}\left|\frac{\Gamma(\nu_{\rm PGW})}{\Gamma\left(\frac{3}{2}\right)}\right|~\nonumber\\
&&~~~~~~~~~~~\times\left\{1-\sqrt{2}\left(\nu_{\rm PGW}-\frac{1}{2}\right)\exp\left(-\frac{i\pi}{4}\right)\right\}\exp\left(-i\left\{\frac{\pi}{2}\left(\nu_{\rm PGW}-\frac{1}{2}\right)-1\right\}\right).\nonumber\\
&&
\\
&&\underline{\textcolor{blue}{\bf B.~Heavy~Hubble~Effective~Mass:\Longrightarrow}}\nonumber\\
 &&\nonumber f'_{\lambda,{\bf k}}(-k\tau_0=1)=i\sqrt{\frac{k}{2}}~2^{-\left(i|\nu_{\rm PGW}|+\frac{3}{2}\right)}\left|\frac{\Gamma(-i|\nu_{\rm PGW}|)}{\Gamma\left(\frac{3}{2}\right)}\right|~\nonumber\\
&&~~~~~~\times \left\{1+\sqrt{2}\left(i|\nu_{\rm PGW}|+\frac{1}{2}\right)\exp\left(-\frac{i\pi}{4}\right)\right\}\underbrace{\exp\left(-\frac{\pi}{2}|\nu_{\rm PGW}|\right)}_{\textcolor{red}{\bf Boltzmann~suppression}}\exp\left(i\left\{\frac{\pi}{4}+1\right\}\right).
\end{eqnarray}
\end{enumerate}

Now,  the canonically conjugate momenta for PGW tensor mode can be computed as:
\bea \underline{\textcolor{red}{\bf PGW~momenta:}}~~~\pi_{\lambda,{\bf k}}(\tau):=\frac{\partial {\cal L}^{(2)}(f_{\lambda,{\bf k}}(\tau),f'_{\lambda,{\bf k}}(\tau),\tau)}{\partial f'_{\lambda,{\bf k}}(\tau)}=\left[v^{'*}_{{\bf k}}(\tau)-\Biggl(\frac{a'(\tau)}{a(\tau)}\Biggr)f_{\lambda,{\bf k}}(\tau)\right].~~~~~~~~\eea 
Then the Hamiltonian for the PGW tensor mode can be further simplified as:
\bea H_{\rm PGW}(\tau)=\frac{1}{2}\int d^3{\bf k}~\Biggl[\left|\pi_{\lambda,{\bf k}}(\tau)+\frac{a'(\tau)}{a(\tau)}f_{\lambda,{\bf k}}(\tau)\right|^2+\mu^2_{\rm PGW}(k,\tau)|f_{\lambda,{\bf k}}(\tau)|^2\Biggr],\eea
where the effective mass $\mu^2_{\rm PGW}(k,\tau)$ of the PGW is defined as:
\bea \mu^2_{\rm PGW}(k,\tau):=\Biggl[k^2-\Biggl(\frac{a'(\tau)}{a(\tau)}\Biggr)^2\Biggr].\eea 
To quantize the Hamiltonian for PGW we introduce the following quantum operators in the Heisenberg picture:
\bea \hat{f}_{\lambda}({\bf x},\tau)&=&{\cal U}^{\dagger}_{\rm PGW}(\tau,\tau_0)\hat{f}_{\lambda}({\bf x},\tau_0){\cal U}_{\rm PGW}(\tau,\tau_0)\nonumber\\
&=&\int \frac{d^3{\bf k}}{(2\pi)^3}~\left[f^{*}_{\lambda,-{\bf k}}(\tau)~\hat{a}_{\bf k}+f_{\lambda,{\bf k}}(\tau)~\hat{a}^{\dagger}_{-{\bf k}}\right]~\exp(i{\bf k}.{\bf x}),\\
\hat{\pi}_{\lambda}({\bf x},\tau)&=&{\cal U}^{\dagger}_{\rm PGW}(\tau,\tau_0)\hat{\pi}_{\lambda}({\bf x},\tau_0){\cal U}_{\rm PGW}(\tau,\tau_0)\nonumber\\
&=&\int \frac{d^3{\bf k}}{(2\pi)^3}~\left[\pi^{*}_{\lambda,-{\bf k}}(\tau)~\hat{a}_{\bf k}+\pi_{\lambda,{\bf k}}(\tau)~\hat{a}^{\dagger}_{-{\bf k}}\right]~\exp(i{\bf k}.{\bf x}).\eea
Then the quantized Hamiltonian for PGW is given by:
\bea \widehat{H}_{\rm PGW}(\tau)&=&\frac{1}{2}\int d^3{\bf k}~\Biggl[\left|\left[f^{*'}_{\lambda,-{\bf k}}(\tau)~\hat{a}_{\bf k}+f^{'}_{\lambda,{\bf k}}(\tau)~\hat{a}^{\dagger}_{-{\bf k}}\right]+\frac{a'(\tau)}{a(\tau)}\left[f^{*}_{\lambda,-{\bf k}}(\tau)~\hat{a}_{\bf k}+f_{\lambda,{\bf k}}(\tau)~\hat{a}^{\dagger}_{-{\bf k}}\right]\right|^2 \nonumber\\
&&~~~~~~~~~~~~~~~~~~~~~~~~~~~~~~~~~~~~~~~~~~~~+\mu^2_{\rm PGW}(k,\tau)|\left[f^{*}_{\lambda,-{\bf k}}(\tau)~\hat{a}_{\bf k}+f_{\lambda,{\bf k}}(\tau)~\hat{a}^{\dagger}_{-{\bf k}}\right]|^2\Biggr]\nonumber\\
&=&\frac{1}{2}\int d^3{\bf k}\Biggl[~\underbrace{\Omega_{\lambda,\bf k}(\tau)\left(\hat{a}^{\dagger}_{\bf k}\hat{a}_{\bf k}+\hat{a}^{\dagger}_{-{\bf k}}\hat{a}_{-{\bf k}}+1\right)}_{\textcolor{red}{\bf Contribution~from~the~free~term}}\nonumber\\
&&~~~~~~~~~~~~~~~~~~~~~~~~+i\underbrace{{\Lambda}_{{\bf k}}(\tau)\Biggl(\exp(-2i\phi_{\lambda,{\bf k}}(\tau))\hat{a}_{\bf k}\hat{a}_{-{\bf k}}-\exp(2i\phi_{\lambda,{\bf k}}(\tau))\hat{a}^{\dagger}_{\bf k}\hat{a}^{\dagger}_{-{\bf k}}\Biggr)}_{\textcolor{red}{\bf Contribution~ from~ the~ Interaction~ term}}~\Biggr],~~~~~~~~~~~~~~\eea
where we define two crucial quantities, the dispersion relation for PGW, $\Omega_{\bf k}(\tau)$ and the time dependent factor $\Lambda_{\bf k}(\tau)$ by:
\bea \Omega_{\lambda,{\bf k}}(\tau):&=&\Biggl\{\left|f^{'}_{\lambda,{\bf k}}(\tau)\right|^2+\mu^2_{\rm PGW}(k,\tau)\left|f_{\lambda,{\bf k}}(\tau)\right|^2\Biggr\}\nonumber\\
&=&\Biggl\{\left|\pi_{\lambda,{\bf k}}(\tau)+\Lambda_{\bf k}(\tau)_{\lambda,{\bf k}}(\tau)\right|^2+\mu^2_{\rm PGW}(k,\tau)\left|f_{\lambda,{\bf k}}(\tau)\right|^2\Biggr\},~~~~~\\
\mu^2_{\rm PGW}(k,\tau):&=&\left(k^2-\Lambda^2_{\bf k}(\tau)\right)=\left(k^2-{\cal H}^2(\tau)\right),\\
\Lambda_{\bf k}(\tau):&=&\Biggl(\frac{a'(\tau)}{a(\tau)}\Biggr)={\cal H}(\tau).~~~~\eea 
For the PGW perturbation we find out that $\Lambda_{\bf k}(\tau)$ actually represents the Hubble parameter in the conformal scale, which makes this computation different from the same study from the quantized scalar modes. Also,  the effective mass and related mass parameters for the scalar and PGW tensor mode perturbations are significantly different, though the classical solutions from the {\it Mukhanov Sasaki equation} for both the mode functions looks similar structure-wise. It would be more clear once we follow the rest of the part of the computation and the physical outcomes that we have obtained in this paper. 
\subsection{Time evolution of quantized PGW }
In this section we will study the time evolution of the quantized PGW tensor mode in presence of different initial conditions.  To study this one need to follow few steps one by one which we discuss in the following subsections in detail.
\textcolor{Sepia}{\subsubsection{\sffamily Fixing the initial condition at horizon crossing}}
Here we fix the intermediate time scale at $\tau=\tau_0$ with the crucial constraint, $-k\tau_0=1$ in such a way that, we get the following normalization in the classical solution of the PGW tensor modes.  This gives us the following results: 
\begin{enumerate}
\item  \underline{\textcolor{red}{\bf Motta-Allen ($\alpha,\gamma$)~vacua:}}\\ 
\bea &&\underline{\textcolor{blue}{\bf A.~Massless~ \& ~Partially ~Massless~Hubble~Effective~Mass:\Longrightarrow}}\nonumber\\
&&f_{\lambda,{\bf k}}(\tau_0)=\frac{2^{\nu_{\rm PGW}-1}}{\sqrt{2k}}\left|\frac{\Gamma(\nu_{\rm PGW})}{\Gamma\left(\frac{3}{2}\right)}\right|~\nonumber\\
&&~~~~~~~~~~~~~~~~~~~~~~~~~~~~\left[\cosh\alpha~\exp\left(-i\left\{\frac{\pi}{2}\left(\nu_{\rm PGW}-2\right)-1\right\}\right)\right.\nonumber\\
&&\left.~~~~~~~~~~~~~~~~~~~~~~~~~~~~~~~~~~~~~~+\sinh\alpha~\exp\left(i\left\{\gamma+\frac{\pi}{2}\left(\nu_{\rm PGW}-2\right)-1\right\}\right)\right],\\
&&\pi_{\lambda,{\bf k}}(\tau_0)=i\sqrt{\frac{k}{2}}~2^{\nu_{\rm PGW}-\frac{3}{2}}\left|\frac{\Gamma(\nu_{\rm PGW})}{\Gamma\left(\frac{3}{2}\right)}\right|~\nonumber\\
&&~~~~~~\times \left[\cosh\alpha~\left\{1-\sqrt{2}\frac{\displaystyle \left(\nu_{\rm PGW}-\frac{1}{2}\right)\left(\nu_{\rm PGW}+\frac{1}{2}+i\right)}{\displaystyle \left(\nu_{\rm PGW}+\frac{1}{2}\right)}\exp\left(-\frac{i\pi}{4}\right)\right\}\right.\nonumber\\
&&\left.~~~~~~~~~~~~~~~~~~~~~~~~~~~~~~~~~~~~~~~~~~~~~\times\exp\left(-i\left\{\frac{\pi}{2}\left(\nu_{\rm PGW}-2\right)-1\right\}\right)\right.\nonumber\\
 &&\left.~~~~-\sinh\alpha~\left\{1-\sqrt{2}\frac{\displaystyle \left(\nu_{\rm PGW}-\frac{1}{2}\right)\left(\nu_{\rm PGW}+\frac{1}{2}+i\right)}{\displaystyle \left(\nu_{\rm PGW}+\frac{1}{2}\right)}\exp\left(\frac{i\pi}{4}\right)\right\}\right.\nonumber\\
&&\left.~~~~~~~~~~~~~~~~~~~~~~~~~~~~~~~~~~~~~~~~~~~~~\times\exp\left(i\left\{\gamma+\frac{\pi}{2}\left(\nu_{\rm PGW}-2\right)-1\right\}\right)\right],~\nonumber\\
 &&\\
&&\underline{\textcolor{blue}{\bf B.~Heavy~Hubble~Effective~Mass:\Longrightarrow}}\nonumber\\ 
&&f_{\lambda,{\bf k}}(\tau_0)=\frac{2^{-(i|\nu_{\rm PGW}|+1)}}{\sqrt{2k}}\left|\frac{\Gamma(-i|\nu_{\rm PGW}|)}{\Gamma\left(\frac{3}{2}\right)}\right|\left[\cosh\alpha~\underbrace{\exp\left(-\frac{\pi}{2}|\nu_{\rm PGW}|\right)}_{\textcolor{red}{\bf Boltzmann~suppression}}~\exp\left(i(\pi+1)\right)\right.\nonumber\\
&&\left.~~~~~~~~~~~~~~~~~~~~~~~~~~~~+\sinh\alpha~\underbrace{\exp\left(\frac{\pi}{2}|\nu_{\rm PGW}|\right)}_{\textcolor{red}{\bf Boltzmann~enhancement}}~\exp\left(i\left(\gamma-(\pi+1)\right)\right)\right],~~~~~~~~~~~ \\
&&\pi_{\lambda,{\bf k}}(\tau_0)=i\sqrt{\frac{k}{2}}~2^{-\left(i|\nu_{\rm PGW}|+\frac{3}{2}\right)}\left|\frac{\Gamma(-i|\nu_{\rm PGW}|)}{\Gamma\left(\frac{3}{2}\right)}\right|~\nonumber\\
&&~~~~~~\times \left[\cosh\alpha~\left\{1+\sqrt{2}\frac{\displaystyle \left(i|\nu_{\rm PGW}|+\frac{1}{2}\right)\left(i|\nu_{\rm PGW}|-\frac{1}{2}-i\right)}{\displaystyle \left(i|\nu_{\rm PGW}|-\frac{1}{2}\right)}\exp\left(-\frac{i\pi}{4}\right)\right\}\right.\nonumber\\
&&\left.~~~~~~~~~~~~~~~~~~~~~~~~~~~~~~~~~~~~~~~~~~~~~\times~\underbrace{\exp\left(-\frac{\pi}{2}|\nu_{\rm PGW}|\right)}_{\textcolor{red}{\bf Boltzmann~suppression}}~\exp\left(i(\pi+1)\right)\right.\nonumber\\
 &&\left.~~~~-\sinh\alpha~\left\{1+\sqrt{2}\frac{\displaystyle \left(i|\nu_{\rm PGW}|+\frac{1}{2}\right)\left(i|\nu_{\rm PGW}|-\frac{1}{2}-i\right)}{\displaystyle \left(i|\nu_{\rm PGW}|-\frac{1}{2}\right)}\exp\left(\frac{i\pi}{4}\right)\right\}\right.\nonumber\\
&&\left.~~~~~~~~~~~~~~~~~~~~~~~~~~~~~~~~~~~~~~~~~~~~~\times\underbrace{\exp\left(\frac{\pi}{2}|\nu_{\rm PGW}|\right)}_{\textcolor{red}{\bf Boltzmann~enhancement}}~\exp\left(i\left(\gamma-(\pi+1)\right)\right)\right].~~~~~~~~~\eea
\item \underline{\textcolor{red}{\bf $\alpha$~vacua:}}\\
\bea &&\underline{\textcolor{blue}{\bf A.~Massless~ \& ~Partially ~Massless~Hubble~Effective~Mass:\Longrightarrow}}\nonumber\\
&&f_{\lambda,{\bf k}}(\tau_0)=\frac{2^{\nu_{\rm PGW}-1}}{\sqrt{2k}}\left|\frac{\Gamma(\nu_{\rm PGW})}{\Gamma\left(\frac{3}{2}\right)}\right|~\nonumber\\
&&~~~~~~~~~~~~~~~~~~~~~~~~~~~~\left[\cosh\alpha~\exp\left(-i\left\{\frac{\pi}{2}\left(\nu_{\rm PGW}-2\right)-1\right\}\right)\right.\nonumber\\
&&\left.~~~~~~~~~~~~~~~~~~~~~~~~~~~~~~~~~~~~~~+\sinh\alpha~\exp\left(i\left\{\frac{\pi}{2}\left(\nu_{\rm PGW}-2\right)-1\right\}\right)\right],\\
&&\pi_{\lambda,{\bf k}}(\tau_0)=i\sqrt{\frac{k}{2}}~2^{\nu_{\rm PGW}-\frac{3}{2}}\left|\frac{\Gamma(\nu_{\rm PGW})}{\Gamma\left(\frac{3}{2}\right)}\right|~\nonumber\\
&&~~~~~~\times \left[\cosh\alpha~\left\{[1-\sqrt{2}\frac{\displaystyle \left(\nu_{\rm PGW}-\frac{1}{2}\right)\left(\nu_{\rm PGW}+\frac{1}{2}+i\right)}{\displaystyle \left(\nu_{\rm PGW}+\frac{1}{2}\right)}\exp\left(-\frac{i\pi}{4}\right)\right\}\right.\nonumber\\
&&\left.~~~~~~~~~~~~~~~~~~~~~~~~~~~~~~~~~~~~~~~~~~~~~\times\exp\left(-i\left\{\frac{\pi}{2}\left(\nu_{\rm PGW}-2\right)-1\right\}\right)\right.\nonumber\\
 &&\left.~~~~-\sinh\alpha~\left\{1-\sqrt{2}\frac{\displaystyle \left(\nu_{\rm PGW}-\frac{1}{2}\right)\left(\nu_{\rm PGW}+\frac{1}{2}+i\right)}{\displaystyle \left(\nu_{\rm PGW}+\frac{1}{2}\right)}\exp\left(\frac{i\pi}{4}\right)\right\}\right.\nonumber\\
&&\left.~~~~~~~~~~~~~~~~~~~~~~~~~~~~~~~~~~~~~~~~~~~~~\times\exp\left(i\left\{\frac{\pi}{2}\left(\nu_{\rm PGW}-2\right)-1\right\}\right)\right],~\nonumber\\
 &&\\
&&\underline{\textcolor{blue}{\bf B.~Heavy~Hubble~Effective~Mass:\Longrightarrow}}\nonumber\\ 
&&f_{\lambda,{\bf k}}(\tau_0)=\frac{2^{-(i|\nu_{\rm PGW}|+1)}}{\sqrt{2k}}\left|\frac{\Gamma(-i|\nu_{\rm PGW}|)}{\Gamma\left(\frac{3}{2}\right)}\right|\left[\cosh\alpha~\underbrace{\exp\left(-\frac{\pi}{2}|\nu_{\rm PGW}|\right)}_{\textcolor{red}{\bf Boltzmann~suppression}}~\exp\left(i(\pi+1)\right)\right.\nonumber\\
&&\left.~~~~~~~~~~~~~~~~~~~~~~~~~~~~+\sinh\alpha~\underbrace{\exp\left(\frac{\pi}{2}|\nu_{\rm PGW}|\right)}_{\textcolor{red}{\bf Boltzmann~enhancement}}~\exp\left(-i\left(\pi+1\right)\right)\right],~~~~~~~~~~~ \\
&&\pi_{\lambda,{\bf k}}(\tau_0)=i\sqrt{\frac{k}{2}}~2^{-\left(i|\nu_{\rm PGW}|+\frac{3}{2}\right)}\left|\frac{\Gamma(-i|\nu_{\rm PGW}|)}{\Gamma\left(\frac{3}{2}\right)}\right|~\nonumber\\
&&~~~~~~\times \left[\cosh\alpha~\left\{1+\sqrt{2}\frac{\displaystyle \left(i|\nu_{\rm PGW}|+\frac{1}{2}\right)\left(i|\nu_{\rm PGW}|-\frac{1}{2}-i\right)}{\displaystyle \left(i|\nu_{\rm PGW}|-\frac{1}{2}\right)}\exp\left(-\frac{i\pi}{4}\right)\right\}\right.\nonumber\\
&&\left.~~~~~~~~~~~~~~~~~~~~~~~~~~~~~~~~~~~~~~~~~~~~~\times~\underbrace{\exp\left(-\frac{\pi}{2}|\nu_{\rm PGW}|\right)}_{\textcolor{red}{\bf Boltzmann~suppression}}~\exp\left(i(\pi+1)\right)\right.\nonumber\\
 &&\left.~~~~-\sinh\alpha~\left\{1+\sqrt{2}\frac{\displaystyle \left(i|\nu_{\rm PGW}|+\frac{1}{2}\right)\left(i|\nu_{\rm PGW}|-\frac{1}{2}-i\right)}{\displaystyle \left(i|\nu_{\rm PGW}|-\frac{1}{2}\right)}\exp\left(\frac{i\pi}{4}\right)\right\}\right.\nonumber\\
&&\left.~~~~~~~~~~~~~~~~~~~~~~~~~~~~~~~~~~~~~~~~~~~~~\times\underbrace{\exp\left(\frac{\pi}{2}|\nu_{\rm PGW}|\right)}_{\textcolor{red}{\bf Boltzmann~enhancement}}~\exp\left(-i\left(\pi+1\right)\right)\right].~~~~~~~~~~~\eea
\item \underline{\textcolor{red}{\bf Bunch-Davies~vacuum:}}\\
\bea &&\underline{\textcolor{blue}{\bf A.~Massless~ \& ~Partially ~Massless~Hubble~Effective~Mass:\Longrightarrow}}\nonumber\\
&&f_{\lambda,{\bf k}}(\tau_0)=\frac{1}{\sqrt{2k}}~2^{\nu_{\rm PGW}-1}\left|\frac{\Gamma(\nu_{\rm PGW})}{\Gamma\left(\frac{3}{2}\right)}\right|~\exp\left(-i\left\{\frac{\pi}{2}\left(\nu_{\rm PGW}-2\right)-1\right\}\right),\\
&&\pi_{\lambda,{\bf k}}(\tau_0)=i\sqrt{\frac{k}{2}}~2^{\nu_{\rm PGW}-\frac{3}{2}}\left|\frac{\Gamma(\nu_{\rm PGW})}{\Gamma\left(\frac{3}{2}\right)}\right|~\exp\left(-i\left\{\frac{\pi}{2}\left(\nu_{\rm PGW}-2\right)-1\right\}\right)~\nonumber\\
 &&~~~~~~~~~~~~~~\left\{1-\sqrt{2}\frac{\displaystyle \left(\nu_{\rm PGW}-\frac{1}{2}\right)\left(\nu_{\rm PGW}+\frac{1}{2}+i\right)}{\displaystyle \left(\nu_{\rm PGW}+\frac{1}{2}\right)}\exp\left(-\frac{i\pi}{4}\right)\right\},~\\
&&\underline{\textcolor{blue}{\bf B.~Heavy~Hubble~Effective~Mass:\Longrightarrow}}\nonumber\\ 
&&f_{\lambda,{\bf k}}(\tau_0)=\frac{1}{\sqrt{2k}}~2^{-(i|\nu_{\rm PGW}|+1)}\left|\frac{\Gamma(-i|\nu_{\rm PGW}|)}{\Gamma\left(\frac{3}{2}\right)}\right|~\underbrace{\exp\left(-\frac{\pi}{2}|\nu_{\rm PGW}|\right)}_{\textcolor{red}{\bf Boltzmann~suppression}}\exp\left(i(\pi+1)\right),~~~~~~~~~~~\\
&&\pi_{\lambda,{\bf k}}(\tau_0)=i\sqrt{\frac{k}{2}}~2^{\nu_{\rm PGW}-\frac{3}{2}}\left|\frac{\Gamma(\nu_{\rm PGW})}{\Gamma\left(\frac{3}{2}\right)}\right|~\underbrace{\exp\left(-\frac{\pi}{2}|\nu_{\rm PGW}|\right)}_{\textcolor{red}{\bf Boltzmann~suppression}}\exp\left(i(\pi+1)\right)~\nonumber\\
 &&~~~~~~~~~~~~~~\left\{1-\sqrt{2}\frac{\displaystyle \left(\nu_{\rm PGW}-\frac{1}{2}\right)\left(\nu_{\rm PGW}+\frac{1}{2}+i\right)}{\displaystyle \left(\nu_{\rm PGW}+\frac{1}{2}\right)}\exp\left(-\frac{i\pi}{4}\right)\right\}.\eea
\end{enumerate}
Hence for the PGW tensor modes we can write the following operators:
\bea \hat{f}_{\lambda,{\bf k}}(\tau)&=&f_{\lambda,{\bf k}}(\tau_0)\Biggl(a_{\bf k}(\tau)+a^{\dagger}_{-{\bf k}}(\tau)\Biggr),\\
\hat{\pi}_{\lambda,{\bf k}}(\tau)&=&-\pi_{\lambda,{\bf k}}(\tau_0)~\Biggl(a_{\bf k}(\tau)-a^{\dagger}_{-{\bf k}}(\tau)\Biggr), \eea
where the creation and the annihilation operators at any arbitrary time scale can be expressed at $\tau=\tau_0$ along with $-k\tau_0=1$ in the Heisenberg picture as:
\bea && a_{\bf k}(\tau):={\cal U}^{\dagger}_{\rm PGW}(\tau,\tau_0)a_{\bf k}{\cal U}_{\rm PGW}(\tau,\tau_0),\\
&& a^{\dagger}_{-{\bf k}}(\tau):={\cal U}^{\dagger}_{\rm PGW}(\tau,\tau_0)a^{\dagger}_{-{\bf k}}{\cal U}_{\rm PGW}(\tau,\tau_0),\eea
where the unitary time evolution operator ${\cal U}_{\rm PGW}(\tau)$ satisfy the following condition:
\bea {\cal U}^{\dagger}_{\rm PGW}(\tau){\cal U}_{\rm PGW}(\tau)=\mathbb{I}.\eea

\textcolor{Sepia}{\subsubsection{\sffamily Squeezed state formalism in PGW dominated Primordial Cosmology}} 

The time evolution of the unitary operator $\mathcal{U}_{\rm PGW}(\tau)$, produced by the full quadratic quantized Hamiltonian which is basically taking care of both the free and interaction contribution, can be understood in a very simplified language of factorization, as given by:
\begin{equation}
\label{eq:unitary}
\textcolor{blue}{\bf PGW~Unitary~Operator:}~~~\mathcal{U}_{\rm PGW}(\tau,\tau_0) =\underbrace{ \underbrace{\hat{\mathcal{S}}_{\rm PGW}(r_{\lambda,{\bf k}}(\tau,\tau_0) ,\phi_{\lambda,{\bf k}}(\tau) )}_{\textcolor{red}{\bf PGW~Squuezing~Operator}}\underbrace{\hat{\mathcal{R}}_{\rm PGW}(\theta_{\lambda,{\bf k}}(\tau) )}_{\textcolor{red}{\bf PGW~Rotation~Operator}}}_{\textcolor{purple}{\bf Simple~factorization}},
\end{equation}
where $\hat{\mathcal{R}}_{\rm PGW}(\theta_{\lambda,{\bf k}}(\tau))$ is a two momentum mode PGW rotation operator, which is defined by the following expression:
\begin{equation}
\label{eq:rotationoperatorPGW}
\textcolor{blue}{\bf PGW ~Rotation~Operator:}~~\hat{\mathcal{R}}_{\rm PGW}(\theta_{\lambda,{\bf k}}(\tau))= \exp\Biggl( -i\theta_{\lambda,{\bf k}}(\tau)\big( \hat{a}_{\bf k}\hat{a}_{\bf k}^{\dagger} + \hat{a}_{-{\bf k}}^{\dagger}\hat{a}_{-{\bf k}} \big) \Biggr),
\end{equation}
and $\hat{\mathcal{S}}_{\rm PGW}(r_{\lambda,{\bf k}}(\tau,\tau_0) ,\phi_{\lambda,{\bf k}}(\tau) )$ is the two momentum mode PGW squeezing operator, which is defined by the following expression:
\bea
\label{eq:SqueezedoperatorPGW}
&&\textcolor{blue}{\bf PGW ~Squeezing~Operator:}~~\nonumber\\&&\hat{\mathcal{S}}_{\rm PGW}(r_{\lambda,{\bf k}}(\tau) ,\phi_{\lambda,{\bf k}}(\tau) )= \exp\Bigg(\frac{\overbrace{r_{\lambda,{\bf k}}(\tau)}^{\textcolor{red}{\bf Amplitude}}}{2} \Big\{ \exp(-2i \underbrace{\phi_{\lambda,{\bf k}}(\tau)}_{\textcolor{red}{\bf Phase}} )\hat{a}_{\bf k}\hat{a}_{-{\bf k}} - \exp(2i \phi_{\lambda,{\bf k}}(\tau))\hat{a}_{-{\bf k}}^{\dagger}\hat{a}_{\bf k}^{\dagger} \Big\}\Bigg).\nonumber\\
&&
\eea
Here the squeezing amplitude for PGW is represented by the parameter, $r_{\lambda, {\bf k}}(\tau)$ and the angle $\phi_{\lambda,{\bf k}}(\tau)$.  Most importantly,  at the level of quantized version of the Hamiltonian functional operator, PGW rotation operator is associated with the free contributions. On the other hand, PGW squeezing operator is specifically capturing the contribution from the interaction part of the Hamiltonian after quantization. But by careful inspection one can point that  the two-mode PGW rotation operator give rise to an irrelevant conformal time dependent phase term, $\exp(i\theta_{\lambda,{\bf k}}(\tau))$ when it operate on the initial quantum mechanical vacuum state. Now since this part completely capturing the information regarding the free part of the PGW perturbation theory, it will not give us any additional significant information after this parametrization and for this reason one can ignore this contribution in our derived results in this paper to avoid the appearance of unnecessary complicated junks. It means that in the rest of the paper we will only concentrate on outcomes of the PGW squeezing operator which is basically playing most significant role to determine the cosmological complexity from PGW in terms of the squeezed parameters which depend upon conformal time and the phase contributions by solving the time evolution equations which we will discuss very soon.

Here it is quite natural that the PGW mode quantization can be visualized in terms of  quantized parametric oscillator,  which is the prime reason for the appearance of a squeezed
quantum mechanical state in the present computation. As a result the whole description can be explicitly described in terms of two-mode two-helicity squeezed state formalism for PGW,  which is the backbone of our present computation of this paper. 
Things will be more clear once we proceed in the later half of this paper. For better understanding, some of the crucial computations are explicitly given in the appendix, which we suggest to the readers to go through.

For the detailed computation we choose the ground state of the Hamiltonian as the initial choice of the helicity and momentum dependent vacuum state:
\bea
\label{eq:initial vacuum}
\hat{a}_{\bf k}\ket{0}_{\lambda,{\bf k},-{\bf k}} = 0 ~~~~~ \forall~~ {\bf k}~\&~\lambda=+,\times,
\eea
which is a {\it Poincare invariant} state.  Obviously,  this is a very general definition of an annihilation operator, which can be further translated in the language of the choice of the initial vacuum states which we have chosen once we have obtained the full classical solution of the PGW mode function. In this paper we are interested in the CPT violating {\it Motta-Allen vacua state} ($|\alpha,\gamma\rangle_{\lambda,{\bf k},-{\bf k}}$) and the CPT preserving $\alpha$-{\it vacua state} ($|\alpha\rangle_{\lambda,{\bf k},-{\bf k}}$), which are related to the well known Euclidean ground state, represented by {\it Bunch-Davies vacuum} state ($|0\rangle_{\lambda,{\bf k},-{\bf k}}$) via {\it Bogoliubov transformation}, which are appended below:
\bea && \underline{\textcolor{red}{\bf Motta~ Allen~ initial~quantum ~vacua ~state:}}\nonumber\\
&&|\alpha,\gamma\rangle_{\lambda,{\bf k},-{\bf k}}=\frac{1}{\sqrt{|\cosh\alpha|}}~\exp\left(-\frac{i}{2}\exp(-i\gamma)~{\rm tanh}\alpha~\hat{a}^{\dagger}_{\bf k}\hat{a}^{\dagger}_{-\bf k}\right)\ket{{\bf BD}}_{\lambda,{\bf k},-{\bf k}},\\
&& \underline{\textcolor{red}{\bf \alpha~ initial~quantum~ vacua~ state:}}\nonumber\\
&& |\alpha\rangle_{\lambda,{\bf k},-{\bf k}}=\frac{1}{\sqrt{|\cosh\alpha|}}~\exp\left(-\frac{i}{2}~{\rm tanh}\alpha~\hat{a}^{\dagger}_{\bf k}\hat{a}^{\dagger}_{-\bf k}\right)\ket{{\bf BD}}_{\lambda,{\bf k},-{\bf k}}.
\eea
Next we use the operator $\hat{\mathcal{S}}_{\rm PGW}(r_{\lambda,{\bf k}}(\tau),\phi_{\lambda,{\bf k}}(\tau))$ which operates on the various initial states and produce a two-mode PGW squeezed vacuum state.  All other initial states will be related to the {\it Bunch-Davies} result via the previously mentioned non-trivial {\it Bogoliubov transformations}, which are appended below:
\bea
&& \underline{\textcolor{red}{\bf Motta~ Allen~ squeezed~quantum ~vacua ~state:}}\nonumber\\
&&\ket{\Psi^{(\alpha,\gamma)}_{\bf sq}}_{\lambda,{\bf k}, -{\bf k}} =\hat{\mathcal{S}}_{\rm PGW}(r_{\lambda,{\bf k}}(\tau),\phi_{\lambda,{\bf k}}(\tau))|\alpha,\gamma\rangle_{\lambda,{\bf k},-{\bf k}}\nonumber\\
&&~~~~~~~~~~~~~= \frac{1}{\cosh r_{\lambda,{\bf k}}(\tau)}\sum_{n=0}^{\infty}(-1)^{n}\exp(-2in~\phi_{\lambda,{\bf k}}(\tau)\tanh^{n}r_{\lambda,{\bf k}}(\tau)\ket{n_{\lambda,{\bf k}}, n_{\lambda,-{\bf k}}}^{(\alpha,\gamma)},~~~~
\\
&& \underline{\textcolor{red}{\bf \alpha~ squeezed~quantum ~vacua ~state:}}\nonumber\\
&&\ket{\Psi^{(\alpha)}_{\bf sq}}_{\lambda,{\bf k}, -{\bf k}} =\hat{\mathcal{S}}_{\rm PGW}(r_{\lambda,{\bf k}}(\tau),\phi_{\lambda,{\bf k}}(\tau))|\alpha,\gamma\rangle_{\lambda,{\bf k},-{\bf k}}\nonumber\\
&&~~~~~~~~~~~~~~~~~= \frac{1}{\cosh r_{\lambda,{\bf k}}(\tau)}\sum_{n=0}^{\infty}(-1)^{n}\exp(-2in~\phi_{\lambda,{\bf k}}(\tau)\tanh^{n}r_{\lambda,{\bf k}}(\tau)\ket{n_{\lambda,{\bf k}}, n_{\lambda,-{\bf k}}}^{(\alpha)},~~~~~~~~~~\\
&& \underline{\textcolor{red}{\bf Bunch~ Davies~ squeezed~quantum ~vacuum ~state:}}\nonumber\\
&&\ket{\Psi^{(\bf BD)}_{\bf sq}}_{\lambda,{\bf k}, -{\bf k}} =\hat{\mathcal{S}}_{\rm PGW}(r_{\lambda,{\bf k}}(\tau),\phi_{\lambda,{\bf k}}(\tau))|\alpha,\gamma\rangle_{\lambda,{\bf k},-{\bf k}}\nonumber\\
&&~~~~~~~~~~~~~~~~~= \frac{1}{\cosh r_{\lambda,{\bf k}}(\tau)}\sum_{n=0}^{\infty}(-1)^{n}\exp(-2in~\phi_{\lambda,{\bf k}}(\tau)\tanh^{n}r_{\lambda,{\bf k}}(\tau)\ket{n_{\lambda,{\bf k}}, n_{\lambda,-{\bf k}}}^{(\bf BD)}.~~~~~~~~~~\eea
Here we define two-mode two helicity dependent excited state (occupation number state) for the three corresponding choices of the initial states,  given by:
\bea 
&& \underline{\textcolor{red}{\bf Motta~ Allen~ squeezed~quantum ~vacua ~state:}}\nonumber\\
&&\label{eq:excitedmode1}
\ket{n_{\lambda,{\bf k}}, n_{\lambda,-{\bf k}}}^{(\alpha,\gamma)}=\frac{1}{n!\sqrt{|\cosh\alpha|}}~\big( \hat{a}_{{\bf k}}^{\dagger} \big)^{n}\big( \hat{a}_{-{\bf k}}^{\dagger} \big)^{n}~\exp\left(-\frac{i}{2}\exp(-i\gamma)~{\rm tanh}\alpha~\hat{a}^{\dagger}_{\bf k}\hat{a}^{\dagger}_{\bf k}\right)\ket{{\bf BD}}_{\lambda,{\bf k},-{\bf k}}\nonumber\\
&&~~~~~~~~~~~~~~~~~~~=\frac{1}{\sqrt{|\cosh\alpha|}}~\exp\left(-\frac{i}{2}\exp(-i\gamma)~{\rm tanh}\alpha~\hat{a}^{\dagger}_{\bf k}\hat{a}^{\dagger}_{\bf k}\right)\ket{n_{\lambda,{\bf k}}, n_{\lambda,-{\bf k}}}^{({\bf BD})}\nonumber\\
&&~~~~~~~~~~~~~~~~~~~=\frac{1}{\sqrt{|\cosh\alpha|}}~\left\{1\right.\nonumber\\
&&\left.~~~~~~~~~~~~~~~~~~~~~~~~~~~~~~~~~+\sum^{\infty}_{m=1}\frac{(-1)^{m}}{m!}\left(\frac{i}{2}\right)^{m}~\exp(-im\gamma)~{\rm tanh}^{m}\alpha~\big(\hat{a}^{\dagger}_{\bf k}\big)^{m}\big(\hat{a}^{\dagger}_{\bf k}\big)^{m}\right\}~\ket{n_{\lambda,{\bf k}}, n_{\lambda,-{\bf k}}}^{({\bf BD})}\nonumber\\
&&~~~~~~~~~~~~~~~~~~~=\frac{1}{\sqrt{|\cosh\alpha|}}~\left\{\ket{n_{\lambda,{\bf k}}, n_{\lambda,-{\bf k}}}^{({\bf BD})}\right.\nonumber\\
&&\left.~~~~~~~~~~~~~~~~~~~~~~~~~~~~~+\sum^{\infty}_{m=1}\frac{(-1)^{m}}{m!}\left(\frac{i}{2}\right)^{m}~\exp(-im\gamma)~{\rm tanh}^{m}\alpha~\ket{(n+m)_{\lambda,{\bf k}}, (n+m)_{\lambda,-{\bf k}}}^{({\bf BD})}\right\},\nonumber\\
&& \\
&& \underline{\textcolor{red}{\bf \alpha~ squeezed~quantum ~vacua ~state:}}\nonumber\\
&&
\label{eq:excitedmode2}
\ket{n_{\lambda,{\bf k}}, n_{\lambda,-{\bf k}}}^{(\alpha)}=\frac{1}{n!\sqrt{|\cosh\alpha|}}~\big( \hat{a}_{{\bf k}}^{\dagger} \big)^{n}\big( \hat{a}_{-{\bf k}}^{\dagger} \big)^{n}~\exp\left(-\frac{i}{2}~{\rm tanh}\alpha~\hat{a}^{\dagger}_{\bf k}\hat{a}^{\dagger}_{\bf k}\right)\ket{{\bf BD}}_{\lambda,{\bf k},-{\bf k}}\nonumber\\
&&~~~~~~~~~~~~~~~~~~~=\frac{1}{\sqrt{|\cosh\alpha|}}~\exp\left(-\frac{i}{2}~{\rm tanh}\alpha~\hat{a}^{\dagger}_{\bf k}\hat{a}^{\dagger}_{\bf k}\right)\ket{n_{\lambda,{\bf k}}, n_{\lambda,-{\bf k}}}^{({\bf BD})}\nonumber\\
&&~~~~~~~~~~~~~~~~~~~=\frac{1}{\sqrt{|\cosh\alpha|}}~\left\{1+\sum^{\infty}_{m=1}\frac{(-1)^{m}}{m!}\left(\frac{i}{2}\right)^{m}~{\rm tanh}^{m}\alpha~\big(\hat{a}^{\dagger}_{\bf k}\big)^{m}\big(\hat{a}^{\dagger}_{\bf k}\big)^{m}\right\}~\ket{n_{\lambda,{\bf k}}, n_{\lambda,-{\bf k}}}^{({\bf BD})}\nonumber\\
&&~~~~~~~~~~~~~~~~~~~=\frac{1}{\sqrt{|\cosh\alpha|}}~\left\{\ket{n_{\lambda,{\bf k}}, n_{\lambda,-{\bf k}}}^{({\bf BD})}\right.\nonumber\\
&&\left.~~~~~~~~~~~~~~~~~~~~~~~~~~~~~~~~~~~~~~~~~~~~+\sum^{\infty}_{m=1}\frac{(-1)^{m}}{m!}\left(\frac{i}{2}\right)^{m}~{\rm tanh}^{m}\alpha~\ket{(n+m)_{\lambda,{\bf k}}, (n+m)_{\lambda,-{\bf k}}}^{({\bf BD})}\right\},\nonumber\\
&& \\
&& \underline{\textcolor{red}{\bf Bunch~ Davies~ squeezed~quantum ~vacuum ~state:}}\nonumber\\
&&
\label{eq:excitedmode3}
\ket{n_{\lambda,{\bf k}}, n_{\lambda,-{\bf k}}}^{({\bf BD})} = \frac{1}{n!}\big( \hat{a}_{{\bf k}}^{\dagger} \big)^{n}\big( \hat{a}_{-{\bf k}}^{\dagger} \big)^{n}\ket{{\bf BD}}_{\lambda,{\bf k},-{\bf k}}.
\eea 
Then the wave function having two-modes ${\bf k}, -{\bf k}$ from the various initial states can be computed as: 
\bea
&& \underline{\textcolor{red}{\bf Motta~ Allen~ squeezed~quantum ~vacua ~state:}}\label{eq:fullwavefunction1}\nonumber\\
&&\ket{\Psi^{(\alpha,\gamma)}_{\bf sq}} = \sum_{\lambda=+,\times}\bigotimes_{{\bf k}}\ket{\Psi^{(\alpha,\gamma)}_{\bf sq}}_{\lambda,{\bf k}, -{\bf k}}\nonumber\\
&&~~~~~~~~~~= \sum_{\lambda=+,\times}\bigotimes_{{\bf k}}\hat{\mathcal{S}}_{\rm PGW}(r_{\lambda,{\bf k}}(\tau),\phi_{\lambda,{\bf k}}(\tau))|\alpha,\gamma\rangle_{\lambda,{\bf k},-{\bf k}}\nonumber\\
&&~~~~~~~~~~= \sum_{\lambda=+,\times}\bigotimes_{{\bf k}} \frac{1}{\cosh r_{\lambda,{\bf k}}(\tau)}\sum_{n=0}^{\infty}(-1)^{n}\exp(-2in~\phi_{\lambda,{\bf k}}(\tau))\tanh^{n}r_{\lambda,{\bf k}}(\tau)\ket{n_{\lambda,{\bf k}}, n_{\lambda,-{\bf k}}}^{(\alpha,\gamma)}\nonumber\\
&&~~~~~~~~~~= \frac{1}{\sqrt{|\cosh\alpha|}}~\sum_{\lambda=+,\times}\bigotimes_{{\bf k}} \frac{1}{\cosh r_{\lambda,{\bf k}}(\tau)}\nonumber\\
&&~~~~~~~~~~~~~~~~~~~~~~~~~~~~~~~~~~\sum_{n=0}^{\infty}\sum_{m=0}^{\infty}\frac{(-1)^{n+m}}{n!m!}\left(\frac{i}{2}\right)^{m}\exp(-i(m\gamma+2n~\phi_{\lambda,{\bf k}}(\tau)))\nonumber\\
&&~~~~~~~~~~~~~~~~~~~~~~~~~~~~~~~~~~~~~~~~~~~~~{\rm tanh}^{m}\alpha\tanh^{n}r_{\lambda,{\bf k}}(\tau)\big( \hat{a}_{{\bf k}}^{\dagger} \big)^{n+m}\big( \hat{a}_{-{\bf k}}^{\dagger} \big)^{n+m}\ket{\bf BD}_{\lambda,{\bf k},-{\bf k}}\nonumber\\
&&~~~~~~~~~~= \frac{1}{\sqrt{|\cosh\alpha|}}~\sum_{\lambda=+,\times}\bigotimes_{{\bf k}} \frac{1}{\cosh r_{\lambda,{\bf k}}(\tau)}\nonumber\\
&&~~~~~~~~~~~~~~~~~~~~~~~~~~~~~~~~~~\sum_{n=0}^{\infty}(-1)^{n}\exp(-2in~\phi_{\lambda,{\bf k}}(\tau)\tanh^{n}r_{\lambda,{\bf k}}(\tau)\left\{\ket{n_{\lambda,{\bf k}}, n_{\lambda,-{\bf k}}}^{({\bf BD})}\right.\nonumber\\
&&\left.~~~~~~~~~~~~~~~~~~~~~~~~~~~~~+\sum^{\infty}_{m=1}\frac{(-1)^{m}}{m!}\left(\frac{i}{2}\right)^{m}~\exp(-im\gamma)~{\rm tanh}^{m}\alpha~\ket{(n+m)_{\lambda,{\bf k}}, (n+m)_{\lambda,-{\bf k}}}^{({\bf BD})}\right\},~~~~\nonumber\\
&&
\\
&& \underline{\textcolor{red}{\bf \alpha~ squeezed~quantum ~vacua ~state:}}\nonumber\\
&&\ket{\Psi^{(\alpha)}_{\bf sq}} = \sum_{\lambda=+,\times}\bigotimes_{{\bf k}}\ket{\Psi^{(\alpha)}_{\bf sq}}_{\lambda,{\bf k}, -{\bf k}}\nonumber\\
&&~~~~~~~~~~= \sum_{\lambda=+,\times}\bigotimes_{{\bf k}}\hat{\mathcal{S}}_{\rm PGW}(r_{\lambda,{\bf k}}(\tau),\phi_{\lambda,{\bf k}}(\tau))|\alpha\rangle_{\lambda,{\bf k},-{\bf k}}\nonumber\\
&&~~~~~~~~~~= \sum_{\lambda=+,\times}\bigotimes_{{\bf k}} \frac{1}{\cosh r_{\lambda,{\bf k}}(\tau)}\sum_{n=0}^{\infty}(-1)^{n}\exp(-2in~\phi_{\lambda,{\bf k}}(\tau))\tanh^{n}r_{\lambda,{\bf k}}(\tau)\ket{n_{\lambda,{\bf k}}, n_{\lambda,-{\bf k}}}^{(\alpha)}\nonumber\\
&&~~~~~~~~~~= \frac{1}{\sqrt{|\cosh\alpha|}}~\sum_{\lambda=+,\times}\bigotimes_{{\bf k}} \frac{1}{\cosh r_{\lambda,{\bf k}}(\tau)}\nonumber\\
&&~~~~~~~~~~~~~~~~~~~~~~~~~~~~~~~~~~\sum_{n=0}^{\infty}\sum_{m=0}^{\infty}\frac{(-1)^{n+m}}{n!m!}\left(\frac{i}{2}\right)^{m}\exp(-2in~\phi_{\lambda,{\bf k}}(\tau))\nonumber\\
&&~~~~~~~~~~~~~~~~~~~~~~~~~~~~~~~~~~~~~~~~~~~~~{\rm tanh}^{m}\alpha\tanh^{n}r_{\lambda,{\bf k}}(\tau)\big( \hat{a}_{{\bf k}}^{\dagger} \big)^{n+m}\big( \hat{a}_{-{\bf k}}^{\dagger} \big)^{n+m}\ket{\bf BD}_{\lambda,{\bf k},-{\bf k}}\nonumber\\
&&~~~~~~~~~~= \frac{1}{\sqrt{|\cosh\alpha|}}~\sum_{\lambda=+,\times}\bigotimes_{{\bf k}} \frac{1}{\cosh r_{\lambda,{\bf k}}(\tau)}\nonumber\\
&&~~~~~~~~~~~~~~~~~~~~~~~~~~~~~~~~~~\sum_{n=0}^{\infty}(-1)^{n}\exp(-2in~\phi_{\lambda,{\bf k}}(\tau)\tanh^{n}r_{\lambda,{\bf k}}(\tau)\left\{\ket{n_{\lambda,{\bf k}}, n_{\lambda,-{\bf k}}}^{({\bf BD})}\right.\nonumber\\
&&\left.~~~~~~~~~~~~~~~~~~~~~~~~~~~~~+\sum^{\infty}_{m=1}\frac{(-1)^{m}}{m!}\left(\frac{i}{2}\right)^{m}~{\rm tanh}^{m}\alpha~\ket{(n+m)_{\lambda,{\bf k}}, (n+m)_{\lambda,-{\bf k}}}^{({\bf BD})}\right\},~~~~\nonumber\\
&&
\\
&& \underline{\textcolor{red}{\bf Bunch~ Davies~ squeezed~quantum ~vacuum ~state:}}\nonumber\\
&&\ket{\Psi^{({\bf BD})}_{\bf sq}} = \sum_{\lambda=+,\times}\bigotimes_{{\bf k}}\ket{\Psi^{({\bf BD})}_{\bf sq}}_{\lambda,{\bf k}, -{\bf k}}\nonumber\\
&&~~~~~~~~~~= \sum_{\lambda=+,\times}\bigotimes_{{\bf k}}\hat{\mathcal{S}}_{\rm PGW}(r_{\lambda,{\bf k}}(\tau),\phi_{\lambda,{\bf k}}(\tau))|{\bf BD}\rangle_{\lambda,{\bf k},-{\bf k}}\nonumber\\
&&~~~~~~~~~~= \sum_{\lambda=+,\times}\bigotimes_{{\bf k}} \frac{1}{\cosh r_{\lambda,{\bf k}}(\tau)}\sum_{n=0}^{\infty}(-1)^{n}\exp(-2in~\phi_{\lambda,{\bf k}}(\tau))\tanh^{n}r_{\lambda,{\bf k}}(\tau)\ket{n_{\lambda,{\bf k}}, n_{\lambda,-{\bf k}}}^{({\bf BD})}\nonumber\\
&&~~~~~~~~~~=\sum_{\lambda=+,\times}\bigotimes_{{\bf k}} \frac{1}{\cosh r_{\lambda,{\bf k}}(\tau)}\sum_{n=0}^{\infty}\frac{(-1)^{n}}{n!}\left(\frac{i}{2}\right)^{m}\exp(-2in~\phi_{\lambda,{\bf k}}(\tau))\nonumber\\
&&~~~~~~~~~~~~~~~~~~~~~~~~~~~~~~~~~~~~~~~~~~~~~~~~~~~~~~~~~~~~~~~~~~\tanh^{n}r_{\lambda,{\bf k}}(\tau)\big( \hat{a}_{{\bf k}}^{\dagger} \big)^{n}\big( \hat{a}_{-{\bf k}}^{\dagger} \big)^{n}\ket{\bf BD}_{\lambda,{\bf k},-{\bf k}}\nonumber\\
&&~~~~~~~~~~=\sum_{\lambda=+,\times}\bigotimes_{{\bf k}} \frac{1}{\cosh r_{\lambda,{\bf k}}(\tau)}\sum_{n=0}^{\infty}(-1)^{n}\exp(-2in~\phi_{\lambda,{\bf k}}(\tau)\tanh^{n}r_{\lambda,{\bf k}}(\tau)\ket{n_{\lambda,{\bf k}}, n_{\lambda,-{\bf k}}}^{({\bf BD})},~~~~\nonumber\\
&&\eea
Here in the squeezed states we have taken the sum over two possible helicities or the polarization of the PGW perturbation, whose contributions in the classical version or in the quantum version of the PGW tensor mode functions looks exactly identical to each other.  Because of this fact after performing the sum over the two helicities we will get an overall contribution of a factor of $2$ which we have not mentioned in the previously mentioned equations, but we will take care of this fact in the rest of the paper.
 
\textcolor{Sepia}{\subsubsection{\sffamily Role of unitary time evolution of PGW fluctuations in squeezed state formalism}}

Here,  our aim is to explore the role of the unitary time evolution of the PGW tensor fluctuations in previously mentioned squeezed state formalism.  Applying the unitary transformation for the PGW perturbation,  one can obtain the following ladder operators:
\bea \hat{a}_{\bf k}(\tau)&=&\hat{\cal U}^{\dagger}_{\rm PGW}(\tau,\tau_0)~\hat{a}_{\bf k}~\hat{\cal U}_{\rm PGW}(\tau,\tau_0)\nonumber\\
&=&\hat{\mathcal{R}}^{\dagger}_{\rm PGW}(\theta_{\lambda,{\bf k}}(\tau))\hat{\mathcal{S}}^{\dagger}_{\rm PGW}(r_{\lambda,{\bf k}}(\tau),\phi_{\lambda,{\bf k}}(\tau))~\hat{a}_{\bf k}~\hat{\mathcal{R}}_{\rm PGW}(\theta_{\lambda,{\bf k}}(\tau))\hat{\mathcal{S}}_{\rm PGW}(r_{\lambda,{\bf k}}(\tau),\phi_{\lambda,{\bf k}}(\tau))\nonumber\\
&=&\cosh r_{\lambda,{\bf k}}(\tau)~\exp(-i\theta_{\lambda,{\bf k}}(\tau))~\hat{a}_{\bf k}-\sinh r_{\lambda,{\bf k}}(\tau)~\exp(i(\theta_{\bf k}(\tau)+2\phi_{\lambda,{\bf k}}(\tau)))~\hat{a}^{\dagger}_{-{\bf k}},~~~~~~~~~~~~~\\
\hat{a}^{\dagger}_{-{\bf k}}(\tau)&=&\hat{\cal U}^{\dagger}_{\rm PGW}((\tau,\tau_0)~\hat{a}^{\dagger}_{-{\bf k}}~\hat{\cal U}_{\rm PGW}((\tau,\tau_0)\nonumber\\
&=&\hat{\mathcal{R}}^{\dagger}_{\rm PGW}((\theta_{\lambda,{\bf k}}(\tau))\hat{\mathcal{S}}^{\dagger}_{\rm PGW}((r_{\lambda,{\bf k}}(\tau),\phi_{\lambda,{\bf k}}(\tau))~\hat{a}^{\dagger}_{-{\bf k}}~\hat{\mathcal{R}}_{\rm PGW}((\theta_{\lambda,{\bf k}}(\tau))\hat{\mathcal{S}}_{\rm PGW}((r_{\lambda,{\bf k}}(\tau),\phi_{\lambda,{\bf k}}(\tau))\nonumber\\ 
&=&\cosh r_{\lambda,{\bf k}}(\tau)~\exp(i\theta_{\lambda,{\bf k}}(\tau))~\hat{a}^{\dagger}_{-{\bf k}}-\sinh r_{\lambda,{\bf k}}(\tau)~\exp(-i(\theta_{\lambda,{\bf k}}(\tau)+2\phi_{\lambda,{\bf k}}(\tau)))~\hat{a}_{{\bf k}}.~~~~\eea

Using these expressions,  the field and momentum operator of the PGW perturbation can be expressed as:

\bea \hat{f}_{\lambda,{\bf k}}(\tau)&=&f_{\lambda,{\bf k}}(\tau_0)\Biggl(\hat{a}_{\bf k}(\tau)+\hat{a}^{\dagger}_{-{\bf k}}(\tau)\Biggr)\nonumber\\
&=&f_{\lambda,{\bf k}}(\tau_0)\Biggl[\hat{a}_{\bf k}\Biggl(\cosh r_{\lambda,{\bf k}}(\tau)~\exp(-i\theta_{\lambda,{\bf k}}(\tau))-\sinh r_{\lambda,{\bf k}}(\tau)~\exp(-i(\theta_{\lambda,{\bf k}}(\tau)+2\phi_{\lambda,{\bf k}}(\tau)))\Biggr)\nonumber\\
&&~~~~~~~+\hat{a}^{\dagger}_{-{\bf k}}\Biggl(\cosh r_{\lambda,{\bf k}}(\tau)~\exp(i\theta_{\lambda,{\bf k}}(\tau))-\sinh r_{\lambda,{\bf k}}(\tau)~\exp(i(\theta_{\lambda,{\bf k}}(\tau)+2\phi_{\lambda,{\bf k}}(\tau)))\Biggr)\Biggr]\nonumber\\
&=&\left[f^{*}_{\lambda,-{\bf k}}(\tau)~\hat{a}_{\bf k}+f_{\lambda,{\bf k}}(\tau)~\hat{a}^{\dagger}_{-{\bf k}}\right],\\
\hat{\pi}_{\lambda,{\bf k}}(\tau)&=&-\pi_{\lambda,{\bf k}}(\tau_0)~\Biggl(a_{\bf k}(\tau)-a^{\dagger}_{-{\bf k}}(\tau)\Biggr)\nonumber\\
&=&-\pi_{\lambda,{\bf k}}(\tau_0)\Biggl[\hat{a}_{\bf k}\Biggl(\cosh r_{\lambda,{\bf k}}(\tau)~\exp(-i\theta_{\lambda,{\bf k}}(\tau))+\sinh r_{\lambda,{\bf k}}(\tau)~\exp(-i(\theta_{\lambda,{\bf k}}(\tau)+2\phi_{\lambda,{\bf k}}(\tau)))\Biggr)\nonumber\\
&&~~~~~~~-\hat{a}^{\dagger}_{-{\bf k}}\Biggl(\cosh r_{\lambda,{\bf k}}(\tau)~\exp(i\theta_{\lambda,{\bf k}}(\tau))+\sinh r_{\lambda,{\bf k}}(\tau)~\exp(i(\theta_{\lambda,{\bf k}}(\tau)+2\phi_{\lambda,{\bf k}}(\tau)))\Biggr)\Biggr]\nonumber\\
&=&\left[\pi^{*}_{\lambda,-{\bf k}}(\tau)~\hat{a}_{\bf k}+\pi_{\lambda,{\bf k}}(\tau)~\hat{a}^{\dagger}_{-{\bf k}}\right], \eea
which are valid for two modes $\pm {\bf k}$ and the two helicities $\pm$ for the PGW tensor modes.  

From above expressions,  the classical counterparts can be written as:
\bea f_{\lambda,{\bf k}}(\tau)&=&f_{\lambda,{\bf k}}(\tau_0)\Biggl(\cosh r_{\lambda,{\bf k}}(\tau)~\exp(i\theta_{\lambda,{\bf k}}(\tau))-\sinh r_{\lambda,{\bf k}}(\tau)~\exp(i(\theta_{\lambda,{\bf k}}(\tau)+2\phi_{\lambda,{\bf k}}(\tau)))\Biggr),~~~~~~~~~~~~\\
\pi_{\lambda,{\bf k}}(\tau)&=&\pi_{\lambda,{\bf k}}(\tau_0)\Biggl(\cosh r_{\lambda,{\bf k}}(\tau)~\exp(i\theta_{\lambda,{\bf k}}(\tau))+\sinh r_{\lambda,{\bf k}}(\tau)~\exp(i(\theta_{\lambda,{\bf k}}(\tau)+2\phi_{\lambda,{\bf k}}(\tau)))\Biggr).\eea 
Further, the time evolution of the quantum operators $\hat{\mathcal{R}}_{\rm PGW}(\theta_{\lambda,{\bf k}}(\tau))$ and $\hat{\mathcal{S}}_{\rm PGW}(r_{\lambda,{\bf k}}(\tau),\phi_{\lambda,{\bf k}}(\tau))$ are described by the Schr\"{o}dinger equation, which gives: 
\begin{align} 
\label{eq:diffneqns1}
&\frac{dr_{\lambda,{\bf k}}(\tau)}{d\tau} = -{\cal H}(\tau)~\cos(2\phi_{\lambda,{\bf k}}(\tau)),\\
\label{eq:diffeqns2}
&\frac{d\phi_{\lambda,{\bf k}}(\tau)}{d\tau} = \Omega_{\lambda,{\bf k}}(\tau) + {\cal H}(\tau)~\coth(2r_{\lambda,{\bf k}}(\tau))\sin(2\phi_{\lambda,{\bf k}}(\tau)),
\end{align}
where $\Omega_{\lambda,{\bf k}}(\tau)$ is defined as:
\bea 
 \Omega_{\lambda,{\bf k}}(\tau):&=&\Biggl\{\left|\pi_{\lambda,{\bf k}}(\tau)+{\cal H}(\tau)~f_{\lambda,{\bf k}}(\tau)\right|^2+\left(k^2-{\cal H}^2(\tau)\right)\left|f_{\lambda,{\bf k}}(\tau)\right|^2\Biggr\}. \eea
It needs to be emphasized that both the conformal time dependent factors are not explicitly dependent on the particular polarization or the helicity of the PGW. Not only that the contribution appearing from both the PGW polarization components become exactly identical to each other. For this reason if we sum over all the helicity components or the polarizations the from both the sides of these evolution equations we will get a factor of $2$ as we have already pointed earlier. Since such factor will appear in the both sides of the evolution equations, it will actually cancel from both the sides and ultimately we have the simplified form of the coupled system which we have to solve in the present context. Additionally, we can see from these above mentioned equations that, the analytical solutions of these equations are not actually possible at all for any arbitrary time scale and for the mentioned classes of the scale factors on which we are interested in this paper. Also, since we are dealing with two first order coupled differential equations, to numerically solve the boundary condition at the cosmological horizon exit play very crucial role.   We fix $-k\tau_0 =1$ condition at the scale $\tau=\tau_0$,  which is chosen to be the boundary condition to solve the evolution in terms of the squeezed amplitude and squeezed angle.  Apart from having the complicated angular dependent contributions in these coupled differential equations the prime and the most significant information is appearing from the dispersion relation of the perturbed PGW or the tensor modes. It is worth to mention that the individual terms appearing in the PGW dispersion relation not explicitly depend on the specific polarization or the helicity of the PGW. However, since two polarization components exist for PGW, so to remind ourself we have kept the helicity index. 

Next, we compute the PGW dispersion relation for the three the quantum vacuum states,  as mentioned earlier in the paper.  More details regarding expressions for these cases for dispersion relation can be found in the Appendix.  The simplified version of the PGW dispersion relation in terms of squeezing parameters can be written as:
\bea 
 \Omega_{\lambda,{\bf k}}(\tau):&=&\Biggl\{\left|\pi_{\lambda,{\bf k}}(\tau_0)\Biggl(\cosh r_{\lambda,{\bf k}}(\tau)~\exp(i\theta_{\lambda,{\bf k}}(\tau))+\sinh r_{\lambda,{\bf k}}(\tau)~\exp(i(\theta_{\lambda,{\bf k}}(\tau)+2\phi_{\lambda,{\bf k}}(\tau)))\Biggr)\nonumber\right.\\
&&\nonumber \left.+{\cal H}(\tau)~f_{\lambda,{\bf k}}(\tau_0)\Biggl(\cosh r_{\lambda,{\bf k}}(\tau)~\exp(i\theta_{\lambda,{\bf k}}(\tau))-\sinh r_{\lambda,{\bf k}}(\tau)~\exp(i(\theta_{\lambda,{\bf k}}(\tau)+2\phi_{\lambda,{\bf k}}(\tau)))\Biggr)\right|^2\\&&+\left(k^2-{\cal H}^2(\tau)\right)\left|f_{\lambda,{\bf k}}(\tau_0)\right|^2\nonumber\\
&&\left|\Biggl(\cosh r_{\lambda,{\bf k}}(\tau)~\exp(i\theta_{\lambda,{\bf k}}(\tau))-\sinh r_{\lambda,{\bf k}}(\tau)~\exp(i(\theta_{\lambda,{\bf k}}(\tau)+2\phi_{\lambda,{\bf k}}(\tau)))\Biggr)\right|^2\Biggr\}.\nonumber\\
&& \eea
\section{Complexity of PGW from squeezed states}
\label{sec:complexityMeasure}
In this section, the complexity will be computed from the squeezed cosmological perturbations which were studied in the preceding section. The wave function method of calculating circuit complexity was introduced in refs.  \cite{Jefferson:2017sdb,Guo:2018kzl} and further used in refs. \cite{Bhattacharyya:2020kgu,Bhattacharyya:2020rpy,Bhattacharyya:2019kvj} has been used here.  Choosing a particular reference and target state is needed to calculate the circuit complexity.   For this purpose the most common choice is the Bunch-Davies state $\ket{0}_{{\bf k},-{\bf k}}$.  The complexity will be studied for three different target quantum states $\ket{\Psi_{\bf sq}}_{{\bf k},-{\bf k}}$: $\alpha$ squeezed quantum vacua state, Bunch-Davies squeezed quantum vacuum states  and Motta-Allen squeezed quantum vacuum state. $\alpha$ initial quantum vacua states and Motta-Allen initial quantum vacua state are obtained from the Bunch-Davies quantum vacua state through the Bogoliubov transformation. Hence, the complexity for these transformations should be a reflection of the complexity of the Bogoliubov transformation and squeezing operator. A variety of methods for calculating the circuit complexity are available in the literature \cite{Ali:2018fcz}. We shall be discussing two such approaches: (1) Nielsen's wave-function method and (2) covariance matrix approach.  

\subsection{Circuit complexity of two mode squeezed states}
Let us start with the two mode squeeze state operator,  which is given by:
\bea
&&\textcolor{blue}{\bf PGW ~Squeezing~Operator:}~~\nonumber\\&&\hat{\mathcal{S}}_{\rm PGW}(r_{\lambda,{\bf k}}(\tau) ,\phi_{\lambda,{\bf k}}(\tau) )= \exp\Bigg(\frac{\overbrace{r_{\lambda,{\bf k}}(\tau)}^{\textcolor{red}{\bf Amplitude}}}{2} \Big\{ \exp(-2i \underbrace{\phi_{\lambda,{\bf k}}(\tau)}_{\textcolor{red}{\bf Phase}} )\hat{a}_{\bf k}\hat{a}_{-{\bf k}} - \exp(2i \phi_{\lambda,{\bf k}}(\tau))\hat{a}_{-{\bf k}}^{\dagger}\hat{a}_{\bf k}^{\dagger} \Big\}\Bigg).\nonumber\\
&&
\eea
Here the squeezing amplitude,  $r_{\lambda, {\bf k}}(\tau)$ and angle for PGW is represented by $\phi_{\lambda,{\bf k}}(\tau)$. The two mode squeezed target state, in this context is given by:
\begin{align}
\nonumber
    \ket{\psi_{\text{sq}}}_{\vec{k},\vec{-k}} &= \hat{\mathcal{S}}_{\rm PGW}(r_{\lambda,{\bf k}}(\tau,\tau_0) ,\phi_{\lambda,{\bf k}}(\tau) ) \ket{0,0} \\
    &=  \exp\Bigg(\frac{\overbrace{r_{\lambda,{\bf k}}(\tau)}^{\textcolor{red}{\bf Amplitude}}}{2} \Big\{ \exp(-2i \underbrace{\phi_{\lambda,{\bf k}}(\tau)}_{\textcolor{red}{\bf Phase}} )\hat{a}_{\bf k}\hat{a}_{-{\bf k}} - \exp(2i \phi_{\lambda,{\bf k}}(\tau))\hat{a}_{-{\bf k}}^{\dagger}\hat{a}_{\bf k}^{\dagger} \Big\}\Bigg)\ket{0,0}
\end{align}
where $\ket{0}_{\vec{k}}\ket{0}_{\vec{-k}} = \ket{0,0}$.  In terms of number states, we can write it as follows:
\begin{equation}
    \ket{\psi_{\text{sq}}}_{\vec{k},\vec{-k}} = \frac{1}{\text{cosh}r_{\lambda,{\bf k}}} \sum_{n= 0}^\infty (-1)^n e^{in\phi_{\lambda,{\bf k}}}(\text{tanh}r_{\lambda,{\bf k}})^n \ket{n_k,n_{-k}})
\end{equation}

We can now calculate circuit complexity for reference and target states.  Here,  we start with the following operators: 
\begin{align}
 \hat{q}_{\vec{k}} &= \frac{1}{\sqrt{2\Omega_k}} \big (\hat{a}^\dagger_{\vec{k}} + \hat{a}_{\vec{k}} \big)\\
 \hat{p}_{\vec{k}} &=i \sqrt{\frac{\Omega_k}{2} }\big (\hat{a}^\dagger_{\vec{k}} - \hat{a}_{\vec{k}} \big)
\end{align}
given that, $[\hat{q}_{\vec{k}},\hat{p}_{\vec{k'}}] = i\delta^3(\vec{k}-\vec{k'})$. 

Reference and target states can be represented as wave functions in position space as follows: 
:\begin{align}
\label{eq:referenceGaussianState}
\nonumber
   \psi_R(q_{\vec{k}},q_{-\vec{k}}) &= \langle q_{\vec{k}}, q_{-\vec{k}} |0\rangle_{\vec{k},-\vec{k}}\\
    &= \left( \frac{\Omega_k}{\pi} \right)^{\frac{1}{4}}\text{exp}\left( -\frac{\Omega_k}{2}\big(q_{\vec{k}}^2+ q_{-\vec{k}}^2 \big)\right) 
\end{align}

\begin{align}
\label{eq:targetGaussianState}
\nonumber
     \psi_{\text{sq}}(q_{\vec{k}},q_{-\vec{k}}) &= \langle q_{\vec{k}}, q_{-\vec{k}} |\psi_{\text{sq}}\rangle_{\vec{k}} \\
    &= \frac{e^{A\big(q_{\vec{k}}^2+ q_{-\vec{k}}^2 \big) - Bq_{\vec{k}}q_{-\vec{k}}}}{\text{cosh}r_{\lambda,{\bf k}}\sqrt{\pi}\sqrt{1-e^{-4i\phi_k}\text{tanh}^2r_{\lambda,{\bf k}}}}   
\end{align}
Here,  $A$ and $B$ denote coefficient which are functions $r_{\rm k}$ and $\phi_{\rm k}$ :
\begin{align}
\begin{split}
    A &= \frac{\Omega_k}{2} \frac{e^{-4i\phi_{\lambda,{\bf k}}}\text{tanh}^2r_{\lambda,{\bf k}}+1}{e^{-4i\phi_{\lambda,{\bf k}}}\text{tanh}^2r_{\lambda,{\bf k}}-1} \\
    B &= 2\Omega_k \frac{e^{-2i\phi_{\lambda,{\bf k}}} \text{tanh}r_{\lambda,{\bf k}}}{e^{-4i\phi_{\lambda,{\bf k}}}\text{tanh}^2r_{\lambda,{\bf k}}-1}
\end{split}
\end{align}
We define following terms:
\begin{align}
\label{eq:omegas}
    \Sigma _{\vec{k}} &= -2A +B,\Sigma _{-\vec{k}} = -2A -B,\omega _{\vec{k}}= \omega_{-\vec{k}}= \frac{\Omega_k}{2}
\end{align}
For computing complexities using different three methods, reader can refer to \cite{Ali:2018fcz}. However, in this paper, we will consider covariance method and Nielsen method for calculating complexity. According to Eq. \eqref{eq:cost-functional} and \eqref{eq:k_cost_functional}, different choice of cost functions will give resultant complexity accordingly. In this paper, we will represent $C_1$ as complexity resulting from linear cost functional, $C_2$ as complexity resulting from to quadratic cost functional and in general, $C_k$ as complexity resulting from $k$ family of functionals $F_k$.

\subsection*{\underline{\textcolor{red}{\bf Covariance Matrix Method:}}}
Reference and Target states \eqref{eq:referenceGaussianState} and \eqref{eq:targetGaussianState}, can be written as covariance matrix as below, because they are in gaussian form as below -

\begin{eqnarray}
    G_k^{s=0}=\displaystyle 
\begin{bmatrix}
\displaystyle \frac{1}{\Omega_k} & 0 & 0 & 0 \\
0 & \displaystyle\Omega_k & 0 & 0\\
0 & 0 &\displaystyle \frac{1}{\Omega_k} & 0 \\
0 & 0 & 0 & \displaystyle\Omega_k
\end{bmatrix}
\end{eqnarray}

\begin{eqnarray}
    G_k^{s=1} = 
\begin{bmatrix}
\displaystyle\frac{1}{\text{Re}(\Sigma _{\bf k})} & \displaystyle-\frac{\text{Im}(\Sigma _{\bf k})}{\text{Re}(\Sigma _{\bf k})} & 0 & 0 \\
\displaystyle-\frac{\text{Im}(\Sigma _{\bf k})}{\text{Re}(\Sigma _{\bf k})} & \displaystyle \frac{|\Sigma_{\bf k}|^2}{\text{Re}(\Sigma _{\bf k})} & 0 & 0 \\
0 & 0 & \displaystyle \frac{1}{\text{Re}(\Sigma _{-\bf k})} &\displaystyle -\frac{\text{Im}(\Sigma _{-\bf k})}{\text{Re}(\Sigma _{-\bf k})} \\
0 & 0 & \displaystyle-\frac{\text{Im}(\Sigma _{-\bf k})}{\text{Re}(-\Sigma _{\bf k})} & \displaystyle\frac{|\Sigma_{-\bf k}|^2}{\text{Re}(\Sigma _{-\bf k})}
\end{bmatrix}\nonumber\\
&&
\end{eqnarray}

Note that $\Sigma _{\bf k}$ and $\Sigma _{-\bf k}$ are already defined \eqref{eq:omegas}. 
Covariance matrix representation is actually equivalent to Wave function representation, as both contain same information in them. 

To make procedure simpler, we will consider two $2 \times 2$ blocks of above $4 \times 4$ covariance matrices $G$ of reference and target states, because other entries in them are zero, as follows - 
\begin{eqnarray}
 G_{k=0}^{s=0} &=& 
\begin{bmatrix}
\displaystyle \frac{1}{\Omega_k} & 0 \\
0 & \displaystyle\Omega_k \\
\end{bmatrix}, 
 G_{k=1}^{s=0} = 
\begin{bmatrix}
\displaystyle\frac{1}{\Omega_k} & 0 \\
0 & \displaystyle\Omega_k \\
\end{bmatrix}
\\
G_{k=0}^{s=1} &=&
\begin{bmatrix}
\displaystyle\frac{1}{\text{Re}(\Sigma _{\bf k})} & \displaystyle-\frac{\text{Im}(\Sigma _{\bf k})}{\text{Re}(\Sigma _{\bf k})} \\
\displaystyle-\frac{\text{Im}(\Sigma _{\bf k})}{\text{Re}(\Sigma _{\bf k})} & \displaystyle\frac{|\Sigma_{\bf k}|^2}{\text{Re}(\Sigma _{\bf k})}
\end{bmatrix}, 
\quad
    G_{k=1}^{s=1} =
\begin{bmatrix}
\displaystyle\frac{1}{\text{Re}(\Sigma _{- \bf k})} &\displaystyle -\frac{\text{Im}(\Sigma _{- \bf k})}{\text{Re}(\Sigma _{- \bf k})} \\
\displaystyle-\frac{\text{Im}(\Sigma _{- \bf k})}{\text{Re}(\Sigma _{- \bf k})} & \displaystyle\frac{|\Sigma_{- \bf k}|^2}{\text{Re}(\Sigma _{- \bf k})}
\end{bmatrix}, 
\end{eqnarray}

After doing this, we will be able to calculate complexity individually for each block and then we can sum it up to find total complexity.

We can now change basis for each block for simplicity of calculation:
\begin{equation}
    \Tilde{G}^{s=1} = SG^{s=1}S^T, \Tilde{G}^{s=0} = SG^{s=0}S^T
\end{equation}
given that,  $\Tilde{G}^{s=0} = 1$.  Here, $S$ is given by:
\begin{eqnarray}
    S =
    \begin{bmatrix}
   \displaystyle \sqrt{\Omega_k} & 0\\
    0 & \displaystyle\frac{1}{\sqrt{\Omega_k} }
    \end{bmatrix}
\end{eqnarray},
 which imply $\Tilde{G}^{s=0} = 1$ and 
\begin{eqnarray}
    \Tilde{G}^{s=1} = 
    \begin{bmatrix}
\displaystyle\frac{\Omega_k}{\text{Re}(\Sigma _{\bf k})} & \displaystyle-\frac{\text{Im}(\Sigma _{\bf k})}{\text{Re}(\Sigma _{\bf k})} \\
 & \\
\displaystyle-\frac{\text{Im}(\Sigma _{\bf k})}{\text{Re}(\Sigma _{\bf k})} & \displaystyle \frac{|\Sigma_{\bf k}|^2}{\Omega_k \text{Re}(\Sigma _{\bf k})}
\end{bmatrix}, 
\end{eqnarray}
The unitary evolution of wave functions with respect to covariance matrices can be expressed as, $\Tilde{G}^s = \Tilde{U}(\tau)\Tilde{G}^{s=0}\Tilde{U}(\tau)^T.$
 The unitary transformations are parametrized with gates satisfying $SL(2,R)$ algebra:

\begin{align}
    \Tilde{U}(\tau) = 
    \begin{bmatrix}
   \displaystyle \text{cos}(\mu(\tau))\text{cosh}(\rho(\tau))-\text{sin}(\theta(\tau))\text{sinh}(\rho(\tau)) & \displaystyle-\text{sin}(\mu(\tau))\text{cosh}(\rho(\tau))+\text{cos}(\theta(\tau))\text{sinh}(\rho(\tau))\\
    &\\
  \displaystyle  \text{sin}(\mu(\tau))\text{cosh}(\rho(\tau))+\text{cos}(\theta(\tau))\text{sinh}(\rho(\tau)) & 
   \displaystyle \text{cos}(\mu(\tau))\text{cosh}(\rho(\tau))+\text{sin}(\theta(\tau))\text{sinh}(\rho(\tau))
    \end{bmatrix}
\end{align}

where, $\mu, \rho, \theta$ are defined in $SL(2,R)$ group.
Considering following boundary conditions,
\begin{equation}
    \begin{aligned}
        \Tilde{G}^{s=1} &= \Tilde{U}(\tau = 1)\Tilde{G}^{s=0}\Tilde{U}(\tau =1)^T \\
        \Tilde{G}^{s=0} &= \Tilde{U}(\tau = 0)\Tilde{G}^{s=0}\Tilde{U}(\tau =0)^T
    \end{aligned}
\end{equation}
Additionally,  we have:
\begin{equation}
\label{eq:boundaryConditions}
    \begin{aligned}
    \left(\text{cosh}(2\rho(1)),\text{tan}(\theta(1)+\mu(1))\right) &= \left( \frac{\Omega_k^2 + |\Sigma_k|^2}{2\Omega \text{Re}(\Sigma_k)}, \frac{\Omega_k^2 - |\Sigma_k|^2}{2\Omega \text{Im}(\Sigma_k)} \right) \\
    \left( \rho(0), \theta(0) + \mu(0) \right) &= \left(0,c\right)
    \end{aligned}
\end{equation}
Again to make calculations simple,  we make following choice:
\begin{enumerate}
\item $\displaystyle \mu(\tau = 1) = \mu(\tau = 0)= 0$.
\item $\displaystyle \theta(\tau = 0) =\theta(\tau = 1) = c = \text{tan}^{-1}\left(\frac{\Omega_k^2 - |\Sigma_k|^2}{2\Omega \text{Im}(\Sigma_k)} \right)$.
\end{enumerate}
Hence the metric for $\Tilde{U}$ can be computed as:
\bea
    ds^2 = d\rho^2 + \text{cosh}(2\rho)\text{cosh}^2\rho d\mu^2 +\text{cosh}(2\rho)\text{sinh}^2\rho d\theta^2
    -\text{sinh}(2\rho)^2d\mu d\theta
\eea
    
The simple geodesic is a straight line on this geometry i.e.  $\rho(\tau) = \rho(1)\tau$.
From, the boundary conditions \eqref{eq:boundaryConditions}, we get:
\begin{equation}
    \rho_k(\tau = 1) = \frac{1}{2}\text{cosh}^{-1} \left[ \frac{\Omega_k^2 + |\Sigma _{\bf k}|^2 }{2\Omega_k\text{Re}(\Sigma _{\bf k})} \right]
\end{equation}
After summing over both the values of momentum mode we finally get: 

\begin{align}
\begin{split}
\label{eq:circuitComplexityCovariance}
C_1(\Omega_k) &=\rho_k(\tau = 1) +\rho_{-k}(\tau = 1) \\ 
&= \frac{1}{2}\Bigg[\text{cosh}^{-1} \left[ \frac{\Omega_k^2 + |\Sigma _{\bf k}|^2 }{2\Omega_k\text{Re}(\Sigma _{\bf k})} \right]  +\text{cosh}^{-1} \left[ \frac{\Omega_{-k}^2 + |\Sigma _{-\bf k}|^2 }{2\Omega_{-k}\text{Re}(\Sigma _{-\bf k})} \right] \Bigg]
\end{split}
\end{align}
\begin{align}
\begin{split}
 C_2(\Omega_k) &= \sqrt{\rho_k(\tau = 1)^2 +\rho_{-k}(\tau = 1)^2} \\ &= \frac{1}{2} \sqrt{\left( \text{cosh}^{-1} \left[ \frac{\Omega_k^2 + |\Sigma _{\bf k}|^2 }{2\Omega_k\text{Re}(\Sigma _{\bf k})} \right]  \right)^2 +\left( \text{cosh}^{-1} \left[ \frac{\Omega_{-k}^2 + |\Sigma _{-\bf k}|^2 }{2\Omega_{-k}\text{Re}(\Sigma _{-\bf k})} \right]  \right)^2 }
 \end{split}
\end{align}
After algebraic simplification we finally get:
\begin{align}
    C_1(\Omega_k) &= 4r_k, \\
    C_2(\Omega_k) &= 2\sqrt{2}r_k,\\
    C_1(\Omega_k) &= \sqrt{2}C_2(\Omega_k)
\end{align}
which is independent of  $\phi_k$.
For,  $r_k \rightarrow 0$,  we have $C_1 \approx 0$ and $C_2 \approx 0$.  We can see that covariance approach is interesting because it is not dependent upon squeezing angle $\phi_k$. 

It is worth mentioning that while calculating entanglement entropy, it will also not depend upon the squeezing angle $\phi_k$. Therefore, this observation will lead us to draw some similarity in behaviour of two quantities - entanglement entropy and Covariance approach of computing complexity.

\subsection*{\underline{\textcolor{red}{\bf Complexity via Nielsen's wave-function method:}}}

 The overall methodology of computing circuit complexity is same as in covariance matrix approach.  The exponent of the target state \eqref{eq:targetGaussianState} can be diagonalized as: 
\begin{eqnarray}
\label{eq:nieslenTarget}
    \psi_{\text{sq}} = \mathcal{N}\text{exp}\left(-\frac{1}{2} \Tilde{\mathcal{M}}^{ab}q_aq_b\right)
\end{eqnarray}
where, $\mathcal{N}$ is the normalization constant.  Here the matrix $\Tilde{\mathcal{M}}$ is defined as:
\begin{eqnarray}
    \Tilde{\mathcal{M}} &=& 
    \begin{bmatrix}
     \displaystyle  -2A+B & 0\\
      & \\
    0 &  \displaystyle  -2A -B
    \end{bmatrix} \nonumber\\
    &=& 
    \begin{bmatrix}
    \displaystyle  \Sigma _{\bf k} & 0\\
      & \\
    0 &  \displaystyle  \Sigma _{- \bf k} 
    \end{bmatrix}
\end{eqnarray}
The reference and target wave function are given by:
\bea
\label{eq:nielsenReference}
        \psi_{\text{R}} &=& \mathcal{N}\text{exp}\left( -\frac{\Omega_k}{2}\big(q_{\bf k}^2+ q_{-\bf k}^2 \big)\right) =  \mathcal{N}\text{exp}\left( \frac{1}{2} \sum_{k,-k}\Omega_k\bf k^2 \right)
\\
    \psi^\tau &=& \mathcal{N}\text{exp}\left( -\frac{1}{2}\left( v_a. \mathcal{A}_{ab}^\tau.v_b  \right) \right)
\eea
where, $v = (q_{\bf k}, q_{- \bf k})$ and $\mathcal{A}^\tau$ is a diagonal matrix.  For the target state eq: \eqref{eq:nieslenTarget},  and reference state eq: \eqref{eq:nielsenReference} we get:
\bea
    \mathcal{A}^{\tau=1} &=& \mathcal{M} = 
    \begin{bmatrix}
   \displaystyle   \Sigma _{\bf k} & 0\\
    &  \\
    0 &  \displaystyle  \Sigma _{- \bf k} 
    \end{bmatrix}
    \quad
    \mathcal{A}^{\tau=0} =
    \begin{bmatrix}
    \displaystyle   \Omega_k & 0\\
         &   \\
    0 &  \displaystyle  \Omega_{-k}
    \end{bmatrix}
\eea
The following unitary transformation is required here,
\begin{equation}
    \mathcal{A}^\tau = \mathcal{U}(\tau).\mathcal{A}^{\tau = 0 }.\mathcal{U}^T(\tau)
\end{equation}
The following constraints are used:
\begin{align}
\begin{split}
     \mathcal{A}^{\tau = 1} &= \mathcal{U}(\tau = 1).\mathcal{A}^{\tau = 0 }.\mathcal{U}^T(\tau = 1) \\
     \mathcal{A}^{\tau = 0} &= \mathcal{U}(\tau = 0).\mathcal{A}^{\tau = 0 }.\mathcal{U}^T(\tau = 0)
\end{split}    
\end{align}
In this case, $\mathcal{U}$ can be parameterized as in eq: \eqref{eq:controlHamiltonian} so that the required goal state is attained at $\tau = 1$.  Elementary gates are limited to $GL(2,C)$ unitaries because $\mathcal{A}^{\tau = 1}$ and $\mathcal{A}^{\tau = 0}$ can both have complex components.  The generators are $\mathcal{O}_I$, and the tangent vector components are complex parameters.  Now,  $Y^I$ can be written as follows:
\begin{equation}
    Y^I = \text{Tr}(\partial _{\tau} U(\tau)U^{-1}(\tau)(\mathcal{O}_I)^T)
\end{equation}
where,  we have note that:
\begin{eqnarray}
\text{Tr}(\mathcal{O}_I.\mathcal{O}_J^T) = \delta^{IJ},
\end{eqnarray} 
and $I,J = 0,1,2,3$. The metric is then given by: 
\begin{eqnarray}
ds^2 = G_{IJ}dY^IdY^{*J}.
\end{eqnarray}
For the sake of ease, we'll use the penalty factors $G_{IJ} = \delta^{IJ}$, where we set it to unity.  As the distance between states grows, the off-diagonal components in $GL(2,C)$ can be set to zero.  As a result of the $U(\tau)$,
\begin{equation}
    U(\tau) = \text{exp}\left(\sum_{i \in (k,-k)}\alpha^i(\tau)\mathcal{O}_i^{diagonal}\right)
\end{equation}
where, $\alpha^i(\tau)$ are complex parameters and $\mathcal{O}_i^{diagonal}$ are generators with identity at $i$ diagonal elements. The metric takes a simple form:
\begin{equation}
    ds^2 = \sum_{i \in (k,-k)} (d\alpha^{i,\text{Re}})^2 + (d\alpha^{i,\text{Im}})^2
\end{equation}
The geodesic is then given by:
\begin{equation}
    \alpha^{i,p}(\tau) = \alpha^{i,p}(\tau =1)+\alpha^{i,p}(\tau =0)
\end{equation}
for each $(i\in k,-k)$ and $(p = \text{Re and Im})$. Given the constraints,  we will get,
\begin{align}
\begin{split}
        \alpha^{i,\text{Re}}(\tau = 0) &= \alpha^{i,\text{Im}}(\tau = 0) = 0 \\
        \alpha^{i,\text{Re}}(\tau = 1) &= \frac{1}{2}\text{ln}\left| \frac{\Sigma _{\vec{i}}}{\omega _{\vec{i}}}\right| \\
        \alpha^{i,\text{Im}}(\tau = 1) &= \frac{1}{2}\text{tan}^{-1}\frac{\text{Im}(\Sigma_{\vec{i}})}{\text{Re}(\Sigma _{\vec{i}})}
\end{split}
\end{align}
for each $(i\in k,-k)$.
Now, the circuit complexity can be computed as:

\begin{align}
\nonumber
  C_1(\Omega_k) &= \alpha^{k,\text{Re}}(\tau = 1) + \alpha^{-k,\text{Re}}(\tau = 1)+ \alpha^{k,\text{Im}}(\tau = 1) + \alpha^{-k,\text{Im}}(\tau = 1)\\
  &=  \frac{1}{2} \Bigg( \text{ln}\left| \frac{\Sigma _{\bf k}}{\omega _{\bf k}}\right| +  \text{ln}\left| \frac{\Sigma _{-\bf k}}{\omega _{-\bf k}}\right| + \text{tan}^{-1}\frac{\text{Im}(\Sigma _{\bf k})}{\text{Re}(\Sigma _{\bf k})} + \text{tan}^{-1}\frac{\text{Im}(\Sigma_{-\bf k})}{\text{Re}(\Sigma _{-\bf k})}\Bigg) 
 \end{align} 
 
\begin{align}
\nonumber
      C_2(\Omega_k) &= \sqrt{(\alpha^{k,\text{Re}}(\tau = 1))^2 + (\alpha^{-k,\text{Re}}(\tau = 1))^2+ (\alpha^{k,\text{Im}}(\tau = 1))^2 + (\alpha^{-k,\text{Im}}(\tau = 1)})^2\\
      &=  \frac{1}{2} \sqrt{ \Bigg(  \text{ln}\left| \frac{\Sigma _{\bf k}}{\omega _{\bf k})}\right|\Bigg)^2  + \Bigg(\text{ln}\left| \frac{\Sigma _{-\bf k}}{\omega _{-\bf k})}\right|  \Bigg)^2 
   + \Bigg(\text{tan}^{-1}\frac{\text{Im}(\Sigma _{\bf k})}{\text{Re}(\Sigma _{\bf k})}  \Bigg)^2 + \Bigg(\text{tan}^{-1}\frac{\text{Im}(\Sigma _{-\bf k})}{\text{Re}(\Sigma _{-\bf k})}  \Bigg)^2}
\end{align}

Using expressions of $\Sigma _{\bf k}$, $\Sigma _{-\bf k}$, $\omega _{\bf k}$ and $\omega _{-\bf k}$ from Eq. \eqref{eq:omegas}, circuit complexity can be obtained as follows:

\begin{align}
 C_1(\Omega_k, \tau) =  \left|\text{ln}\left| \frac{1+\text{exp}(-2i\phi_k(\tau))\text{tanh}r_k(\tau)}{1-\text{exp}(-2i\phi_k(\tau))\text{tanh}r_k(\tau)}\right|\right| + \left| \text{tanh}^{-1}(\text{sin}(2\phi_k(\tau))\text{sinh}(2r_k(\tau))) \right|
\end{align}
\begin{align}
C_2(\Omega_k, \tau)  = \frac{1}{\sqrt{2}}\sqrt{\left(\text{ln}\left| \frac{1+\text{exp}(-2i\phi_k(\tau))\text{tanh}r_k(\tau)}{1-\text{exp}(-2i\phi_k(\tau))\text{tanh}r_k(\tau)}\right|\right)^2 + \left( \text{tanh}^{-1}(\text{sin}(2\phi_k(\tau))\text{sinh}(2r_k(\tau))) \right)^2}  
\end{align}
 As we can see, this approach for calculating circuit complexity of two mode squeezed states depends upon both squeezed parameters: $r_k$ and $\phi_k$
For the large squeezing parameter $r_k$ and $\phi_k \rightarrow -\frac{\pi}{2}$ we get:
\bea
    C_1(\Omega_k) \approx \sqrt{2}C_2(\Omega_k) \approx \left| \text{ln} \left( \frac{1-\text{tanh}r_k}{1+\text{tanh}r_k} \right) \right| \approx r_k.
\eea

\subsection{Complexity in PGW}
The evolution of the unitary operator $\mathcal{U}_{\rm PGW}(\tau)$ in eq: \ref{eq:unitary}, produced by the full quadratic quantized Hamiltonian functional is given by:
\begin{equation}
\textcolor{blue}{\bf PGW~Unitary~Operator:}~~~\mathcal{U}_{\rm PGW}(\tau,\tau_0) =\underbrace{ \underbrace{\hat{\mathcal{S}}_{\rm PGW}(r_{\lambda,{\bf k}}(\tau,\tau_0) ,\phi_{\lambda,{\bf k}}(\tau) )}_{\textcolor{red}{\bf PGW~Squuezing~Operator}}\underbrace{\hat{\mathcal{R}}_{\rm PGW}(\theta_{\lambda,{\bf k}}(\tau) )}_{\textcolor{red}{\bf PGW~Rotation~Operator}}}_{\textcolor{purple}{\bf Simple~factorization}},
\end{equation}
where $\hat{\mathcal{R}}_{\rm PGW}(\theta_{\lambda,{\bf k}}(\tau))$ is a two momentum mode PGW rotation operator given in eq: \ref{eq:rotationoperatorPGW} while $\hat{\mathcal{S}}_{\rm PGW}(r_{\lambda,{\bf k}}(\tau,\tau_0) ,\phi_{\lambda,{\bf k}}(\tau) )$ is the two momentum mode PGW squeezing operator in eq: \ref{eq:SqueezedoperatorPGW}. Since two mode rotation operator correspond to the free terms, we will mostly be interested in the squeezing operator. The complexity of the squeezing transformation has been calculated using covariance and Nielsen's approach in previous section. In the context of time evolution of PGW, one can choose different initial vacua states. We want to calculate quantum circuit complexity for CPT violating {\it Motta-Allen vacua state} ($|\alpha,\gamma\rangle_{\lambda,{\bf k},-{\bf k}}$) and the CPT preserving $\alpha$-{\it vacua state} ($|\alpha\rangle_{\lambda,{\bf k},-{\bf k}}$), which are related to the well known Euclidean ground state, represented by {\it Bunch-Davies vacuum} state ($|0\rangle_{\lambda,{\bf k},-{\bf k}}$) via {\it Bogoliubov transformation}, which are appended below:
\bea && \underline{\textcolor{red}{\bf Motta~ Allen~ initial~quantum ~vacua ~state:}}\nonumber\\
\label{MottaAllenIQVS}
&&|\alpha,\gamma\rangle_{\lambda,{\bf k},-{\bf k}}=\frac{1}{\sqrt{|\cosh\alpha|}}~\exp\left(-\frac{i}{2}\exp(-i\gamma)~{\rm tanh}\alpha~\hat{a}^{\dagger}_{\bf k}\hat{a}^{\dagger}_{-\bf k}\right)\ket{{\bf BD}}_{\lambda,{\bf k},-{\bf k}},\\
&& \underline{\textcolor{red}{\bf \alpha~ initial~quantum~ vacua~ state:}}\nonumber\\
&& |\alpha\rangle_{\lambda,{\bf k},-{\bf k}}=\frac{1}{\sqrt{|\cosh\alpha|}}~\exp\left(-\frac{i}{2}~{\rm tanh}\alpha~\hat{a}^{\dagger}_{\bf k}\hat{a}^{\dagger}_{-\bf k}\right)\ket{{\bf BD}}_{\lambda,{\bf k},-{\bf k}}.
\eea
So, we will individually compute quantum circuit complexity for all these three different vacua. In order to capture terms from Bogoliubov transformation and two-mode squeezing operator, we will choose {\it Bunch-Davies } initial state as the reference state and {\it Motta-Allen squeezed state} , $\alpha$-{\it squeezed  state}, and {\it Bunch-Davies squeezed state} as the target state respectively.

\subsubsection*{\underline{\textcolor{red}{\bf Motta~ Allen~ squeezed~quantum ~vacua ~state:}}}
The initial state $|\alpha,\gamma\rangle_{\lambda,{\bf k},-{\bf k}}$ is obtained by applying Bogoliubov transformation on the Bunch-Davies initial state:
\bea \nonumber\\
&&|\alpha,\gamma\rangle_{\lambda,{\bf k},-{\bf k}}=\frac{1}{\sqrt{|\cosh\alpha|}}~\exp\left(-\frac{i}{2}\exp(-i\gamma)~{\rm tanh}\alpha~\hat{a}^{\dagger}_{\bf k}\hat{a}^{\dagger}_{-\bf k}\right)\ket{{\bf BD}}_{\lambda,{\bf k},-{\bf k}}
\eea
while the target state is given by applying squeezing operator in the initial state:
$\ket{\psi_{\text{sq}}}^{\text{MA}}_{\vec{k},\vec{-k}} = \sum_{\lambda=+,\times}\bigotimes_{{\bf k}}\hat{\mathcal{S}}_{\rm PGW}(r_{\lambda,{\bf k}}(\tau),\phi_{\lambda,{\bf k}}(\tau))|\alpha,\gamma\rangle_{\lambda,{\bf k},-{\bf k}}$. The full target state is given in \ref{eq:fullwavefunction1}.
\bea
&& \ket{\Psi^{(\alpha,\gamma)}_{\bf sq}} = \frac{1}{\sqrt{|\cosh\alpha|}}~\sum_{\lambda=+,\times}\bigotimes_{{\bf k}} \frac{1}{\cosh r_{\lambda,{\bf k}}(\tau)}\nonumber\\
&&~~~~~~~~~~~~~~~~~~~~~~~~~~~~~~~~~~\sum_{n=0}^{\infty}(-1)^{n}\exp(-2in~\phi_{\lambda,{\bf k}}(\tau))\tanh^{n}r_{\lambda,{\bf k}}(\tau)\left\{\ket{n_{\lambda,{\bf k}}, n_{\lambda,-{\bf k}}}^{({\bf BD})}\right.\nonumber\\
&&\left.~~~~~~~~~~~~~~~~~~~~~~~~~~~~~+\sum^{\infty}_{m=1}\frac{(-1)^{m}}{m!}\left(\frac{i}{2}\right)^{m}~\exp(-im\gamma)~{\rm tanh}^{m}\alpha~\ket{(n+m)_{\lambda,{\bf k}}, (n+m)_{\lambda,-{\bf k}}}^{({\bf BD})}\right\}~~~~\nonumber \\
\eea

Because the initial state is obtained from the Bunch-Davies vacuum state, we will chose our reference state to be Bunch-Davies and the target state is obtained by applying two unitary transformations $U_2U_1$ where $U_1$ is the Bogoliubov transformation:
\bea \nonumber\\
&& U_1 =\frac{1}{\sqrt{|\cosh\alpha|}}~\exp\left(-\frac{i}{2}\exp(-i\gamma)~{\rm tanh}\alpha~\hat{a}^{\dagger}_{\bf k}\hat{a}^{\dagger}_{-\bf k}\right)
\eea
and $U_2$ is time-dependent squeezing transformation:
\bea
&&U_2 = \hat{\mathcal{S}}_{\rm PGW}(r_{\lambda,{\bf k}}(\tau) ,\phi_{\lambda,{\bf k}}(\tau) )= \exp\Bigg(\frac{\overbrace{r_{\lambda,{\bf k}}(\tau)}^{\textcolor{red}{\bf Amplitude}}}{2} \Big\{ \exp(-2i \underbrace{\phi_{\lambda,{\bf k}}(\tau)}_{\textcolor{red}{\bf Phase}} )\hat{a}_{\bf k}\hat{a}_{-{\bf k}} - \exp(2i \phi_{\lambda,{\bf k}}(\tau))\hat{a}_{-{\bf k}}^{\dagger}\hat{a}_{\bf k}^{\dagger} \Big\}\Bigg).\nonumber\\
&&
\eea
Now, because of the triangle inequality, the complexity from reference state to target state, $C$, is bounded by $C \leq C[U_1] + C[U_2]$ where $C[U_1]$ is the complexity of unitary transformation $U_1$ and $C[U_2]$ is the complexity of unitary transformation $U_2$. The complexity of $U_2$, $C[U_2]$, is same as for the two mode squeezed operator in previous section. One can also relate the Bogoliubov transformation, $U_1$ in \ref{MottaAllenIQVS} with the squeezing operator where the squeezing parameter $r$ is $r = \alpha $ and the squeezing angle is $\phi= \gamma/2$. Then using the result of circuit complexity for two-mode squeezed operator from previous section, we can write the expression for $C[U_1]$ too. Then, one can compute the complexity using both covariance matrix and Nielsen's wave function approaches: 

\subsubsection*{\underline{\textcolor{blue}{\bf Covariance Matrix approach:}}}
The complexity from Covariance Matrix approach is:
\begin{align}
    C_1(\Omega_k) &= C_1(U_1)+ C_1(U_2) = 4(r_k(\tau) + \alpha) \\
    C_2(\Omega_k) &= C_2(U_1)+ C_2(U_2) = 2\sqrt{2}(r_k(\tau) + \alpha)
\end{align}
Interestily circuit complexity using covariance matrix approach is independent of the angular terms like $\phi$ and $\gamma$.
\subsubsection*{\underline{\textcolor{blue}{\bf Nielsen's wave function approach:}}}
The total complexity from Nielsen's approach, up to the bound is given by:
\bea
&& C_1(\Omega_k) = C_1(U_1)+ C_1(U_2) \nonumber \\
&&~~~~~~~~~~=\left|\text{ln}\left| \frac{1+\text{exp}(-i\gamma)\text{tanh}\alpha}{1-\text{exp}(-i\gamma)\text{tanh}\alpha}\right|\right| + \left| \text{tanh}^{-1}(\text{sin}(\gamma)\text{sinh}(2\alpha) \right| \nonumber \\
&&~~~~~~~~~~~~+ \left|\text{ln}\left| \frac{1+\text{exp}(-2i\phi_k(\tau))\text{tanh}r_k(\tau)}{1-\text{exp}(-2i\phi_k(\tau))\text{tanh}r_k(\tau)}\right|\right| + \left| \text{tanh}^{-1}(\text{sin}(2\phi_k(\tau))\text{sinh}(2r_k(\tau))) \right|~~~~\nonumber\\
&&
\eea
\bea
&&  C_2(\Omega_k) = C_2(U_1)+ C_2(U_2) \nonumber \\
&&~~~~~~~~~~=\frac{1}{\sqrt{2}}\sqrt{\left( \left|\text{ln}\left| \frac{1+\text{exp}(-i\gamma)\text{tanh}\alpha}{1-\text{exp}(-i\gamma)\text{tanh}\alpha}\right|\right|\right)^2 + \left( \left| \text{tanh}^{-1}(\text{sin}(\gamma)\text{sinh}(2\alpha) \right|\right)^2} \nonumber \\
&&~~~~~~~~~~~~+\frac{1}{\sqrt{2}}\sqrt{\left(\text{ln}\left| \frac{1+\text{exp}(-2i\phi_k(\tau))\text{tanh}r_k(\tau)}{1-\text{exp}(-2i\phi_k(\tau))\text{tanh}r_k(\tau)}\right|\right)^2 + \left( \text{tanh}^{-1}(\text{sin}(2\phi_k(\tau))\text{sinh}(2r_k(\tau))) \right)^2}~~~~\nonumber\\
&&
\eea

This complexity $C_1(\Omega_k)$ and $C_2(\Omega_k)$ captures the effect from both squeezing operator and Bogoliubov transformation. $C(U_1)$ shows the complexity necessary for creating Motta-Allen Initial state which is as expected independent of time. $C(U_2)$ is the complexity of the squeezing operator applied on the initial state which is now dependent on time. 

\subsubsection*{\underline{\textcolor{red}{\bf $\alpha$ squeezed~quantum ~vacua ~state:}}}
The initial state $|\alpha \rangle_{\lambda,{\bf k},-{\bf k}}$ is obtained by applying Bogoliubov transformation on the Bunch-Davies initial state:
\bea
 |\alpha\rangle_{\lambda,{\bf k},-{\bf k}}=\frac{1}{\sqrt{|\cosh\alpha|}}~\exp\left(-\frac{i}{2}~{\rm tanh}\alpha~\hat{a}^{\dagger}_{\bf k}\hat{a}^{\dagger}_{-\bf k}\right)\ket{{\bf BD}}_{\lambda,{\bf k},-{\bf k}}.
\eea
while the target state is given by applying squeezing operator in the initial state:
$\ket{\psi_{\text{sq}}}^{\text{Alpha}}_{\vec{k},\vec{-k}} = \sum_{\lambda=+,\times}\bigotimes_{{\bf k}}\hat{\mathcal{S}}_{\rm PGW}(r_{\lambda,{\bf k}}(\tau),\phi_{\lambda,{\bf k}}(\tau)) |\alpha \rangle_{\lambda,{\bf k},-{\bf k}}$. The full target state is given by:
\bea
&& \ket{\psi_{\text{sq}}}^{\text{Alpha}}_{\vec{k},\vec{-k}} = \frac{1}{\sqrt{|\cosh\alpha|}}~\sum_{\lambda=+,\times}\bigotimes_{{\bf k}} \frac{1}{\cosh r_{\lambda,{\bf k}}(\tau)}\nonumber\\
&&~~~~~~~~~~~~~~~~~~~~~~~~~~~~~~~~~~\sum_{n=0}^{\infty}(-1)^{n}\exp(-2in~\phi_{\lambda,{\bf k}}(\tau))\tanh^{n}r_{\lambda,{\bf k}}(\tau)\left\{\ket{n_{\lambda,{\bf k}}, n_{\lambda,-{\bf k}}}^{({\bf BD})}\right.\nonumber\\
&&\left.~~~~~~~~~~~~~~~~~~~~~~~~~~~~~+\sum^{\infty}_{m=1}\frac{(-1)^{m}}{m!}\left(\frac{i}{2}\right)^{m}~{\rm tanh}^{m}\alpha~\ket{(n+m)_{\lambda,{\bf k}}, (n+m)_{\lambda,-{\bf k}}}^{({\bf BD})}\right\},~~~~\nonumber\\
&&
\eea

Like in Motta-Allen case, we will choose our reference state to be Bunch davies vacuum state, and our target state is obtained by applying two unitary transformations $U_2U_1$ where $U_1$ is the Bogoliubov transformation:
\bea \nonumber\\
&& U_1 =\frac{1}{\sqrt{|\cosh\alpha|}}~\exp\left(-\frac{i}{2}~{\rm tanh}\alpha~\hat{a}^{\dagger}_{\bf k}\hat{a}^{\dagger}_{-\bf k}\right)
\eea

and $U_2$ is time-dependent squeezing transformation:
\bea
&&U_2 = \hat{\mathcal{S}}_{\rm PGW}(r_{\lambda,{\bf k}}(\tau) ,\phi_{\lambda,{\bf k}}(\tau) )= \exp\Bigg(\frac{\overbrace{r_{\lambda,{\bf k}}(\tau)}^{\textcolor{red}{\bf Amplitude}}}{2} \Big\{ \exp(-2i \underbrace{\phi_{\lambda,{\bf k}}(\tau)}_{\textcolor{red}{\bf Phase}} )\hat{a}_{\bf k}\hat{a}_{-{\bf k}} - \exp(2i \phi_{\lambda,{\bf k}}(\tau))\hat{a}_{-{\bf k}}^{\dagger}\hat{a}_{\bf k}^{\dagger} \Big\}\Bigg).\nonumber\\
&&
\eea

The operator $U_1$ is also a squeezing operator with squeezing angle $\phi = 0$. The overall framework to compute complexity is already discussed in Motta-Allen case, so will now directly write down the result.
\subsubsection*{\underline{\textcolor{blue}{\bf Covariance Matrix approach:}}}
The complexity from Covariance Matrix approach is:
\begin{align}
    C_1(\Omega_k) &= C_1(U_1)+ C_1(U_2) = 4(r_k(\tau) + \alpha) \\
    C_2(\Omega_k) &= C_2(U_1)+ C_2(U_2) = 2\sqrt{2}(r_k(\tau) + \alpha)
\end{align}
Final results in this case are independent of phases $\phi_{\bf k}$ and $\gamma$.
\subsubsection*{\underline{\textcolor{blue}{\bf Nielsen's wave function approach:}}}
The complexity from Nielsen's approach, up to the bound is given by:
\bea
&& C_1(\Omega_k) = C_1(U_1)+ C_1(U_2) \nonumber \\
&&~~~~~~~~~~=\left|\text{ln}\left| \frac{1+\text{tanh}\alpha}{1-\text{tanh}\alpha}\right|\right| + \left| \text{tanh}^{-1}(\text{sinh}(2\alpha) \right| \nonumber \\
&&~~~~~~~~~~~~+ \left|\text{ln}\left| \frac{1+\text{exp}(-2i\phi_k(\tau))\text{tanh}r_k(\tau)}{1-\text{exp}(-2i\phi_k(\tau))\text{tanh}r_k(\tau)}\right|\right| + \left| \text{tanh}^{-1}(\text{sin}(2\phi_k(\tau))\text{sinh}(2r_k(\tau))) \right|~~~~\nonumber\\
&&
\eea
\bea
&&  C_2(\Omega_k) = C_2(U_1)+ C_2(U_2) \nonumber \\
&&~~~~~~~~~~=\frac{1}{\sqrt{2}}\sqrt{\left( \left|\text{ln}\left| \frac{1+\text{tanh}\alpha}{1-\text{tanh}\alpha}\right|\right|\right)^2 + \left( \left| \text{tanh}^{-1}(\text{sinh}(2\alpha) \right|\right)^2} \nonumber \\
&&~~~~~~~~~~~~+\frac{1}{\sqrt{2}}\sqrt{\left(\text{ln}\left| \frac{1+\text{exp}(-2i\phi_k(\tau))\text{tanh}r_k(\tau)}{1-\text{exp}(-2i\phi_k(\tau))\text{tanh}r_k(\tau)}\right|\right)^2 + \left( \text{tanh}^{-1}(\text{sin}(2\phi_k(\tau))\text{sinh}(2r_k(\tau))) \right)^2}~~~~\nonumber\\
&&
\eea

\subsubsection*{\underline{\textcolor{red}{\bf Bunch~ Davies~ squeezed~quantum ~vacua ~state:}}}
The reference state is $\ket{0}_{{\bf k},-{\bf k}}$ while the target state is given by:
\begin{equation}
   \ket{\Psi_{\bf sq}}^{\text{BD}}_{{\bf k},-{\bf k}} = \sum_{\lambda=+,\times}\bigotimes_{{\bf k}}\hat{\mathcal{S}}_{\rm PGW}(r_{\lambda,{\bf k}}(\tau),\phi_{\lambda,{\bf k}}(\tau))\ket{0}_{{\bf k},-{\bf k}} 
\end{equation}
Since the complexity of two-mode squeezed transformation has already been obtained in previous sections and it corresponds to the Bunch-Davies case, we will directly quote down the result. 
\subsubsection*{\underline{\textcolor{blue}{\bf Covariance Matrix approach:}}}
The complexity from Covariance Matrix approach is:
\begin{align}
    C_1(\Omega_k) &= 4r_k \\
    C_2(\Omega_k) &= 2\sqrt{2}r_k
\end{align}
Final results in this case are independent of phase $\phi_{\bf k}$.
\subsubsection*{\underline{\textcolor{blue}{\bf Nielsen's wave function approach:}}}
The complexity from Nielsen's wave function approach is:
\begin{align}
 C_1(\Omega_k, \tau) =  \left|\text{ln}\left| \frac{1+\text{exp}(-2i\phi_k(\tau))\text{tanh}r_k(\tau)}{1-\text{exp}(-2i\phi_k(\tau))\text{tanh}r_k(\tau)}\right|\right| + \left| \text{tanh}^{-1}(\text{sin}(2\phi_k(\tau))\text{sinh}(2r_k(\tau))) \right|
\end{align}
\begin{align}
C_2(\Omega_k, \tau)  = \frac{1}{\sqrt{2}}\sqrt{\left(\text{ln}\left| \frac{1+\text{exp}(-2i\phi_k(\tau))\text{tanh}r_k(\tau)}{1-\text{exp}(-2i\phi_k(\tau))\text{tanh}r_k(\tau)}\right|\right)^2 + \left( \text{tanh}^{-1}(\text{sin}(2\phi_k(\tau))\text{sinh}(2r_k(\tau))) \right)^2}  
\end{align}
One can also directly obtain the expression of quantum circuit complexity for Bunch-Davies squeezed quantum vacua state by directly substituting parameters like $\alpha$ and $\gamma$ in Motta-Allen and $\alpha$ squeezed cases to be zero. This gives the consistency check for our lengthy calculations.

\section{Entanglement Entropy of PGW from squeezed states}
\label{sec:entanglementEntropy}
In this section,  our object is to compute entanglement entropy using the squeezed states for PGW.  In terms of number states,  the reduced density matrix for the two modes are given by:
\bea
    \hat{\rho}_k &=& \sum_{n = 0}^\infty \frac{1}{(\text{cosh }r_k)^2}(\text{tanh }r_k)^{2n}\bra{n}_{\text{k k}}\ket{n},\quad
    \hat{\rho}_{-k} = \sum_{n = 0}^\infty \frac{1}{(\text{cosh }r_{-k})^2}(\text{tanh }r_{-k})^{2n}\bra{n}_{\text{-k -k}}\ket{n}.\quad\quad\quad
\eea
The corresponding probability of having $n$ photons in a single mode having momenta $k$ or $-k$ is given by the following expression:
\begin{equation}
    P_n^{(i)} = \frac{(\text{tanh}r_k)^{2n}}{(\text{cosh}r_k)^2}, \textit{}{ i = a,b}
\end{equation}
There are different measures exist for entropy in the context of quantum information theory.  Von-Neumann and Renyi Entanglment measures for entropies are commonly used in this context.  In the present context Von-Neumann entanglement entropy can be computed as:
\begin{align}
\begin{split}
\label{eq:entanglementEntropy}
S(\hat{\rho}_k) &= - \sum_{n = 0}^\infty P_n \text{ ln}P_n = S(\hat{\rho}_{-k})\\
    &= - \sum_{n = 0}^\infty \frac{\text{tanh}^{2n}r_k}{\text{cosh}^2r_k} \text{ln}\frac{\text{tanh}^{2n}r_k}{\text{cosh}^2r_k} \\
    &= -\sum_{n = 0}^\infty \frac{\text{tanh}^{2n}r_k}{\text{cosh}^2r_k}  \big({\text{ln}(\text{tanh}^{2n}r_k) - \text{ln}(\text{cosh}^{2}r_k)  \big) }\\
    &= \text{ln(cosh}^2r_k) \text{cosh}^2r_k - \text{ln(sinh}^2r_k) \text{sinh}^2r_k
\end{split}
\end{align}
The rise in entropy with increasing $r _k$ is then evident. Since the squeezed states are pure states, we didn't calculate the entropy associated with them because it will be zero. Instead, entropy for the reduced density matrix has been computed.  Further using the formalism Renyi-entropy can be computed as:
\begin{align}
\begin{split}
   \label{eq:renyi-entropy eqn}
    S_\mu  &= \frac{1}{1-\mu} \text{ln} \sum_{n= 1}^d P_n  = \frac{2\mu \text{ ln coshr}_k + \text{ln}(1- \text{tanh}^{2\mu}r_k)}{\mu -1}   
\end{split}
\end{align}
where $\mu \geq 0$ is the Renyi Parameter and $d$ is the Schmidt rank. Again, we can see that Renyi entropy increases with increasing squeezing parameter $r_k$. For large values of $r_k$, we get:
\begin{equation}
    S_\mu(r_k \rightarrow \infty) \approx \frac{2 \mu r_k}{(\mu - 1)}
\end{equation}

If we take the limit $\mu \rightarrow 1$, we get the Von-Neumann entropy \ref{eq:entanglementEntropy}. Meanwhile, Renyi-2 entropy is given by $S_2(r_k) = \text{ln cosh}2r_k$. 

We can also compute effective temperature in thermal distribution
$\langle\hat{n}_i\rangle = \text{sinh}^2 r_k$. The average photon number is: 
\begin{equation}
    \langle\hat{n}_i\rangle = \Bar{n} = \frac{1}{\text{exp}(\hbar\omega/k_BT)-1}
\end{equation}
Then, effective temperature is given by:
\begin{align}
\nonumber
    T &= \frac{\hbar\omega_i}{k_B} \text{ln}\left( \frac{\langle\hat{n}_i\rangle}{ \langle\hat{n}_i\rangle +1 } \right) \\ \nonumber
    &=  \frac{\hbar\omega_i}{k_B} \text{ln}\left( \frac{\text{sinh}^2 r_k }{\text{sinh}^2 r_k +1} \right) \\
    &= \frac{\hbar\omega_i}{2k_B\text{ln(coth}r_k)}
\end{align}
where, $\omega_i = i/c$ is the frequency of the mode and $i \in (k,-k)$.

In the context of time evolution of PGW, one can choose different initial vacua states. In this paper we are interested in calculating quantum circuit complexity for CPT violating {\it Motta-Allen vacua state} ($|\alpha,\gamma\rangle_{\lambda,{\bf k},-{\bf k}}$) and the CPT preserving $\alpha$-{\it vacua state} ($|\alpha\rangle_{\lambda,{\bf k},-{\bf k}}$), which are related to the well known Euclidean ground state, represented by {\it Bunch-Davies vacuum} state ($|0\rangle_{\lambda,{\bf k},-{\bf k}}$) via {\it Bogoliubov transformation}, which are appended below:
\bea && \underline{\textcolor{red}{\bf Motta~ Allen~ initial~quantum ~vacua ~state:}}\nonumber\\
&&|\alpha,\gamma\rangle_{\lambda,{\bf k},-{\bf k}}=\frac{1}{\sqrt{|\cosh\alpha|}}~\exp\left(-\frac{i}{2}\exp(-i\gamma)~{\rm tanh}\alpha~\hat{a}^{\dagger}_{\bf k}\hat{a}^{\dagger}_{-\bf k}\right)\ket{{\bf BD}}_{\lambda,{\bf k},-{\bf k}},\\
&& \underline{\textcolor{red}{\bf \alpha~ initial~quantum~ vacua~ state:}}\nonumber\\
\label{alphaIQVS}
&& |\alpha\rangle_{\lambda,{\bf k},-{\bf k}}=\frac{1}{\sqrt{|\cosh\alpha|}}~\exp\left(-\frac{i}{2}~{\rm tanh}\alpha~\hat{a}^{\dagger}_{\bf k}\hat{a}^{\dagger}_{-\bf k}\right)\ket{{\bf BD}}_{\lambda,{\bf k},-{\bf k}}.
\eea
So, we will individually compute entanglement entropy for all these three different vacua.
The initial state $|\alpha,\gamma\rangle_{\lambda,{\bf k},-{\bf k}}$ is obtained by applying Bogoliubov transformation on the Bunch-Davies initial vacuum state
\bea \nonumber\\
&&|\alpha,\gamma\rangle_{\lambda,{\bf k},-{\bf k}}=\frac{1}{\sqrt{|\cosh\alpha|}}~\exp\left(-\frac{i}{2}\exp(-i\gamma)~{\rm tanh}\alpha~\hat{a}^{\dagger}_{\bf k}\hat{a}^{\dagger}_{-\bf k}\right)\ket{{\bf BD}}_{\lambda,{\bf k},-{\bf k}}
\eea
while the target state is given by applying squeezing operator in the initial state:
$\ket{\psi_{\text{sq}}}^{\text{MA}}_{\vec{k},\vec{-k}} = \sum_{\lambda=+,\times}\bigotimes_{{\bf k}}\hat{\mathcal{S}}_{\rm PGW}(r_{\lambda,{\bf k}}(\tau),\phi_{\lambda,{\bf k}}(\tau))|\alpha,\gamma\rangle_{\lambda,{\bf k},-{\bf k}}$. The full target state is given in \ref{eq:fullwavefunction1}.

\subsubsection*{\underline{\textcolor{red}{\bf Motta~ Allen~ squeezed~quantum ~vacua ~state:}}}
The initial state $|\alpha,\gamma\rangle_{\lambda,{\bf k},-{\bf k}}$ is obtained by applying Bogoliubov transformation on the Bunch-Davies initial vacuum state
\bea \nonumber\\
&&|\alpha,\gamma\rangle_{\lambda,{\bf k},-{\bf k}}=\frac{1}{\sqrt{|\cosh\alpha|}}~\exp\left(-\frac{i}{2}\exp(-i\gamma)~{\rm tanh}\alpha~\hat{a}^{\dagger}_{\bf k}\hat{a}^{\dagger}_{-\bf k}\right)\ket{{\bf BD}}_{\lambda,{\bf k},-{\bf k}}
\eea
while the target state is given by applying squeezing operator in the initial state:
$\ket{\psi_{\text{sq}}}^{\text{MA}}_{\vec{k},\vec{-k}} = \sum_{\lambda=+,\times}\bigotimes_{{\bf k}}\hat{\mathcal{S}}_{\rm PGW}(r_{\lambda,{\bf k}}(\tau),\phi_{\lambda,{\bf k}}(\tau))|\alpha,\gamma\rangle_{\lambda,{\bf k},-{\bf k}}$. The Motta-Allen squeezed quantum vacua state is given in \ref{eq:fullwavefunction1}:
\bea
&& \ket{\Psi^{(\alpha,\gamma)}_{\bf sq}} =  \frac{1}{\sqrt{|\cosh\alpha|}}~\sum_{\lambda=+,\times}\bigotimes_{{\bf k}} \frac{1}{\cosh r_{\lambda,{\bf k}}(\tau)}\nonumber\\
&&~~~~~~~~~~~~~~~~~~~~~~~~~~~~~~~~~~\sum_{n=0}^{\infty}\sum_{m=0}^{\infty}\frac{(-1)^{n+m}}{n!m!}\left(\frac{i}{2}\right)^{m}\exp(-i(m\gamma+2n~\phi_{\lambda,{\bf k}}(\tau)))\nonumber\\
&&~~~~~~~~~~~~~~~~~~~~~~~~~~~~~~~~~~~~~~~~~~~~~{\rm tanh}^{m}\alpha\tanh^{n}r_{\lambda,{\bf k}}(\tau)\big( \hat{a}_{{\bf k}}^{\dagger} \big)^{n+m}\big( \hat{a}_{-{\bf k}}^{\dagger} \big)^{n+m}\ket{\bf BD}_{\lambda,{\bf k},-{\bf k}}\nonumber\\
&&
\eea 
Now we can compute Von-Neumann entanglement entropy for Motta-Allen  squeezed quantum  vacua  state as:
\bea
\label{eq:mottaAllenEntanglementEntropy}
&&S(\hat{\rho}_k) =  S(\hat{\rho}_{-k}) \nonumber \\
&&~~~~~~~~~ = \sum_{n = 0}^\infty \sum_{m = 0}^\infty \frac{\text{tanh}^{2n}r_k\text{tanh}^{2m}\alpha}{\text{cosh}^2r_k\text{cosh}\alpha} \text{ln}\frac{\text{tanh}^{2n}r_k \text{tanh}^{2m}\alpha}{\text{cosh}^2r_k \text{cosh}\alpha} \nonumber \\
&&~~~~~~~~~ = \sum_{n = 0}^\infty \sum_{m = 0}^\infty \Biggr\{ \biggr[\frac{\text{tanh}^{2n}r_k\text{tanh}^{2m}\alpha}{\text{cosh}^2r_k\text{cosh}\alpha} \biggr] \big\{ \text{ln}[\text{cosh}^2r_k \text{cosh}\alpha] - \text{ln} \text[(\text{tanhr}_k)^{2n}] - \text{ln}\text[(\text{tanh}\alpha)^{2m}] \big\} \Biggr \} \nonumber \\
&&~~~~~~~~~ =  \text{cosh}\alpha \big\{(1 + \text{sinh}^2r_k) \text{ln(cosh}^2r_k) - \text{sinh}^2r_k \text{ln(sinh}^2r_k)   \nonumber \\
&&~~~~~~~~~~~~~~~~~~ + (\frac{1}{2} + \text{sinh}^2\alpha) \text{ln(cosh}^2\alpha)  -  \text{sinh}^2\alpha \text{ln(sinh}^2\alpha) \big\} \nonumber \\
&&
\eea

\subsubsection*{\underline{\textcolor{red}{\bf $\alpha$ squeezed~quantum ~vacua ~state:}}}
The initial state $|\alpha \rangle_{\lambda,{\bf k},-{\bf k}}$ is obtained by applying Bogoliubov transformation on the Bunch-Davies initial vacuum state
\bea
 |\alpha\rangle_{\lambda,{\bf k},-{\bf k}}=\frac{1}{\sqrt{|\cosh\alpha|}}~\exp\left(-\frac{i}{2}~{\rm tanh}\alpha~\hat{a}^{\dagger}_{\bf k}\hat{a}^{\dagger}_{-\bf k}\right)\ket{{\bf BD}}_{\lambda,{\bf k},-{\bf k}}.
\eea
while the target state is given by applying squeezing operator in the initial state:
$\ket{\psi_{\text{sq}}}^{\text{Alpha}}_{\vec{k},\vec{-k}} = \sum_{\lambda=+,\times}\bigotimes_{{\bf k}}\hat{\mathcal{S}}_{\rm PGW}(r_{\lambda,{\bf k}}(\tau),\phi_{\lambda,{\bf k}}(\tau)) |\alpha \rangle_{\lambda,{\bf k},-{\bf k}}$. The full target state is given by:
\bea
&& \ket{\psi_{\text{sq}}}^{\text{Alpha}}_{\vec{k},\vec{-k}} = \frac{1}{\sqrt{|\cosh\alpha|}}~\sum_{\lambda=+,\times}\bigotimes_{{\bf k}} \frac{1}{\cosh r_{\lambda,{\bf k}}(\tau)}\nonumber\\
&&~~~~~~~~~~~~~~~~~~~~~~~~~~~~~~~~~~\sum_{n=0}^{\infty}\sum_{m=0}^{\infty}\frac{(-1)^{n+m}}{n!m!}\left(\frac{i}{2}\right)^{m}\exp(-2in~\phi_{\lambda,{\bf k}}(\tau))\nonumber\\
&&~~~~~~~~~~~~~~~~~~~~~~~~~~~~~~~~~~~~~~~~~~~~~{\rm tanh}^{m}\alpha\tanh^{n}r_{\lambda,{\bf k}}(\tau)\big( \hat{a}_{{\bf k}}^{\dagger} \big)^{n+m}\big( \hat{a}_{-{\bf k}}^{\dagger} \big)^{n+m}\ket{\bf BD}_{\lambda,{\bf k},-{\bf k}}\nonumber\\
&&
\eea
Now we can compute Von-Neumann entanglement entropy for $\alpha$ squeezed quantum vacua state as:
\bea
&&S(\hat{\rho}_k) =  S(\hat{\rho}_{-k}) \nonumber \\
&&~~~~~~~~~ = \sum_{n = 0}^\infty \sum_{m = 0}^\infty \frac{\text{tanh}^{2n}r_k\text{tanh}^{2m}\alpha}{\text{cosh}^2r_k\text{cosh}\alpha} \text{ln}\frac{\text{tanh}^{2n}r_k \text{tanh}^{2m}\alpha}{\text{cosh}^2r_k \text{cosh}\alpha} \nonumber \\
&&~~~~~~~~~ = \sum_{n = 0}^\infty \sum_{m = 0}^\infty \Biggr\{ \biggr[\frac{\text{tanh}^{2n}r_k\text{tanh}^{2m}\alpha}{\text{cosh}^2r_k\text{cosh}\alpha} \biggr] \big\{ \text{ln}[\text{cosh}^2r_k \text{cosh}\alpha] - \text{ln} \text[(\text{tanhr}_k)^{2n}] - \text{ln}\text[(\text{tanh}\alpha)^{2m}] \big\} \Biggr \} \nonumber \\
&&~~~~~~~~~ =  \text{cosh}\alpha \big\{(1 + \text{sinh}^2r_k) \text{ln(cosh}^2r_k) - \text{sinh}^2r_k \text{ln(sinh}^2r_k)   \nonumber \\
&&~~~~~~~~~~~~~~~~~~ + (\frac{1}{2} + \text{sinh}^2\alpha) \text{ln(cosh}^2\alpha)  -  \text{sinh}^2\alpha \text{ln(sinh}^2\alpha) \big\} \nonumber \\
&&
\eea

\subsubsection*{\underline{\textcolor{red}{\bf Bunch~ Davies~ squeezed~quantum ~vacua ~state:}}}
The reference state is $\ket{0}_{{\bf k},-{\bf k}}$ while the target state is given by:
\begin{equation}
   \ket{\Psi_{\bf sq}}^{\text{BD}}_{{\bf k},-{\bf k}} = \sum_{\lambda=+,\times}\bigotimes_{{\bf k}}\hat{\mathcal{S}}_{\rm PGW}(r_{\lambda,{\bf k}}(\tau),\phi_{\lambda,{\bf k}}(\tau))\ket{0}_{{\bf k},-{\bf k}} 
\end{equation}
Since the entanglement entropy of two-mode squeezed transformation has already been obtained and it corresponds to the Bunch-Davies case, we will directly quote down the result. 
\begin{align}
\begin{split}
S(\hat{\rho}_k) &= - \sum_{n = 0}^\infty P_n \text{ ln}P_n = S(\hat{\rho}_{-k})\\
    &= - \sum_{n = 0}^\infty \frac{\text{tanh}^{2n}r_k}{\text{cosh}^2r_k} \text{ln}\frac{\text{tanh}^{2n}r_k}{\text{cosh}^2r_k} \\
    &= -\sum_{n = 0}^\infty \frac{\text{tanh}^{2n}r_k}{\text{cosh}^2r_k}  \big({\text{ln}(\text{tanh}^{2n}r_k) - \text{ln}(\text{cosh}^{2}r_k)  \big) }\\
    &= \text{ln(cosh}^2r_k) \text{cosh}^2r_k - \text{ln(sinh}^2r_k) \text{sinh}^2r_k
\end{split}
\end{align}

\section{Numerical results and interpretation}
\label{sec:Numeric}

The primary goal of this section is to numerically resolve the time-dependent squeezed state parameter $r_{\bf k}(\tau)$ and squeezed angle $\phi_{\bf k}(\tau)$, which are provided in the equations \Cref{eq:diffneqns1} and \Cref{eq:diffeqns2}. The scale factor $a(\tau)$ has been selected as the dynamical variable instead of the conformal time $\tau$. As a result, the calculation is made to be simpler and easy to physically support. The differential operator in the aforementioned evolution equations must be replaced with the following one using the chain rule in order to transform the variable from $\tau\rightarrow a(\tau)$:
\bea \frac{d}{d\tau}=\frac{d}{da(\tau)}\frac{da(\tau)}{d\tau}=a'(\tau)\frac{d}{da(\tau)}.\eea
In terms of the new variable the time evolution equations of the squeezing parameters are given by:
\begin{align} \label{eq:diffeqnswa1}
&\frac{dr_{\bf k}(a)}{da} = -\frac{\lambda_{\bf k}(a)}{a'} \cos 2\phi_{\bf k}(a),\\
\label{eq:diffeqnswa2}
&\frac{d\phi_{\bf k}(a)}{da} = \frac{\Omega_{\bf k}}{a'} -\frac{\lambda_{\bf k}(a)}{a'} \coth2 r_{\bf k}(a)\sin 2\phi_{\bf k}(a) 
\end{align} 
Here the dispersion relation is derived in Appendix \ref{sec:dispersionRelation}, for general initial quantum states.

We will now briefly comment on the choice of cost functions for circuit complexity used in our calculations for circuit complexity.
Circuit complexity $C_1$ is close to the counting of gates in quantum computation while circuit complexity $C_2$ is the geodesic distance in the manifold of unitaries. We have computed the circuit complexity for both: linear $C_1$ and quadratic circuit complexity $C_2$ of using Covariance and Nielsen's approach. These two different approaches give the different structure of circuit complexity.
 These differences are discussed below:
\begin{itemize}
   \item Covariance measure of circuit complexity is not sensitive to the squeezing angle $\phi_k$ while circuit complexity obtained via Nielsen's approach is. Since, entanglement entropy is also independent of the the squeezing angle $\phi_k$ it is easier to make comparison of entropy with Covariance measure rather than Nielsen's ones.
   \item
   The circuit complexity via covariance approach is always linearly dependent on the squeezing parameter $r_k$ while this is not true for Nielsen's measure.  Furthermore on different limiting conditions,   structure of the Nielsen's measure of complexity  can be very different. In contrast, we always have one limiting condition in Covariance measure: $C_1(\Omega_k) = \sqrt{2}C_2(\Omega_k) = 4\sqrt{2}r_k$.
    \item Nielsen's measure of circuit complexity is sensitive to the details of evolution of the wave function while covariance measure is not. This is because Nielsen's measure is dependent on both parameters: squeezing angle $\phi_k$ and squeezing parameter $r_k$.
\end{itemize}

\subsection{De Sitter}
In Figure (\ref{fig:deSitterPlots}), we have numerically plotted the squeezing parameters, derived circuit complexity measures for Bunch-Davies Vacua, $\alpha$ Vacua and Motta-Allen Vacua, comparison of entanglement entropy with circuit complexity measure and quantum chaos in de Sitter Model on super-horizon scales for the parameters $k = 0.1$, $\alpha = 0.2$ and $ \gamma = 0.4$. We have set $r_k(a = 1) = 1, \phi_k(a = 1) = 1$ as our initial conditions. 
\subsubsection*{Squeezing Parameters}
In Figure (\ref{fig:deSitterPlots}.a), we have plotted squeezing parameters $r_k$ and $\phi_K$ for each vacua using the parameters listed above. Both $r_k$ and $\phi_k$ are oscillating for each vacua before $a= 0.5$. However, interestingly after that point these squeezing parameters starts to converge at each other and stops to oscillate. Since both circuit complexity and entanglement entropy are strongly dependent on the values of these parameters, the behaviour of these parameters are significant to understand the dynamics of the system.  
\subsubsection*{Complexity Measure}
\begin{figure}[h!]
	\centering
	\includegraphics[width=\textwidth,height=\textheight,keepaspectratio]{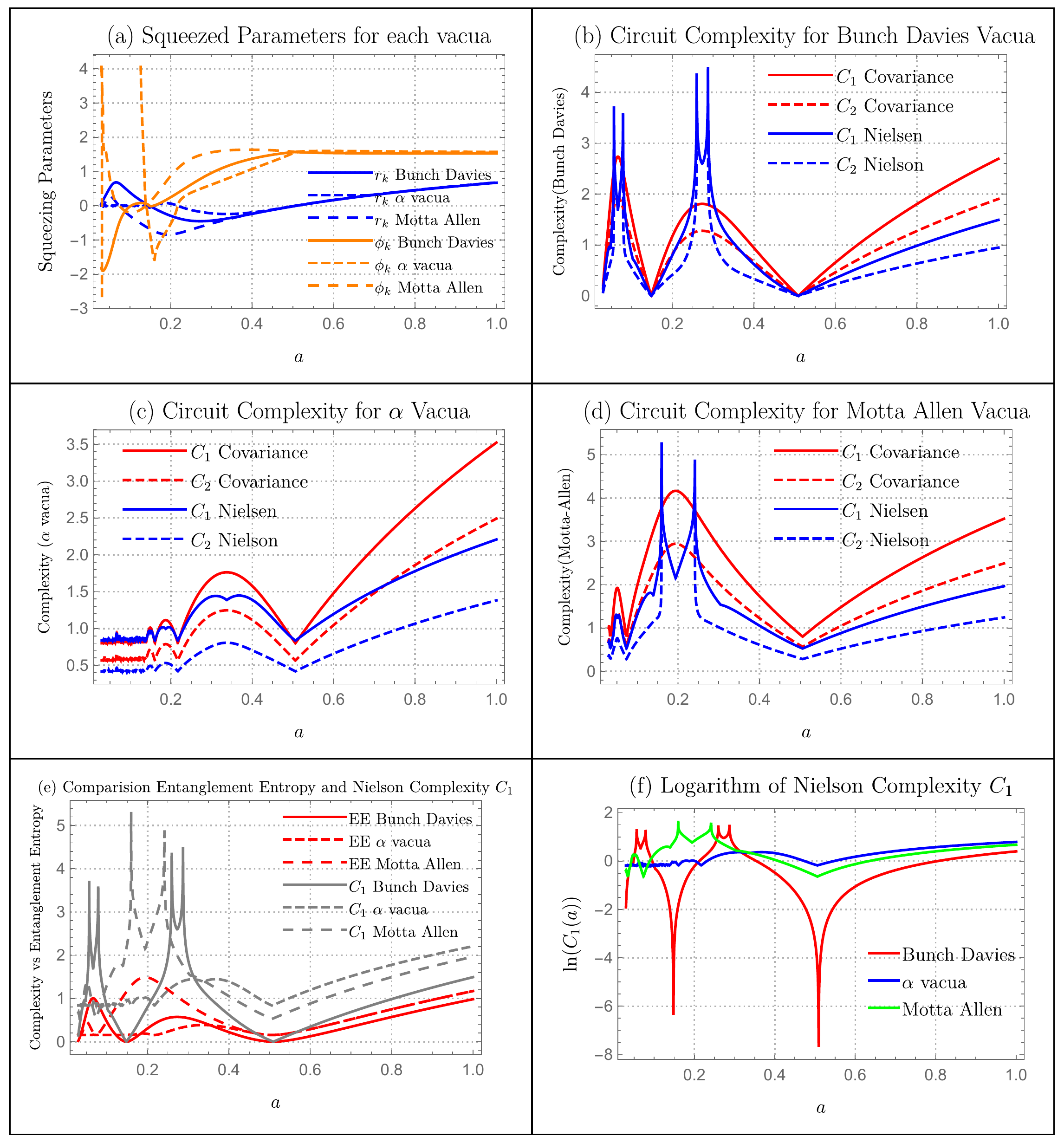}
	\caption{\textbf{De Sitter Model }(a.) Behaviour of the squeezing parameters in De Sitter Model for $k = 0.1$, $\alpha = 0.2$ and $ \gamma = 0.4$ for Bunch-Davies, $\alpha$ and Motta-Allen Vacua (b,c,d) Circuit complexity $C_1$ and $C_2$ for Bunch-Davies Vacua, $\alpha$ Vacua and Motta-Allen Vacua respectively using both Nielsen and Covariance approach for same parameters as in fig.a (e.) Comparison of entanglement entropy with Nielsen circuit complexity $C_1$ (f.) Logarithm of Nielsen circuit complexity $C_1$.}
	\label{fig:deSitterPlots}
\end{figure}

In Figure (\ref{fig:deSitterPlots}.b ,\ref{fig:deSitterPlots}.c, \ref{fig:deSitterPlots}.d), we have plotted the circuit complexity measures for Bunch-Davies Vacua, $\alpha$ Vacua and Motta-Allen Vacua respectively using both Covariance as well as Nielsen Approach using calculations from Section \ref{sec:complexityMeasure}. Before discussing each Vacua individually, let us discuss the overall features of these complexity measures. Clearly both Covariance and Nielsen complexity measures have similar growth pattern. However Nielsen's complexity measure has more details on it. The reason for this is that covariance approach of computing complexity is independent of the squeezing angle $\phi_k$ while Nielsen's approach depends on it too. Still to have the similar growth pattern shows that circuit complexity could indeed be an useful tool to probe the dynamics of the system. After scale factor $a = 0.5$, both Covariance and Nielsen's complexity measures starts to grow linearly while before that point, they show the oscillating behaviour. The reason for this linear growth of complexity is that squeezing parameters $r_k$ and $\phi_k$ starts to grow linearly after $a = 0.5$.

Since the dispersion relations Sec: \ref{sec:superHubbleDispersion} for each vacua, squeezing parameters are also different. This has a direct effect in the circuit complexity measure. For $\alpha$ vacua, we have set $\alpha = 0.2$ while for Motta-Allen Vacua, we have set $\alpha = 0.2$ and $\gamma = 0.4$. Because these measures are oscillating before the point $a= 0.5$, it is simpler to compare them after it. Circuit complexity measure of Motta-Allen vacua is largest followed by $\alpha$ vacua and then by Bunch-Davies vacua. One can understand this in the following way. After $a= 0.5$, the squeezing parameters starts to merge at each other, so if $\alpha$ and $\gamma$ would be zero, the circuit complexity measure would be same. However, with increase of the number of parameters the circuit complexity is expected to increase which is then reflected in plot. 
\subsubsection*{Comparison of entanglement entropy with Complexity}
Entanglement entropy is a very popular probe to study the dynamics of quantum systems. We would like to see if circuit complexity can also be a similar candidate. In Figure (\ref{fig:deSitterPlots}.e), we have plotted a comparison of Nielsen's circuit complexity $C_1$ and entanglement entropy.
We have chosen Nielsen's measure of complexity because, both Covariance complexity and entanglement entropy are independent of squeezing angle and depends linearly on squeezing parameter $r_k$. So, the pattern obtained from covariance measure of complexity will be similar as to entanglement entropy except that both complexity measure $C_1$ and $C_2$ is greater than entanglement entropy. However with Nielsen's measure of complexity, we can obtain more interesting details of the system. For example, the nature of oscillation obtained in complexity measure is not visible in the entanglement entropy. After the point $a = 0.5$, both entanglement entropy and complexity starts to grow linearly with entanglement entropy being bounded by complexity. So, given the complexity plot, we can say that entanglement entropy is bounded by it.
\subsubsection*{Quantum Lyapunov Exponent}
Circuit complexity has also been proposed as a tool to measure quantum chaos. In particular, low growth of complexity indicates less chaotic system while higher growth of complexity indicates highly chaotic system. So, slope of the complexity could be a measure of quantum chaos. As a crude approximation, we will call the Lyapunov exponent $\lambda_i$ to be:
\begin{equation}
    \lambda_i = \frac{\text{ln}C_i(\text{point of saturation})-\text{ln}C_i(\text{point of rise})}{a(\text{point of saturation})-a(\text{point of rise})}
\end{equation}
where $i$ indicates the choice of vacua. For simplicity, we will restrict to Nielsen complexity $C_1$ and we will obtain Lyapunov exponent $\lambda_i$ to be:
\begin{equation}
\begin{aligned}
  \lambda_{Motta-Allen} &= 3.3 \\
  \lambda_{\alpha} &= 3.1 \\
  \lambda_{Bunch-Davies} &= 4.1
\end{aligned}
\end{equation}
Bunch-Davies Vacua has the largest Lyapunov exponent, so it is the most chaotic cosmological model followed by Motta-Allen and $\alpha$ vacua.
\subsection{Inflation/ Quasi De Sitter}
In Figure (\ref{fig:inflationPlots}), we have numerically plotted the squeezing parameters, derived circuit complexity measures for Bunch-Davies Vacua, $\alpha$ Vacua and Motta-Allen Vacua, comparison of entanglement entropy with circuit complexity measure and quantum chaos in de Sitter Model on super-horizon scales for the parameters $k = 0.1, \epsilon_* = 0.5, \tau_* = 1$, $\alpha = 0.2$ and $ \gamma = 0.4$. We have set $r_k(a = 1) = 1, \phi_k(a = 1) = 1$ as our initial conditions. 
\subsubsection*{Squeezing Parameters}
In Figure (\ref{fig:inflationPlots}.a), we have plotted squeezing parameters $r_k$ and $\phi_K$ for each vacua using the parameters listed above. The behavior for this model is similar to the one for de Sitter Model. Both $r_k$ and $\phi_k$ are oscillating for each vacua before $a= 0.5$. However, interestingly after that point these squeezing parameters starts to converge at each other and stops to oscillate. Since both circuit complexity and entanglement entropy are strongly dependent on the values of these parameters, the behavior of these parameters are significant to understand the dynamics of the system. 

\subsubsection*{Complexity Measure}
\begin{figure}[h!]
	\centering
	\includegraphics[width=\textwidth,height=\textheight,keepaspectratio]{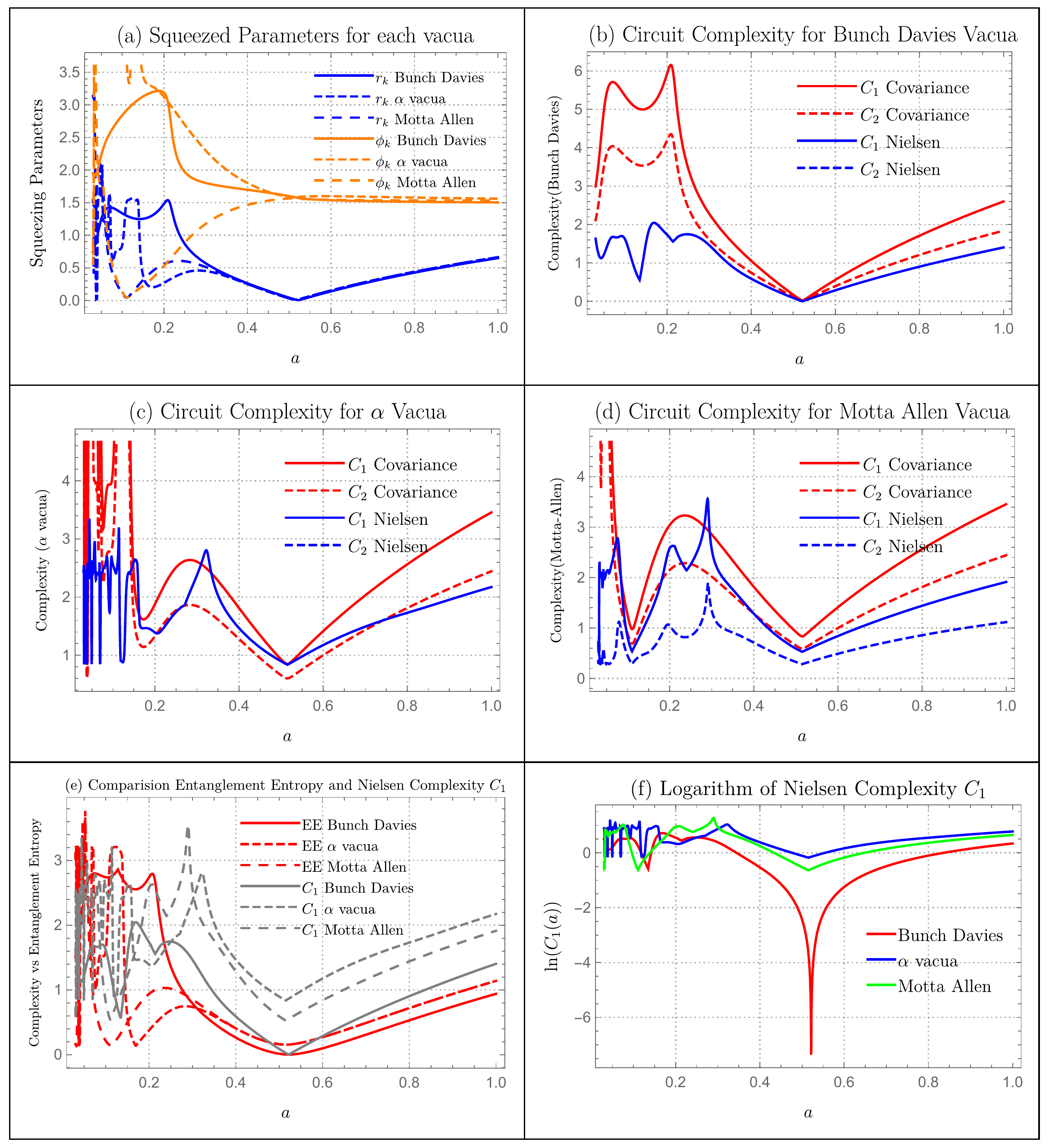}
	\caption{\textbf{Inflation/ Quasi De Sitter Model }(a.) Behaviour of the squeezing parameters in Inflation/ Quasi De Sitter Model for $k = 0.1, \epsilon_* = 0.5, \tau_* = 1$, $\alpha = 0.2$ and $ \gamma = 0.4$ for Bunch-Davies, $\alpha$ and Motta-Allen Vacua (b,c,d) Circuit complexity $C_1$ and $C_2$ for Bunch-Davies Vacua, $\alpha$ Vacua and Motta-Allen Vacua respectively using both Nielsen and Covariance approach for same parameters as in fig.a (e.) comparison of entanglement entropy with Nielsen circuit complexity $C_1$ (f.) Logarithm of Nielsen circuit complexity $C_1$.}
	\label{fig:inflationPlots}
\end{figure}

In Figure (\ref{fig:inflationPlots}.b ,\ref{fig:inflationPlots}.c, \ref{fig:inflationPlots}.d), we have plotted the circuit complexity measures for Bunch-Davies Vacua, $\alpha$ Vacua and Motta-Allen Vacua respectively using both Covariance as well as Nielsen Approach using calculations from Section \ref{sec:complexityMeasure}. Before discussing each Vacua individually, let us discuss the overall features of these complexity measures. Clearly both Covariance and Nielsen complexity measures have similar growth pattern. However Nielsen's complexity measure has more details on it. The reason for this is that covariance approach of computing complexity is independent of the squeezing angle $\phi_k$ while Nielsen's approach depends on it too. Still to have the similar growth pattern shows that circuit complexity could indeed be an useful tool to probe the dynamics of the system. After scale factor $a = 0.5$, both Covariance and Nielsen's complexity measures starts to grow linearly while before that point, they show the oscillating behavior. The reason for this linear growth of complexity is that squeezing parameters $r_k$ and $\phi_k$ starts to grow linearly after $a = 0.5$. The gap between the magnitude of complexity between covariance approach and Nielsen's approach is highest in the case of Bunch-Davies Vacua. This difference drops as we go to Motta-Allen and Alpha Vacua.

Since the dispersion relations Sec: \ref{sec:superHubbleDispersion} for each vacua, squeezing parameters are also different. This has a direct effect in the circuit complexity measure. For $\alpha$ vacua, we have set $\alpha = 0.2$ while for Motta-Allen Vacua, we have set $\alpha = 0.2$ and $\gamma = 0.4$. Because these measures are oscillating before the point $a= 0.5$, it is simpler to compare them after it. Circuit complexity measure of Motta-Allen vacua is largest followed by $\alpha$ vacua and then by Bunch-Davies vacua. One can understand this in the following way. After $a= 0.5$, the squeezing parameters starts to merge at each other, so if $\alpha$ and $\gamma$ would be zero, the circuit complexity measure would be same. However, with increase of the number of parameters the circuit complexity is expected to increase which is then reflected in plot. 

\subsubsection*{comparison of entanglement entropy with Complexity}
Entanglement entropy is a very popular probe to study the dynamics of quantum systems. We would like to see if circuit complexity can also be a similar candidate. In Figure (\ref{fig:inflationPlots}.e), we have plotted a comparison of Nielsen's circuit complexity $C_1$ and entanglement entropy.
We have chosen Nielsen's measure of complexity because, both Covariance complexity and entanglement entropy are independent of squeezing angle and depends linearly on squeezing parameter $r_k$. So, the pattern obtained from covariance measure of complexity will be similar as to entanglement entropy except that both complexity measure $C_1$ and $C_2$ is greater than entanglement entropy. However with Nielsen's measure of complexity, we can obtain more interesting details of the system. For example, we can see more details regarding the evolution of system in the complexity graph compared to entanglement entropy. After the point $a = 0.5$, both entanglement entropy and complexity starts to grow linearly with entanglement entropy being bounded by complexity. So, given the complexity plot, we can say that entanglement entropy is bounded by it.
\subsubsection*{Quantum Lyapunov Exponent}
Circuit complexity has also been proposed as a tool to measure quantum chaos. In particular, low growth of complexity indicates less chaotic system while higher growth of complexity indicates highly chaotic system. So, slope of the complexity could be a measure of quantum chaos. As a crude approximation, we will call the lyapunov exponent $\lambda_i$ to be:
\begin{equation}
    \lambda_i = \frac{\text{ln}C_i(\text{point of saturation})-\text{ln}C_i(\text{point of rise})}{a(\text{point of saturation})-a(\text{point of rise})}
\end{equation}
where $i$ indicates the choice of vacua. For simplicity, we will restrict to Nielsen complexity $C_1$ and we will obtain Lyapunov exponent $\lambda_i$ to be:
\begin{equation}
\begin{aligned}
  \lambda_{Motta-Allen} &= 2.375 \\
  \lambda_{\alpha} &= 2.25 \\
  \lambda_{Bunch-Davies} &= 3.75
\end{aligned}
\end{equation}
Bunch-Davies Vacua has the largest Lyapunov exponent, so it is the most chaotic cosmological model followed by Motta-Allen and $\alpha$ vacua.

\subsection{Reheating}
In Figure (\ref{fig:reheatingPlots}), we have numerically plotted the squeezing parameters, derived circuit complexity measures for Bunch-Davies Vacua, $\alpha$ Vacua and Motta-Allen Vacua, comparison of entanglement entropy with circuit complexity measure and quantum chaos in de Sitter Model on super-horizon scales for the parameters $k = 0.1, w=1/4$, $\alpha = 0.2$ and $ \gamma = 0.4$. We have set $r_k(a = 1) = 1, \phi_k(a = 1) = 1$ as our initial conditions. 
\subsubsection*{Squeezing Parameters}
In Figure (\ref{fig:reheatingPlots}.a), we have plotted squeezing parameters $r_k$ and $\phi_K$ for each vacua using the parameters listed above. The value of these squeezing parameters keep growing for all vacua. Since both circuit complexity and entanglement entropy are strongly dependent on the values of these parameters, the behaviour of these parameters are significant to understand the dynamics of the system.

\subsubsection*{Complexity Measure}
\begin{figure}[h!]
	\centering
	\includegraphics[width=\textwidth,height=\textheight,keepaspectratio]{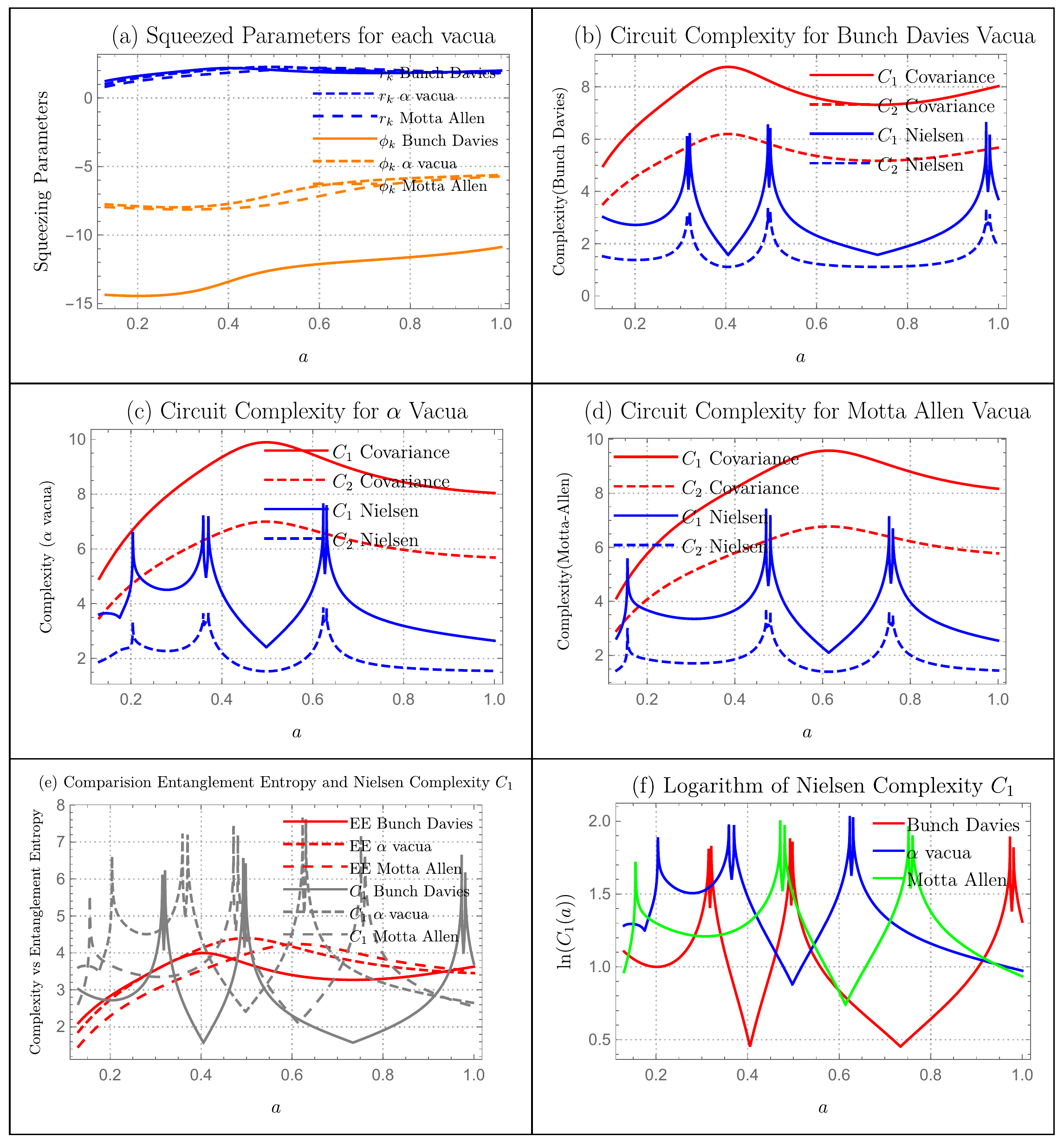}
	\caption{\textbf{Reheating Model }(a.) Behaviour of the squeezing parameters in Reheating Model for $k = 0.1, w=1/4$, $\alpha = 0.2$ and $ \gamma = 0.4$ for Bunch-Davies, $\alpha$ and Motta-Allen Vacua (b,c,d) Circuit complexity $C_1$ and $C_2$ for Bunch-Davies Vacua, $\alpha$ Vacua and Motta-Allen Vacua respectively using both Nielsen and Covariance approach for same parameters as in fig.a (e.) comparison of entanglement entropy with Nielsen circuit complexity $C_1$ (f.) Logarithm of Nielsen circuit complexity $C_1$.}
	\label{fig:reheatingPlots}
\end{figure}

In Figure (\ref{fig:reheatingPlots}.b ,\ref{fig:reheatingPlots}.c, \ref{fig:reheatingPlots}.d), we have plotted the circuit complexity measures for Bunch-Davies Vacua, $\alpha$ Vacua and Motta-Allen Vacua respectively using both Covariance as well as Nielsen Approach using calculations from Section \ref{sec:complexityMeasure}. Before discussing each Vacua individually, let us discuss the overall features of these complexity measures. Clearly both Covariance and Nielsen complexity measures have similar growth pattern. However Nielsen's complexity measure has more details on it. The complexity from Nielsen's method starts to grow linearly and then starts to oscillate. In contrast, the complexity obtained using Covariance method grows and show saturating behaviour. The reason that we don't see oscillating behaviours in covariance is that covariance approach of computing complexity is independent of the squeezing angle $\phi_k$ while Nielsen's approach depends on it too. 

Since the dispersion relations Sec: \ref{sec:superHubbleDispersion} for each vacua, squeezing parameters are also different. This has a direct effect in the circuit complexity measure. For $\alpha$ vacua, we have set $\alpha = 0.2$ while for Motta-Allen Vacua, we have set $\alpha = 0.2$ and $\gamma = 0.4$. In the complexity computed using Nielsen's approach, the gap between oscillating peaks are different for each vacua. While for Covariance approach, there is not much difference. 

\subsubsection*{comparison of entanglement entropy with Complexity}
Entanglement entropy is a very popular probe to study the dynamics of quantum systems. We would like to see if circuit complexity can also be a similar candidate. In Figure (\ref{fig:reheatingPlots}.e), we have plotted a comparison of Nielsen's circuit complexity $C_1$ and entanglement entropy.
We have chosen Nielsen's measure of complexity because, both Covariance complexity and entanglement entropy are independent of squeezing angle and depends linearly on squeezing parameter $r_k$. So, the pattern obtained from covariance measure of complexity will be similar as to entanglement entropy except that both complexity measure $C_1$ and $C_2$ is greater than entanglement entropy. However with Nielsen's measure of complexity, we can obtain more interesting details of the system. For example, we can see more details regarding the evolution of system in the complexity graph compared to entanglement entropy.

\subsubsection*{Quantum Lyapunov Exponent}
Circuit complexity has also been proposed as a tool to measure quantum chaos. In particular, low growth of complexity indicates less chaotic system while higher growth of complexity indicates highly chaotic system. So, slope of the complexity could be a measure of quantum chaos. As a crude approximation, we will call the Lyapunov exponent $\lambda_i$ to be:
\begin{equation}
    \lambda_i = \frac{\text{ln}C_i(\text{point of saturation})-\text{ln}C_i(\text{point of rise})}{a(\text{point of saturation})-a(\text{point of rise})}
\end{equation}
where $i$ indicates the choice of vacua. The logarithm of complexity plot for Nielsen's measure of complexity is very oscillatory and it doesn't reach a saturation point. So, it is not able to measure the chaotic component here. 

\subsection{Radiation}
In Figure (\ref{fig:radiationPlots}), we have numerically plotted the squeezing parameters, derived circuit complexity measures for Bunch-Davies Vacua, $\alpha$ Vacua and Motta-Allen Vacua, comparison of entanglement entropy with circuit complexity measure and quantum chaos in de Sitter Model on super-horizon scales for the parameters $k = 0.1, \alpha = 0.2$ and $ \gamma = 0.4$. We have set $r_k(a = 1) = 1, \phi_k(a = 1) = 1$ as our initial conditions. 

\subsubsection*{Squeezing Parameters}
In Figure (\ref{fig:radiationPlots}.a), we have plotted squeezing parameters $r_k$ and $\phi_K$ for each vacua using the parameters listed above. The overall growth pattern is similar to the reheating model. The value of these squeezing parameters keep growing for all vacua. Since both circuit complexity and entanglement entropy are strongly dependent on the values of these parameters, the behaviour of these parameters are significant to understand the dynamics of the system. 

\subsubsection*{Complexity Measure}
\begin{figure}[h!]
	\centering
	\includegraphics[width=\textwidth,height=\textheight,keepaspectratio]{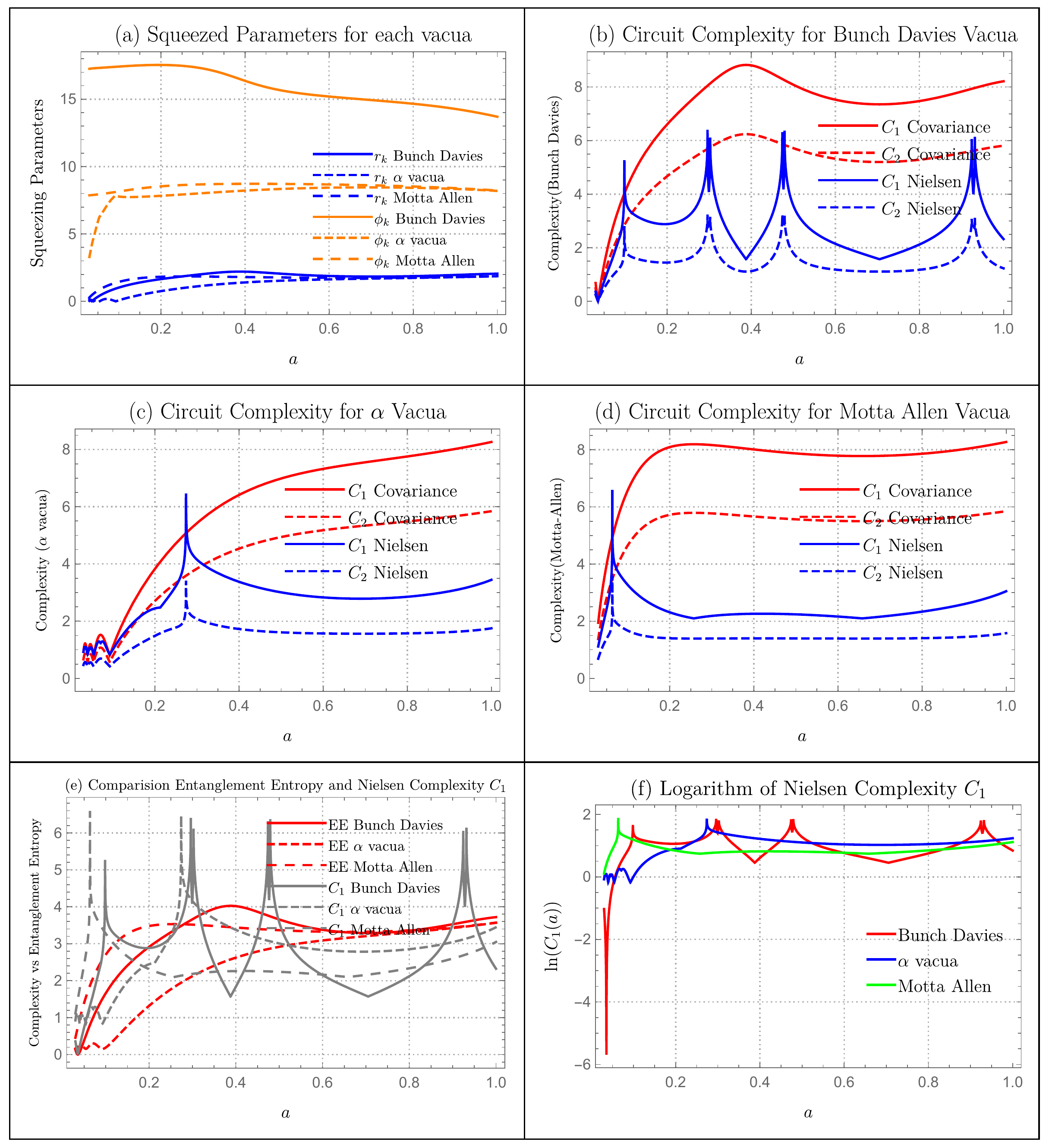}
	\caption{\textbf{Radiation Model }(a.) Behaviour of the squeezing parameters in Radiation Model for $k = 0.1$, $\alpha = 0.2$ and $ \gamma = 0.4$ for Bunch-Davies, $\alpha$ and Motta-Allen Vacua (b,c,d) Circuit complexity $C_1$ and $C_2$ for Bunch-Davies Vacua, $\alpha$ Vacua and Motta-Allen Vacua respectively using both Nielsen and Covariance approach for same parameters as in fig.a (e.) comparison of entanglement entropy with Nielsen circuit complexity $C_1$ (f.) Logarithm of Nielsen circuit complexity $C_1$.}
	\label{fig:radiationPlots}
\end{figure}

In Figure (\ref{fig:radiationPlots}.b ,\ref{fig:radiationPlots}.c, \ref{fig:radiationPlots}.d), we have plotted the circuit complexity measures for Bunch-Davies Vacua, $\alpha$ Vacua and Motta-Allen Vacua respectively using both Covariance as well as Nielsen Approach using calculations from Section \ref{sec:complexityMeasure}. Before discussing each Vacua individually, let us discuss the overall features of these complexity measures. Clearly both Covariance and Nielsen complexity measures have similar growth pattern with Nielsen's complexity measure is bounded by Covariance's complexity measure. However Nielsen's complexity measure has more details on it. The complexity from Nielsen's method starts to grow linearly and then sometimes starts to oscillate. In contrast, the complexity obtained using Covariance method grows and show saturating behaviour. The reason that we don't see oscillating behaviours in covariance is that covariance approach of computing complexity is independent of the squeezing angle $\phi_k$ while Nielsen's approach depends on it too. 

Since the dispersion relations Sec: \ref{sec:superHubbleDispersion} for each vacua, squeezing parameters are also different. This has a direct effect in the circuit complexity measure. For $\alpha$ vacua, we have set $\alpha = 0.2$ while for Motta-Allen Vacua, we have set $\alpha = 0.2$ and $\gamma = 0.4$. For the bunch davies vacua, Nielsen's complexity is oscillatory while for Alpha and Motta-Allen Vacua, complexity grows to a peak and then saturates. Covariance's complexity measure for all three vacua converges at some point of $\alpha$.

\subsubsection*{comparison of entanglement entropy with Complexity}
Entanglement entropy is a very popular probe to study the dynamics of quantum systems. We would like to see if circuit complexity can also be a similar candidate. In Figure (\ref{fig:radiationPlots}.e), we have plotted a comparison of Nielsen's circuit complexity $C_1$ and entanglement entropy.
We have chosen Nielsen's measure of complexity because, both Covariance complexity and entanglement entropy are independent of squeezing angle and depends linearly on squeezing parameter $r_k$. So, the pattern obtained from covariance measure of complexity will be similar as to entanglement entropy except that both complexity measure $C_1$ and $C_2$ is greater than entanglement entropy. However with Nielsen's measure of complexity, we can obtain more interesting details of the system. For example, we can see more details regarding the evolution of system in the complexity graph compared to entanglement entropy. The complexity grows and then saturates after certain point while complexity starts to oscillating behaviour. 

\subsubsection*{Quantum Lyapunov Exponent}
Circuit complexity has also been proposed as a tool to measure quantum chaos. In particular, low growth of complexity indicates less chaotic system while higher growth of complexity indicates highly chaotic system. So, slope of the complexity could be a measure of quantum chaos. As a crude approximation, we will call the Lyapunov exponent $\lambda_i$ to be:
\begin{equation}
    \lambda_i = \frac{\text{ln}C_i(\text{point of saturation})-\text{ln}C_i(\text{point of rise})}{a(\text{point of saturation})-a(\text{point of rise})}
\end{equation}
where $i$ indicates the choice of vacua. For simplicity, we will restrict to Nielsen complexity $C_1$ and we will obtain lyapunov exponent $\lambda_i$ to be:
\begin{equation}
\begin{aligned}
  \lambda_{Motta-Allen} &= 0.52 \\
  \lambda_{\alpha} &= 1.05 \\
  \lambda_{Bunch-Davies} &= 3.15
\end{aligned}
\end{equation}
Bunch-Davies Vacua has the largest lyapunov exponent, so it is the most chaotic cosmological model followed by $\alpha$ vacua and Motta-Allen Vacua respectively.

\subsection{Matter}
In Figure (\ref{fig:matterPlots}), we have numerically plotted the squeezing parameters, derived circuit complexity measures for Bunch-Davies Vacua, $\alpha$ Vacua and Motta-Allen Vacua, comparison of entanglement entropy with circuit complexity measure and quantum chaos in de Sitter Model on super-horizon scales for the parameters $k = 0.1, \alpha = 0.2$ and $ \gamma = 0.4$. We have set $r_k(a = 1) = 1, \phi_k(a = 1) = 1$ as our initial conditions. 

\subsubsection*{Squeezing Parameters}
In Figure (\ref{fig:matterPlots}.a), we have plotted squeezing parameters $r_k$ and $\phi_K$ for each vacua using the parameters listed above. The value of these squeezing parameters $r_k$ keeps constant while $\phi_k$ keep growing for all vacua. Since both circuit complexity and entanglement entropy are strongly dependent on the values of these parameters, the behaviour of these parameters are significant to understand the dynamics of the system. 

\subsubsection*{Complexity Measure}
\begin{figure}[h!]
	\centering
	\includegraphics[width=\textwidth,height=\textheight,keepaspectratio]{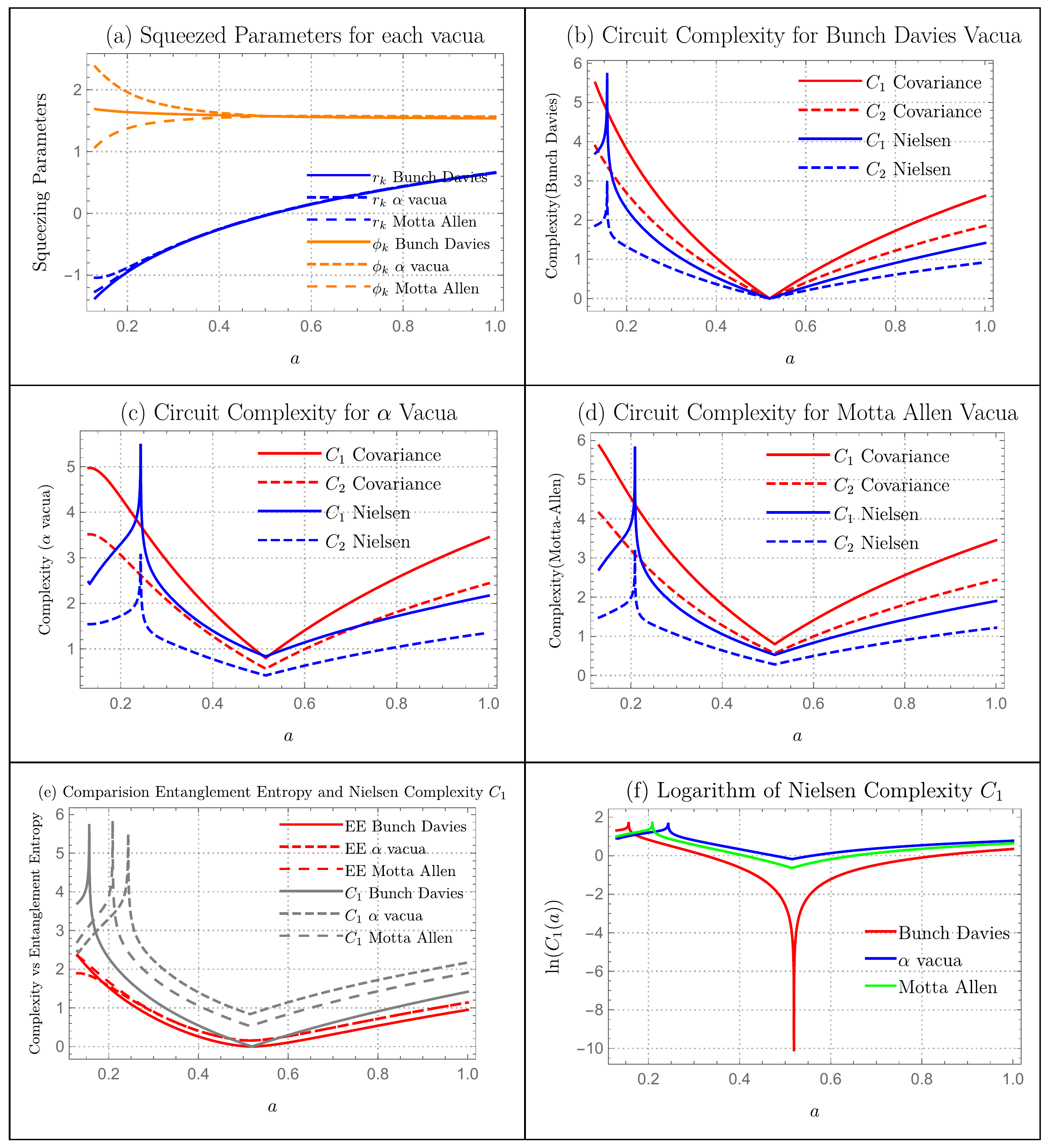}
	\caption{\textbf{Matter Model }(a.) Behaviour of the squeezing parameters in Matter Model for $k = 0.1$, $\alpha = 0.2$ and $ \gamma = 0.4$ for Bunch-Davies, $\alpha$ and Motta-Allen Vacua (b,c,d) Circuit complexity $C_1$ and $C_2$ for Bunch-Davies Vacua, $\alpha$ Vacua and Motta-Allen Vacua respectively using both Nielsen and Covariance approach for same parameters as in fig.a (e.) comparison of entanglement entropy with Nielsen circuit complexity $C_1$ (f.) Logarithm of Nielsen circuit complexity $C_1$.}
	\label{fig:matterPlots}
\end{figure}

In Figure (\ref{fig:matterPlots}.b ,\ref{fig:matterPlots}.c, \ref{fig:matterPlots}.d), we have plotted the circuit complexity measures for Bunch-Davies Vacua, $\alpha$ Vacua and Motta-Allen Vacua respectively using both Covariance as well as Nielsen Approach using calculations from Section \ref{sec:complexityMeasure}. Before discussing each Vacua individually, let us discuss the overall features of these complexity measures. Clearly both Covariance and Nielsen complexity measures have similar growth pattern with Nielsen's complexity measure is bounded by Covariance's complexity measure. However Nielsen's complexity measure has more details on it. The complexity using Nielsen's approach shows some oscillatory behaviour while after reaching value $a= 0.5$, it starts to grow linearly. While Covariance's complexity grows at the beginning, then takes a dip. After the dip it starts to grow linearly. The reason that we don't see oscillating behaviours in covariance is that covariance approach of computing complexity is independent of the squeezing angle $\phi_k$ while Nielsen's approach depends on it too. 

\subsubsection*{comparison of entanglement entropy with Complexity}
Entanglement entropy is a very popular probe to study the dynamics of quantum systems. We would like to see if circuit complexity can also be a similar candidate. In Figure (\ref{fig:matterPlots}.e), we have plotted a comparison of Nielsen's circuit complexity $C_1$ and entanglement entropy.
We have chosen Nielsen's measure of complexity because, both Covariance complexity and entanglement entropy are independent of squeezing angle and depends linearly on squeezing parameter $r_k$. So, the pattern obtained from covariance measure of complexity will be similar as to entanglement entropy except that both complexity measure $C_1$ and $C_2$ is greater than entanglement entropy. However with Nielsen's measure of complexity, we can obtain more interesting details of the system. For example, we can see some oscillatory details regarding the evolution of system in the complexity graph compared to entanglement entropy. After $a=0.5$, both complexity and entanglement entropy starts to grow linearly with entanglement entropy being bounded by complexity growth.

\subsubsection*{Quantum Lyapunov Exponent}
Circuit complexity has also been proposed as a tool to measure quantum chaos. In particular, low growth of complexity indicates less chaotic system while higher growth of complexity indicates highly chaotic system. So, slope of the complexity could be a measure of quantum chaos. As a crude approximation, we will call the Lyapunov exponent $\lambda_i$ to be:
\begin{equation}
    \lambda_i = \frac{\text{ln}C_i(\text{point of saturation})-\text{ln}C_i(\text{point of rise})}{a(\text{point of saturation})-a(\text{point of rise})}
\end{equation}
where $i$ indicates the choice of vacua. For simplicity, we will restrict to Nielsen complexity $C_1$ and we will obtain Lyapunov exponent $\lambda_i$ to be:
\begin{equation}
\begin{aligned}
  \lambda_{Motta-Allen} &= 3.12 \\
  \lambda_{\alpha} &= 2.5 \\
  \lambda_{Bunch-Davies} &= 3.75
\end{aligned}
\end{equation}
Bunch-Davies Vacua has the largest lyapunov exponent, so it is the most chaotic cosmological model followed by Motta-Allen and $\alpha$ vacua.

\subsection{Bouncing model}
From Figure (\ref{fig:contractionPlots}) - (\ref{fig:expansionPlots}) , we have numerically plotted the squeezing parameters, derived circuit complexity measures for Bunch-Davies Vacua, $\alpha$ Vacua and Motta-Allen Vacua, comparison of entanglement entropy with circuit complexity measure and quantum chaos for various Bouncing models on super-horizon scales for the parameters $k = 0.1, \alpha = 0.2$ and $ \gamma = 0.4$. We have set $r_k(a = 1) = 1, \phi_k(a = 1) = 1$ as our initial conditions for exponential bounce, sechyperbolic bounce, cosechyperbolic bounce models  and  $r_k(a = 2) = 2, \phi_k(a = 2) = 2$ for Expansion bounce, polynomial bounce, power law bounce, cosinehyperbolic bounce, sinehyperbolic bounce, contraction bounce, matter bounce models . The extra parameters we have selected for each bouncing model will be highlighted under it's own headings.

\subsubsection*{Squeezing Parameters}
\begin{figure}[h!]
	\centering
	\includegraphics[width=\textwidth,height=\textheight,keepaspectratio]{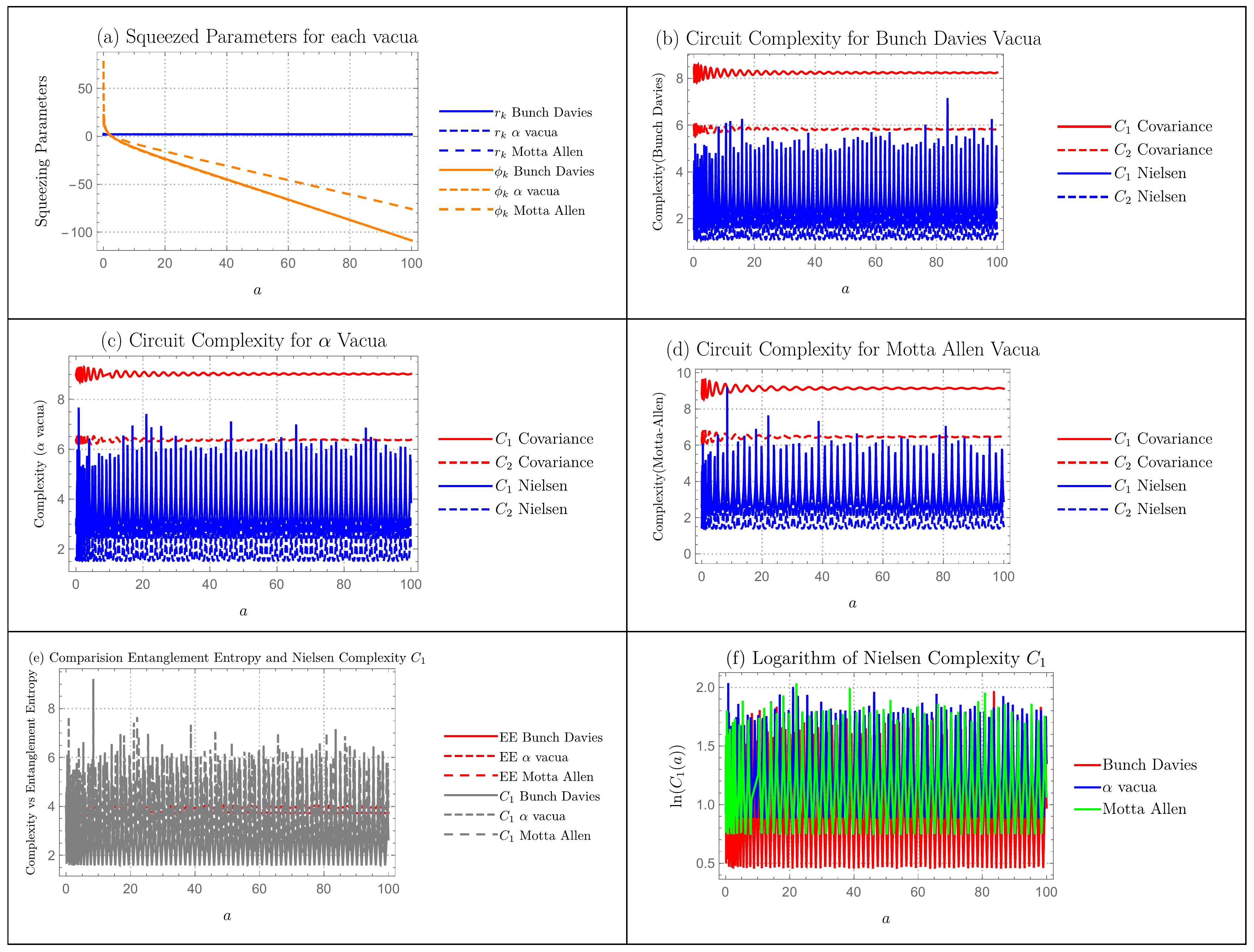}
	\caption{\textbf{Contraction Bounce Model }(a) Behaviour of the squeezing parameters in Contraction Bounce Model for $k = 0.1$, $\alpha = 0.2$ and $ \gamma = 0.4$ for Bunch-Davies, $\alpha$ and Motta-Allen Vacua (b,c,d) Circuit complexity $C_1$ and $C_2$ for Bunch-Davies Vacua, $\alpha$ Vacua and Motta-Allen Vacua respectively using both Nielsen and Covariance approach for same parameters as in fig.a (e) comparison of entanglement entropy with Nielsen circuit complexity $C_1$ (f) Logarithm of Nielsen circuit complexity $C_1$.}
	\label{fig:contractionPlots}
\end{figure}

\begin{figure}[h!]
	\centering
	\includegraphics[width=\textwidth,height=\textheight,keepaspectratio]{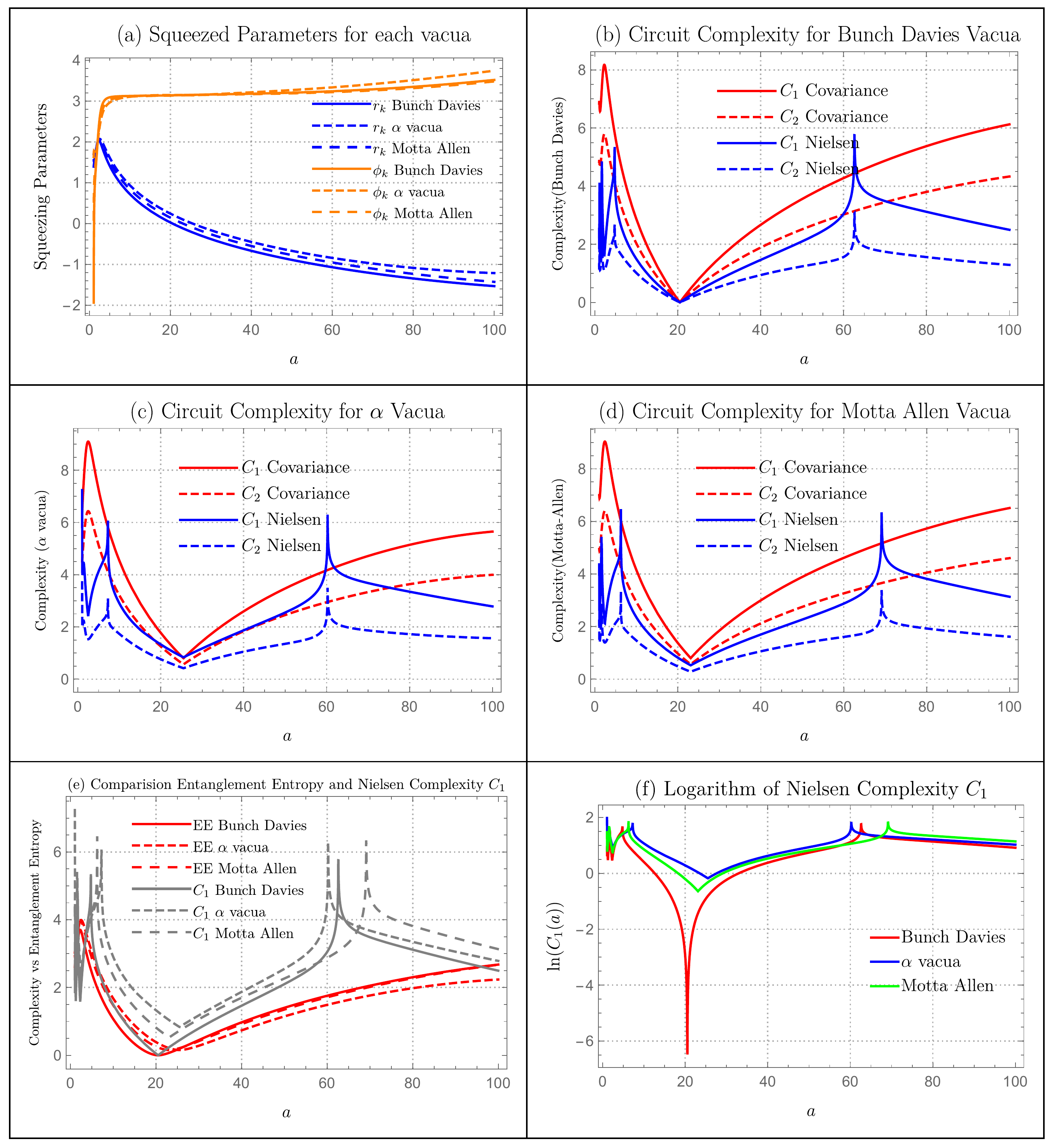}
	\caption{\textbf{Matter Bounce Model }(a) Behaviour of the squeezing parameters in Matter Bounce Model for $k = 0.1$, $\alpha = 0.2$ and $ \gamma = 0.4$ for Bunch-Davies, $\alpha$ and Motta-Allen Vacua (b,c,d) Circuit complexity $C_1$ and $C_2$ for Bunch-Davies Vacua, $\alpha$ Vacua and Motta-Allen Vacua respectively using both Nielsen and Covariance approach for same parameters as in fig.a (e) comparison of entanglement entropy with Nielsen circuit complexity $C_1$ (f) Logarithm of Nielsen circuit complexity $C_1$.}
	\label{fig:matterBouncePlots}
\end{figure}

\begin{figure}[h!]
	\centering
	\includegraphics[width=\textwidth,height=\textheight,keepaspectratio]{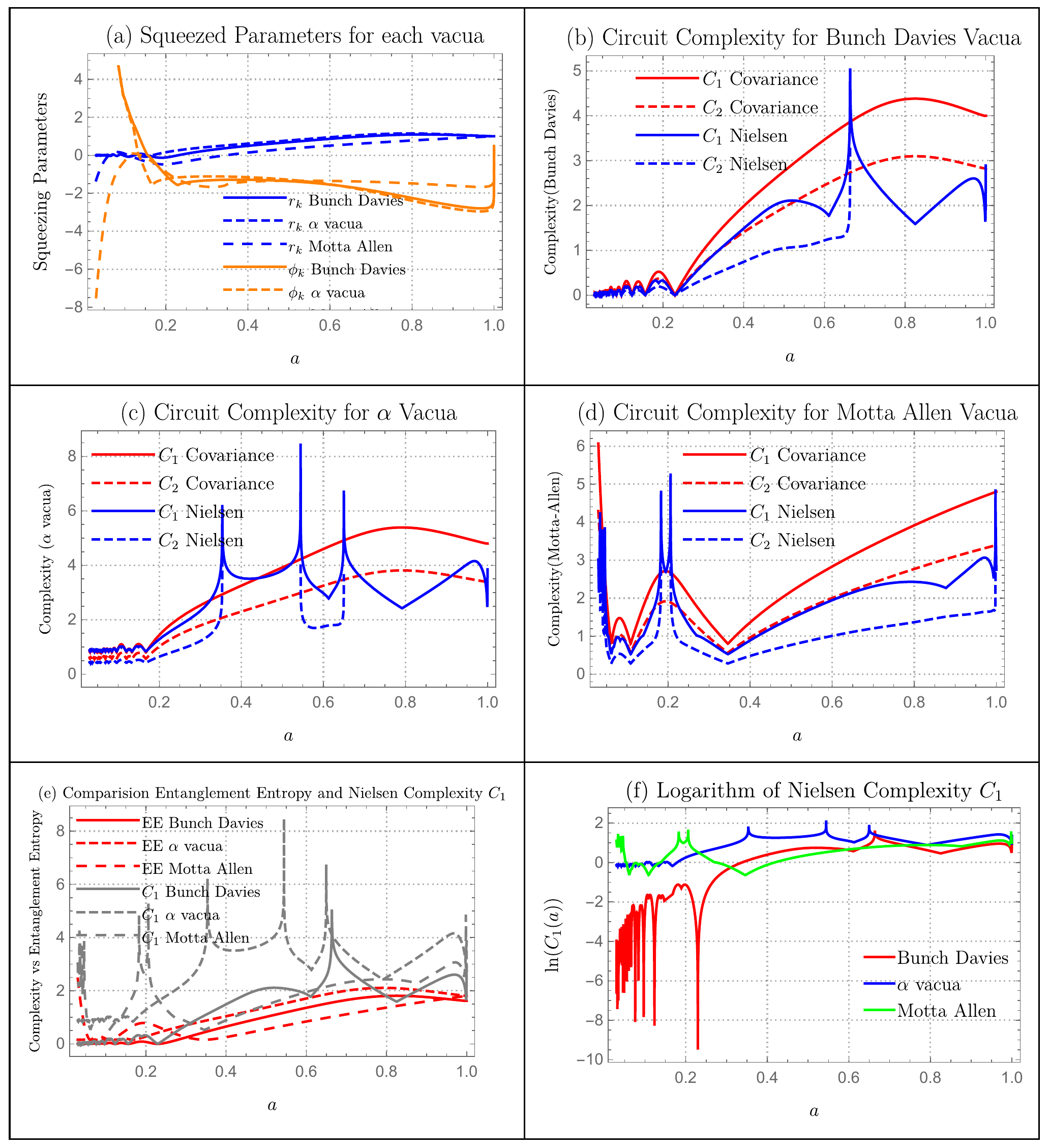}
	\caption{\textbf{Sechyperbolic Bounce Model }(a) Behaviour of the squeezing parameters in Sechyperbolic Bounce Model for $k = 0.1$, $\alpha = 0.2$ and $ \gamma = 0.4$ for Bunch-Davies, $\alpha$ and Motta-Allen Vacua (b,c,d) Circuit complexity $C_1$ and $C_2$ for Bunch-Davies Vacua, $\alpha$ Vacua and Motta-Allen Vacua respectively using both Nielsen and Covariance approach for same parameters as in fig.a (e) comparison of entanglement entropy with Nielsen circuit complexity $C_1$ (f) Logarithm of Nielsen circuit complexity $C_1$.}
	\label{fig:sechPlots}
\end{figure}

\begin{figure}[h!]
	\centering
	\includegraphics[width=\textwidth,height=\textheight,keepaspectratio]{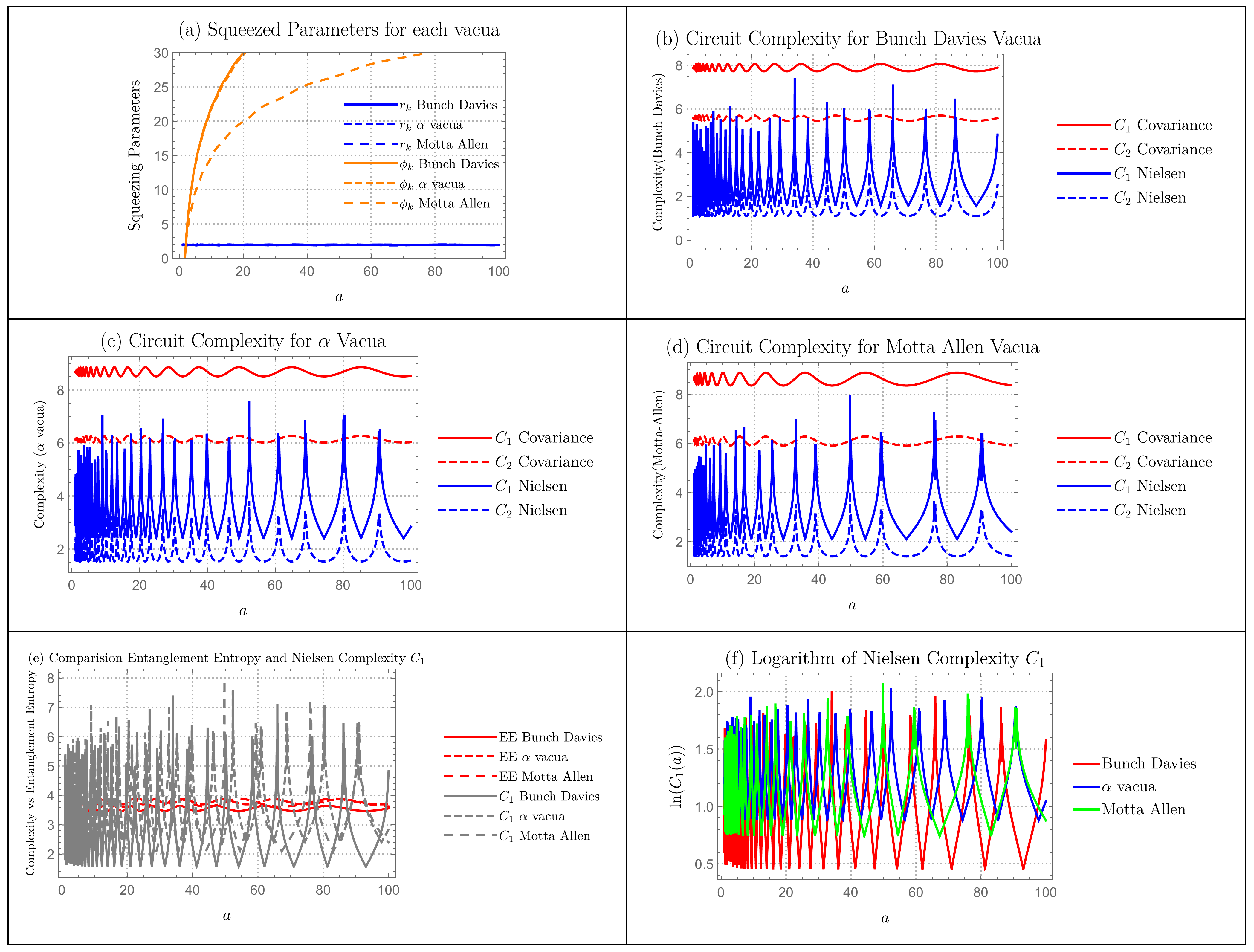}
	\caption{\textbf{Cosinehyperbolic Bounce Model }(a) Behaviour of the squeezing parameters in Cosinehyperbolic Bounce Model for $k = 0.1$, $\alpha = 0.2$ and $ \gamma = 0.4$ for Bunch-Davies, $\alpha$ and Motta-Allen Vacua (b,c,d) Circuit complexity $C_1$ and $C_2$ for Bunch-Davies Vacua, $\alpha$ Vacua and Motta-Allen Vacua respectively using both Nielsen and Covariance approach for same parameters as in fig.a (e) comparison of entanglement entropy with Nielsen circuit complexity $C_1$ (f) Logarithm of Nielsen circuit complexity $C_1$.}
	\label{fig:coshPlots}
\end{figure}

\begin{figure}[h!]
	\centering
	\includegraphics[width=\textwidth,height=\textheight,keepaspectratio]{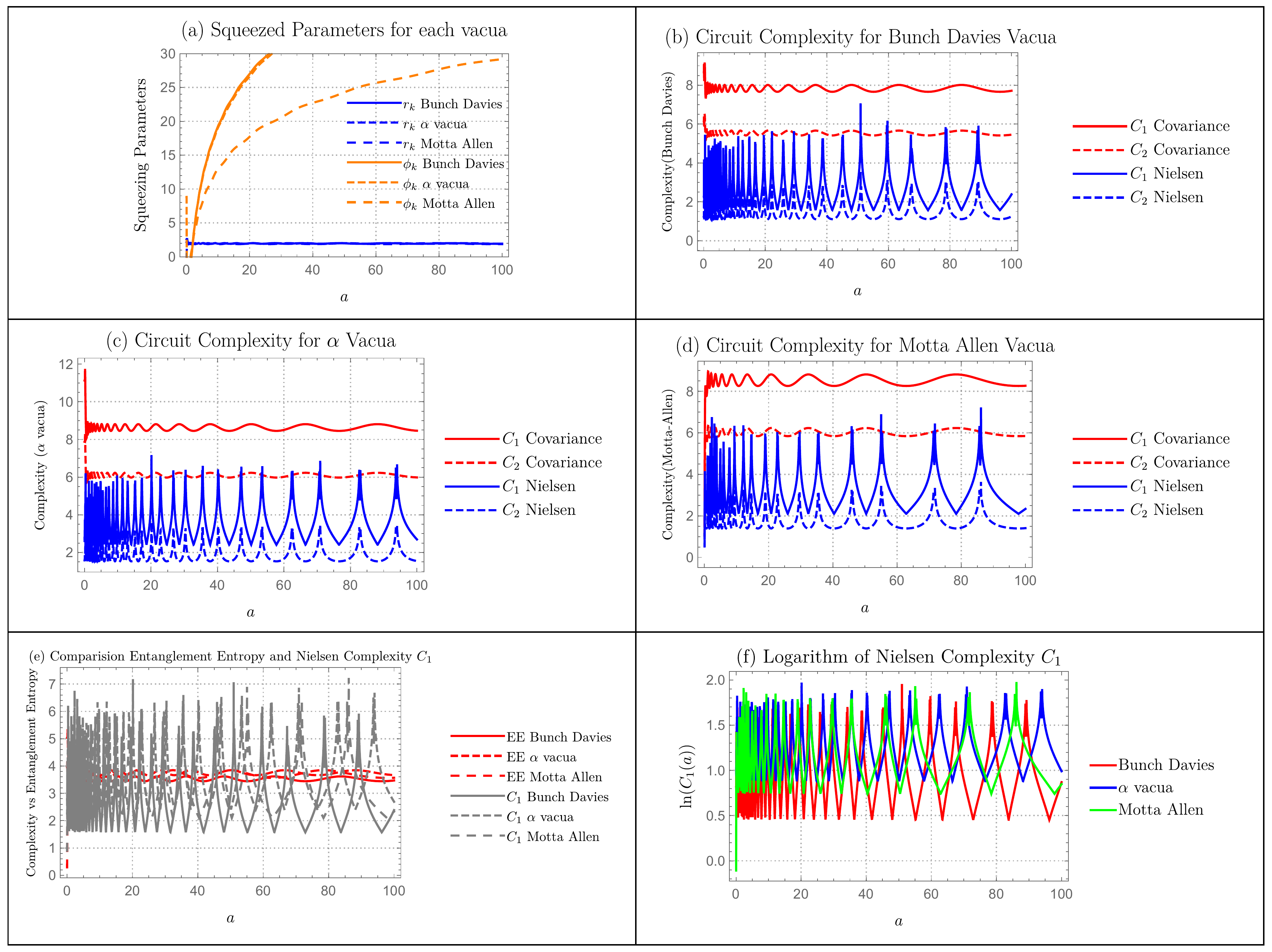}
	\caption{\textbf{Sinehyperbolic Bounce Model }(a) Behaviour of the squeezing parameters in Sinehyperbolic Bounce Model for $k = 0.1$, $\alpha = 0.2$ and $ \gamma = 0.4$ for Bunch-Davies, $\alpha$ and Motta-Allen Vacua (b,c,d) Circuit complexity $C_1$ and $C_2$ for Bunch-Davies Vacua, $\alpha$ Vacua and Motta-Allen Vacua respectively using both Nielsen and Covariance approach for same parameters as in fig.a (e) comparison of entanglement entropy with Nielsen circuit complexity $C_1$ (f) Logarithm of Nielsen circuit complexity $C_1$.}
	\label{fig:sinhPlots}
\end{figure}

\begin{figure}[h!]
	\centering
	\includegraphics[width=\textwidth,height=\textheight,keepaspectratio]{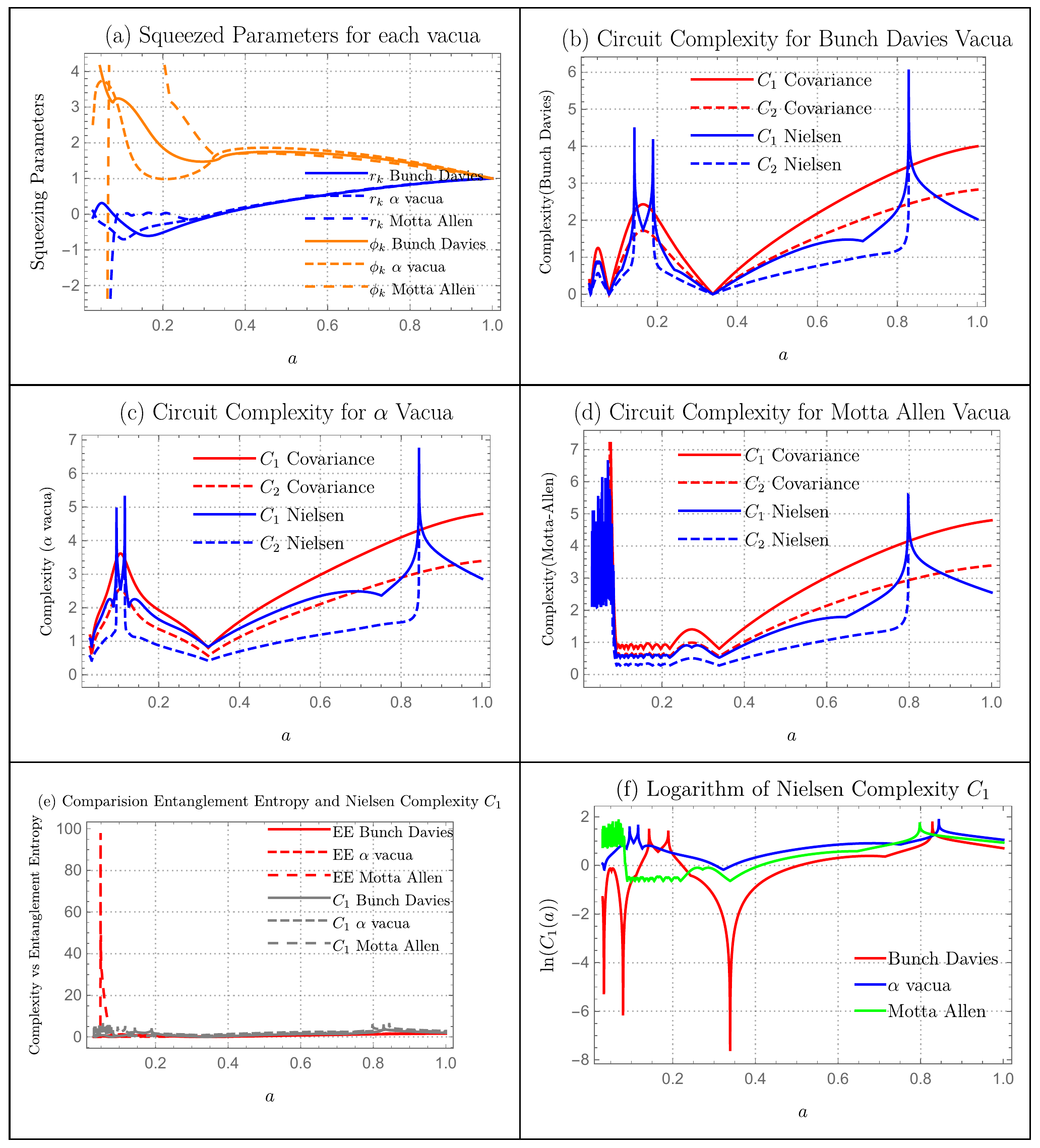}
	\caption{\textbf{Cosechyperbolic Bounce Model }(a) Behaviour of the squeezing parameters in Cosechyperbolic Bounce Model for $k = 0.1$, $\alpha = 0.2$ and $ \gamma = 0.4$ for Bunch-Davies, $\alpha$ and Motta-Allen Vacua (b,c,d) Circuit complexity $C_1$ and $C_2$ for Bunch-Davies Vacua, $\alpha$ Vacua and Motta-Allen Vacua respectively using both Nielsen and Covariance approach for same parameters as in fig.a (e) comparison of entanglement entropy with Nielsen circuit complexity $C_1$ (f) Logarithm of Nielsen circuit complexity $C_1$.}
	\label{fig:bouncingCschPlots}
\end{figure}

\begin{figure}[h!]
	\centering
	\includegraphics[width=\textwidth,height=\textheight,keepaspectratio]{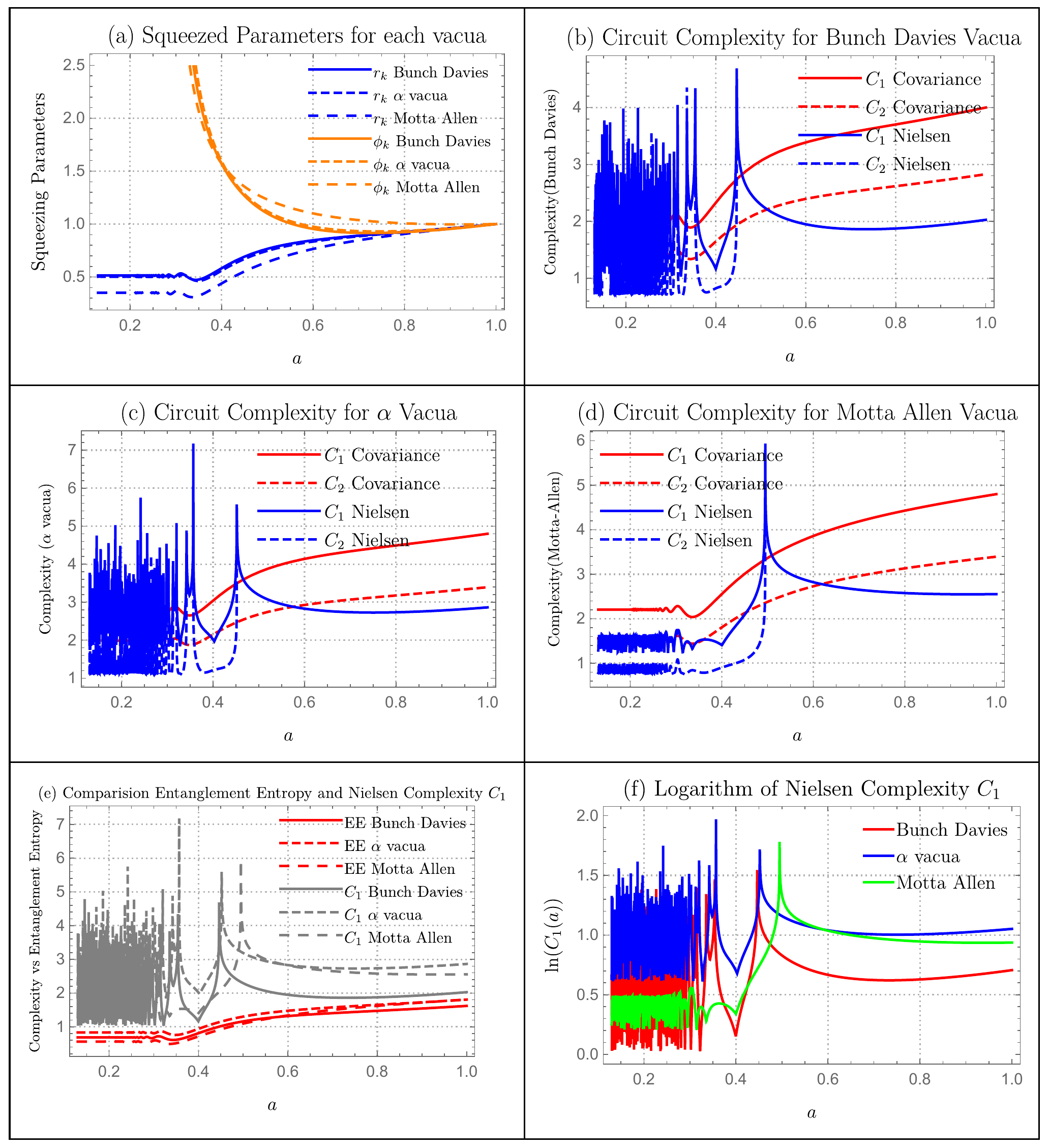}
	\caption{\textbf{Exponential Bounce Model }(a) Behaviour of the squeezing parameters in Exponential Bounce Model for $k = 0.1$, $\alpha = 0.2$ and $ \gamma = 0.4$ for Bunch-Davies, $\alpha$ and Motta-Allen Vacua (b,c,d) Circuit complexity $C_1$ and $C_2$ for Bunch-Davies Vacua, $\alpha$ Vacua and Motta-Allen Vacua respectively using both Nielsen and Covariance approach for same parameters as in fig.a (e) comparison of entanglement entropy with Nielsen circuit complexity $C_1$ (f) Logarithm of Nielsen circuit complexity $C_1$.}
	\label{fig:BounceInverseErfPlots}
\end{figure}

\begin{figure}[h!]
	\centering
	\includegraphics[width=\textwidth,height=\textheight,keepaspectratio]{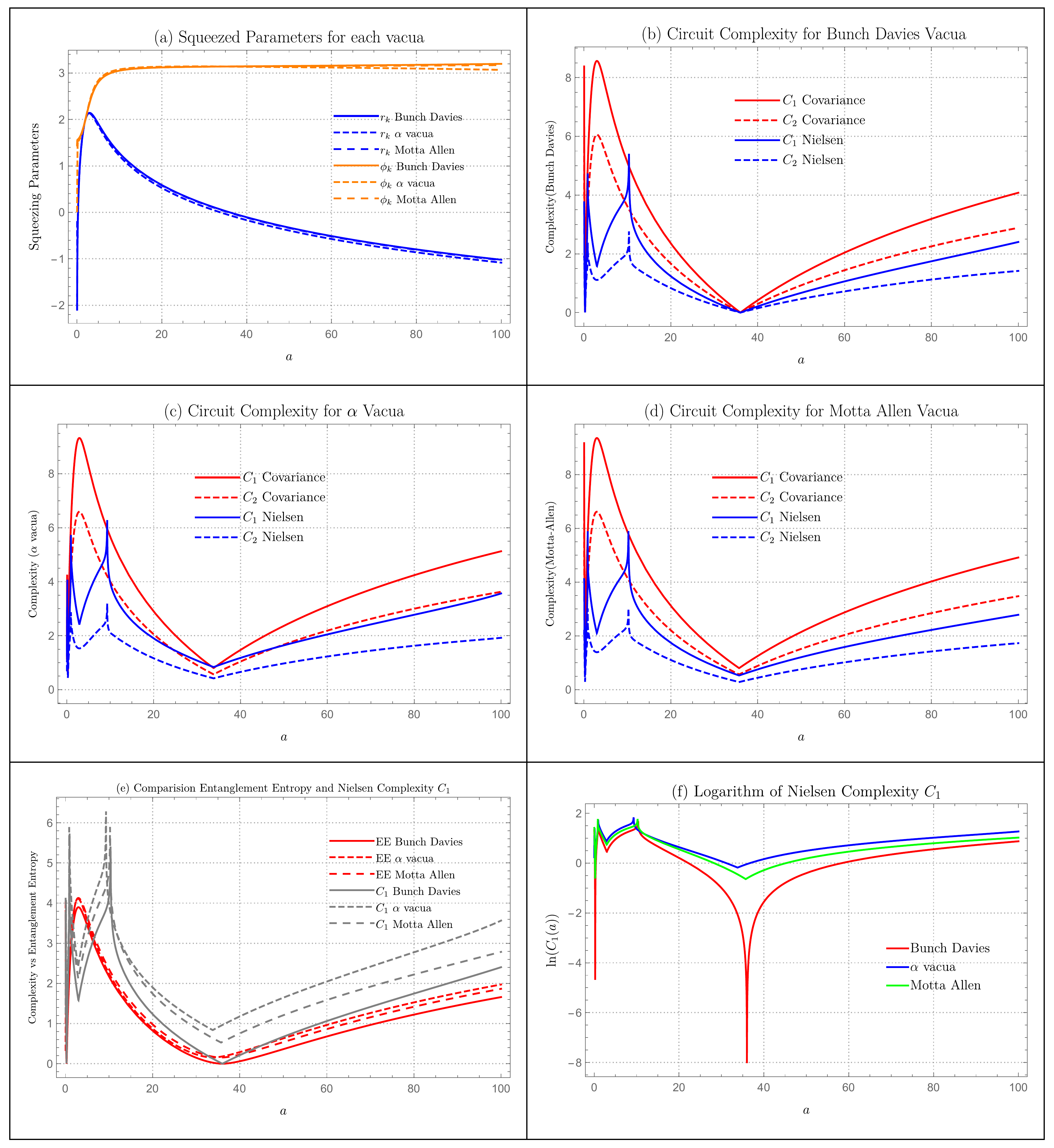}
	\caption{\textbf{Power Law Bounce Model }(a) Behaviour of the squeezing parameters in Power Law Bounce Model for $k = 0.1,~ \alpha_*=0.9,~\tau_*=0.2,~a_*=1$, $\alpha = 0.2$ and $ \gamma = 0.4$ for Bunch-Davies, $\alpha$ and Motta-Allen Vacua (b,c,d) Circuit complexity $C_1$ and $C_2$ for Bunch-Davies Vacua, $\alpha$ Vacua and Motta-Allen Vacua respectively using both Nielsen and Covariance approach for same parameters as in fig.a (e) comparison of entanglement entropy with Nielsen circuit complexity $C_1$ (f) Logarithm of Nielsen circuit complexity $C_1$.}
	\label{fig:PowerLawBouncePlots}
\end{figure}

\begin{figure}[h!]
	\centering
	\includegraphics[width=\textwidth,height=\textheight,keepaspectratio]{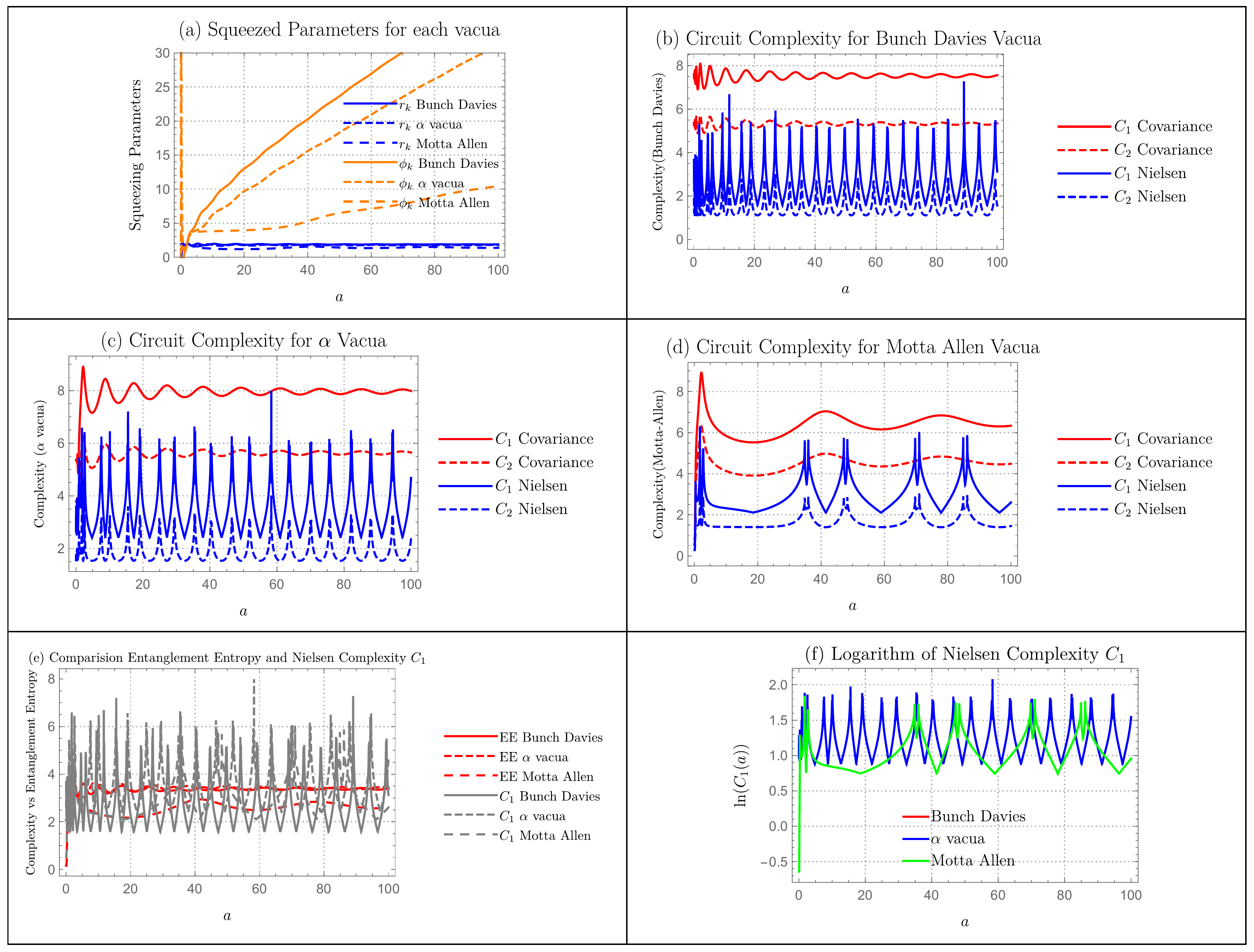}
	\caption{\textbf{Polynomial Bounce Model }(a) Behaviour of the squeezing parameters in Polynomial Bounce Model for $k = 0.1,~\tau_*=0.2, \gamma=0.5,~\delta=0.5,~a_*=1$, $\alpha = 0.2$ and $ \gamma = 0.4$ for Bunch-Davies, $\alpha$ and Motta-Allen Vacua (b,c,d) Circuit complexity $C_1$ and $C_2$ for Bunch-Davies Vacua, $\alpha$ Vacua and Motta-Allen Vacua respectively using both Nielsen and Covariance approach for same parameters as in fig.a (e) comparison of entanglement entropy with Nielsen circuit complexity $C_1$ (f) Logarithm of Nielsen circuit complexity $C_1$.}
	\label{fig:polynomialBouncePlots}
\end{figure}

\begin{figure}[h!]
	\centering
	\includegraphics[width=\textwidth,height=\textheight,keepaspectratio]{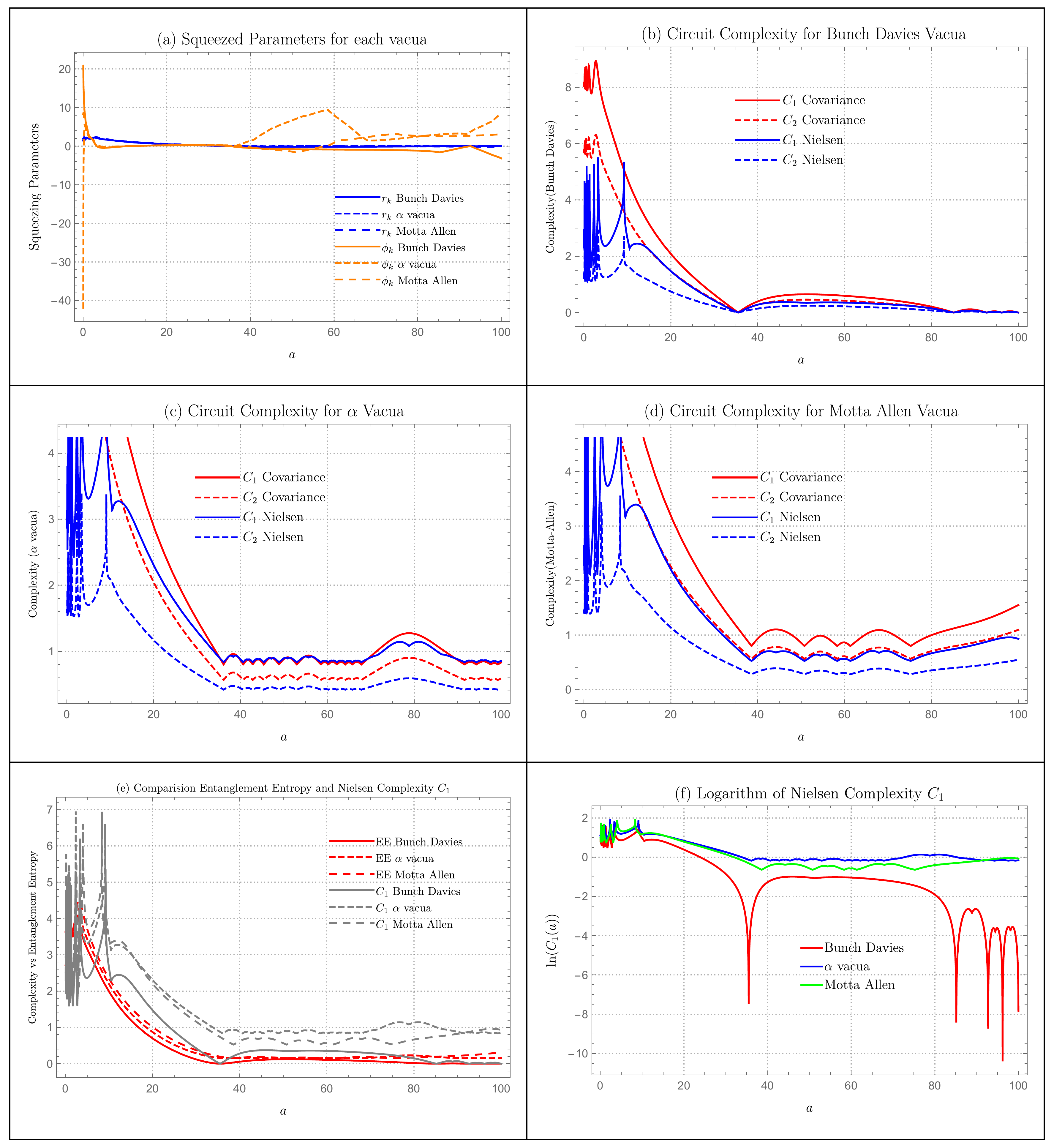}
	\caption{\textbf{Expansion (Post-Bounce) Model }(a) Behaviour of the squeezing parameters in Expansion (Post-Bounce) Model for $k = 0.1,~\tau_*=0.2, \gamma=0.5,~\delta=0.5,~a_*=1$, $\alpha = 0.2$ and $ \gamma = 0.4$ for Bunch-Davies, $\alpha$ and Motta-Allen Vacua (b,c,d) Circuit complexity $C_1$ and $C_2$ for Bunch-Davies Vacua, $\alpha$ Vacua and Motta-Allen Vacua respectively using both Nielsen and Covariance approach for same parameters as in fig.a (e) comparison of entanglement entropy with Nielsen circuit complexity $C_1$ (f) Logarithm of Nielsen circuit complexity $C_1$.}
	\label{fig:expansionPlots}
\end{figure}

For each figures from (\ref{fig:contractionPlots}) - (\ref{fig:expansionPlots}), in sub figure a, we have plotted squeezing parameters $r_k$ and $\phi_K$ for each vacua using the parameters that will be given below. Since all measures of interest such as Complexity, Entropy and Chaos are dependent on these parameters, it is crucial to understand their behaviour.

\begin{enumerate}
    \item \textit{Contraction Bounce Model: } In (\ref{fig:contractionPlots}).a, we have plotted squeezing parameters $r_k$ and $\phi_k$ for each vacua
 using the parameters $k = 0.1, \alpha = 0.2$ and $ \gamma = 0.4$. The growth of $r_k$ is very small for all vacua and
 they merge at each other. The evolution of $\phi_k$ is different to Motta-Allen vacua than $\alpha$ and Bunch-Davies Vacua.

    \item \textit{Matter Bounce Model: } In (\ref{fig:matterBouncePlots}).a, we have plotted squeezing parameters $r_k$ and $\phi_k$ for each vacua
 using the parameters $k = 0.1, \alpha = 0.2$ and $ \gamma = 0.4$. For all three vacua, $r_k$ grows at the early values of scale factor $a$, and then
keeps dropping. While for all three vacua, $\phi_k$ grows and then saturates.

\item \textit{Sechyperbolic Bounce Model: } In (\ref{fig:sechPlots}).a, we have plotted squeezing parameters $r_k$ and $\phi_k$ for each vacua
 using the parameters $k = 0.1, \alpha = 0.2$ and $ \gamma = 0.4$. For all three vacua, $r_k$ keeps growing with respect to the scale factor $a$.
While the behavior of $\phi_k$ is different for each vacua at early $a$. Then, they starts to merge after some time.

\item \textit{Cosinehyperbolic Bounce Model: } In (\ref{fig:coshPlots}).a, we have plotted squeezing parameters $r_k$ and $\phi_k$ for each vacua
 using the parameters $k = 0.1, \alpha = 0.2$ and $ \gamma = 0.4$. The growth of $r_k$ is very small for all vacua and
 they merge at each other. At early time, $\phi_k$ starts to grow together with the evolution of $\phi_k$ being different to Motta-Allen vacua than 
$\alpha$ and Bunch-Davies Vacua at later $a$. 

\item \textit{Sinehyperbolic Bounce Model: } In (\ref{fig:sinhPlots}).a, we have plotted squeezing parameters $r_k$ and $\phi_k$ for each vacua
 using the parameters $k = 0.1, \alpha = 0.2$ and $ \gamma = 0.4$. The growth of $r_k$ is very small for all vacua and
 they merge at each other. At early time, $\phi_k$ starts to grow together with the evolution of $\phi_k$ being different to Motta-Allen vacua than 
$\alpha$ and Bunch-Davies Vacua at later $a$. 

\item \textit{Cosechyperbolic Bounce Model: } In (\ref{fig:bouncingCschPlots}).a, we have plotted squeezing parameters $r_k$ and $\phi_k$ for each vacua
 using the parameters $k = 0.1, \alpha = 0.2$ and $ \gamma = 0.4$. The growth of $r_k$ takes a dip around $a=0.15$ and then keeps a linear growth. This holds
for all vacua. While value of $\phi_k$ are different for each vacua until $a=0.3$, and then they merge. After that, they keep on dropping until it merge with
$r_k$.

\item \textit{Exponential Bounce Model: } In (\ref{fig:BounceInverseErfPlots}).a, we have plotted squeezing parameters $r_k$ and $\phi_k$ for each vacua
 using the parameters $k = 0.1, \alpha = 0.2$ and $ \gamma = 0.4$. The value of $r_k$ is constant up to $a=0.35$, and then they starts to grow linearly.
The value of $\phi_k$ keeps dropping until it merges with $r_k$. This holds for all vacua.

\item \textit{Power Law Bounce Model: } In (\ref{fig:PowerLawBouncePlots}).a, we have plotted squeezing parameters $r_k$ and $\phi_k$ for each vacua
 using the parameters $k = 0.1, \alpha = 0.2$ and $ \gamma = 0.4$. The value of $r_k$ takes a sharp growth and starts to drop. While $\phi_k$ grows and then
saturates.  

\item \textit{Polynomial Bounce Model: } In (\ref{fig:polynomialBouncePlots}).a, we have plotted squeezing parameters $r_k$ and $\phi_k$ for each vacua
 using the parameters $k = 0.1, \alpha = 0.2$ and $ \gamma = 0.4$. The growth of $r_k$ is constant and almost same for each vacua. Bunch-Davies vauca has the largest growth
of $\phi_k$ followed by Alpha and then Motta-Allen vacua. 

 \item \textit{Expansion (Post-Bounce) Model: } In (\ref{fig:expansionPlots}).a, we have plotted squeezing parameters $r_k$ and $\phi_k$ for each vacua
 using the parameters $k = 0.1, \alpha = 0.2$ and $ \gamma = 0.4$. The growth of $r_k$ is constant and almost same for each vacua. The growth of $\phi_k$ is 
similar in the beginning for each vacua but they starts to diverge around $a=50$.
  
\end{enumerate}

\subsubsection*{Complexity Measure}
For each figures from (\ref{fig:contractionPlots}) - (\ref{fig:expansionPlots}), in sub figure b,c and d, we have plotted the circuit complexity measures for Bunch-Davies Vacua, $\alpha$ Vacua and Motta-Allen Vacua respectively using both Covariance as well as Nielsen Approach using calculations from Section \ref{sec:complexityMeasure}. Before discussing each case individually, let us discuss the overall features of these complexity measures. Covariance measure of complexity has similar pattern to the growth of squeezing parameter $r_k$. However, Nielsen's complexity measure has more details on it. This is reasonable as Nielsen's complexity measure is sensitive to both $r_k$ and $\phi_k$.

\begin{enumerate}
    \item \textit{Contraction Bounce Model: } Circuit complexity plots for each vacua is identical in overall pattern except for the magnitude.
 Overall, Motta-Allen Vacua has the largest magnitude with Bunch-Davies being lowest which is due to Motta-Allen Vacua having largest number of
 external parameters. Covariance measure of complexity is similar to the growth of the squeezing parameter $r_k$ with increase in magnitude.
 However, Nielsen's measure is telling a different story. Nielsen's measure of complexity is highly oscillatory for each vacua and for both cost function
 $C_1$ and $C_2$.

\item \textit{Matter Bounce Model: } Circuit complexity plots for each vacua is identical in overall pattern except for the magnitude. The plot for the 
covariance measure of complexity is smooth with a growth at early times and then taking a dip around $a=20$. After the dip, both $C_1$ and $C_2$ growing 
linearly. Nielsen measure of complexity has similar pattern but with more details on it. Around $a=60$, we can observe a pick in both $C_1$ and $C_2$. Even 
in early values of $a$, we can observe some oscillatory behaviour. 

\item \textit{Sechyperbolic Bounce Model: } Interestingly, we can observe different behaviour in evolution of complexity for each vacua. For Bunch-Davies case,
both $C_1$ and $C_2$ Covariance's complexity is oscillatory before $a=0.2$, and then they keep growing with $C_2$ being bounded by $C_1$. While this pattern
is visible for Nielsen's approach too, there is a oscillatory around $a=0.65$. For $\alpha$ vacua, more such peaks are visible in Nielsen's measure of
complexity. In Motta-Allen Vacua, both covariance and Nielsen's measure of complexity is oscillatory before $a=0.35$ after which they starts to take a smooth
rise.

\item \textit{Cosinehyperbolic Bounce Model: } Circuit complexity plots for each vacua is identical in overall pattern except for the magnitude.
 Overall, Motta-Allen Vacua has the largest magnitude with Bunch davies being lowest which is due to Motta-Allen Vacua having largest number of
 external parameters. Covariance measure of complexity is similar to the growth of the squeezing parameter $r_k$ with increase in magnitude.
 However, Nielsen's measure is telling a different story. Nielsen's measure of complexity is highly oscillatory for each vacua and for both cost function
 $C_1$ and $C_2$. In all cases, $C_2$ is being bounded by $C_1$.

\item \textit{Sinehyperbolic Bounce Model: } Circuit complexity plots for each vacua is identical in overall pattern except for the magnitude.
 Overall, Motta-Allen Vacua has the largest magnitude with Bunch davies being lowest which is due to Motta-Allen Vacua having largest number of
 external parameters. Covariance measure of complexity is similar to the growth of the squeezing parameter $r_k$ with increase in magnitude.
 However, Nielsen's measure is telling a different story. Nielsen's measure of complexity is highly oscillatory for each vacua and for both cost function
 $C_1$ and $C_2$. In all cases, $C_2$ is being bounded by $C_1$.

\item \textit{Cosechyperbolic Bounce Model: } Interestingly, we can observe different behavior in evolution of complexity for each vacua. For Bunch-Davies case,
both $C_1$ and $C_2$ Covariance's complexity is oscillatory before $a=0.35$, and then they keep growing with $C_2$ being bounded by $C_1$. While this pattern
is visible for Nielsen's approach too, there is a oscillatory peak around $a=0.8$. Before $a=0.35$, the complexity measure is oscillatory just like for Bunch-
Davies case but with more sharp peaks visible. In all cases, $C_2$ is being bounded by $C_1$.

\item \textit{Exponential Bounce Model: } The overall pattern of complexity measure is identical for each vacua expect for the change in magnitude. Covariance
measure of complexity is similar to the growth of $r_k$ with $C_2$ being bounded by $C_1$. Nielsen's measure of complexity is very oscillatory before $a=0.5$
after which it takes a growth and saturates.

\item \textit{Power Law Bounce Model: } The overall pattern of complexity measure is identical for each vacua expect for the change in magnitude. Covariance
measure of complexity takes a sharp growth and takes a dip until $a= 36$ and then keep a linear growth with $C_2$ being bounded by $C_1$. While this is also
true for Nielsen's measure, there are some peaks visible around $a=10$.

\item \textit{Polynomial Bounce Model: } Circuit complexity plots patter for Bunch-Davies and $\alpha$ vacua is identical in overall pattern except for the 
magnitude. For both vacua, covariance measure has the growth pattern similar to the squeezing parameter $r_k$. However, Nielsen's measure of complexity is
very oscillatory. For the case of Motta-Allen vacua, there are more gaps in the oscillatory peaks.

\item \textit{Expansion (Post-Bounce) Model: } For each vacua, circuit complexity drops until $a=35$ and then saturates around that point. In all vacua, 
Nielsen's measure of complexity is quite chaotic and oscillatory before $a=10$. For $\alpha$ and Motta-Allen vacua, there are also more oscillatory bumps
visible after the point $a=35$.
    
\end{enumerate}

\subsubsection*{comparison of entanglement entropy with Complexity}
Entanglement entropy is a very popular probe to study the dynamics of quantum systems. We would like to see if circuit complexity can also be a similar candidate. 
In Figure e, we have plotted a comparison of Nielsen's circuit complexity $C_1$ and entanglement entropy.
We have chosen Nielsen's measure of complexity because, both Covariance complexity and entanglement entropy are independent of 
squeezing angle and depends linearly on squeezing parameter $r_k$. So, the pattern obtained from covariance measure of complexity 
will be similar as to entanglement entropy except that both complexity measure $C_1$ and $C_2$ is greater than entanglement entropy.
 However with Nielsen's measure of complexity, we can obtain more interesting details of the system.
 
\begin{enumerate}

\item \textit{Contraction Bounce Model: } Complexity is highly oscillatory compared to the entanglement entropy.

\item \textit{Matter Bounce Model: } Both entanglement entropy and complexity measure has similar growth pattern. However, complexity measure has more peaks
visible around $a=60$.

\item \textit{Sechyperbolic Bounce Model: } Again complexity measure has more oscillatory peaks compared to the entanglement entropy.This allows us to understand
the system in more detail.

\item \textit{Cosinehyperbolic Bounce Model: } Complexity is highly oscillatory compared to the entanglement entropy.

\item \textit{Sinehyperbolic Bounce Model: } Complexity is highly oscillatory compared to the entanglement entropy.

\item \textit{Cosechyperbolic Bounce Model: } Both entanglement entropy and complexity has similar growth pattern but entanglement entropy has a very huge peak
at early values of $a$.

\item \textit{Exponential Bounce Model: } Nielsen's measure of complexity is very oscillatory before $a=0.5$ after which it takes a growth and saturates while
entanglement entropy is not. But, entanglement entropy also takes a growth and saturates just like complexity measure.

\item \textit{Power Law Bounce Model: } Both complexity and entanglement entropy has a similar growth pattern, but there are some peaks visible around $a=10$
for the case of complexity measure.

\item \textit{Polynomial Bounce Model: } Complexity is highly oscillatory compared to the entanglement entropy.

\item \textit{Expansion (Post-Bounce) Model: } Nielsen's measure of complexity is quite chaotic and oscillatory before $a=10$ but the overall pattern is similar 
to of entanglement entropy.

\end{enumerate}

\subsubsection*{Quantum Lyapunov Exponent}
Circuit complexity has also been proposed as a tool to measure quantum chaos. In particular, low growth of complexity indicates less chaotic system while higher growth of complexity indicates highly chaotic system. So, slope of the complexity could be a measure of quantum chaos. As a crude approximation, we will call the Lyapunov exponent $\lambda_i$ to be:
\begin{equation}
    \lambda_i = \frac{\text{ln}C_i(\text{point of saturation})-\text{ln}C_i(\text{point of rise})}{a(\text{point of saturation})-a(\text{point of rise})}
\end{equation}
where $i$ indicates the choice of vacua. For simplicity, we will restrict to Nielsen complexity $C_1$ and we will obtain Lyapunov exponent $\lambda_i$.

\begin{enumerate}

\item \textit{Contraction Bounce Model: }
The log. complexity plot is too oscillatory to get the chaotic measure.

\item \textit{Matter Bounce Model: }
\begin{equation}
\begin{aligned}
  \lambda_{Motta-Allen} &= 0.008 \\
  \lambda_{\alpha} &= 0.007 \\
  \lambda_{Bunch-Davies} &= 0.008
\end{aligned}
\end{equation}
All three vacua has similar lyapunov exponent, so they have similar chaotic properties.

\item \textit{Sechyperbolic Bounce Model: }
\begin{equation}
\begin{aligned}
  \lambda_{Motta-Allen} &= 1 \\
  \lambda_{\alpha} &= 0.01\\
  \lambda_{Bunch-Davies} &= 0.16
\end{aligned}
\end{equation}
Motta-Allen Vacua has the largest lyapunov exponent, so it is the most chaotic cosmological model followed by Bunch-Davies and $\alpha$ vacua.

\item \textit{Cosinehyperbolic Bounce Model: } The log. complexity plot is too oscillatory to get the chaotic measure.

\item \textit{Sinehyperbolic Bounce Model: } The log. complexity plot is too oscillatory to get the chaotic measure.

\item \textit{Cosechyperbolic Bounce Model: } 
\begin{equation}
\begin{aligned}
  \lambda_{Motta-Allen} &= 1.25 \\
  \lambda_{\alpha} &= 0.41\\
  \lambda_{Bunch-Davies} &= 2.91
\end{aligned}
\end{equation}
Bunch-Davies Vacua has the largest Lyapunov exponent, so it is the most chaotic cosmological model followed by Motta-Allen and $\alpha$ vacua.

\item \textit{Exponential Bounce Model: } 
\begin{equation}
\begin{aligned}
  \lambda_{Motta-Allen} &= 0.83 \\
  \lambda_{\alpha} &= 0.75\\
  \lambda_{Bunch-Davies} &= 0.75
\end{aligned}
\end{equation}
All three vacua has similar Lyapunov exponent, so they have similar chaotic properties.

\item \textit{Power Law Bounce Model: }
\begin{equation}
\begin{aligned}
  \lambda_{Motta-Allen} &= 0.02 \\
  \lambda_{\alpha} &= 0.01\\
  \lambda_{Bunch-Davies} &= 0.03
\end{aligned}
\end{equation}
All three vacua has similar Lyapunov exponent, so they have similar chaotic properties.

\item \textit{Polynomial Bounce Model: } The log. complexity plot is too oscillatory to get the chaotic measure.

\item \textit{Expansion (Post-Bounce) Model: } There is no growth or saturation point of complexity, so we will not be able to get the chaotic measure.

\end{enumerate}

\subsection{Cyclic Models}

In Figure (\ref{fig:matterCyclicPlots}) and (\ref{fig:radiationCyclicPlots}), we have numerically plotted the squeezing parameters,
 derived circuit complexity measures for Bunch-Davies Vacua, $\alpha$ Vacua and Motta-Allen Vacua, comparison of entanglement entropy
 with circuit complexity measure and quantum chaos for Matter cyclic and Radiation cyclic model on super-horizon scales for the parameters
 $k = 0.1, \alpha = 0.2$ and $ \gamma = 0.4$. We have set $r_k(a = 1) = 1, \phi_k(a = 1) = 1$ as our initial conditions. 
The extra parameters we have selected for each bouncing model will be highlighted under it's own headings.

\subsubsection*{Squeezing Parameters}
\begin{figure}[h!]
	\centering
	\includegraphics[width=\textwidth,height=\textheight,keepaspectratio]{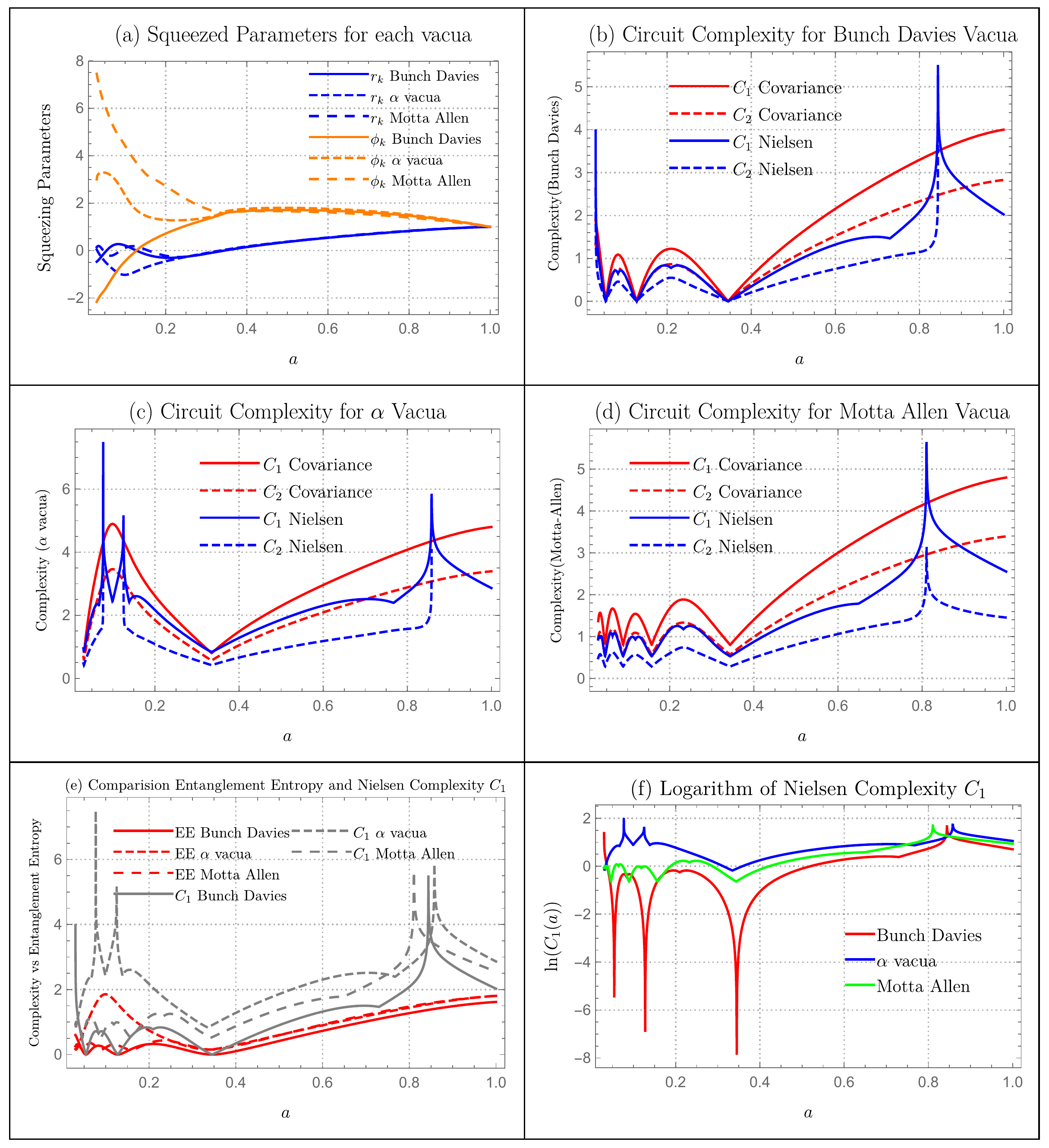}
	\caption{\textbf{Matter Cyclic Model }(a) Behaviour of the squeezing parameters in Matter Cyclic Model for $k = 0.1,~\tau_*=0.2, \gamma=0.5,~\delta=0.5,~a_*=1$, $\alpha = 0.2$ and $ \gamma = 0.4$ for Bunch-Davies, $\alpha$ and Motta-Allen Vacua (b,c,d) Circuit complexity $C_1$ and $C_2$ for Bunch-Davies Vacua, $\alpha$ Vacua and Motta-Allen Vacua respectively using both Nielsen and Covariance approach for same parameters as in fig.a (e) comparison of entanglement entropy with Nielsen circuit complexity $C_1$ (f) Logarithm of Nielsen circuit complexity $C_1$.}
	\label{fig:matterCyclicPlots}
\end{figure}

\begin{figure}[h!]
	\centering
	\includegraphics[width=\textwidth,height=\textheight,keepaspectratio]{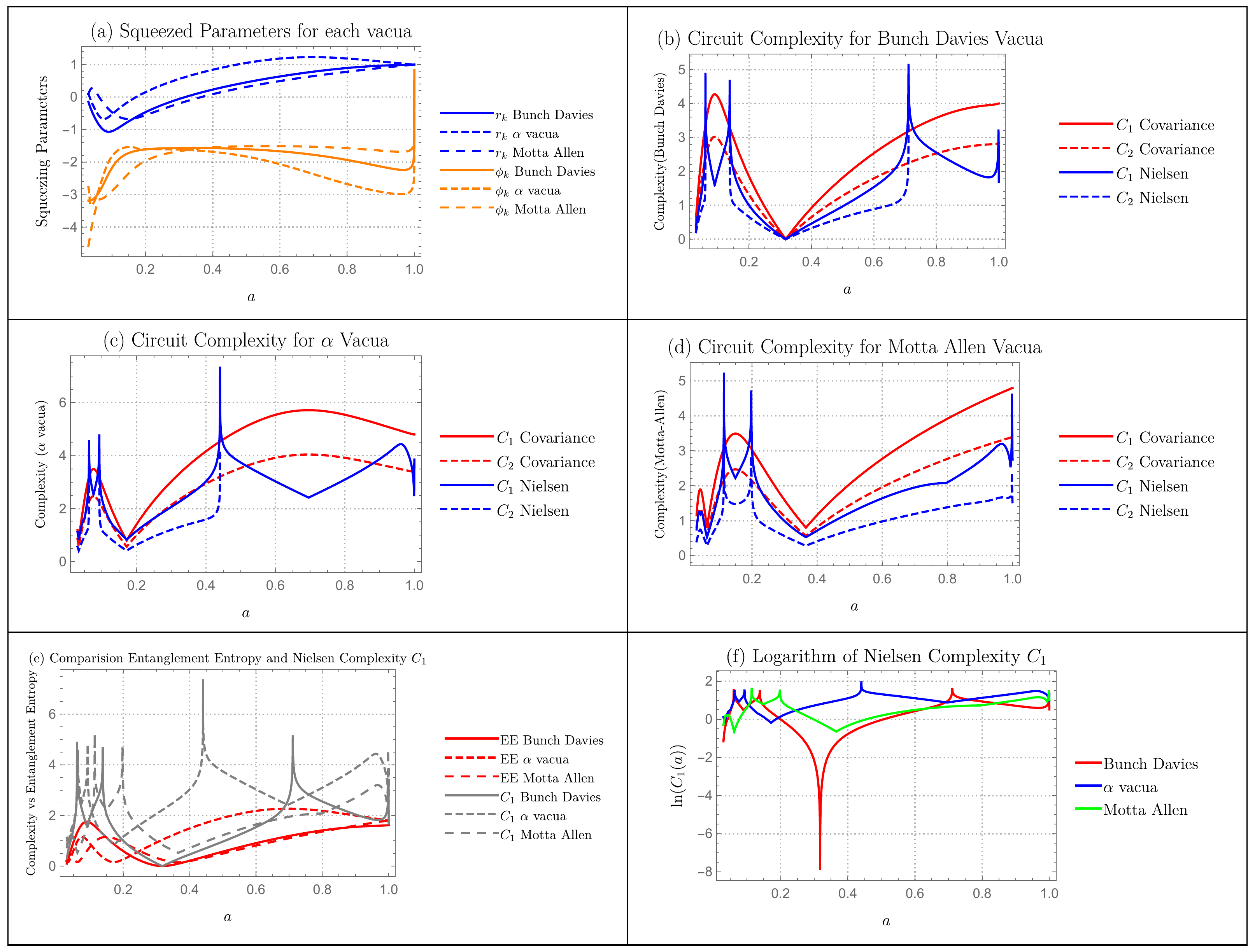}
	\caption{\textbf{Radiation Cyclic Model }(a) Behaviour of the squeezing parameters in Radiation Cyclic Model for $k = 0.1,~\tau_*=0.2,  \gamma=0.5,~\delta=0.5,~a_*=1$, $\alpha = 0.2$ and $ \gamma = 0.4$ for Bunch-Davies, $\alpha$ and Motta-Allen Vacua (b,c,d) Circuit complexity $C_1$ and $C_2$ for Bunch-Davies Vacua, $\alpha$ Vacua and Motta-Allen Vacua respectively using both Nielsen and Covariance approach for same parameters as in fig.a (e) comparison of entanglement entropy with Nielsen circuit complexity $C_1$ (f) Logarithm of Nielsen circuit complexity $C_1$.}
	\label{fig:radiationCyclicPlots}
\end{figure}

For each figures from (\ref{fig:matterCyclicPlots}) and (\ref{fig:radiationCyclicPlots}), in sub figure a, we have plotted squeezing parameters
 $r_k$ and $\phi_K$ for each vacua using the parameters that will be given below. Since all measures of interest such as Complexity, 
Entropy and Chaos are dependent on these parameters, it is crucial to understand their behavior.

\begin{enumerate}
    \item \textit{Matter cyclic model: } In (\ref{fig:matterCyclicPlots}).a, we have plotted squeezing parameters $r_k$ and $\phi_k$ for each vacua
 using the parameters $k = 0.1, \alpha = 0.2$ and $ \gamma = 0.4$. Before $a=0.2$,  $r_k$ is different for each vacua. However, after that point, they
merge and starts to grow linearly. While before $a=0.3$, $\phi_k$ is different for each vacua, after that point they merge and saturates.

\item \textit{Radiation cyclic model: } In (\ref{fig:radiationCyclicPlots}).a, we have plotted squeezing parameters $r_k$ and $\phi_k$ for each vacua
 using the parameters $k = 0.1, \alpha = 0.2$ and $ \gamma = 0.4$. The value of $r_k$ grows for each vacua but with a different magnitude. $\alpha$ vacua
has the largest magnitude while Bunch-Davies and Motta-Allen have comparable values of $r_k$. At early values of $a$, $\phi_k$ grows for each vacua and saturates.
However at around $a=1$, all three vacua merges and takes a sharp growth.

\end{enumerate}

\subsubsection*{Complexity Measure}
For each figures from (\ref{fig:matterCyclicPlots}) and (\ref{fig:radiationCyclicPlots}), in sub figure b,c and d, we have plotted the circuit complexity measures
 for Bunch-Davies Vacua, $\alpha$ Vacua and Motta-Allen Vacua respectively using both Covariance as well as Nielsen Approach using calculations from
 Section \ref{sec:complexityMeasure}. Before discussing each case individually, let us discuss the overall features of these complexity measures.
 Covariance measure of complexity has similar pattern to the growth of squeezing parameter $r_k$. However, Nielsen's complexity measure has more details
 on it. This is reasonable as Nielsen's complexity measure is sensitive to both $r_k$ and $\phi_k$.

\begin{enumerate}

\item \textit{Matter cyclic model: } For Bunch-Davies and Motta-Allen Vacua, covariance measure of complexity has a similar pattern but with different magnitude.
Until $a=0.35$, it is oscillatory in nature with small magnitude. After which, it takes a linear growth. For $\alpha$ vacua, complexity grows at early values of $a$, then 
take a dip at $a=0.35$, and then grows linearly. Nielsen's measure of complexity also has a similar pattern as covariance but has a peak around $a=0.8$.

    \item \textit{Radiation cyclic model: } Covariance measure of complexity has the similar pattern for all three vacua. It grows at early values of $a$, then 
take a dip and then grows linearly. The point of dip is different for each vacua. However, Nielsen's measure of complexity has more details coming from peaks
at different values of $a$.

\end{enumerate}

\subsubsection*{comparison of entanglement entropy with Complexity}
Entanglement entropy is a very popular probe to study the dynamics of quantum systems. We would like to see if circuit complexity can also be a similar candidate. 
In Figure e, we have plotted a comparison of Nielsen's circuit complexity $C_1$ and entanglement entropy.
We have chosen Nielsen's measure of complexity because, both Covariance complexity and entanglement entropy are independent of 
squeezing angle and depends linearly on squeezing parameter $r_k$. So, the pattern obtained from covariance measure of complexity 
will be similar as to entanglement entropy except that both complexity measure $C_1$ and $C_2$ is greater than entanglement entropy.
 However with Nielsen's measure of complexity, we can obtain more interesting details of the system.
 
\begin{enumerate}

\item \textit{Matter Cyclic Model: } Both entanglement entropy and complexity measure has similar growth pattern. However, complexity measure has more peaks
visible around $a=0.8$.

\item \textit{Radiation Cyclic Model: } Both entanglement entropy and complexity measure has similar growth pattern. However, complexity measure has more peaks.
This shows that complexity measure is able to carry more details regarding the system compared to matter cyclic model.
\end{enumerate}
\subsubsection*{Quantum Lyapunov Exponent}
Circuit complexity has also been proposed as a tool to measure quantum chaos. In particular, low growth of complexity indicates less chaotic system while higher growth of complexity indicates highly chaotic system. So, slope of the complexity could be a measure of quantum chaos. As a crude approximation, we will call the lyapunov exponent $\lambda_i$ to be:
\begin{equation}
    \lambda_i = \frac{\text{ln}C_i(\text{point of saturation})-\text{ln}C_i(\text{point of rise})}{a(\text{point of saturation})-a(\text{point of rise})}
\end{equation}
where $i$ indicates the choice of vacua. For simplicity, we will restrict to Nielsen complexity $C_1$ and we will obtain lyapunov exponent $\lambda_i$.

\begin{enumerate}
\item \textit{Matter Cyclic Model: }
\begin{equation}
\begin{aligned}
  \lambda_{Motta-Allen} &= 1.83 \\
  \lambda_{\alpha} &= 1.33 \\
  \lambda_{Bunch-Davies} &= 3.16
\end{aligned}
\end{equation}
Bunch-Davies Vacua has the largest lyapunov exponent, so it is the most chaotic cosmological model followed by Motta-Allen and $\alpha$ vacua.

\item \textit{Radiation Cyclic Model: }
\begin{equation}
\begin{aligned}
  \lambda_{Motta-Allen} &= 2\\
  \lambda_{\alpha} &= 0.33 \\
  \lambda_{Bunch-Davies} &= 2.5
\end{aligned}
\end{equation}
\end{enumerate}
Bunch-Davies Vacua has the largest lyapunov exponent, so it is the most chaotic cosmological model followed by Motta-Allen and $\alpha$ vacua.

\subsection{Black Hole Gas}
In Figure (\ref{fig:blackHolePlots}), we have numerically plotted the squeezing parameters, derived circuit complexity measures for Bunch-Davies Vacua, $\alpha$ Vacua and Motta-Allen Vacua, comparison of entanglement entropy with circuit complexity measure and quantum chaos in de Sitter Model on super-horizon scales for the parameters $k = 0.1$, $\alpha = 0.2$ and $ \gamma = 0.4$. We have set $r_k(a = 1) = 1, \phi_k(a = 1) = 1$ as our initial conditions. 
\subsubsection*{Squeezing Parameters}
In Figure (\ref{fig:blackHolePlots}.a), we have plotted squeezing parameters $r_k$ and $\phi_K$ for each vacua using the parameters listed above. $r_k$ takes a constant growth while $\phi_k$ grows very quickly. Since both circuit complexity and entanglement entropy are strongly dependent on the values of these parameters, the behavior of these parameters are significant to understand the dynamics of the system.  
\subsubsection*{Complexity Measure}
\begin{figure}[h!]
	\centering
	\includegraphics[width=\textwidth,height=\textheight,keepaspectratio]{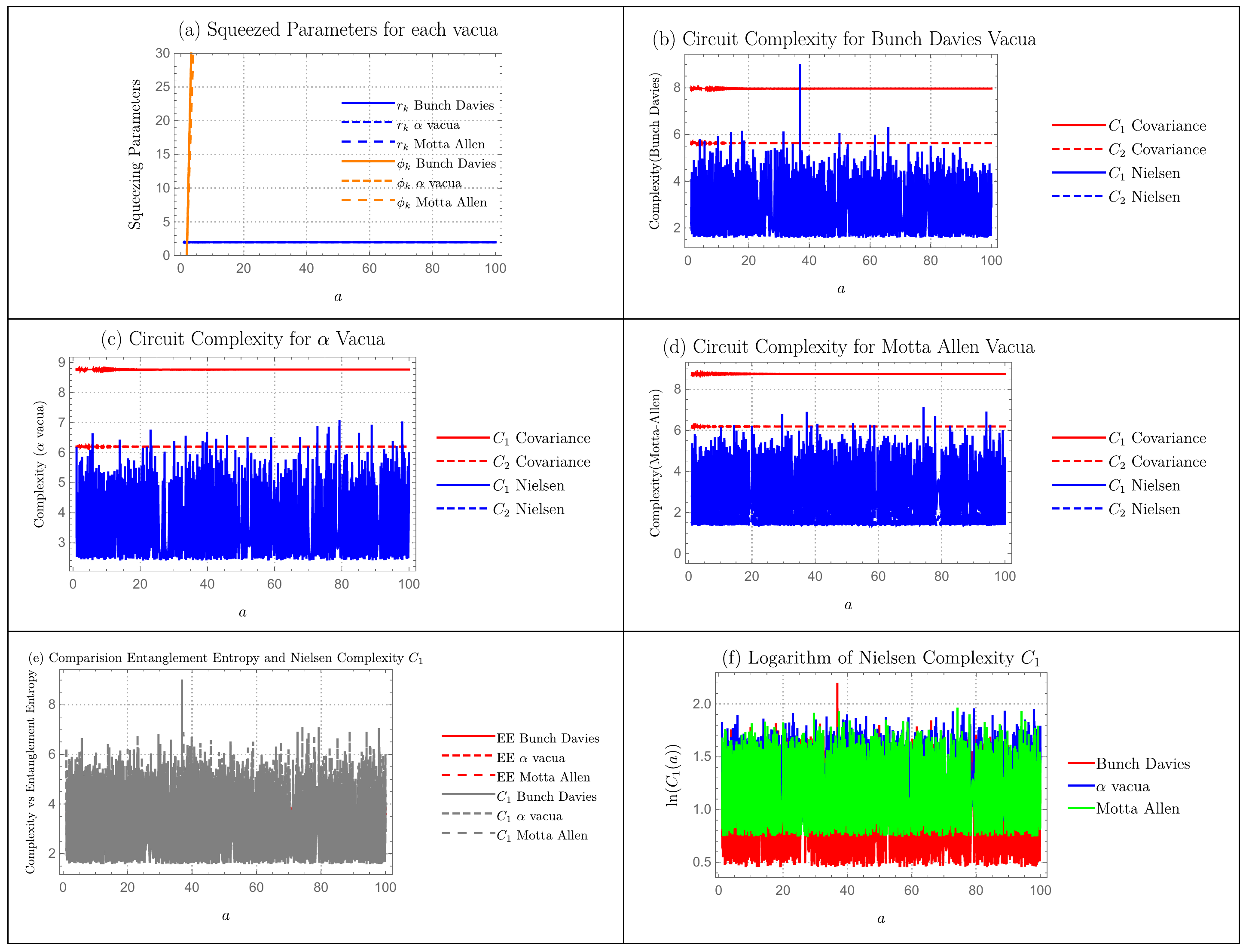}
	\caption{\textbf{Black Hole Gas Model }(a) Behaviour of the squeezing parameters in Black Hole gas model for $a_*=1
	$ for Bunch-Davies, $\alpha$ and Motta-Allen Vacua (b,c,d) Circuit complexity $C_1$ and $C_2$ for Bunch-Davies Vacua, $\alpha$ Vacua and Motta-Allen Vacua respectively using both Nielsen and Covariance approach for same parameters as in fig.a (e) comparison of entanglement entropy with Nielsen circuit complexity $C_1$ (f) Logarithm of Nielsen circuit complexity $C_1$.}
	\label{fig:blackHolePlots}
\end{figure}

In Figure (\ref{fig:blackHolePlots}.b ,\ref{fig:blackHolePlots}.c, \ref{fig:blackHolePlots}.d), we have plotted the circuit complexity measures for Bunch-Davies Vacua, $\alpha$ Vacua and Motta-Allen Vacua respectively using both Covariance as well as Nielsen Approach using calculations from Section \ref{sec:complexityMeasure}. 
Nielsen's measure of complexity is very oscillatory while Covariance measure grows similar to $r_k$ but with larger magnitude.
\subsubsection*{comparison of entanglement entropy with Complexity}
Entanglement entropy is a very popular probe to study the dynamics of quantum systems. We would like to see if circuit complexity can also be a similar candidate. In Figure (\ref{fig:blackHolePlots}.e), we have plotted a comparison of Nielsen's circuit complexity $C_1$ and entanglement entropy.
We have chosen Nielsen's measure of complexity because, both Covariance complexity and entanglement entropy are independent of squeezing angle and depends linearly on squeezing parameter $r_k$. So, the pattern obtained from covariance measure of complexity will be similar as to entanglement entropy except that both complexity measure $C_1$ and $C_2$ is greater than entanglement entropy. However with Nielsen's measure of complexity, we can obtain more interesting details of the system such as the oscillatory behavior of complexity which is absent in entanglement entropy.
\subsubsection*{Quantum Lyapunov Exponent}
Circuit complexity has also been proposed as a tool to measure quantum chaos. In particular, low growth of complexity indicates less chaotic system while higher growth of complexity indicates highly chaotic system. So, slope of the complexity could be a measure of quantum chaos. Because Nielsen's measure of complexity is too oscillatory and doesn't reach a saturation point, we will not be able to compute the lyapunov exponent term. 

\section{Conclusions }
\label{sec:conclusion}
From our study we have the following final remarks:
 
\begin{itemize}
\item Compared to that calculated using the covariance matrix method, the circuit complexity computed using Nielsen's wave function approach offers a much better understanding because it relies on the squeezing angle and the squeezing parameter and can therefore be linked to the entanglement entropy.

    \item We have computed the circuit complexity using both Nielsen's wave function approach and Covariance matrix approach. From the study of both approaches to compute complexity in different cosmological models, it is clear that the overall pattern one obtain is similar in both cases. However, the circuit complexity computed from Nielsen's wave function approach gives a much detailed understanding of the evolution of the system. The reason is that covariance measure of complexity is insensitive to the squeezing angle.
    
    \item The behaviour of the squeezing parameter is also different for each vacua: Bunch-Davies, $\alpha$ vacua and Motta-Allen Vacua. The reason for this is that dispersion relation for each vacua is different. So, while solving the set of differential equations, we will get different solutions for the squeezing parameter.
    
    \item We have computed the circuit complexity for all three vacua: Bunch-Davies, $\alpha$ vacua and Motta-Allen Vacua. Because the dispersion relation for each vacua is different, the solutions for the squeezing parameter is also different. This has the direct consequence in the properties of complexity measure. In addition, Motta-Allen and $\alpha$ vacua have more parameters than the simple Bunch-Davies vacua. This is also reflected in the complexity measure. Most of the times, Motta-Allen vacua has the largest complexity magnitude followed by $\alpha$ vacua and then Bunch-Davies vacua.
    
    \item The behavior of the entanglement entropy is also different for each vacua. Again, this is the result of having different set of solutions for the squeezing parameter. The values of entanglement entropy is bounded by the covariance measure of complexity $C_1$ and $C_2$. Entanglement entropy has the similar pattern as the covariance measure of complexity because both are independent of the squeezing angle $\phi_k$. However, entanglement entropy has the different behaviour than the Nielsen's measure of complexity. We are able to see more details in the complexity that is not visible in entanglement entropy growth. This shows that, complexity could also be a probe to detect important properties of a quantum system just like entanglement entropy. Since entanglement entropy is sometimes ridiculously difficult to compute, complexity could be a better alternative at those cases.
    
    \item The behaviour of circuit complexity is very dependent on the model and scale factor involved. For example: Nielsen's measure of complexity is extremely oscillatory for models like Polynomial bounce and Sinehyperbolic bounce while very smooth for models like de-Sitter. 
    
\end{itemize}

The future prospects of the work are:
\begin{itemize}
    \item We have made qualitative comments on the relation between growth of complexity and quantum chaos. It would be interesting to see if it can be rigorously proven. One way could be to use the quantum speed limit properties to show the bound on complexity. Using Random Matrix theory, one can associate the bound on quantum speed limit to quantum chaos.
    
    \item For the framework of two-mode squeezed states, quantum complexity computed using Covariance matrix show similar feature that of entanglement entropy. It would be interesting to see if this holds true for other models.
    
    \item The relation between entanglement entropy and complexity can be studied in even more detail.
    
    \item In this paper, we have only studied the complexity measure for various cosmological models. It would be interesting to apply the lesson from these complexity plots to study the physical evolution of these models and observe to what extent it actually matches. Furthermore, concrete experimental prediction would be much desirable.

\end{itemize}



	\subsection*{Acknowledgements}
The research fellowship of SC is supported by the J.  C.  Bose National Fellowship of Sudhakar Panda.  SC also would line to thank School of Physical Sciences, National Institute for Science Education and Research (NISER),  Bhubaneswar for providing the work friendly environment.  SC also thank all the members of our newly formed virtual international non-profit consortium Quantum Structures of the Space-Time \& Matter (QASTM) for elaborative discussions.  Kiran Adhikari would like to thank TTK, RWTH and JARA, Institute of Quantum Information for fellowships. KA would also like to thank Dr. David Di Vincenzo for his help in understanding quantum information theoretic concepts such as entanglement entropy and complexity.
  Last but not least,  we would like to acknowledge our debt to the people belonging to the various part of the world for their generous and steady support for research in natural sciences.

\clearpage
	\appendix

\section{Role of squeezing in the dispersion relation for PGW }
\label{sec:dispersionRelation}
The general relation for the dispersion relation for PGW is given by:
\bea 
 \Omega_{\lambda,{\bf k}}(\tau):&=&\Biggl\{\left|\pi_{\lambda,{\bf k}}(\tau_0)\Biggl(\cosh r_{\lambda,{\bf k}}(\tau)~\exp(i\theta_{\lambda,{\bf k}}(\tau))+\sinh r_{\lambda,{\bf k}}(\tau)~\exp(i(\theta_{\lambda,{\bf k}}(\tau)+2\phi_{\lambda,{\bf k}}(\tau)))\Biggr)\nonumber\right.\\
&&\nonumber \left.+{\cal H}(\tau)~f_{\lambda,{\bf k}}(\tau_0)\Biggl(\cosh r_{\lambda,{\bf k}}(\tau)~\exp(i\theta_{\lambda,{\bf k}}(\tau))-\sinh r_{\lambda,{\bf k}}(\tau)~\exp(i(\theta_{\lambda,{\bf k}}(\tau)+2\phi_{\lambda,{\bf k}}(\tau)))\Biggr)\right|^2\\&&+\left(k^2-{\cal H}^2(\tau)\right)\left|f_{\lambda,{\bf k}}(\tau_0)\right|^2\nonumber\\
&&\left|\Biggl(\cosh r_{\lambda,{\bf k}}(\tau)~\exp(i\theta_{\lambda,{\bf k}}(\tau))-\sinh r_{\lambda,{\bf k}}(\tau)~\exp(i(\theta_{\lambda,{\bf k}}(\tau)+2\phi_{\lambda,{\bf k}}(\tau)))\Biggr)\right|^2\Biggr\}.\nonumber\\
&=&\Biggl\{\left|\pi_{\lambda,{\bf k}}(\tau_0)\right|^2\left|\Biggl(\cosh r_{\lambda,{\bf k}}(\tau)~\exp(i\theta_{\lambda,{\bf k}}(\tau))+\sinh r_{\lambda,{\bf k}}(\tau)~\exp(i(\theta_{\lambda,{\bf k}}(\tau)+2\phi_{\lambda,{\bf k}}(\tau)))\Biggr)\right|^2\nonumber\\
&&+k^2\left|f_{\lambda,{\bf k}}(\tau_0)\right|^2\nonumber\\
&&\left|\Biggl(\cosh r_{\lambda,{\bf k}}(\tau)~\exp(i\theta_{\lambda,{\bf k}}(\tau))-\sinh r_{\lambda,{\bf k}}(\tau)~\exp(i(\theta_{\lambda,{\bf k}}(\tau)+2\phi_{\lambda,{\bf k}}(\tau)))\Biggr)\right|^2\nonumber \\  
&& +{\cal H}(\tau)\left(\pi^{*}_{\lambda,{\bf k}}(\tau_0)f_{\lambda,{\bf k}}(\tau_0)+\pi_{\lambda,{\bf k}}(\tau_0)f^{*}_{\lambda,{\bf k}}(\tau_0)\right)\nonumber\\
&& +i{\cal H}(\tau)\sinh 2r_{\lambda,{\bf k}}(\tau)\sin 2\phi_{\lambda,{\bf k}}(\tau)\left(\pi_{\lambda,{\bf k}}(\tau_0)f^{*}_{\lambda,{\bf k}}(\tau_0)-\pi^{*}_{\lambda,{\bf k}}(\tau_0)f_{\lambda,{\bf k}}(\tau_0)\right)\nonumber\Biggr\}.\nonumber\\ 
&& \eea

The dispersion relation for individual vacua are given by:
\begin{enumerate}
    \item  \underline{\textcolor{red}{\bf Motta-Allen ($\alpha,\gamma$)~vacua:}}\\
    \bea &&\underline{\textcolor{blue}{\bf A.~Massless~ \& ~Partially ~Massless~Hubble~Effective~Mass:\Longrightarrow}}~\nonumber\\
    &&\Omega_{\lambda,{\bf k}}(\tau) = 4^{\nu_{\rm PGW}-2} \left|\frac{\Gamma(\nu_{\rm PGW})}{\Gamma\left(\frac{3}{2}\right)}\right|^2~\nonumber\\
   &&~~~~~~~~~~~~~~~~~~~~~~\left[2\sqrt{2}H(\sinh(2\alpha)\cos(2-\gamma)-\sinh(2r)\sin(2\phi) \right.~\nonumber\\
   &&~~~~~~~~~~~~~~~~~~~~~~ \left.-k\sinh(2r)\cos(2\phi)(3\sinh(2\alpha)\sin(2-\gamma)+\cosh(2\alpha))\right.~\nonumber\\
   &&~~~~~~~~~~~~~~~~~~~~~~ \left.+k\cosh(2r)(\sinh(2\alpha)\sin(2-\gamma)+3\cosh(2\alpha))\right]\\
    &&~\nonumber\\
  &&\underline{\textcolor{blue}{\bf B.~Heavy~Hubble~Effective~Mass:\Longrightarrow}}~\nonumber\\
  &&\Omega_{\lambda,{\bf k}}(\tau) = \frac{1}{16} e^{-\pi /2}\left|\frac{\Gamma(-i|\nu_{\rm PGW}|)}{\Gamma\left(\frac{3}{2}\right)}\right| \left|\frac{\Gamma(i|\nu_{\rm PGW}|)}{\Gamma\left(\frac{3}{2}\right)}\right|~\nonumber\\
  &&~~~~~~~~~~~~~~~~~~~~~~ \left[-e^{\pi /2} \sinh (2 \alpha ) \left(\sqrt{2} H (\sin(2-\gamma )+\cos (2-\gamma ))\right.\right.~\nonumber\\
   &&~~~~~~~~~~~~~~~~~~~~~~ +\sinh (2 r) \left(\sqrt{2} H \sin (2 \phi ) (\cos
   (2-\gamma )-\sin (2-\gamma ))\right.~\nonumber\\
   &&~~~~~~~~~~~~~~~~~~~~~~\left. \left. +2 k \cos (2-\gamma ) \cos (2 \phi )\right)\right)-2
   \sqrt{2} H \cosh ^2(\alpha ) (\sinh (2 r) \sin (2 \phi )+1)~\nonumber\\
    &&~~~~~~~~~~~~~~~~~~~~~~ +k \cosh (2 r) \left(2
   e^{\pi /2} \sinh (2 \alpha ) \cos (2-\gamma )+2 e^{\pi } \sinh ^2(\alpha )+2 \cosh
   ^2(\alpha ) \right.~\nonumber\\
    &&~~~~~~~~~~~~~~~~~~~~~~\left. \left.+\cosh (2 \alpha ) +1\right)-2 e^{\pi } k \sinh ^2(\alpha ) \sinh (2 r)
   \cos (2 \phi )\right]~\nonumber\\
 && \eea
 \item \underline{\textcolor{red}{\bf $\alpha$~vacua:}}\\
\bea &&\underline{\textcolor{blue}{\bf A.~Massless~ \& ~Partially ~Massless~Hubble~Effective~Mass:\Longrightarrow}}\nonumber\\
 &&\Omega_{\lambda,{\bf k}}(\tau) = 2^{2\nu_{\rm PGW}-\frac{9}{2}}\left|\frac{\Gamma(\nu_{\rm PGW})}{\Gamma\left(\frac{3}{2}\right)}\right|^2~\nonumber\\
&&~~~~~~~~~~~~~~~~~~~~~~\left(4H\cos(2)\sinh(2\alpha)-4H\sinh(2r)\sin(2\phi) \right. ~\nonumber\\
&&~~~~~~~~~~~~~~~~~~~~~~+\sqrt{2}k(\cosh(2r)(3\cosh(2\alpha) ~\nonumber\\
&&~~~~~~~~~~~~~~~~~~~~~~+\sin(2)\sinh(2\alpha))-\sinh(2r)\cos(2\phi)(\cosh(2\alpha)~\nonumber\\
&&~~~~~~~~~~~~~~~~~~~~~~\left.+3\sin(2)\sinh(2\alpha)))\right)~\nonumber\\
&& \\
 &&\underline{\textcolor{blue}{\bf B.~Heavy~Hubble~Effective~Mass:\Longrightarrow}}~\nonumber\\
  &&\Omega_{\lambda,{\bf k}}(\tau) = \frac{1}{16} e^{-\pi /2}\left|\frac{\Gamma(-i|\nu_{\rm PGW}|)}{\Gamma\left(\frac{3}{2}\right)}\right| \left|\frac{\Gamma(i|\nu_{\rm PGW}|)}{\Gamma\left(\frac{3}{2}\right)}\right|~\nonumber\\
  &&~~~~~~~~~~~~~~~~~~~~~~ \left(-e^{\pi /2} \sinh (2 \alpha ) \left(\sqrt{2} H (\sin
   (2-\gamma )+\cos (2-\gamma )) \right. \right. ~\nonumber\\
   &&~~~~~~~~~~~~~~~~~~~~~~ +\sinh (2 r) \left(\sqrt{2} H \sin (2 \phi ) (\cos
   (2-\gamma )-\sin (2-\gamma ))\right. ~\nonumber\\
   &&~~~~~~~~~~~~~~~~~~~~~~\left. \left.+2 k \cos (2-\gamma ) \cos (2 \phi )\right)\right)-2
   \sqrt{2} H \cosh ^2(\alpha ) ~\nonumber\\
   &&~~~~~~~~~~~~~~~~~~~~~~(\sinh (2 r) \sin (2 \phi )+1)~\nonumber\\
   &&~~~~~~~~~~~~~~~~~~~~~~+k \cosh (2 r) \left(2
   e^{\pi /2} \sinh (2 \alpha ) \cos (2-\gamma ) \right.~\nonumber\\
  &&~~~~~~~~~~~~~~~~~~~~~~ \left.+2 e^{\pi } \sinh ^2(\alpha )+2 \cosh
   ^2(\alpha )+\cosh (2 \alpha )+1\right)~\nonumber\\
   &&~~~~~~~~~~~~~~~~~~~~~~ \left.-2 e^{\pi } k \sinh ^2(\alpha ) \sinh (2 r)
   \cos (2 \phi )\right)~\nonumber\\
   &&~\nonumber \\
 && \eea
 \item \underline{\textcolor{red}{\bf Bunch-Davies~vacuum:}}\\
\bea &&\underline{\textcolor{blue}{\bf A.~Massless~ \& ~Partially ~Massless~Hubble~Effective~Mass:\Longrightarrow}}\nonumber\\
  &&\Omega_{\lambda,{\bf k}}(\tau) = 2^{2 \nu_{\rm PGW}-\frac{9}{2}}\left|\frac{\Gamma(\nu_{\rm PGW})}{\Gamma\left(\frac{3}{2}\right)}\right|^2 \left(\sqrt{2} k (3 \cosh (2 r)-\sinh (2 r) \cos (2 \phi ))\right.~\nonumber\\
  &&~~~~~~~~~~~~~~~~~~~~~~ \left.-4 H\sinh (2 r) \sin (2 \phi )\right)~\nonumber\\
&& \\   
&&\underline{\textcolor{blue}{\bf B.~Heavy~Hubble~Effective~Mass:\Longrightarrow}}~\nonumber\\
 &&\Omega_{\lambda,{\bf k}}(\tau) =\frac{1}{8} e^{-\pi /2} \left|\frac{\Gamma(-i|\nu_{\rm PGW}|)}{\Gamma\left(\frac{3}{2}\right)}\right| \left|\frac{\Gamma(i|\nu_{\rm PGW}|)}{\Gamma\left(\frac{3}{2}\right)}\right|\left(2 k \cosh (2 r)-\sqrt{2} H (\sinh (2 r) \sin (2 \phi
   )+1)\right)~\nonumber\\
 && \eea
\end{enumerate}
\subsection{Sub-Hubble limiting result}
In the sub-Hubble limit,$-k\tau \gg 1$, it is expected to have very small contribution from
the squeezed parameter, $r_k(\tau )$ for which one can use the following approximations:
\begin{equation}
    \cosh r_k(\tau) \approx 1, \sinh r_k(\tau) \approx r_k(\tau) 
\end{equation}
Consequently, in the limit $r_k(\tau ) \rightarrow 0$, we get the following result for the dispersion relation
in the sub-Hubble region:
\begin{enumerate}
    \item  \underline{\textcolor{red}{\bf Motta-Allen ($\alpha,\gamma$)~vacua:}}\\
    \bea &&\underline{\textcolor{blue}{\bf A.~Massless~ \& ~Partially ~Massless~Hubble~Effective~Mass:\Longrightarrow}}~\nonumber\\
    &&\Omega_{\lambda,{\bf k}}^{\text{sub}}(\tau) = 4^{\nu_{\rm PGW}-2}\left|\frac{\Gamma(\nu_{\rm PGW})}{\Gamma\left(\frac{3}{2}\right)}\right|^2 \left(\sinh (2 \alpha ) \right.~\nonumber\\
    &&~~~~~~~~~~~~~~~~~~~~~~  \left(2 \sqrt{2} H \cos (2-\gamma ) \right.~\nonumber\\
    &&~~~~~~~~~~~~~~~~~~~~~~\left. \left. +k \sin (2-\gamma
   )\right)+3 k \cosh (2 \alpha )\right)~\nonumber\\
    && \\
    &&\underline{\textcolor{blue}{\bf B.~Heavy~Hubble~Effective~Mass:\Longrightarrow}}~\nonumber\\
    &&\Omega_{\lambda,{\bf k}}^{\text{sub}}(\tau) =\frac{1}{16}\left|\frac{\Gamma(-i|\nu_{\rm PGW}|)}{\Gamma\left(\frac{3}{2}\right)}\right| \left|\frac{\Gamma(i|\nu_{\rm PGW}|)}{\Gamma\left(\frac{3}{2}\right)}\right|  ~\nonumber\\
    &&~~~~~~~~~~~~~~~~~~~~~~ e^{-\pi/2}\left(-e^{\pi/2}\sinh(2\alpha)\left(\sqrt{2}H(\sin(2-\gamma)+\cos(2-\gamma)) \right. \right.~\nonumber\\
    &&~~~~~~~~~~~~~~~~~~~~~~ \left. -2k\cos(2-\gamma)\right)+~\nonumber\\
    &&~~~~~~~~~~~~~~~~~~~~~~ \left.\cosh^2(\alpha)\left(4k-2\sqrt{2}H\right)+2e^{\pi}k\sinh^2(\alpha)\right)~\nonumber\\
    &&\eea
     \item \underline{\textcolor{red}{\bf $\alpha$~vacua:}}\\
\bea &&\underline{\textcolor{blue}{\bf A.~Massless~ \& ~Partially ~Massless~Hubble~Effective~Mass:\Longrightarrow}}~\nonumber\\
 &&\Omega_{\lambda,{\bf k}}^{\text{sub}}(\tau) = 2^{2\nu_{\rm PGW}-\frac{9}{2}}\left|\frac{\Gamma(\nu_{\rm PGW})}{\Gamma\left(\frac{3}{2}\right)}\right|^2~\nonumber\\
 &&~~~~~~~~~~~~~~~~~~~~~~\left(4 H \cos (2) \sinh (2 \alpha )+\sqrt{2} k (3 \cosh (2
   \alpha ) +\sin (2) \sinh (2 \alpha ))\right)~\nonumber\\
 && \\
    &&\underline{\textcolor{blue}{\bf B.~Heavy~Hubble~Effective~Mass:\Longrightarrow}}~\nonumber\\  
 &&\Omega_{\lambda,{\bf k}}^{\text{sub}}(\tau) =   \frac{1}{16} \left|\frac{\Gamma(-i|\nu_{\rm PGW}|)}{\Gamma\left(\frac{3}{2}\right)}\right| \left|\frac{\Gamma(i|\nu_{\rm PGW}|)}{\Gamma\left(\frac{3}{2}\right)}\right|  ~\nonumber\\
 &&~~~~~~~~~~~~~~~~~~~~~~  \left(-2 e^{-\pi /2} \cosh ^2(\alpha ) \left(\sqrt{2} H-2 k\right) \right.~\nonumber\\
  &&~~~~~~~~~~~~~~~~~~~~~~ +\sinh(2 \alpha ) \left(2 k \cos (2)-\sqrt{2} H (\sin (2)+\cos (2))\right)~\nonumber\\
  &&~~~~~~~~~~~~~~~~~~~~~~\left. +2 e^{\pi /2}k \sinh ^2(\alpha )\right)~\nonumber\\
&&\eea
 \item \underline{\textcolor{red}{\bf Bunch-Davies~vacuum:}}\\
\bea &&\underline{\textcolor{blue}{\bf A.~Massless~ \& ~Partially ~Massless~Hubble~Effective~Mass:\Longrightarrow}}\nonumber\\
&&\Omega_{\lambda,{\bf k}}^{\text{sub}}(\tau) = 3 k 4^{\nu_{\rm PGW}-2}\left|\frac{\Gamma(\nu_{\rm PGW})}{\Gamma\left(\frac{3}{2}\right)}\right|^2~\nonumber\\
 && \\
    &&\underline{\textcolor{blue}{\bf B.~Heavy~Hubble~Effective~Mass:\Longrightarrow}}~\nonumber\\  
    &&\Omega_{\lambda,{\bf k}}^{\text{sub}}(\tau) =-\frac{1}{8} e^{-\pi /2} \left(\sqrt{2} H-2 k\right)\left|\frac{\Gamma(-i|\nu_{\rm PGW}|)}{\Gamma\left(\frac{3}{2}\right)}\right| \left|\frac{\Gamma(i|\nu_{\rm PGW}|)}{\Gamma\left(\frac{3}{2}\right)}\right|~\nonumber\\
&&\eea
\end{enumerate}

\subsection{Super-Hubble limiting result}
In the super-Hubble limit,$-k\tau \ll 1$, the dispersion relation takes the following form:

\label{sec:superHubbleDispersion}
\begin{enumerate}
    \item  \underline{\textcolor{red}{\bf Motta-Allen ($\alpha,\gamma$)~vacua:}}\\
    \bea &&\underline{\textcolor{blue}{\bf A.~Massless~ \& ~Partially ~Massless~Hubble~Effective~Mass:\Longrightarrow}}~\nonumber\\
     &&\Omega_{\lambda,{\bf k}}^{\text{sub}}(\tau) =4^{\nu_{\rm PGW}-2}\left|\frac{\Gamma(\nu_{\rm PGW})}{\Gamma\left(\frac{3}{2}\right)}\right|^2  \left(\sinh (2 \alpha ) \left(2 \sqrt{2} H \cos (2-\gamma ) \right. \right.~\nonumber\\
     &&~~~~~~~~~~~~~~~~~~~~~~\left.+k \sin (2-\gamma) (\cosh (2 r)-3 \sinh (2 r))\right)~\nonumber\\
     &&~~~~~~~~~~~~~~~~~~~~~~ \left. +k \cosh (2 \alpha ) (3 \cosh (2 r)-\sinh (2
   r))\right)~\nonumber\\
   &&\\
    &&\underline{\textcolor{blue}{\bf B.~Heavy~Hubble~Effective~Mass:\Longrightarrow}}~\nonumber\\
    &&\Omega_{\lambda,{\bf k}}^{\text{sub}}(\tau) =\frac{1}{16} 
\left|\frac{\Gamma(-i|\nu_{\rm PGW}|)}{\Gamma\left(\frac{3}{2}\right)}\right| \left|\frac{\Gamma(i|\nu_{\rm PGW}|)}{\Gamma\left(\frac{3}{2}\right)}\right|\nonumber\\
&&~~~~~~~~~~~~~~~~~~~~~~e^{-\pi /2} \left(-2 \sqrt{2} H \cosh ^2(\alpha ) \right.\nonumber\\
&&~~~~~~~~~~~~~~~~~~~~~~+e^{\pi /2} \sinh (2\alpha ) \left(2 k e^{-2 r} \cos (2-\gamma ) \right.\nonumber\\
&&~~~~~~~~~~~~~~~~~~~~~~\left. -\sqrt{2} H (\sin (2-\gamma )+\cos
   (2-\gamma ))\right)+2 k e^{\pi -2 r} \sinh ^2(\alpha ) \nonumber\\
&&~~~~~~~~~~~~~~~~~~~~~~ \left. +4 k \cosh ^2(\alpha ) \cosh(2 r)\right) \nonumber\\
    &&\eea
     \item \underline{\textcolor{red}{\bf $\alpha$~vacua:}}\\
\bea &&\underline{\textcolor{blue}{\bf A.~Massless~ \& ~Partially ~Massless~Hubble~Effective~Mass:\Longrightarrow}}~\nonumber\\
&&\Omega_{\lambda,{\bf k}}^{\text{sub}}(\tau) = 2^{2 \nu_{\rm PGW}-\frac{9}{2}} \left|\frac{\Gamma(\nu_{\rm PGW})}{\Gamma\left(\frac{3}{2}\right)}\right|^2\nonumber\\
&&~~~~~~~~~~~~~~~~~~~~~~\left(4 H \cos (2) \sinh (2 \alpha )+\sqrt{2} k (\cosh (2 \alpha)\right.\nonumber\\
&&~~~~~~~~~~~~~~~~~~~~~~\left. (3 \cosh (2 r)-\sinh (2 r))+\sin (2) \sinh (2 \alpha ) (\cosh (2 r)-3 \sinh (2r)))\right)\nonumber\\
&&\\
&&\underline{\textcolor{blue}{\bf B.~Heavy~Hubble~Effective~Mass:\Longrightarrow}}~\nonumber\\
&&\Omega_{\lambda,{\bf k}}^{\text{sub}}(\tau) =\frac{1}{16}\left|\frac{\Gamma(-i|\nu_{\rm PGW}|)}{\Gamma\left(\frac{3}{2}\right)}\right| \left|\frac{\Gamma(i|\nu_{\rm PGW}|)}{\Gamma\left(\frac{3}{2}\right)}\right|\nonumber\\
&&~~~~~~~~~~~~~~~~~~~~~~\left(-\sqrt{2} H \sinh (2 \alpha ) (\sin (2)+\cos (2))+e^{-\pi /2}
   \cosh ^2(\alpha ) \left(4 k \cosh (2 r)-2 \sqrt{2} H\right) \right.\nonumber\\
&&~~~~~~~~~~~~~~~~~~~~~~ \left. +2 k e^{-2 r} \sinh
   (\alpha ) \left(e^{\pi /2} \sinh (\alpha )+2 \cos (2) \cosh (\alpha
   )\right)\right)\nonumber\\
&&\eea
 \item \underline{\textcolor{red}{\bf Bunch-Davies~vacuum:}}\\
\bea &&\underline{\textcolor{blue}{\bf A.~Massless~ \& ~Partially ~Massless~Hubble~Effective~Mass:\Longrightarrow}}\nonumber\\
&&\Omega_{\lambda,{\bf k}}^{\text{sub}}(\tau) =k\left|\frac{\Gamma(\nu_{\rm PGW})}{\Gamma\left(\frac{3}{2}\right)}\right|^2 4^{\nu_{\rm PGW}-2} (3 \cosh (2 r)-\sinh (2 r))
 \nonumber\\
 &&\\
&&\underline{\textcolor{blue}{\bf B.~Heavy~Hubble~Effective~Mass:\Longrightarrow}}~\nonumber\\
&&\Omega_{\lambda,{\bf k}}^{\text{sub}}(\tau) = -\left|\frac{\Gamma(-i|\nu_{\rm PGW}|)}{\Gamma\left(\frac{3}{2}\right)}\right| \left|\frac{\Gamma(i|\nu_{\rm PGW}|)}{\Gamma\left(\frac{3}{2}\right)}\right|\frac{1}{8} e^{-\pi /2} \left(\sqrt{2} H-2 k \cosh (2 r)\right)\nonumber\\
&&\eea
\end{enumerate}

\newpage
\phantomsection
\addcontentsline{toc}{section}{References}
\bibliographystyle{utphys}
\bibliography{references}

\end{document}